\PassOptionsToPackage{final}{graphicx}
\documentclass[aps,prd,amsmath,twocolumn,final]{revtex4-2}

\usepackage{graphicx}
\usepackage{xfrac}

\begin{document}

\title{General Expressions for Measurable Parameters in Curved Spacetime}

\author{Dmitri Lebedev and Kayll Lake}
\email[]{dmitri.lebedev@queensu.ca, lakek@queensu.ca}
\affiliation{Department of Physics, Queen's University, Kingston, Ontario, Canada}

\date{\today}

\begin{abstract}
General covariant expressions for measurable angles, distances, velocities, and accelerations are provided in terms of fundamental parameters that can be applied in any setup. The relativistic aberration of light relationship is presented in full generality, which is applicable to any orientation of observers and light rays. An expansion for the geometrical exponential map is established and used to form an expression for the physical distance between an observer and a nearby object within its extended local frame. Curvature effects on measurable distances, velocities, and accelerations are made explicit and appear in general tensorial form. The concepts of Fermi frames on timelike worldlines and the Fermi-Walker derivative are discussed in detail and used throughout; and in examining the meaning of relative stationarity between timelike observers, the Fermi-Walker derivative is established from first principles through physically meaningful consideration. A generalized type of Taylor expansion is provided for tensors of any rank in a covariant form. Expressions for the optically based angular diameter distance and luminosity distance are provided in general forms, and the reciprocity theorem is discussed and verified. A generalized version of the geodesic deviation equation, applicable to extreme relative motion, is provided as well.
\end{abstract}

\maketitle

\section{\label{sec:intro}Introduction}

In a curved spacetime, measurable parameters such as angles, distances, and relative velocities require extra care in their definitions and interpretations. Much effort has been put into studying measurable parameters since the establishment of the theory, yet disagreements on the definitions of measurable angles and distances have led to the well known debate on the role of the cosmological constant in gravitational lensing; see \cite{ri,park,ll,ll2} and the references within. Concepts such as relative velocity and acceleration are of central importance in astrophysics and cosmology, yet it is clear that such parameters are not uniquely defined, even in the simplest cases, and their definitions are often model dependent. In 2000, an effort was made by the IAU (International Astronomical Union) to properly define `radial velocity' in astronomy, given the ongoing improvement in measurement precision and the recognition of inconsistencies in past definitions, \cite{iau}. To this end, two definitions were proposed based on observable and geometric reasoning. The definitions are not equivalent, but both properly capture the concept. Based on these initial propositions, in 2003, the concept of radial velocity was extended to three different definitions, which were designated `kinematic radial velocity', `astrometric radial velocity', and `barycentric radial velocity', \cite{lindrav}. The definitions rely on the fact that for `radial velocity' the displacement is along the line of sight. However, obtaining general expressions for these velocities for any spacetime and motion was left as an open topic due to the complexity of the task. In 2007, the concept of `relative velocity' was outlined and manifested through four different definitions in a fairly rigorous pioneering work on the subject, \cite{bolos}. These four definitions are based on the previous effort in \cite{lindrav}, but extend beyond the radial case. The author calls them \textit{kinematic} relative velocity, \textit{Fermi} relative velocity, \textit{spectroscopic} relative velocity, and \textit{astrometric} relative velocity. Expressions for these velocities in special cases were obtained in \cite{bolos} and later works (see, for example, \cite{bolos2,bolos3,bolos4,bolos,klein,klein2,klein3,klein4}), but truly general expressions were established only for \textit{kinematic} and \textit{spectroscopic} relative velocities. As we shall see, only the Fermi relative velocity of the four above can be considered measurable. Furthermore, Synge introduced his own definition of relative velocity in \cite{synge2}, which we will consider as well.

In this work, by focusing on physically measurable parameters, we clarify which definitions of velocity are actually useful. Starting from fundamental concepts, we investigate measurable angles, distances, velocities, and accelerations. First, we provide a general formula to calculate measurable angles and a generalized relativistic aberration relationship. We then arrive at different possible definitions for velocity, which rely entirely on measurements, and compare our results to the definitions in the literature. For each definition of measurable velocity, we provide a general expression that allows it's calculation in any setting. While general expressions for the so called \textit{kinematic} and \textit{spectroscopic} relative velocities are given in \cite{bolos} and other sources, currently, no such expressions can be found for the `Fermi' and \textit{astrometric} relative velocities. In fact, in \cite{bolos3}, the author states: ``There is an open problem that consists on finding intrinsic expressions (in a coordinate-free language) for the Fermi and \textit{astrometric} relative velocities... It is a hard geometric problem, but it would be very useful for the interpretation and computation of these relative velocities." We provide such expressions and extend our analysis to acceleration.

A large portion of the analysis is dedicated to distances and velocities that are measurable through purely optical observations. The concepts of angular diameter distance and luminosity distance are studied in detail and used to define proper optical velocity. General expressions for these optical velocities and distances are provided. Some interesting phenomena regarding peculiar observations suggested by the outcomes are investigated.

The fundamental question we are dealing with can be stated as follows: For a general spacetime and two given worldlines for an observer and an object, what possible measurements of distance, velocity, and acceleration can be made by the observer, and how can the outcomes of measurement be expressed in a fully general way in terms of the parameters of the system? Specifically, we are seeking coordinate independent expressions involving the given curves, the event of measurement on the observer's worldline, the corresponding 4-velocities and their derivatives, and metric components representing the curvature. We do not want to have any restrictions on the form of the metric or the relative motion of the observer and the object.

The concepts of Fermi coordinates, local Fermi frames, and the Fermi-Walker derivative are natural to this topic and are discussed in detail. In our effort to investigate physical properties of relatively non-moving objects, we end up deriving the Fermi-Walker derivative purely from physical considerations based on possible measurements. This reveals important physical aspects of the Fermi-Walker transport and gives clear measurable criteria for identifying a non-rotating frame. A thorough investigation and utilization of these concepts were done by Synge in \cite{synge2}, with which we compare our results as well.

The Fermi frame with corresponding Fermi coordinates could be thought of as a natural representation of an observer's extended local space. For two nearby observers in general motion, we ask the following fundamental questions: If one has constant Fermi coordinates in the other's frame, will it, in general, also see the other at constant Fermi coordinates in its own frame? Further, if one is held at constant Fermi coordinates in the other's frame, what 4-velocity must it have with respect to that of the other? Or rather, what 4-velocity must one have with respect to the other to be stationary in its Fermi frame? As part of our initial analysis, we provide this 4-velocity as well as a clear answer to the first question, in the negative. Throughout this work, we casually use the terms static, stationary, and non-moving as description of objects having attributes of not changing in distance and not rotating relative to an observer (in a sense that will be fully clarified).

We begin our investigation of measurable relative velocity by exposing the asymmetry with the Fermi frame mentioned above. This fact together with other unexpected consequences of what should be considered as a relatively non-moving neighboring object leads us to a detailed discussion of the physical requirements for relative stationarity. Specifically, we ask what physical conditions would we expect to be satisfied by two observers that are relatively stationary, and what mathematical conditions would it translate to. With what was discovered, we proceed to provide definitions and expressions for relative velocity. The idea of proper optical velocity is then introduced as a natural continuation of the investigation.

To support the analysis, we derive some more general mathematical results and put them to use. Among these, we present a generalized Taylor expansion for tensors defined on curves in an arbitrary metric space. We also derive general relationships for subvolumes under a transformation that is confined to a subspace. Additionally, we propose a generalized geodesic deviation equation and provide an expansion for the geometric exponential map.\\

This work is structured as follows: In section \ref{sec:ang} we discuss measurable intersection angles and relativistic aberration of light. In section \ref{sec:vel} we develop the concept of relative velocity. A discussion of stationarity and the derivation of the Fermi-Walker derivative take place in section \ref{sec:rot}. Precise definitions of relative velocity are provided in section \ref{sec:relvel}. Section \ref{sec:opt} is dedicated to optically based distances and proper optical velocities. A thorough discussion of optical distances is provided in section \ref{sec:optdis} in the framework of geometrical ray bundles and possible measurements on them. In section \ref{sec:defopt} we define and express proper optical velocities, and in section \ref{sec:acc} we define and provide expressions for acceleration with respect to each definition of velocity. We then summarize and discuss our main results in section \ref{sec:dis}, and make a final comparison of our definitions to the literature. Many important derivations are done in the appendix, which has been divided into three sections.

\section{\label{sec:ang}Measurable Angles}

\subsection{\label{sec:gen}The General Formulas}

Consider a Lorentzian manifold with metric $g_{\alpha \beta}$ of positive signature, and two arbitrary curves that intersect at an event. In general, for a pseudo-Riemannian manifold the intersection angle between the tangents of the two curves is not defined. However, if we pick a subspace with a timelike normal (purely Riemannian subspace) and project the curves onto it in the region of the intersection event, then the resulting intersection angle is well defined and unique. It is the measurable angle by an observer having the 4-velocity normal to the subspace, with the subspace being its local ``laboratory frame". That is, given two arbitrary intersecting curves and an observer at an event, we can clearly define a measurable intersection angle in a Lorentzian manifold. Let $K^\alpha$ and $W^\alpha$ be the tangent vectors of the two intersecting curves being considered, and let $U^\alpha$ be the 4-velocity of an observer at the event. We find that the observable intersection angle, $\theta_U$, between the tangents of the curves projected onto the observer's frame is given by
\begin{equation}
	\cos(\theta_U)=\frac{K_\alpha W^\alpha + (U_\alpha K^\alpha)(U_\beta W^\beta)}{\sqrt{K_\alpha K^\alpha + (U_\alpha K^\alpha)^2} \sqrt{W_\alpha W^\alpha + (U_\alpha W^\alpha)^2}}. \label{genform}
\end{equation}
This formula can be derived as follows: Let $K^\beta h^\alpha_\beta$ and $W^\beta h^\alpha_\beta$ be the projections of $K^\alpha$ and $W^\alpha$ onto the space of the observer, so that $K^\beta h^\alpha_\beta$ and $W^\beta h^\alpha_\beta$ are spacelike and perpendicular to $U^\alpha$ with respect to the metric $g_{\alpha \beta}$. We easily find that the projection tensor $h^\alpha_\beta$ is given by
\begin{equation}
	h^\alpha_\beta=\delta^\alpha_\beta+U^\alpha U_\beta, \label{temp6}
\end{equation}
so
\begin{align}
	&K^\beta h^\alpha_\beta=K^\alpha+K^\beta U_\beta U^\alpha,\\
	&W^\beta h^\alpha_\beta=W^\alpha+W^\beta U_\beta U^\alpha.
\end{align}
The Riemannian metric on the subspace is the confinement of $g_{\alpha \beta}$ to the subspace, which is equivalent to the projection of the metric onto the subspace, $g_{\gamma\epsilon}h^\gamma_\alpha h^\epsilon_\beta = h_{\alpha\beta}=g_{\alpha\beta}+U_\alpha U_\beta$. Therefore the angle, $\theta_U$, between the vectors $K^\beta h^\alpha_\beta$ and $W^\beta h^\alpha_\beta$ that takes place in the observer's space is given by
\begin{equation}
	\cos(\theta_U) = \frac{h_{\alpha \beta} \left(K^\gamma h^\alpha_\gamma\right) \left(W^\gamma h^\beta_\gamma\right)}{\sqrt{h_{\alpha \beta} (K^\gamma h^\alpha_\gamma) (K^\gamma h^\beta_\gamma)} \sqrt{h_{\alpha \beta} (W^\gamma h^\alpha_\gamma) (W^\gamma h^\beta_\gamma)}}.
\end{equation}
Using the expressions for $K^\beta h^\alpha_\beta$, $W^\beta h^\alpha_\beta$, and $h^\alpha_\beta$, we promptly get equation \eqref{genform}, which we were not able to find in the literature on the topic or elsewhere. In the special case where the tangent vectors are null and future pointing, equation \eqref{genform} simplifies to the following important angle formula, (for more details see \cite{ll,ll2}, and compare with equation (11) in \cite{ri})
\begin{equation}
	\cos(\theta_U)=1+\frac{K^\alpha W_\alpha}{(U^\alpha K_\alpha)(U^\alpha W_\alpha)}. \label{angleformula}
\end{equation}
This formula is fundamental to the topic of measurements involving optical signals. In particular, its relevance to the field of gravitational lensing is clear; yet before it was introduced in \cite{ll}, it did not appear in any of the discussions regarding the influence of the cosmological constant on gravitational lensing, or related areas. In fact, it took much effort to find this formula or a variation of it in any available sources. The formula is known to be used by the GAIA (Global Astrometric Interferometer for Astrophysics) team. We found it in an explicit form in \cite{firstform,tlplform}, and in implicit form in \cite{defe1,defe2}. We were not able to find the formula in books on differential geometry and related areas, except for very special cases appearing in \cite{oneill,kriele} (also see \cite{ellis2,cohen,ehlersform}).

It is important to note that the formulas given by \eqref{genform} and \eqref{angleformula} are both signature dependent. Furthermore, for the given positive signature, equation \eqref{angleformula} remains true for past pointing null vectors, but obtains an overall negative sign when one is past and one is future pointing. This is due to the way the root is dealt with in equation \eqref{genform} for the null case.

In case the angle is very small, we can rewrite the above formula for future or past orientation as follows:
\begin{align}
	\cos(\theta_U)&=1+\frac{K^\alpha}{(U^\alpha K_\alpha)}\frac{W_\alpha}{(U^\alpha W_\alpha)}\nonumber\\
	1-\frac{1}{2}\theta_U^2+\mathcal{O}(\theta_U^4)&=1-\frac{1}{2}\left|\frac{K^\alpha}{(U^\alpha K_\alpha)}-\frac{W^\alpha}{(U^\alpha W_\alpha)}\right|^2\nonumber\\
	\theta_U&=\left|\frac{K^\alpha}{(U^\alpha K_\alpha)}-\frac{W^\alpha}{(U^\alpha W_\alpha)}\right| +\mathcal{O}(|\;\;|^3). \label{smallangform}
\end{align}
Now, if the vectors are of mixed orientation, then without loss of generality suppose $K^\alpha$ is future and $W^\alpha$ is past pointing. The angle expression turns out to be
\begin{equation}
	\cos(\theta_U)=-1-\frac{K^\alpha W_\alpha}{(U^\alpha K_\alpha)(U^\alpha W_\alpha)},
\end{equation}
but we can always convert $W^\alpha$ to a future pointing $\tilde W^\alpha$ with respect to the frame of $U^\alpha$, such that $K^\alpha$ and $\tilde W^\alpha$ give the same angle. This future pointing version of $W^\alpha$ is given by
\begin{equation}
	\tilde W^\alpha= W^\alpha+2(W^\beta U_\beta)U^\alpha,
\end{equation}
and when we substitute it in the angle formula we get
\begin{align}
	\cos(\theta_U)&=-1-\frac{K^\alpha \tilde W_\alpha-2(W^\beta U_\beta)K^\alpha U_\alpha}{(U^\alpha K_\alpha)(U^\alpha W_\alpha)}\nonumber\\
	&=1-\frac{K^\alpha \tilde W_\alpha}{(U^\alpha K_\alpha)(U^\alpha W_\alpha)}\nonumber\\
	&=1+\frac{K^\alpha \tilde W_\alpha}{(U^\alpha K_\alpha)(U^\alpha \tilde W_\alpha)},
\end{align}
as expected, and from which we obtain
\begin{equation}
	\theta_U=\left|\frac{K^\alpha}{(U^\alpha K_\alpha)}-\frac{\tilde W^\alpha}{(U^\alpha \tilde W_\alpha)}\right| +\mathcal{O}(|\;\;|^3).
\end{equation}
Replacing $\tilde W^\alpha$ with the original $W^\alpha$ we get
\begin{equation}
	\theta_U=\left|\frac{K^\alpha}{(U^\alpha K_\alpha)}+\frac{W^\alpha}{(U^\alpha W_\alpha)}+2U^\alpha\right| +\mathcal{O}(|\;\;|^3), \label{smallangform2}
\end{equation}
which is the mixed orientation version of \eqref{smallangform}. Note that there are various technicalities to consider when dealing with small angles. When the null vectors differ by a small parameter, then blindly using the formula \eqref{angleformula} may lead to problems. The formulas for small measurable angles \eqref{smallangform} and \eqref{smallangform2} are designed to prevent such issues and produce the correct result to the given order of accuracy.

\subsection{Relativistic Aberration of Light}

Consider two observers with 4-velocities $U^\alpha$ and $V^\alpha$ at an event where two null geodesics intersect. Assuming the null tangents $K^\alpha$ and $W^\alpha$ have the same (future or past) orientation then \eqref{angleformula} can be used to express the measurable intersection angle for each observer. Taking the ratio of these expressions we find
\begin{equation}
	\frac{1-\cos(\theta_V)}{1-\cos(\theta_U)}=\frac{(U^\alpha K_\alpha)(U^\alpha W_\alpha)}{(V^\alpha K_\alpha)(V^\alpha W_\alpha)}, \label{relab}
\end{equation}
which can be considered the fully general relativistic aberration formula. It can be used in any orientation of light rays and relatively moving observers, as well as in any coordinates and background metric.

Recall the basic version of the aberration relationship one encounters early in studying special relativity (see, for example, \cite{rindler}, or any popular introductory text),
\begin{equation}
	\cos(\theta_V)=\frac{\cos(\theta_U)-v}{1-v\cos(\theta_U)},
\end{equation}
where $v$ is the relative speed between the observers. In this relationship, it is assumed that each observer sees the other moving in a parallel direction to one of the light rays. This assumption greatly limits the applicability of the familiar aberration relationship. A fully general relationship between the measurable angles for any orientation is surprisingly missing from textbooks. Clearly, the above must be a special case of \eqref{relab}. We will show this by expressing \eqref{relab} in terms of measurable angles in the space frame of $U^\alpha$, and consider the special case of alignment.

Let $\hat V^\alpha$, $\hat K^\alpha$, and $\hat W^\alpha$ be the spacelike unit vectors corresponding to the projections onto the space frame of $U^\alpha$. Then,
\begin{equation} \label{temp297}
    \begin{gathered}
        V^\alpha = \gamma U^\alpha + \gamma v \hat V^\alpha, \\
        K^\alpha = -(U^\alpha K_\alpha)(U^\alpha+\hat K^\alpha), \\
        W^\alpha = -(U^\alpha W_\alpha)(U^\alpha+\hat W^\alpha),
    \end{gathered}
\end{equation}
where $\gamma=-U^\alpha V_\alpha$ is the relativistic $\gamma$ factor. (We assume here that null vectors are future pointing.) Let $\alpha$ and $\beta$ respectively be the angles between $\hat V^\alpha$ and $\hat K^\alpha$, and $\hat V^\alpha$ and $\hat W^\alpha$, in the frame of $U^\alpha$. Then,
\begin{equation}
    \begin{gathered}
        V^\alpha K_\alpha = \gamma U^\alpha K_\alpha - (U^\alpha K_\alpha) \gamma v \hat V^\alpha \hat K_\alpha, \\
        V^\alpha W_\alpha = \gamma U^\alpha W_\alpha - (U^\alpha W_\alpha) \gamma v \hat V^\alpha \hat W_\alpha,
    \end{gathered}
\end{equation}
and
\begin{equation}
	\frac{V^\alpha K_\alpha}{U^\alpha K_\alpha} = \gamma\left(1-v\cos(\alpha)\right), \;\; \frac{V^\alpha W_\alpha}{U^\alpha W_\alpha} = \gamma\left(1-v\cos(\beta)\right). \label{temp299}
\end{equation}
Inserting these in \eqref{relab} and solving for $\cos(\theta_V)$, we have
\begin{equation}
	\cos(\theta_V)=1-\frac{\left(1-\cos(\theta_U)\right)\left(1-v^2\right)}{\left(1-v\cos(\alpha)\right)\left(1-v\cos(\beta)\right)}. \label{relab2}
\end{equation}
This is a version of the general aberration relationship expressed entirely in terms of measurable quantities by the observer $U^\alpha$. Thus, an observer that can measure the angle between two intersecting light rays as well as the angles between the rays and the trajectory of another observer in motion, will be able to calculate the angle that the moving observer will measure between the light rays with the above formula. For the special case where the motion of the moving observer in the frame of $U^\alpha$ is aligned with one of the rays, say $W^\alpha$, we must have $\beta = 0$ or $\pi$, and $\alpha=\theta_U$ or $\pi-\theta_U$ respectively, so
\begin{align}
	\cos(\theta_V)&=1-\frac{\left(1-\cos(\theta_U)\right)\left(1-v^2\right)}{\left(1\mp v\cos(\theta_U)\right)\left(1\mp v\right)}\nonumber\\
	&=\frac{1\mp v\cos(\theta_U)-\left(1-\cos(\theta_U)\right)\left(1\pm v\right)}{1\mp v\cos(\theta_U)}\nonumber\\
	&=\frac{\cos(\theta_U) \mp v}{1\mp v\cos(\theta_U)},
\end{align}
as required. Again, while the aberration relationship for this special case of orientation is well known, the general expression that is applicable to any orientation as the one given by \eqref{relab2} or \eqref{relab} is not. We notice that while the angle expression \eqref{angleformula} is signature dependent, the aberration relationship \eqref{relab} derived from it is not.\\

When the measurable angles are very small, so that both $\theta_U \ll 1$ and $\theta_V \ll 1$, the aberration relationship \eqref{relab} becomes
\begin{equation}
	\frac{\theta_V^2}{\theta_U^2}=\frac{(U^\alpha K_\alpha)(U^\alpha W_\alpha)}{(V^\alpha K_\alpha)(V^\alpha W_\alpha)}. \label{temp1}
\end{equation}
It can be further simplified to
\begin{equation}
	\frac{\theta_V}{\theta_U}=\frac{(U^\alpha K_\alpha)}{(V^\alpha K_\alpha)}=\frac{(U^\alpha W_\alpha)}{(V^\alpha W_\alpha)},  \label{temp2}
\end{equation}
where only one of the null vectors is used. The easy way to see this outcome is to recognize that $\beta = \alpha + \mathcal{O}(\theta_U)$ in \eqref{temp299}, which makes the above true to the lowest order in $\theta_U$. In terms of measurable parameters the above becomes
\begin{equation}
	\frac{\theta_V}{\theta_U}=\frac{1}{\gamma\left(1-v\cos(\alpha)\right)}+ \mathcal{O}(\theta_U). \label{temp2a}
\end{equation}
In case of alignment, with $\alpha=0+ \mathcal{O}(\theta_U)$, the above takes the well known form
\begin{equation}
	\frac{\theta_V}{\theta_U}=\sqrt{\frac{1+v}{1-v}}+ \mathcal{O}(\theta_U). \label{relabsb}
\end{equation}
It is important to note that for small but not infinitesimal measurable angles going from \eqref{temp1} to \eqref{temp2} is not straightforward, and close attention must be paid to the conditions under which this can be done. A closer inspection of the process of replacing $\beta$ with $\alpha$ in this approximation reveals that for orientations of near alignment, where $0<\alpha \ll \theta_U$ (with $\theta_U$ small but finite), higher order terms in $\theta_U$ could have coefficients that are very large and cannot be neglected. We discuss this situation in detail in appendix \ref{app:relab}, and derive the required condition for \eqref{temp2} and \eqref{temp2a} to hold true. In terms of the measurable quantities defined in this section, the requirement on the motion of $V^\alpha$ is given by (see equation \eqref{cond}, and appendix \ref{app:relab} for details)
\begin{equation}
	\frac{v\theta_U^2}{1-v\cos(\alpha)} \ll 1.
\end{equation}
This condition is also necessary for \eqref{relabsb} to hold true, which in the literature is justified by the assumption that the angles are infinitesimal. (See for example \cite{rindler,gibbons}.)

Furthermore, the well known aberration relationship for small solid angles is given by
\begin{equation}
	\frac{d\Omega_V}{d\Omega_U}=\frac{(U^\alpha K_\alpha)^2}{(V^\alpha K_\alpha)^2}, \label{temp3}
\end{equation}
where the null vector $K^\alpha$ is tangent to any member ray of the bundle that makes up the small solid angle. It is clear that the above expression is valid whenever \eqref{temp2} is valid, but the derivation of \eqref{temp3} always assumes infinitesimal angles, which avoids any issues of alignment and extreme relative motion (see \cite{gibbons,sef}). In case where the solid angle $d\Omega_U$ is set small but finite, the validity of \eqref{temp3} is also subject to the condition on $V^\alpha$ that we derive in appendix \ref{app:relab}, which is an important aspect that is not discussed in the literature.

In section \ref{sec:solid} we make an intuitive definition of a measurable solid angle in the process of defining optical distances. Our definition does not rely on any local coordinates as is usually the case (see \cite{gibbons,sef}) and is based on the limit of a well defined measurable area (see equation \eqref{temp61}). As a quick consequence of our approach, we arrive at the aberration relationship given by \eqref{temp3} (see equation \eqref{temp141}), with the condition for its validity arising from the derivation process.

\section{\label{sec:vel}The Challenges in Defining Relative Velocity for General Motion and Curvature}

Consider two timelike trajectories parametrized with proper times $t$ and $\tau$, with coordinates $x^{\alpha}(t)$ and $x^{\alpha}(\tau)$, and with tangents $U^\alpha(t) = \frac{dx^{\alpha}}{dt}$ and $V^\alpha(\tau) = \frac{dx^{\alpha}}{d\tau}$. In a situation when the curves are sufficiently close (neighboring), how can we express the velocity that one observer would measure of the other? In this section, the concept of measurable velocity is discussed in detail, leading to specific definitions and coordinate independent expressions in terms of the involved trajectories and metric components.

\subsection{\label{sec:phys}Physical Position Vector and its Derivatives}

For some $t$ and (appropriately chosen, as will be made clear below) $\tau$, that correspond to particular events $x^{\alpha}(t)$ and $x^{\alpha}(\tau)$, consider the unique geodesic connecting the two events. We will analyze the motion of an object moving along the curve $x^{\alpha}(\tau)$ within the local space of an observer moving along $x^{\alpha}(t)$. To this end, assume that $\tau$ is set with respect to the given $t$, so that the connecting geodesic is spacelike and the tangent to the geodesic at $x^{\alpha}(t)$ is in the space frame of the observer, being orthogonal to $U^\alpha(t)$ with respect to the metric. In a neighborhood where the curves $x^{\alpha}(t)$ and $x^{\alpha}(\tau)$ are close enough for a range of $t$ and $\tau$, this construction yields a unique connecting geodesic at each event on $x^{\alpha}(t)$, while also establishing a connection between $t$ and $\tau$. The tangent to the connecting geodesic at $x^{\alpha}(t)$ is given by (see appendix \ref{app:deri}, equation \eqref{LofA})
\begin{equation}
	D^\alpha = A^\alpha + \frac{1}{2}\Gamma^{\alpha}_{\beta\gamma}A^\beta A^\gamma + \frac{1}{6}(\partial_\beta\Gamma^{\alpha}_{\gamma\epsilon} + \Gamma^{\alpha}_{\beta\rho}\Gamma^{\rho}_{\gamma\epsilon})A^\beta A^\gamma A^\epsilon + ...\,, \label{DofA}
\end{equation}
where $A^\alpha=x^{\alpha}(\tau)-x^{\alpha}(t)$, and $|D|=D$ is the metric distance between the events. In the above expression it is implicit that the connecting geodesic has a particular parametrization, which sets the magnitude $D$. This is the `normalized affine parametrization', for which the affine parameter increment between the events is unity. (See \cite{synge2,hawk} and the derivation of \eqref{LofA} in appendix \ref{app:deri}.) We will use this parametrization frequently.

Requiring that $D^{\alpha}U_{\alpha}=0$ sets a restriction on $\tau$, giving a relationship $\tau(t)$. $D^{\alpha}(t)$ is thus constructed along the trajectory $x^\alpha(t)$ in the neighborhood of interest. $D^{\alpha}(t)$ represents the direction and a distance to the passing object at $x^{\alpha}(\tau)$ from the point of view of the observer at $x^{\alpha}(t)$. $D$ is the distance to the object in the extended Fermi frame of the observer. We will refer to it as the Fermi distance, as is common in the literature. The first derivative is found to be (see appendix \ref{app:deri}, equations \eqref{ddotap},\eqref{taudotap})
\begin{equation}
	\dot D^\alpha = \dot \tau \bar V^\alpha - U^\alpha + \frac{1}{6}\left(\dot \tau \bar V^\beta + 2U^\beta\right)R^\alpha_{D\beta D} + \mathcal{O}(D^3), \label{ddot}
\end{equation}
where $\dot \tau=\frac{d\tau}{dt}$, and is given by
\begin{equation}
	\dot\tau \gamma = 1+ D^\alpha \dot U_\alpha+ \frac{1}{6}\left(\frac{1}{\gamma} \bar V^\alpha + 2U^\alpha\right)R_{\alpha DUD}+\mathcal{O}(D^3), \label{taudot}
\end{equation}
$\bar V^\alpha$ is the parallel transport of $V^\alpha$ from $x^\alpha(\tau(t))$ to $x^\alpha(t)$ along the connecting geodesic (see \eqref{partrans}), and $\gamma=-\bar V^\alpha U_\alpha$ is a generalized relativistic $\gamma$ factor. $R^\alpha_{\beta\gamma\epsilon}$ is the Riemann curvature tensor at $x^\alpha(t)$, given by equation \eqref{riem}. We use $\dot{(\;\;)}$ to represent the proper derivative of any tensor or scalar quantity along the curve on which it is defined; in the above $\dot D^\alpha$ and $\dot U^\alpha$ are covariant derivatives in the direction of $U^\alpha$.

The second derivative is (see appendix \ref{app:deri}, equations \eqref{ddotdotap},\eqref{taudotdotap})
\begin{align}
	\ddot D^\alpha =& \ddot\tau \bar V^\alpha + \dot\tau^2\bar{\dot V}^\alpha - \dot U^\alpha +R^\alpha_{UUD} \nonumber\\
    &\!\!\!+\frac{2}{3}(\dot \tau \bar V^\beta-U^\beta)(\dot \tau \bar V^\gamma+2U^\gamma)R^\alpha_{\beta\gamma D}+\mathcal{O}(D^2), \label{ddotdot}
\end{align}
where
\begin{multline}
	\ddot \tau \gamma = 2\dot \tau\bar V^\alpha \dot U_\alpha+\dot \tau^2\bar{\dot V}^\alpha U_\alpha+D^\alpha \ddot U_\alpha \\
    +\frac{2}{3}\dot \tau (\dot \tau \bar V^\alpha+2U^\alpha)R_{\alpha DU\bar V}+\mathcal{O}(D^2), \label{taudotdot}
\end{multline}
and $\bar{\dot V}^\alpha$ is the parallel transport of $\dot V^\alpha$ from $x^\alpha(\tau(t))$ to $x^\alpha(t)$ along the connecting geodesic. Since $V^\alpha$ is defined on the curve $x^\alpha(\tau)$, in our notation $\dot V^\alpha$ is the covariant derivative of $V^\alpha$ with respect to $\tau$, in the direction of $V^\alpha$; so $\bar{\dot V}^\alpha$ is the proper 4-acceleration vector of the object parallelly transported to the location of the observer. (See equation \eqref{partransdot} and its derivation for more details.)

$D^\alpha(t)$ has the qualities of the familiar position vector. It points in the direction of the `straight line' to the passing object, and its magnitude is the metric distance to the object. We can visualize a geodesic wire in the observer space (or rather embedded in its extended laboratory frame), stretched from the observer to the object, which can change in length over time. The wire will point in the direction of $D^\alpha(t)$ and will have length $D(t)$. We will refer to $D^\alpha(t)$ as the direction of the connecting wire to remind us of this physical interpretation and also to distinguish it from other significant directions within the space of the observer. The goal now is to connect the change in $D^\alpha(t)$ to what we can call an observable velocity of the moving object.

\subsection{\label{sec:rot}Observable Motion and Rotation}

In this section we conduct an analysis of what it means for nearby objects to be relatively static, which is necessary for a clear definition of relative velocity. Throughout the process we pay close attention to the expected physical attributes of non-moving objects and the possible measurements that can be done to determine stationarity.

\subsubsection{Fermi Relative Velocity and the Bouncing Photon}

Let the timelike curves $x^{\alpha}(t)$ and $x^{\alpha}(\tau)$ be given, which allows the construction of $D^\alpha(t)$ and its derivatives in the region of interest as described in section \ref{sec:phys}. Naively, we may suppose that $\dot D^\alpha$ is the \textit{observed} velocity, but while it is guaranteed by construction that $D^\alpha(t)$ is within the space frame of $U^\alpha$ there is no reason to expect $\dot D^\alpha(t)$ to be perpendicular to $U^\alpha$ (or even to be spacelike), as one would expect from the \textit{observable} velocity vector. In fact, from $D^{\alpha}U_{\alpha}=0$ we have that $\dot D^{\alpha} U_{\alpha} = -D^{\alpha} \dot U_{\alpha}$, which means that if $U^\alpha$ is accelerating $\dot D^\alpha$ may have a component in the direction of $U^{\alpha}$ itself. This observation suggests exploring the projection of $\dot D^\alpha$ onto the space of $U^\alpha$, given by $\dot D^\beta h^\alpha_\beta$, as a possible definition for the \textit{observed} velocity, with $h^\alpha_\beta$ given by \eqref{temp6}.

This definition of velocity turns out to be the Fermi-Walker derivative of $D^\alpha(t)$, as will be made clear below. It represents the rate of change of the vector $D^\alpha(t)$ within the space of $U^\alpha$, and for this reason it has been used as a definition of velocity in some sources (see for example \cite{bolos,bolos2,bolos3,bolos4,bolos5,klein,klein2,klein3,klein4}). In this definition, trajectories $x^{\alpha}(\tau)$ that satisfy $\dot D^\beta h^\alpha_\beta=0$ have zero velocity and therefore represent objects that do not move with respect to the observer, and $D^\alpha(t)$ is said to be Fermi-Walker transported along $x^{\alpha}(t)$.

In general, the Fermi-Walker transport of a vector $A^\alpha$ from $x^{\alpha}(t)$ to $x^{\alpha}(t+\delta t)$ is given by $A^\beta \left(\delta^\alpha_\beta + \dot U_\beta U^\alpha \delta t - U_\beta \dot U^\alpha \delta t - \Gamma^\alpha_{\beta\gamma}U^\gamma \delta t \right)$; and the covariant derivative of a vector field $A^\alpha(t)$ generated from the Fermi-Walker transport satisfies $\dot A^\alpha = A^\beta \dot U_\beta U^\alpha - A^\beta U_\beta \dot U^\alpha$. The latter is often viewed in itself as the Fermi-Walker transport rule. The Fermi-Walker derivative of any vector field $B^\alpha(t)$ on the curve is given by $\dot B^\alpha - B^\beta \dot U_\beta U^\alpha + B^\beta U_\beta \dot U^\alpha$, and it represents the rate of change of the vector field relative to the instantaneous spacetime frame of the observer (which itself may change in time). In the case where the vector field also satisfies $B^\alpha U_\alpha=0$, the Fermi-Walker derivative reduces to $\dot B^\alpha - B^\beta \dot U_\beta U^\alpha$, which represents the rate of change of the vector field within the instantaneous space frame. Such frames are assumed to be non-rotating in the sense that was addressed by Synge, \cite{synge2}, and is a concept that is thoroughly investigated in this section.

For the spacelike vector field $D^\alpha(t)$ in the above construction, the equality $\dot D^{\alpha} U_{\alpha} = -D^{\alpha} \dot U_{\alpha}$ can be used to establish that the projection of $\dot D^\alpha(t)$ onto the space of the observer is in fact the Fermi-Walker derivative of $D^\alpha(t)$ as stated above, $\dot D^\beta h^\alpha_\beta = \dot D^\alpha + \dot D^\beta U_\beta U^\alpha = \dot D^\alpha - D^\beta \dot U_\beta U^\alpha$. There are various appealing arguments in support of this definition of velocity, and it is often referred to as Fermi relative velocity (\cite{bolos,bolos2,bolos3,bolos4,bolos5,klein,klein2,klein3,klein4}). If an observer could identify neighboring objects that have zero Fermi relative velocity, then the observer could use these objects as reference in measuring velocities of moving objects. It is therefore important to identify the physical properties of non-moving neighboring objects in accordance with this definition. It appears that Synge was the first to investigate the physical significance of the Fermi-Walker transport and of the Fermi frame of an observer, \cite{synge2}. He considered a bouncing photon, shot and received by an observer after reflecting from a nearby object. Synge summarizes: ``if $\tau$ is the trip-time and $\theta$ the angle through which the photon gun must be turned, then the limit of $\theta / \tau$ as $\tau$ tends to zero, is zero for Fermi frames and for them alone." He thereby explained that if a photon is emitted by an observer in arbitrary motion in any direction and very shortly later received after being reflected, then the direction in which it is received is the Fermi-Walker transport of the direction in which it was emitted. In this sense, a local frame composed of directions that undergo Fermi-Walker transport is non-rotating; and we will sometimes refer to such vector fields on the worldline of an observer as ones that `pass the bouncing photon test'. Also importantly, Synge points out that for a finite trip time there is a developing angle between the direction in which the photon is received and the Fermi-Walker transported direction in which the photon was emitted. He finds this angle implicitly, reveals that it depends only on the second order (and higher) of the trip time, and explains how measuring this deviation would yield information about the motion of the observer and the local curvature. It is noteworthy that Synge's definition of relative velocity in his book (\cite{synge2}) is not the Fermi relative velocity that we have here. In the following analysis some of Synge's key observations will be made fully clear and his definition of velocity will be discussed and compared to others.

The above property of the Fermi-Walker transport indicates that for a zero Fermi relative velocity object the direction of the connecting geodesic is non-rotating in a local sense. It is also clear that such an object will remain at constant geodesic distance from the observer, since $\dot D^\beta h^\alpha_\beta=0 \implies \dot D^\alpha=D^\beta \dot U_\beta U^\alpha \implies \dot D^\alpha D_\alpha = 0$, as one would expect of a non-moving object. However, such non-moving objects that are not extremely close to the observer would not pass the bouncing photon test if the photons are reflected by the objects themselves (also recognized by Synge). Surprisingly, further investigation reveals that zero Fermi relative velocity objects do not have a 4-velocity that equals the observer's 4-velocity when parallel transported along the connecting geodesic, as opposed to what one may expect of a relatively static object. Indeed, from \eqref{ddot} and \eqref{taudot} with $\dot D^\alpha=D^\beta \dot U_\beta U^\alpha$, we get
\begin{equation}
	\bar V^\alpha = U^\alpha-\frac{1}{2}R^\beta_{DUD}h^\alpha_\beta + \mathcal{O}(D^3) \label{vbarzerofermi}.
\end{equation}
The above makes it evident that the requirement for two observers to be at rest with respect to one another in the Fermi sense is not symmetric. This statement will be further clarified below. A suspicion of this asymmetry was brought up in \cite{bolos}.

\begin{figure}
  \centering
  \includegraphics[width=\columnwidth]{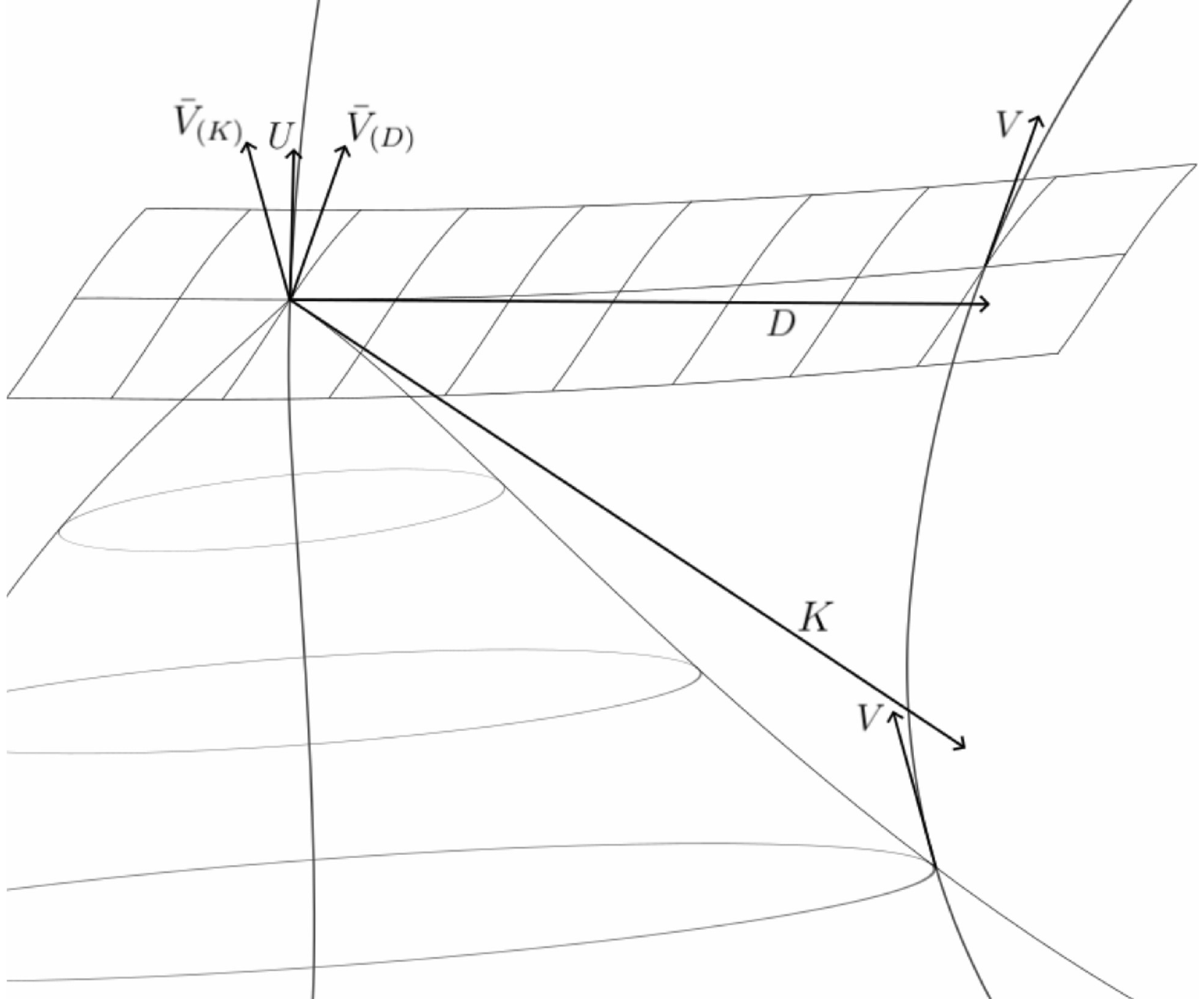}
  \caption{An observer on a timelike curve with 4-velocity $U^\alpha$, a neighboring object on a timelike curve with 4-velocity $V^\alpha$, and two connecting geodesics that are spacelike and null with tangents $D^\alpha$ and $K^\alpha$, respectively. The geodesics are parametrized with normalized affine parametrization (see appendix \ref{app:deri}), which is reflected in their lengths on the diagram. The spacelike geodesic belongs to the space simultaneity slice, also referred to as Fermi simultaneity. The null geodesic belongs to the past light cone, also referred to as lightlike simultaneity. $\bar V^\alpha_{(D)}$ and $\bar V^\alpha_{(K)}$ are parallel transports of the 4-velocity $V^\alpha$ along the spacelike and null geodesics, respectively, from the corresponding events on the object's trajectory. The \textit{kinematic} and \textit{spectroscopic} relative velocities are defined with respect to $\bar V^\alpha_{(D)}$ and $\bar V^\alpha_{(K)}$, respectively.} \label{fig:sim}
\end{figure}

Another common approach of defining relative velocity in the literature is by comparing the parallel transport of the 4-velocity of an object to that of the observer. This clearly differs from the method involving the position vector discussed above. In this approach a non-moving object must have $\bar V^\alpha = U^\alpha$ with respect to an observer. The velocity is found from $\bar V^\alpha$ as if the object is at the same event as the observer (given by $\frac{1}{\gamma}h^\alpha_\beta\bar V^\beta$), without accounting for curvature. The vector $\bar V^\alpha$ can either be the parallel transport of $V^\alpha$ along the connecting spacelike geodesic as described in the construction above, or it can be the transport along a null geodesic that connects the two trajectories. Relative velocities defined in this way are referred to as \textit{kinematic} and \textit{spectroscopic}, see Figure \ref{fig:sim} and \cite{lindrav,bolos}. The definition given by Synge in \cite{synge2} has to do with transporting $V^\alpha$ along the past null cone, and it corresponds to the \textit{spectroscopic} relative velocity defined in \cite{bolos}, but with an extra factor of $\gamma$.

However, there are various shortcomings with this way of defining relative velocity. The most important is that these velocities cannot be directly determined from physical measurements; these quantities serve as purely mathematical measures of relative motion. What we want is to construct expressions for a realistically measurable rate of change in position within a given model in terms of fundamental system parameters, and these velocities are not directly tied to physically observable positions. Additionally, we see by \eqref{ddot} that the connecting vectors $D^\alpha$, which connect to zero \textit{kinematic} velocity objects (with $\bar V^\alpha = U^\alpha$) will not undergo Fermi-Walker transport, so they will not in general satisfy the bouncing photon condition. This means that the position vector $D^\alpha$ of a non-moving object (in this definition) will be rotating with respect to a (non-rotating) Fermi frame. Indeed, from \eqref{ddot} and \eqref{taudot} with $\bar V^\alpha = U^\alpha$, we have
\begin{equation}
	\dot D^\alpha = D^\beta \dot U_\beta U^\alpha +\frac{1}{2}R^\beta_{DUD}h^\alpha_\beta + \mathcal{O}(D^3),
\end{equation}
which makes clear that $D^\alpha$ does not have a zero Fermi-Walker derivative. This is an important observation that is not discussed in the literature; two objects that satisfy $\bar V^\alpha = U^\alpha$, will in general see each other rotate (but not drift away, since $\dot D^\alpha D_\alpha = 0$).

These issues of inconsistency, non-symmetry, and failing the bouncing photon test of objects at rest with respect to one another under the above definitions reveal the need of having clear physical and mathematical requirements on non-moving objects before attempting to define a measurable relative velocity. Such a definition should be rooted in an intuitive physical interpretation and also allow for an experimental test. Whichever reasonable definition of velocity one may come up with, the physical attributes of zero-velocity objects should be intuitive and consistent. As a first step in the process to establish a requirement for stationarity, we revisit the analysis of the bouncing photon pioneered by Synge, but rather than assuming the existence of a Fermi frame a priori, we will construct a general propagation rule for the direction of incoming reflected photons (which has its own applications) and arrive at the Fermi-Walker transport from first principles as a limiting case. In particular, the Fermi-Walker derivative and Fermi frames are natural geometrical constructions in analyzing the space around an observer throughout its path; and while Synge demonstrated their physical attributes and provided a clear interpretation, we present a fundamental derivation of the Fermi-Walker derivative that is fully based on physical principles and observations, fully demonstrating the non-rotation aspect of the Fermi frame.\\

\begin{figure}
  \centering
  \includegraphics[width=\columnwidth]{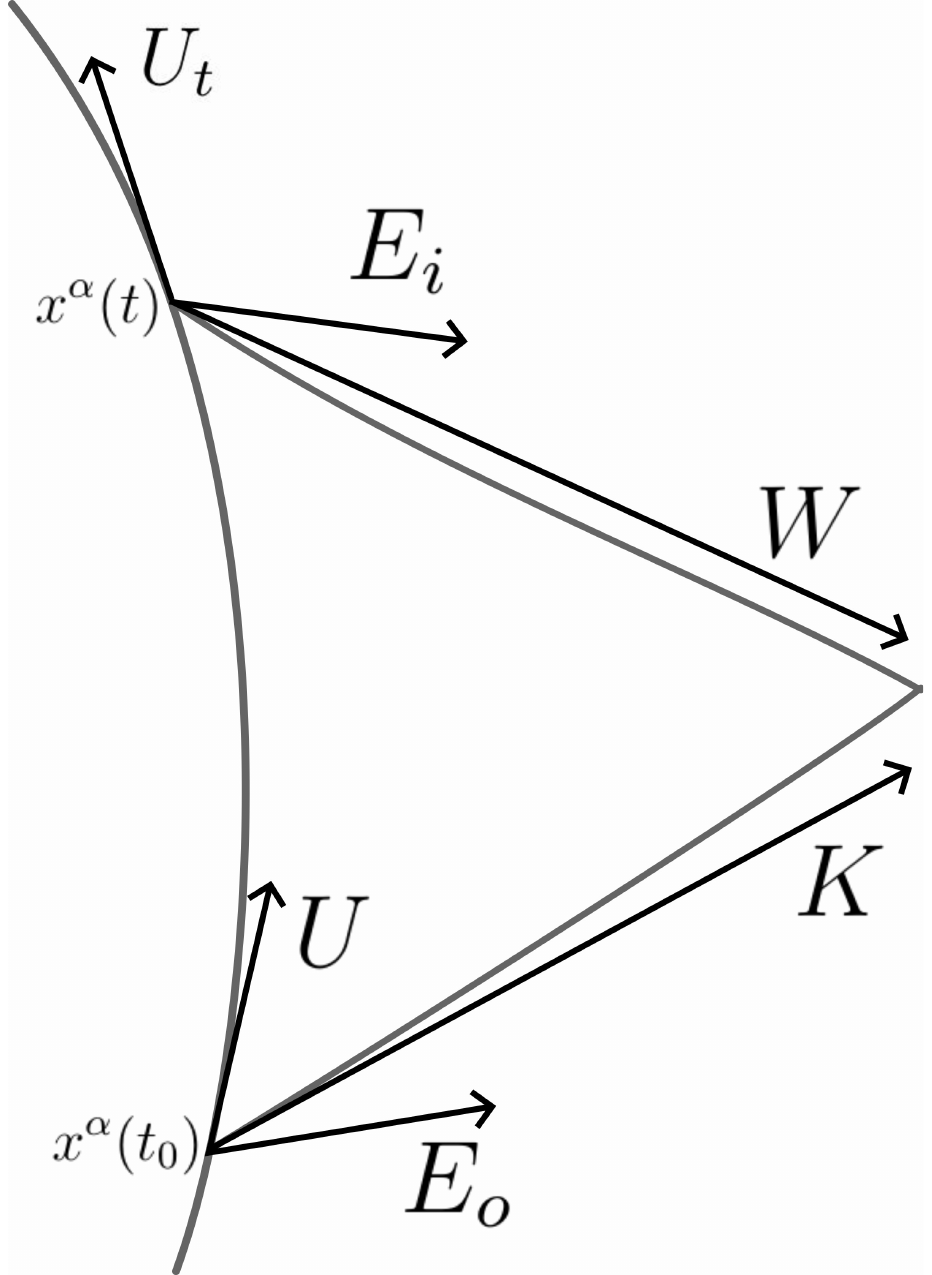}
  \caption{An observer on a timelike curve with 4-velocity $U^\alpha$, and the null trajectory of a bouncing photon that leaves the observer at $t_0$ and arrives at a later proper time $t$. $U^\alpha_t$ is the 4-velocity at the time of reception. The null vectors $K^\alpha$ and $W^\alpha$ are tangent to the outgoing and incoming photon trajectories at the events of emission and reception, respectively. They are parametrized with normalized affine parametrization (see appendix \ref{app:deri}), which is reflected in their lengths on the diagram. $E^\alpha_o$ and $E^\alpha_i$ are the unit vectors in the direction of the outgoing and incoming photon within the space frame of $U^\alpha$ and $U^\alpha_t$, respectively.} \label{fig:bp}
\end{figure}

Consider Figure \ref{fig:bp} and the definitions in the description. The direction of the outgoing photon at time $t_0$ is $E^{\alpha}_o$ and that of the incoming photon at time $t$ is $E^{\alpha}_i$. Each is a spacelike unit vector at its event, pointing towards the reflection event. Let
\begin{equation}
    \begin{gathered}
        \bar W^{\alpha}=W^{\beta}[t \to t_0]^\alpha_\beta, \\
        \bar E^{\alpha}_i=E^{\beta}_i[t \to t_0]^\alpha_\beta, \\
        \bar U^{\alpha}_t=U^{\beta}_t [t \to t_0]^\alpha_\beta,
    \end{gathered}
\end{equation}
where $[t \to t_0]^\alpha_\beta$ is the parallel transport operator from $x^\alpha(t)$ to $x^\alpha(t_0)$ along the curve, and is given by equation \eqref{partransu} in appendix \ref{app:deri}. The goal is to express $\bar E^{\alpha}_i$ in terms of $ E^{\alpha}_o$, the Cauchy data of the curve and the metric components at $x^\alpha(t_0)$, as a function of time lapse $\delta t = t - t_0$. In this setup it is clear that given the reflection event and the curve, the null vectors $K^{\alpha}$ and $W^{\alpha}$ are fully determined up to a constant, which is set by parametrization of the connecting geodesics (see appendix \ref{app:deri}). With the normalized affine parametrization for both null vectors, given some $K^\alpha$ and $U^\alpha$ (and its derivatives) at $x^\alpha(t_0)$, we can uniquely determine $W^\alpha$ and therefore $\bar W^\alpha$. The relationship among the vectors at the event $x^\alpha(t_0)$ is found to be (see appendix \ref{app:deri}, equation \eqref{omegaoflambdau})
\begin{multline}
	\bar W^{\alpha}=K^{\alpha}-U^{\alpha}\delta t-\frac{1}{2}\dot U^{\alpha}\delta t^2-\frac{1}{6}\ddot U^{\alpha}\delta t^3\\
    +\frac{1}{3}R^{\alpha}_{K U K}\delta t+ \frac{1}{6}R^{\alpha}_{U K U}\delta t^2 + ...\,, \label{wbar}
\end{multline}
where $R^{\alpha}_{\beta \gamma \epsilon}$ is the Riemann tensor at $x^\alpha(t_0)$. The vector $K^\alpha$ at $x^\alpha(t_0)$ fully determines the event of reflection (through equation \eqref{AofL}), and the time lapse $\delta t$ is set by the condition that $W^\alpha$ is null. We decompose the null vectors as follows,
\begin{equation}
	K^\alpha = -K^\beta U_\beta \left( U^\alpha + E^\alpha_o \right), \label{kueo}
\end{equation}
and
\begin{align}
	\nonumber \bar W^\alpha = W^\beta [t \to t_0]^\alpha_\beta &= W^\gamma U_{t\gamma}\left( -U^\beta_t + E^\beta_i \right)[t \to t_0]^\alpha_\beta\\
	&= -\bar W^\beta \bar U_{t\beta}\left( \bar U^\alpha_t - \bar E^\alpha_i \right). \label{wuei}
\end{align}
With the normalized affine parametrization, $-K^\alpha U_\alpha$ (and $W^\alpha U_{t\alpha}$) may be interpreted as a measure of $K^\alpha$ (and $W^\alpha$), in the sense that the larger this quantity is the further the reflection event, and the longer the bouncing trip takes. By means of equations \eqref{wbar}, \eqref{kueo}, and the null condition on $W^\alpha$, we can establish a relationship between $K^\alpha U_\alpha$ and $\delta t$ in terms of $E^\alpha_o$. Further, $\bar U^\alpha_t$ depends on the Cauchy data at $x^\alpha(t_0)$ and $\delta t$ only, and is given by (see appendix \ref{app:deri}, equation \eqref{abartdeltat})
\begin{equation}
	\bar U^\alpha_t = U^\alpha +\dot U^\alpha \delta t + \frac{1}{2}\ddot U^\alpha \delta t^2 +... \label{ubar}
\end{equation}
The above can be combined with \eqref{wbar} and \eqref{kueo} to express $\bar W^\alpha \bar U_{t\alpha}$ in terms of $K^\alpha U_\alpha$, $\delta t$, and $E^\alpha_o$, which then allows forming the following ratios,
\begin{equation}
	\frac{K^\alpha U_\alpha}{\bar W^\alpha \bar U_{t\alpha}}=-1+\frac{1}{6}E^\alpha_o \ddot U_\alpha \delta t^2 + \mathcal{O}(\delta t^3), \label{rat1}
\end{equation}
and
\begin{multline}
	\frac{\delta t}{\bar W^\alpha \bar U_{t\alpha}}=2-E^\alpha_o \dot U_\alpha \delta t +\frac{1}{6}\dot U^2 \delta t^2-\frac{2}{3}E^\alpha_o \ddot U_\alpha\delta t^2 \\
    - \frac{1}{6}R_{EUEU}\delta t^2 + \mathcal{O}(\delta t^3). \label{rat2}
\end{multline}
Inserting the expressions \eqref{kueo} and \eqref{wuei} into \eqref{wbar}, and solving for $\bar E^\alpha_i$ by making use of \eqref{rat1}, \eqref{rat2}, and \eqref{ubar}, we find
\begin{align}
	\bar E^\alpha_i \!&= E^\alpha_o + E^\beta_o \dot U_\beta U^\alpha \delta t \nonumber\\
    &\;+ \frac{1}{2}\!\left( \! E^\beta_o \ddot U_\beta U^\alpha \!+\! E^\beta_o \dot U_\beta \dot U^\alpha \!+\! \frac{1}{3}\left( \ddot U^\beta \!+\! R^\beta_{EUE} \right)\!H^\alpha_\beta \!\right)\!\delta t^2 \nonumber\\
    &\hspace{5.5cm}+ \mathcal{O}(\delta t^3), \label{eibar}
\end{align}
where
\begin{align}
	H^\alpha_\beta = h^\alpha_\gamma \left( \delta^\gamma_\beta-E^\gamma_o E_{o \beta} \right) &= \left( \delta^\alpha_\gamma-E^\alpha_o E_{o \gamma} \right) h^\gamma_\beta \nonumber\\
    &= \delta^\alpha_\beta+U^\alpha U_\beta  -E^\alpha_o E_{o \beta}, \label{H}
\end{align}
is a combined projection operator that projects onto the space of $U^\alpha$ and onto the subspace normal to $E^\alpha_o$. (This operator can also be expressed as $H^\alpha_\beta=\delta^\alpha_\beta + K^\alpha L_\beta + K_\beta L^\alpha$ for some null $L^\alpha$ that satisfies $L^\alpha K_\alpha =-1$, see \cite{poisson}; in our case $L^\alpha=\frac{-1}{2K^\beta U_\beta}\left(U^\alpha-E^\alpha_o\right)$.)

Equation \eqref{eibar} completes our derivation. It is a general evolution rule for $\bar E^\alpha_i$, expressed to second order in $\delta t$. Given a particular direction of an outgoing photon $E^\alpha_o$ at an event $x^\alpha(t_0)$, a vector field $E^\alpha_i$ on the curve is generated by means of \eqref{eibar}, which represents the direction of a returning reflected photon at the corresponding event. This evolution rule is not stated in \cite{synge2} or other sources, and we will put it to use in the coming investigation.

We first recognize that the evolution rule for $\bar E^\alpha_i$ is not a consistent differential evolution rule in the sense that it does not yield a differential equation that governs the evolution. If we were to have a differentiable vector field $E^\alpha(t)$ on the curve such that $E^\alpha(t_0)=E^\alpha_o$, then $E^\alpha(t_0+\delta t)$ parallel transported back to $x^\alpha(t_0)$ would be given by (see equation \eqref{abartdeltat}) $\bar E^\alpha = E^\alpha_o +\dot E^\alpha \delta t + \frac{1}{2}\ddot E^\alpha \delta t^2+...$, where the coefficients of $\delta t$ and $\delta t^2$ are related by differentiation. The coefficients of $\delta t$ and $\delta t^2$ in \eqref{eibar} clearly do not satisfy the same differential relation, since there is an extra term due to curvature and jerk (the 4-jerk, or simply the jerk is the second derivative of the 4-velocity). What this essentially means is that while it is possible to populate $E^\alpha_i$ over a portion of the curve by means of \eqref{eibar}, the resulting vector field would directly depend on the $E^\alpha_o$ vector and the other terms at the starting event in such a way that if any other vector of the resulting field was used as a new $E^\alpha_o$ to repopulate the $E^\alpha_i$ field according to the same evolution rule, then the new $E^\alpha_i$ field will be different. A consistent differential evolution rule, such as the Fermi-Walker rule, produces a vector field that can be reconstructed into itself from any of its vectors and the rule. We will clarify the above statements and physical consequence of this observation on equation \eqref{eibar} in what follows.
\begin{figure}
  \centering
  \includegraphics[width=\columnwidth]{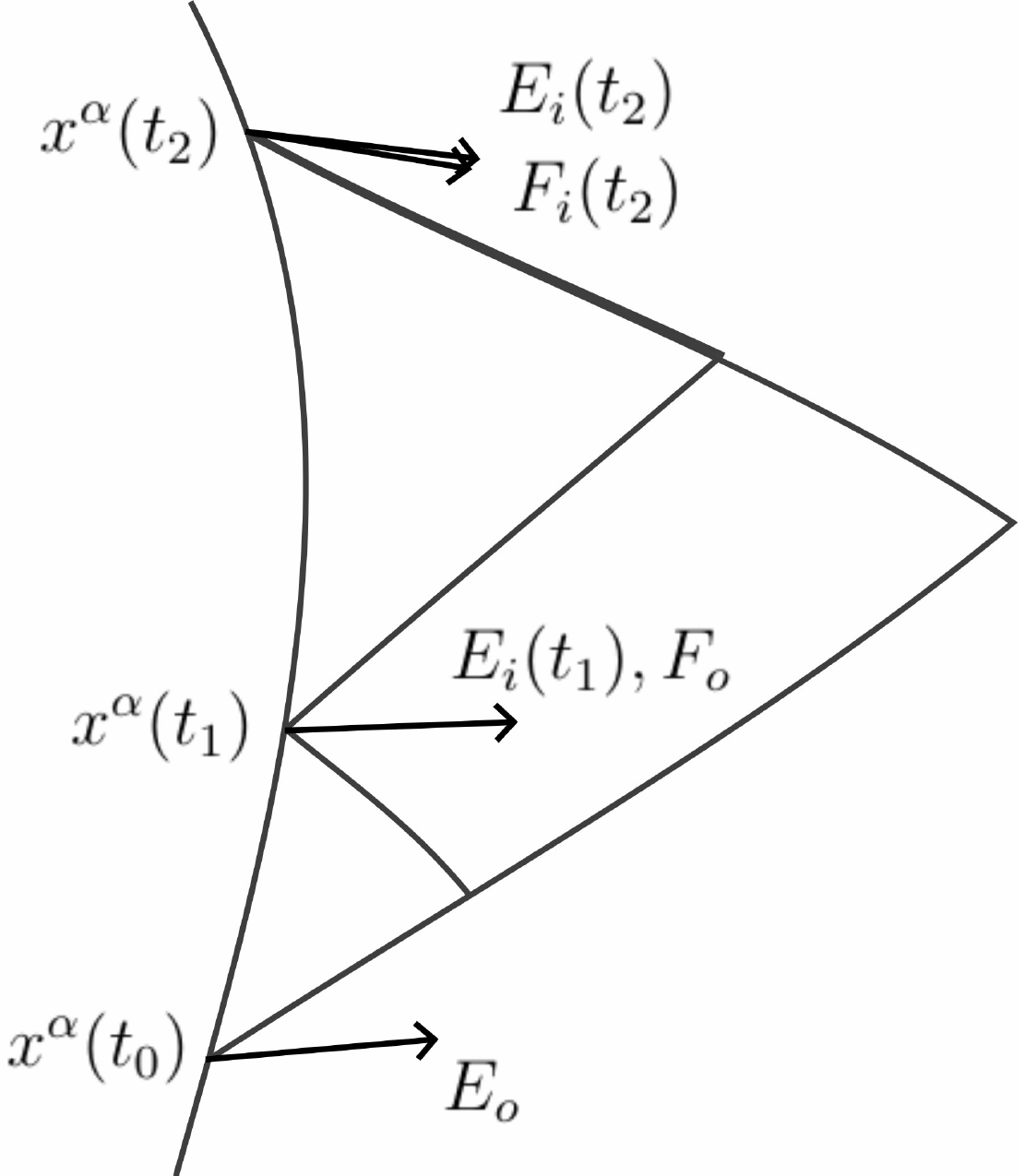}
  \caption{A burst of photons at $x^\alpha(t_0)$ generates $E^\alpha_i(t>t_0)$ by reflecting and coming back. A new burst at $x^\alpha(t_1)$ in the direction of $E^\alpha_i(t_1)$ generates another future field $F^\alpha_i(t>t_1)$ with respect to the new incoming reflected photons. $E^\alpha_i(t>t_1)$ and $F^\alpha_i(t>t_1)$ are compared at $x^\alpha(t_2)$.} \label{fig:tb}
\end{figure}

Consider a burst of photons at time $t_0$ of which reflected photons are received at all later times $t>t_0$, forming the field $E^\alpha_i(t)$. Consider another burst at time $t_1$ in the same direction as the received photons from the original burst, which are then also reflected and received at consequent times, forming another field $F^\alpha_i(t)$. We compare the two vector fields at a later time $t_2$, using the event $x^\alpha(t_0)$ as reference, see Figure \ref{fig:tb}. $E^\alpha_i(t_2)$ transported to $x^\alpha(t_0)$ is simply given by \eqref{eibar} with $\delta t= t_2-t_0$. $F^\alpha_i(t_2)$ transported to $x^\alpha(t_0)$ is obtained by using \eqref{eibar} with $\delta t= t_2-t_1$ to get $\bar F^\alpha_i(t_2)$ at $x^\alpha(t_1)$, and then transporting it to $x^\alpha(t_0)$. In the expression for $\bar F^\alpha_i(t_2)$ the outgoing direction must be set to $E^\alpha_i(t_1)$, and the rest of the terms evaluated at $x^\alpha(t_1)$. Relating all terms to the event $x^\alpha(t_0)$ and comparing the resulting expressions for $E^\beta_i(t_2)[t_2 \to t_0]^\alpha_\beta$ and $F^\beta_i(t_2)[t_2 \to t_0]^\alpha_\beta$, we get
\begin{align}
	E^\beta_i&(t_2)[t_2 \to t_0]^\alpha_\beta - F^\beta_i(t_2)[t_2 \to t_0]^\alpha_\beta \nonumber\\
    &= \bar E^\alpha_i(t_2)-\bar F^\beta_i(t_2)[t_1 \to t_0]^\alpha_\beta\nonumber\\
	&= \frac{1}{3}\left( \ddot U^\beta +R^\beta_{EUE} \right)H^\alpha_\beta(t_2-t_1)(t_1-t_0)+\mathcal{O}(\delta t^3). \label{fibar}
\end{align}
The above demonstrates the difference between the vector fields explicitly. It is due to the ($H$) term in the coefficient of $\delta t^2$ in \eqref{eibar}, which prevents the coefficients of the $\delta t$ and $\delta t^2$ terms from being related by differentiation. Clearly, a field generated by \eqref{eibar} fully depends on the starting event, and this transport rule for non-infinitesimal $\delta t$ cannot be converted into a differential equation. It is noteworthy that the difference between the above two vectors at $x^\alpha(t_2)$ is physically measurable, as well as the time intervals in \eqref{fibar}; so the ($H$) term in \eqref{fibar} is in principle measurable.

However, when $\delta t$ is infinitesimally small, \eqref{eibar} reduces to the Fermi-Walker transport for vectors perpendicular to $U^\alpha$, $\bar E^\alpha = E^\alpha + E^\beta \dot U_\beta U^\alpha \delta t$; and in the limit $\delta t \to 0$, the definition of covariant differentiation yields
\begin{equation}
	\dot E^\alpha = E^\beta \dot U_\beta U^\alpha, \label{ftrans}
\end{equation}
the (zero) Fermi-Walker derivative for vectors that are perpendicular to $U^\alpha$. This limiting case is interpreted as local observations on bouncing photons, and sets a clear criterion that allows physical determination of Fermi-Walker frames. It is clear now in what sense Fermi-Walker transported vectors and Fermi-Walker frames are non-rotating, and what the limitations are. The linear appearance of the unit vector $E^\alpha$ in \eqref{ftrans} means that the expression can be generalized to vectors of any magnitude. In fact, it is trivial to demonstrate that any vector field $A^\alpha(t)$ on the curve that has constant magnitude, and a projected component ($A^\beta h^\alpha_\beta(t)$) that satisfies \eqref{ftrans}, will itself satisfy
\begin{equation}
	\dot A^\alpha = A^\beta \dot U_\beta U^\alpha - A^\beta U_\beta \dot U^\alpha, \label{fwtrans}
\end{equation}
which is the full differential expression of the Fermi-Walker transport. Apparently Fermi introduced \eqref{ftrans} and the corresponding transport implicitly, while Walker made it explicit and extended it to \eqref{fwtrans} and the full transport, see \cite{fermi,walker,bija}. This concludes the derivation of the Fermi-Walker transport from first principles based on a physically relevant requirement.\\

Finally, consider a photon shot at event $x^\alpha(t_0)$ in the direction of $E^\alpha_o$, and the Fermi-Walker transport of $E^\alpha_o$ into future events, call it $E^\alpha(t)$. This field will satisfy \eqref{ftrans}, and therefore the parallel transport, $\bar E^\alpha(t_0,t)$, of $E^\alpha(t)$ from $x^\alpha(t)$ to $x^\alpha(t_0)$ along the curve is given by (equation \eqref{abartdeltat})
\begin{align}
	\nonumber \bar E^\alpha(t_0,t) &= E^\alpha(t_0) + \dot E^\alpha(t_0) \delta t + \frac{1}{2}\ddot E^\alpha(t_0) \delta t^2 +...\\
	&=E^\alpha_o + E^\beta_o \dot U_\beta U^\alpha \delta t \nonumber\\
    &\;\;\;\;+ \frac{1}{2} \left(E^\beta_o \ddot U_\beta U^\alpha + E^\beta_o \dot U_\beta \dot U^\alpha\right) \delta t^2 + \mathcal{O}(\delta t^3).
\end{align}
On the other hand, if the outgoing photon gets reflected and returned to the observer at time $t$, then the direction in which it is received, $E^\alpha_i(t)$, transported back to the event $x^\alpha(t_0)$ is given by \eqref{eibar}. The difference between the two vectors is
\begin{equation}
	\bar E^\alpha_i-\bar E^\alpha= \frac{1}{6}\left( \ddot U^\beta +R^\beta_{EUE} \right)H^\alpha_\beta\delta t^2+\mathcal{O}(\delta t^3),
\end{equation}
and it represents the deviation of $E^\alpha_i(t)$ from the Fermi-Walker frame over time $\delta t$. The angle between these vectors is given by
\begin{align}
	\nonumber \cos\delta\theta &= E^\alpha_i E_\alpha = \bar E^\alpha_i \bar E_\alpha\\
	\nonumber 1-\frac{1}{2}\delta\theta^2 +... &= 1-\frac{1}{2}|\bar E^\alpha_i-\bar E^\alpha|^2\\
	\nonumber \implies \;\;\;\;\;\; \delta\theta &= |\bar E^\alpha_i-\bar E^\alpha| + \mathcal{O}(|\bar E^\alpha_i-\bar E^\alpha|^3)\\
	&= \frac{1}{6}\left|\left(\ddot U^\beta +R^\beta_{EUE}\right)H^\alpha_\beta\right|\delta t^2 + \mathcal{O}(\delta t^3). \label{deltatheta}
\end{align}
The above is an explicit expression for the developing deviation angle. Compare the above result to equation (98) in the bouncing photon section of \cite{synge2}, and also note its dependence on $\delta t^2$ as lowest order term. In principle, the angle $\delta \theta$ can be physically measured (as well as $\delta t$) to experimentally determine the term in \eqref{deltatheta}; and clearly, the ratio $\delta \theta / \delta t$ goes to zero as $\delta t$ goes to zero, as Synge deduced. The vector within the absolute value sign in \eqref{deltatheta} will appear quite often in the following analysis.

\subsubsection{\label{sec:relsta}Relatively Static Objects}

Now consider an observer in arbitrary motion with a Fermi-Walker frame transported along its worldline, which can be thought of as a grid created from a collection of non-rotating directions. A constant vector in this frame will pass the local bouncing photon test, as described above. A nearby object with zero Fermi relative velocity could be connected to the observer with a geodesic wire, which will not change in length throughout the motion, and its tangent direction at the observer will be constant with respect to the frame. However, on the other end of the wire a second observer may also set a Fermi-Walker frame by bouncing local photons, and for this observer the tangent of the connecting wire will no longer be constant with respect to the frame, as we demonstrate below.

If $D^\alpha$ is the position vector at the first observer (tangent to the connecting wire, and having magnitude equal the length of the wire) then it satisfies $\dot D^\alpha = D^\beta \dot U_\beta U^\alpha$ as it is constant in the frame, for which case the 4-velocities of the observers are related through \eqref{vbarzerofermi}. On the other side of the wire, we can define $B^\alpha$ as the position vector of the first observer, and form a similar expression for its derivative as \eqref{ddot}. The relationship between the 4-velocities, \eqref{vbarzerofermi}, then reduces the derivative of $B^\alpha$ to
\begin{equation}
	\dot B^\alpha = B^\beta \dot V_\beta V^\alpha + R^\beta_{BVB}h^\alpha_\beta + \mathcal{O}(D^3),
\end{equation}
where the terms are defined at the second observer and the projection is onto its space. The above makes it clear that the tangent direction will not pass the bouncing photon test. In fact, over a period of proper time $\delta\tau$ the developing angle between the connecting wire and the local Fermi-Walker frame is
\begin{align}
	\nonumber \delta\theta &= |\bar {\hat B}^\alpha - \bar {\hat B}^\alpha_{FW}| + \mathcal{O}(|\bar {\hat B}^\alpha - \bar {\hat B}^\alpha_{FW}|^3)\\
	\nonumber &= |\dot {\hat B}^\alpha\delta\tau - \dot {\hat B}^\alpha_{FW}\delta\tau| + \mathcal{O}(\delta\tau^2)\\
	&= |R^\beta_{\hat B V \hat B}h^\alpha_\beta|D\delta\tau + \mathcal{O}(\delta\tau^2),
\end{align}
where $D$ is the constant distance between the observers, and $\hat B^\alpha$ is the instantaneous unit direction of the connecting wire. ($\bar {\hat B}^\alpha_{FW}$ is the Fermi-Walker propagated copy of $\hat B^\alpha$, the expression for the angle is obtained as in \eqref{deltatheta}, and \eqref{abartdeltat} was used to express the transported vectors.) Interestingly, this developing angle is completely independent of the acceleration and jerk associated with the motions of the observers, and it reveals yet another way of measuring the curvature. Clearly in practice these terms are extremely small, but in the presence of curvature and with the ability to make sensitive measurements it is important to be aware of this asymmetry in the definition of the Fermi relative velocity.

While the Fermi-Walker frame identifies locally non-rotating directions, extending the concept to velocities of distant objects leads to the inconsistency of relatively static observers described above. Furthermore, while distant objects of zero Fermi relative velocity may be connected to the observer with geodesic wires in directions that pass the local bouncing photon test, the actual photons that are bounced off the object itself must in general be aimed at and received from different directions with respect to the Fermi-Walker frame; and it is not yet clear how these two directions are related to the direction of the wire and whether they will rotate with respect to the frame. As already pointed out, these issues with the Fermi relative velocity suggest that we should revisit the definition of velocity in the presence of curvature, for which relatively static objects satisfy some intuitive physical requirements. It is not guarantied that a better definition can be formulated, but the investigation is necessary before settling on the shortcomings of Fermi velocity, and useful for what it reveals. The appeal in the definition of Fermi relative velocity is that zero velocity objects have constant Fermi coordinates, constant metric distance to the observer, and the direction to the object is locally non-rotating. We should always require the constancy of distance for zero velocity objects under any definition, but the concept of non-rotation and its physical meaning must be properly developed from the perspective of possible measurements and physical requirements.

Notice that of the four velocities mentioned in the introduction, we already see that \textit{kinematic} and \textit{spectroscopic} (defined in Figure \ref{fig:sim}) are not directly measurable. Also, finding expressions for them is trivial since their definition is based on having an analogous expression to the local case but with a transported 4-velocity. As we shall see, \textit{astrometric} relative velocity it is also not measurable and happens to be of the optical kind, which we will discuss in detail later. Thus, for now we challenge the usefulness of the definition of Fermi velocity alone, being the only one of the four that is directly measurable within an extended local frame; and starting from first principles we will look for possible alternatives.\\

As the first step, it is intuitive to expect that a relatively static object will appear to an observer in the same direction of space where photons should be emitted to illuminate the object over time, and that this direction should correspond to the direction of the connecting geodesic wire to the object. Further, we would also expect that the bouncing photons travel within the wire, at least in the limiting case of geodesic motion and no curvature. As we shall see, these intuitive expectations will have to be abandoned for the case of general motion and curvature, but our investigation reveals the reasons, as well as the needed logical modifications to the physical expectations of relatively static neighbors. We translate the physical requirements suggested above into mathematical conditions as follows.

Let $E^\alpha_o$ be the direction (unit vector) of an outgoing photon released by an observer in arbitrary motion at a given time, and let $D^\alpha$ be the position vector of the photon from the point of view of the observer at later times. With $\hat D^\alpha$ being the unit vector in the direction of $D^\alpha$ and $\bar {\hat D}^\alpha$ its parallel transport back to the emission event, a similar method that led to \eqref{eibar} yields the following relationship,
\begin{multline}
	\bar{\hat D}^\alpha = E^\alpha_o + \left( E^\beta_o \dot U_\beta U^\alpha - \frac{1}{2}\dot U^\beta H^\alpha_\beta \right)\delta t\\
	+ \frac{1}{2} \!\left( \!\left( \!(E^\beta_o \dot U_\beta)^2 \!-\! \dot U^2 \!\!+\!\! E^\beta_o \ddot U_\beta\! \right)U^\alpha \!+\! \left(\!\frac{5}{4}(E^\beta_o \dot U_\beta)^2 \!-\! \frac{1}{4}\dot U^2\! \right)E^\alpha_o \right. \\
	\left. + \left( \frac{1}{2}E^\gamma_o \dot U_\gamma \dot U^\beta - \frac{1}{3}\ddot U^\beta +\frac{2}{3}R^\beta_{EUE}-\frac{1}{3}R^\beta_{UEU} \right) H^\alpha_\beta \right)\delta t^2 \\
    + \mathcal{O}(\delta t^3). \label{outphoton}
\end{multline}
If we imagine a geodesic wire coming out of the observer in the direction of the outgoing photon at the event of emission, then the direction of the wire must obey the above evolution rule so that the photon remains confined within it. Replacing $E^\alpha_o$ with $\hat D^\alpha$ and multiplying by $D$, at least to first order in time we must have
\begin{equation}
	\bar D^\alpha = D^\alpha + \left( D^\beta \dot U_\beta U^\alpha - \frac{1}{2}D\dot U^\beta H^\alpha_\beta \right)\delta t + \mathcal{O}(\delta t^2) \label{outwire}
\end{equation}
as an evolution rule for the position vector of an object which receives photons that are aimed in the direction of the connecting wire to it, while also ensuring that the photon does not stray from the wire on its way to the object.

Likewise, let $D^\alpha$ be the position vector of an incoming photon from the point of view of an observer in arbitrary motion, and let $E^\alpha_i$ be the direction from which the incoming photon is received by the observer at a later event. With $\bar E^\alpha_i$ being the parallel transport of $E^\alpha_i$ to the event of $D^\alpha$, we get
\begin{multline}
	\bar E^\alpha_i = \hat D^\alpha + \left( \hat D^\beta \dot U_\beta U^\alpha + \frac{1}{2}\dot U^\beta H^\alpha_\beta \right)\delta t\\
	+ \frac{1}{2} \left( \left( -(\hat D^\beta \dot U_\beta)^2 + \dot U^2 + \hat D^\beta \ddot U_\beta \right)U^\alpha \right. \\
    + \left(\frac{5}{4}(\hat D^\beta \dot U_\beta)^2 - \frac{1}{4}\dot U^2 \right)\hat D^\alpha \\
	\left. + \left( \hat D^\gamma \dot U_\gamma \dot U^\beta + \frac{2}{3}\ddot U^\beta +\frac{2}{3}R^\beta_{\hat D U \hat D} + \frac{1}{3}R^\beta_{U \hat D U} \right) H^\alpha_\beta \right)\delta t^2 \\
    + \mathcal{O}(\delta t^3).
\end{multline}
(Here the projection $H^\alpha_\beta=\delta^\alpha_\beta+U^\alpha U_\beta -\hat D^\alpha\hat D_\beta$.) Again, if we imagine a geodesic wire which always contains the photon as it is moving towards the observer, then the direction of the wire at the observer must obey the above evolution rule, with $E^\alpha_i$ replaced by $\hat D^\alpha$ at the event of reception. Thus, making the replacement and multiplying by $D$, we get that for nearby incoming photons, at least to first order in time,
\begin{equation}
	\bar D^\alpha = D^\alpha + \left( D^\beta \dot U_\beta U^\alpha + \frac{1}{2}D\dot U^\beta H^\alpha_\beta \right)\delta t + \mathcal{O}(\delta t^2). \label{incwire}
\end{equation}
This is the evolution rule for the position vector of a nearby object that is visible from the same direction as that of the connecting geodesic wire to it, while also ensuring that photons coming from the object remain confined to the wire on their way to the observer.

Clearly the evolution rules given by \eqref{outwire} and \eqref{incwire} are inconsistent in the presence of an acceleration component perpendicular to $D^\alpha$, and we can also see that even without acceleration there are curvature terms in the next order. (We want our investigation of relatively static objects to encompass accelerating frames.) The restriction that the bouncing photons between the observer and object remain confined to the connecting wire is clearly too strong and cannot be satisfied for general motion. In fact, this investigation reveals a fundamental issue with the bouncing photon test. Even for extremely short distance bounces, the photon direction deviates from the Fermi-Walker frame as it travels away, and then deviates again but in an opposite manner as it travels back, as evident from equations \eqref{outwire} and \eqref{incwire}. This deviation cancels out through the short bounce motion, which explains the absence of a similar term in \eqref{eibar}. However, while the test identifies non-rotating directions, even extremely short geodesic wires that are constant in the Fermi-Walker sense will not in general contain bouncing photons throughout their motion. Therefore the reflection in such a test will in general happen from an object outside the geodesic wire. Of course, the discussion applies only to infinitely thin wires and point-like photons, but this is another issue that arises when considering sensitive measurements in the context of identifying non-rotation.\\

Abandoning the restrictive requirement, we now demand only that a non-rotating object will still be visible from the direction of its connecting geodesic wire, and that photons shot in that direction reach the object as well, while the motion of the wire is no longer subject to contain the photons as they travel. With the object being at constant distance $D$, from the process of establishing the evolution rule \eqref{outphoton} the relationship between the travel time $\delta t$ and the distance is found to be
\begin{multline}
	\delta t = D\left( 1-\frac{1}{2}D^\alpha \dot U_\alpha + \frac{1}{8}(D^\alpha\dot U_\alpha)^2+ \frac{5}{24}\dot U^2 D^2 \right. \\
    \left. - \frac{1}{3}D^\alpha \ddot U_\alpha D - \frac{1}{6}R_{UDUD} \right) + \mathcal{O}(D^4),
\end{multline}
where the terms are evaluated at the event of emission. Thus, if $\hat D^\alpha$ is the direction of emission as well as the direction of the wire at a given event, then after a period $\delta t$ given by the above (for the fixed $D$) the direction to the object will have to coincide with the one given by \eqref{outphoton}, since the photon arrives at the object. This means that we can multiply \eqref{outphoton} by $D$ and set $\delta t$ to equal the travel time given by the above to get a relationship between the parallel transported position vector of the object from the event of arrival and the position vector of the object at the event of emission, in terms of the distance $D$. Note that only when $\delta t$ is set to the above will the relationship be true, since the wire does not need to coincide with the photon before it arrives at the object. Now if a differentiable vector field $D^\alpha(t)$ is to be constructed from the established condition on the position vector and its parallel transport as described above, then the relationship between one of its values and its parallel transport separated by $\delta t$ is given by \eqref{abartdeltat}. In particular, for this $\delta t$ being set to the travel time for the distance $D$ to the object given by the above, we get another relationship between the transported position vector at the time of arrival and the position vector at the time of emission. Combining the two relationships yields a simple differential equation for $\dot D^\alpha$, the solution to which is given by
\begin{multline}
	\dot D^\alpha \!\!=\!\! D^\beta \dot U_\beta U^\alpha - \frac{1}{2}\dot U^\beta H^\alpha_\beta D + \!\left( \!\!- \frac{1}{8} D^\gamma \dot U_\gamma \dot U^\beta D \!+\! \frac{1}{12} \ddot U^\beta D^2 \right. \\
    \left. + \frac{1}{3}R^\beta_{DUD} - \frac{1}{6}R^\beta_{UDU} D \right) H^\alpha_\beta + \mathcal{O}(D^3). \label{outnotwire}
\end{multline}
(Again with $H^\alpha_\beta=\delta^\alpha_\beta+U^\alpha U_\beta -\hat D^\alpha\hat D_\beta$.) Hence, the above is the differential condition on the position vector of an object that always receives the photons which are emitted in the direction of its connecting geodesic wire. Notice that we were able to construct the differential condition only after relaxing the restrictive requirement on the wire (otherwise we would have a propagation rule that fully depends on the event of emission, not yielding a differential equation).

As for the incoming photons from the object, through the same process we find the travel period $\delta t$ in terms of the constant distance $D$.
\begin{multline}
	\delta t = D\left( 1-\frac{1}{2}D^\alpha \dot U_\alpha + \frac{3}{8}(D^\alpha\dot U_\alpha)^2- \frac{1}{24}\dot U^2 D^2 \right. \\
    \left. + \frac{1}{6}D^\alpha \ddot U_\alpha D + \frac{1}{6}R_{UDUD} \right) + \mathcal{O}(D^4),
\end{multline}
where the terms are evaluated at the time of emission (in accordance to the simultaneity slice of the observer). Analogous reasoning that lead to \eqref{outnotwire} for objects receiving outgoing photons, leads to 
\begin{multline}
	\dot D^\alpha \!=\! D^\beta \dot U_\beta U^\alpha + \frac{1}{2}\dot U^\beta H^\alpha_\beta D +\! \left( \frac{1}{8} D^\gamma \dot U_\gamma \dot U^\beta D \!+\! \frac{1}{12} \ddot U^\beta D^2 \right. \\
    \left. + \frac{1}{3}R^\beta_{DUD} + \frac{1}{6}R^\beta_{UDU} D \right) H^\alpha_\beta + \mathcal{O}(D^3). \label{incnotwire}
\end{multline}
for objects emitting incoming photons. The above is the differential condition on the position vector of an object that is always visible from the direction of the connecting geodesic wire. Evidently the conditions on $\dot D^\alpha$ given by \eqref{outnotwire} and \eqref{incnotwire} are not achievable simultaneously given general motion and curvature. This means that the physical requirements imposed on the motion of the object are still too restrictive in general. The outcome is not surprising considering the example of a constantly accelerating frame discussed by Rindler, \cite{rindler}, in which case projected light trajectories (onto the local space frame of an observer) are not spacelike geodesics; and objects that are at rest with respect to one another are in general visible to each other from directions that are not the same as connecting geodesic wires. We shall therefore abandon the requirement that photons which bounce back and forth between an observer and a neighboring static object must be sent at and received from the direction of the connecting spacelike geodesic. In our next step will allow the possibility that the relatively static object is visible to the observer from any direction without any prior restrictions, but still require that photons which are shot in the visibility direction also arrive at the object throughout the motion; a reasonable requirement.\\

\begin{figure}
  \centering
  \includegraphics[width=\columnwidth]{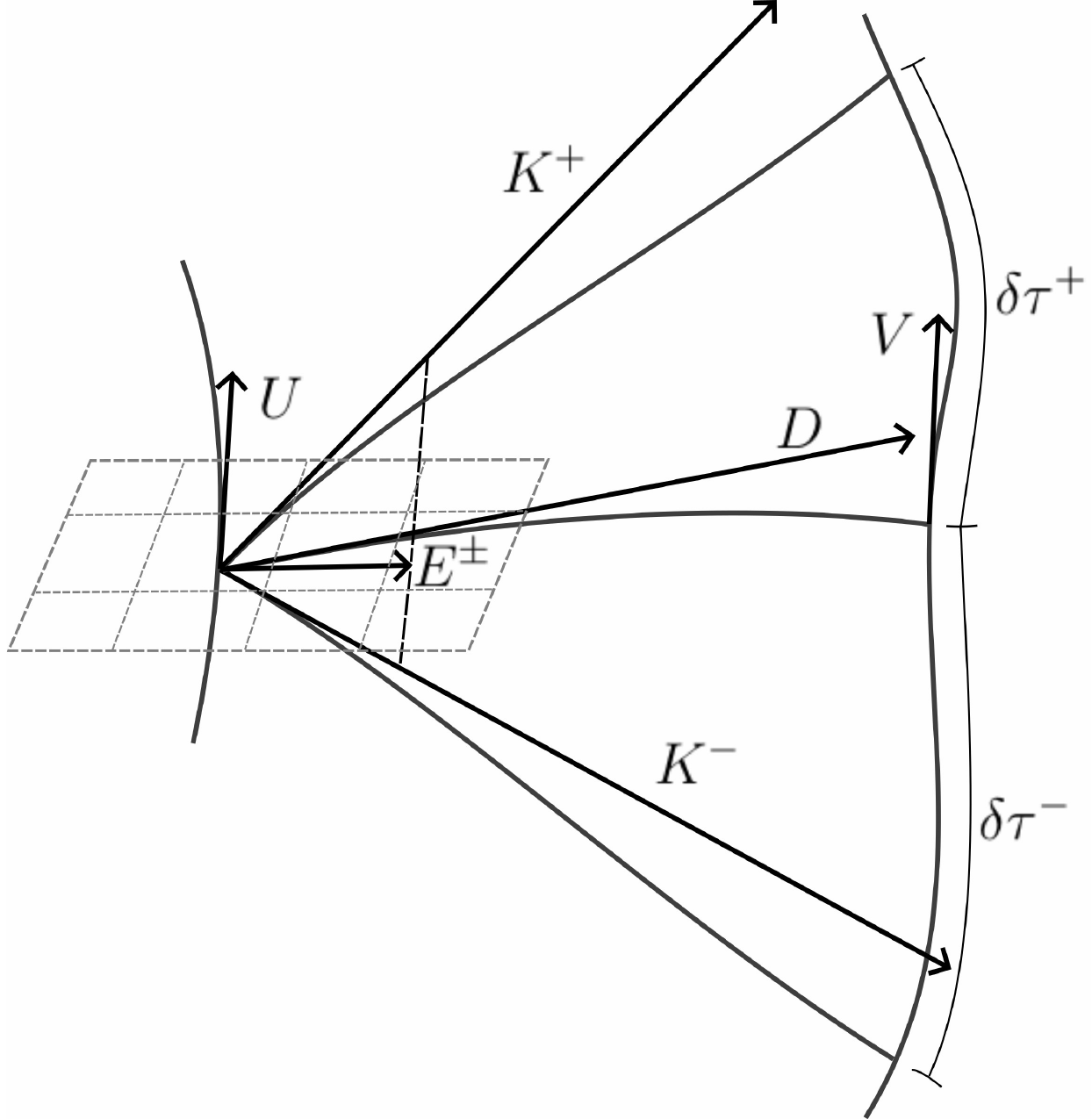}
  \caption{Two timelike curves, not necessarily geodesics, with 4-velocities $U^\alpha$ and $V^\alpha$, connected by two null geodesics that intersect at the event depicted. One represents the light incoming from past $V^\alpha$ to $U^\alpha$ and has tangent $K^{-\alpha}$, and the other represents the light outgoing from $U^\alpha$ at the given event that reaches a future $V^\alpha$ and has tangent $K^{+\alpha}$. Both null geodesics are parametrized with normalized affine parametrization as indicated by the magnitudes of their tangents on the diagram. $D^\alpha$ is the Fermi position vector of $V^\alpha$ in the frame of $U^\alpha$ at the event of emission/reception of light from/to $V^\alpha$. $E^{\pm\alpha}$ is the unit vector in the space frame of $U^\alpha$ pointing in the direction of the projections of $K^{\pm\alpha}$. In this particular case, the projections have the same direction in the observer's frame, but this need not be true in general. $\delta \tau^\pm=\tau^\pm-\tau$ is the lapse in proper time on the curve of $V^\alpha$, where $x^\alpha(\tau)$ belongs to the spacelike Fermi simultaneity slice with respect to the observer $U^\alpha$ and $x^\alpha(\tau^\pm)$ belong to the future and past null light cones with respect to the event of $U^\alpha$.} \label{fig:sd}
\end{figure}

For the case described in Figure \ref{fig:sd} and the terms defined within, the tangents to the future and past null geodesics connecting the emission/reception event on the observer's curve to the corresponding events on the neighboring object's curve are given by (see appendix \ref{app:deri}, equation \eqref{temp62})
\begin{multline}
	K^{\pm \alpha}=D^{\alpha}+\bar V^{\alpha}\delta \tau^\pm+\frac{1}{2}\bar{\dot V}^{\alpha}\delta \tau^{\pm 2}+\frac{1}{6}\bar{\ddot V}^{\alpha}\delta \tau^{\pm 3} \\
    +\frac{1}{6}R^{\alpha}_{D \bar V D}\delta \tau^\pm -\frac{1}{3}R^{\alpha}_{\bar V D \bar V}\delta \tau^{\pm 2}+ ... \label{kpm}
\end{multline}
The aim is to set a condition on the neighboring curve with 4-velocity $V^\alpha$ under the given physical restrictions, in terms of the observer's curve with 4-velocity $U^\alpha$, the position vector $D^\alpha$, and the curvature. An object with a 4-velocity satisfying this condition would be considered static relative to $U^\alpha$ under this definition. The time intervals $\delta \tau^\pm$ in the above are set through the requirement that $K^{\pm\alpha}$ are null, and under the condition that the distance $D=|D^\alpha|$ is constant throughout the motion. In fact, the constancy of $D$ sets fundamental restrictions on the trajectory of the neighboring object, which affects any candidate definition of being relatively static. We find (see appendix \ref{app:neigh}, equation \eqref{temp21})
\begin{multline}
	\delta \tau^\pm=\pm\frac{D}{\gamma}\left( 1+\frac{1}{2}D^\alpha\dot U_\alpha-\frac{1}{8}(D^\alpha\dot U_\alpha)^2-\frac{1}{24}\dot U^2 D^2 \right. \\
    \left. +\frac{1}{3} R_{DUDU} \right)+\frac{D^2}{6}D^\alpha \ddot U_\alpha+\mathcal{O}(D^4). \label{deltataupm}
\end{multline}

The small angle measured by the observer between the space directions of $K^{+\alpha}$ and $K^{-\alpha}$ in its frame is given by (see section \ref{sec:gen}, equation \eqref{smallangform2})
\begin{equation}
	\theta_U = \left| \frac{K^{+\alpha}}{(K^{+\alpha}U_\alpha)} + \frac{K^{-\alpha}}{(K^{-\alpha}U_\alpha)} + 2U^\alpha \right| + \mathcal{O}(|\;\;|^3),
\end{equation}
and with equations \eqref{kpm}, \eqref{deltataupm}, \eqref{dconstcondv}, \eqref{dconstcondvdot} and \eqref{temp78} we get
\begin{multline}
	\theta_U = \left| \dot U^\beta \bar V_\beta D^\alpha-2\left( \bar V^\beta +\frac{1}{6}D^2\ddot U^\beta +\frac{1}{6}R^\beta_{DUD} \right)H^\alpha_\beta \right| \\
    + \mathcal{O}(D^3), \label{thetau}
\end{multline}
where $H^\alpha_\beta$ projects with respect to $U^\alpha$ and $\hat D^\alpha$ ($H^\alpha_\beta=\delta^\alpha_\beta+U^\alpha U_\beta -\hat D^\alpha\hat D_\beta$). Requiring $\theta_U = 0$ forces both terms in the above to be zero, yielding the following condition on $V^\alpha$, (using $\bar V^\alpha D_\alpha=0$, since $|D|$ is constant)
\begin{equation}
	\bar V^\alpha = U^\alpha-\frac{1}{6}\left( D^2\ddot U^\beta + R^\beta_{DUD} \right)H^\alpha_\beta + \mathcal{O}(D^3). \label{temp38}
\end{equation}
Finally, by means of \eqref{ddot} and \eqref{taudot} we get the following condition on $\dot D^\alpha$ for an object that remains at constant distance, and is visible from the same direction in which emitted photons must travel to reach it,
\begin{equation}
	\dot D^\alpha = D^\beta \dot U_\beta U^\alpha - \frac{1}{6}\left( D^2 \ddot U^\beta -2R^\beta_{DUD} \right)H^\alpha_\beta + \mathcal{O}(D^3). \label{temp39}
\end{equation}
This is the first consistent condition that we obtain, given our requirements for being relatively static, as an alternative to the zero Fermi velocity. It becomes the same as zero Fermi velocity for very small distances. Notice the modification appearing in second order of $D$ and the effects of curvature and jerk on the required development of $D^\alpha$. With these results we can go back to \eqref{kpm} and find $E^{\pm \alpha}$, the corresponding unit projections of $K^{\pm \alpha}$ onto the space of $U^\alpha$,
\begin{equation}
	K^{\pm\alpha}=\mp K^{\pm\beta} U_\beta\left( \pm U^\alpha+E^{\pm\alpha} \right),
\end{equation}
\begin{equation}
	E^{\pm\alpha}=\mp \left( \frac{ K^{\pm\alpha}}{K^{\pm\beta} U_\beta} + U^\alpha \right), \label{temp41a}
\end{equation}
which simplifies to
\begin{multline}
	E^{\pm\alpha} = \left(1 -\frac{D^2}{8}\dot U^\beta \dot U^\gamma H_{\beta\gamma}\right)\hat D^\alpha \\
    + \left(\frac{D}{2}\dot U^\beta - \frac{D}{4}D^\gamma \dot U_\gamma \dot U^\beta + \frac{D}{6}R^\beta_{UDU} \right)H^\alpha_\beta \\
    + \mathcal{O}(D^3). \label{temp40}
\end{multline}
As expected the two are the same in this construction, so we may drop the $\pm$ superscript for now. (Indeed, the equality of the two could have been used as the condition leading to \eqref{temp38}, but the angle expression is more versatile for the analysis.) From the above we see that $E^\alpha-\hat D^\alpha = \frac{1}{2}D\dot U^\beta H^\alpha_\beta + \mathcal{O}(D^2)$, so the main difference between the direction of visibility and the physical position vector (wire) comes from the acceleration. In case of geodesic motion the difference is in second order and is due to curvature only. From \eqref{temp40} and \eqref{temp39} we get the time derivative of $E^\alpha$,
\begin{equation}
	\dot E^\alpha = E^\beta \dot U_\beta U^\alpha + \frac{D}{3}\left(\ddot U^\beta+ R^\beta_{EUE} \right)H^\alpha_\beta + \mathcal{O}(D^2). \label{temp41}
\end{equation}
The above expressions are the same whether $H^\alpha_\beta$ is with respect to $D^\alpha$ or $E^\alpha$.\\

We can test the established conditions for the motion of the object (\eqref{temp38}, \eqref{temp39}) by considering a round trip of a bouncing photon to the object and back. At a given event $x^\alpha(t)$, $E^\alpha(t)$ is the direction of an outgoing photon to the object and is given by \eqref{temp40}. After a time interval $\delta t_{RT}$ that corresponds to a round trip, the photon should be received from a future direction $E^\alpha(t+\delta t_{RT})$, and we will now check if this is indeed the case. The parallel transport of $E^\alpha(t+\delta t_{RT})$ from $x^\alpha(t+\delta t_{RT})$ back to $x^\alpha(t)$ is given by $\bar E^\alpha(t+\delta t_{RT})=E^\alpha(t)+\dot E^\alpha(t)\delta t_{RT}+\frac{1}{2}\ddot E^\alpha \delta t^2_{RT}+...$ (eq. \eqref{abartdeltat}), where the derivatives of $E^\alpha$ are given through \eqref{temp41}. In this way we relate the directions of the outgoing photon and the parallel transport of what should be of the incoming photon in this construction, and compare it to the general relationship derived earlier given by \eqref{eibar}. While $\delta t_{RT}$ can easily be expressed more accurately, for our purpose all we need is its lowest order in $D$, which is clearly $\delta t_{RT} = 2D+\mathcal{O}(D^2)$. Therefore,
\begin{align}
    \bar E^\alpha &= E^\alpha +\dot E^\alpha \delta t_{RT} +\frac{1}{2}\ddot E^\alpha \delta t^2_{RT} + \mathcal{O}(\delta t^3_{RT}) \nonumber\\
	& =E^\alpha +\left(E^\beta \dot U_\beta U^\alpha +\frac{D}{3}\left( \ddot U^\beta+R^\beta_{EUE} \right)H^\alpha_\beta \right) \delta t_{RT} \nonumber\\
    &\hspace{1cm}+ \frac{1}{2}\left( E^\beta \ddot U_\beta U^\alpha+E^\beta \dot U_\beta \dot U^\alpha \right) \delta t^2_{RT} + \mathcal{O}(\delta t^3_{RT}) \nonumber\\
	&=E^\alpha +E^\beta \dot U_\beta U^\alpha \delta t_{RT} + \frac{1}{2}\bigg(E^\beta \ddot U_\beta U^\alpha+E^\beta \dot U_\beta \dot U^\alpha \nonumber\\
    &\hspace{1.5cm} +\frac{1}{3}\left( \ddot U^\beta+R^\beta_{EUE} \right)H^\alpha_\beta \bigg) \delta t^2_{RT} + \mathcal{O}(\delta t^3_{RT}),
\end{align}
and we find agreement with equation \eqref{eibar}, which means that if photons are ever shot in the direction of the vector field $E^\alpha$ constructed above, they will always come back in the direction of a future $E^\alpha$ after a round trip to the object, as required. Notice that if the photon bounces from a distance shorter or longer than the distance to the object, then we would not arrive at a similar conclusion; it only works for the particular constant distance associated with the object, $D$, that appears in the differential development rule for $E^\alpha$, \eqref{temp41}.\\

\begin{figure}
  \centering
  \includegraphics[width=\columnwidth]{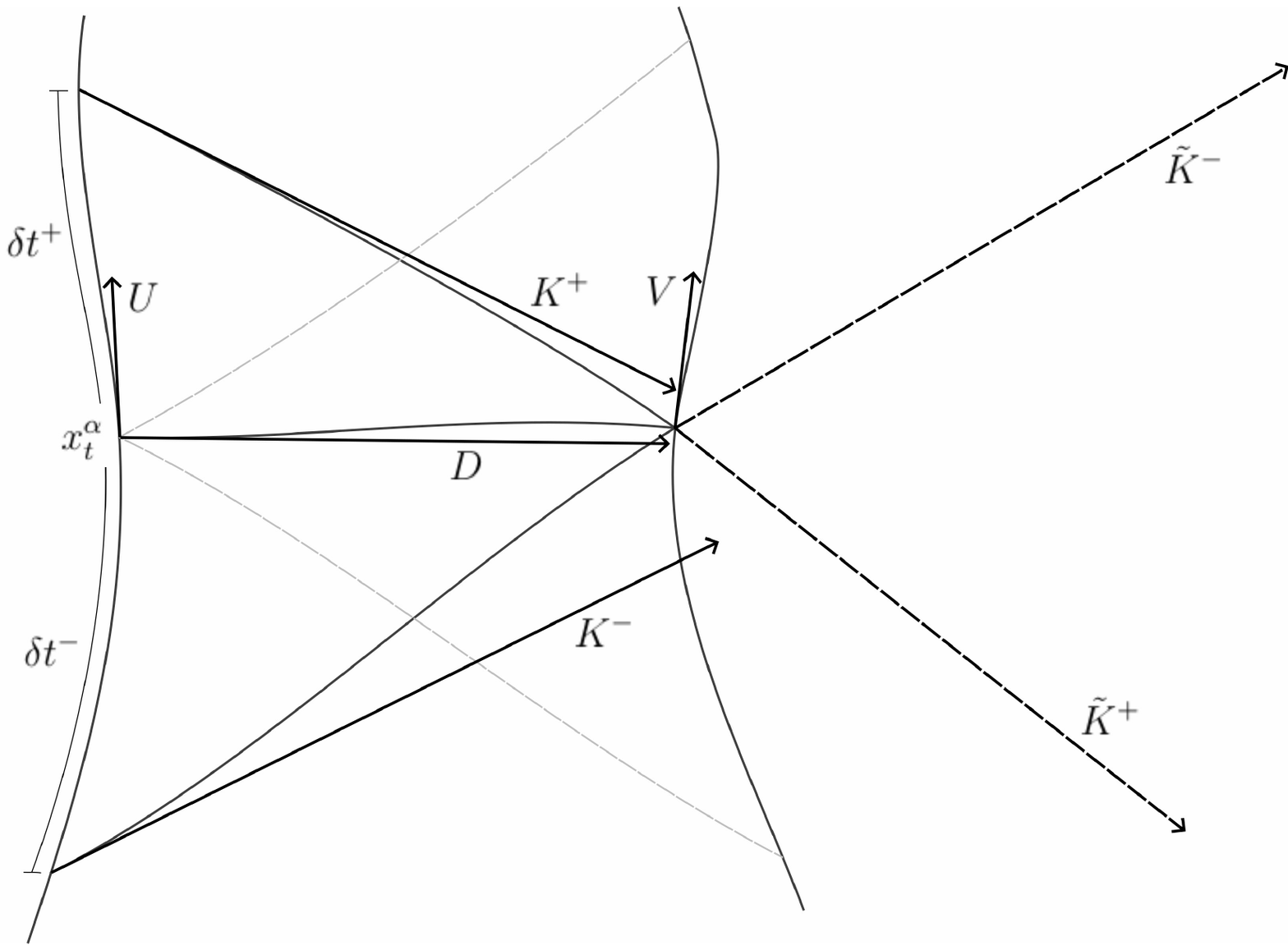}
  \caption{A similar situation as the one depicted in Figure \ref{fig:sd}, but with emphasis on the measurements done by $V^\alpha$ on the incoming and outgoing photons at an event on its curve. Here $K^{\pm\alpha}$ are tangents to the null geodesics at events on the curve of $U^\alpha$, and $\tilde{K}^{\pm\alpha}$ are their parallel transports to the event of $V^\alpha$, which are also tangent to the intersecting null geodesics at that event. (Not to be confused with $\bar{K}^{\pm\alpha}$, which are the parallel transports of $K^{\pm\alpha}$ along the timelike curve to $x^\alpha_t$.) $x^\alpha_t$ is the event on $U^\alpha$ that determines the spacelike simultaneity slice which contains the event of intersection on $V^\alpha$ and the spacelike connecting geodesic with tangent $D^\alpha$. $\delta t^\pm$ are proper time lapses on the curve of $U^\alpha$, similarly defined as $\delta \tau^\pm$ in Figure \ref{fig:sd}.} \label{fig:av}
\end{figure}

While the condition is fully consistent, evidently, the visibility direction $E^\alpha$ itself does not satisfy the Fermi-Walker condition. This means that given a local non-rotating Fermi frame, the optical direction $E^\alpha$ will rotate relative to the frame for this motion of the object, as would $D^\alpha$. We can also see that the angle between $D^\alpha$ and $E^\alpha$ can change over time as well, another undesirable behavior of a static object. Thus, to satisfy our requirements for being relatively static we pay a high price for the condition to not be the same as zero Fermi velocity, and, therefore, for the object to not have constant Fermi-Walker coordinates. This goes against the intuitive understanding of the Fermi frame and does not rid the resulting zero velocity from undesirable consequences. However, the biggest shortcoming of using \eqref{temp39} as a condition for relatively static objects is that it does not in any way help us resolve the asymmetry issue that we encountered with zero Fermi velocity, which was the primary reason for seeking an alternative. (Compare \eqref{temp38} to \eqref{vbarzerofermi}). The asymmetry suggests that while the observer will conclude that its neighbor is non-rotating, the neighbor may not arrive at the same conclusion. With reference to Figure \ref{fig:av} and the definitions within, the small angle $\theta_V$ measured by the observer with 4-velocity $V^\alpha$ between the incoming and outgoing bouncing photons is given by (see equation \eqref{smallangform2})
\begin{align}
	\nonumber \theta_V &= \left| \frac{\tilde K^{+\alpha}}{(\tilde K^{+\alpha}V_\alpha)} + \frac{\tilde K^{-\alpha}}{(\tilde K^{-\alpha}V_\alpha)} + 2V^\alpha \right| + \mathcal{O}(|\;\;|^3)\\
	&=\left| \frac{\bar{\tilde K}^{+\alpha}}{(\bar{\tilde K}^{+\alpha}\bar V_\alpha)} + \frac{\bar{\tilde K}^{-\alpha}}{(\bar{\tilde K}^{-\alpha}\bar V_\alpha)} + 2\bar V^\alpha \right| + \mathcal{O}(|\;\;|^3), \label{temp45}
\end{align}
where $\bar{\tilde K}^{\pm\alpha}$ are the parallel transports of $\tilde K^{\pm\alpha}$ to $x^\alpha_t$ along the connecting spacelike geodesic, and $\tilde K^{\pm\alpha}$ are the tangents of the null geodesics at the observer with 4-velocity $V^\alpha$. The parallel transport performed on the above will allow us to express the angle in terms of quantities at the event $x^\alpha_t$ for proper comparison with previous results. With the help of equation \eqref{omegaoflambdau} we have
\begin{multline}
	\bar K^{\pm\alpha}=D^{\alpha}-U^{\alpha}\delta t^\pm-\frac{1}{2}\dot U^{\alpha}\delta t^{\pm 2}-\frac{1}{6}\ddot U^{\alpha}\delta t^{\pm 3} \\
    +\frac{1}{3}R^{\alpha}_{D U D}\delta t^\pm+\frac{1}{6}R^{\alpha}_{U D U}\delta t^{\pm 2} + ...\,,
\end{multline}
where $\bar K^{\pm\alpha}$ are the parallel transports of $K^{\pm\alpha}$ to $x^\alpha_t$ along the timelike curve. By equation \eqref{tritransa} and the configuration in our setup, $\bar{\tilde K}^{\pm\alpha}=\bar K^{\pm\beta}\left( \delta^\alpha_\beta + \frac{1}{2}R^\alpha_{\beta DU}\delta t^\pm \right)+...$, therefore
\begin{multline}
	\bar{\tilde K}^{\pm\alpha}=D^{\alpha}-U^{\alpha}\delta t^\pm-\frac{1}{2}\dot U^{\alpha}\delta t^{\pm 2}-\frac{1}{6}\ddot U^{\alpha}\delta t^{\pm 3} \\
    -\frac{1}{6}R^{\alpha}_{D U D}\delta t^\pm-\frac{1}{3}R^{\alpha}_{U D U}\delta t^{\pm 2} + ...\,,
\end{multline}
where, as with \eqref{kpm}, the time intervals $\delta t^\pm$ are set by the condition that $K^{\pm \alpha}$ are null. Thus, we have all the ingredients to express $\theta_V$ given by \eqref{temp45},
\begin{multline}
	\theta_V = \bigg| \dot U^\beta \bar V_\beta D^\alpha-2\left( \bar V^\beta +\frac{1}{6}D^2\ddot U^\beta +\frac{1}{6}R^\beta_{DUD} \right)H^\alpha_\beta \\
    +4\left( \bar V^\alpha - U^\alpha \right) \bigg| + \mathcal{O}(D^3). \label{thetav}
\end{multline}
For the $\bar V^\alpha$ satisfying the latest condition of non-rotation, \eqref{temp38}, the measured angle between the outgoing and incoming bouncing photons between it and the observer is found to be
\begin{equation}
	\theta_V = \frac{2}{3}\left| \left( D^2\ddot U^\beta + R^\beta_{DUD} \right)H^\alpha_\beta \right| + \mathcal{O}(D^3).
\end{equation}
If we were to require that $\theta_V=0$, then \eqref{thetav} would lead to the following condition on $V^\alpha$,
\begin{equation}
	\bar V^\alpha = U^\alpha+\frac{1}{6}\left( D^2\ddot U^\beta + R^\beta_{DUD} \right)H^\alpha_\beta + \mathcal{O}(D^3),
\end{equation}
which is clearly inconsistent (and opposite in a sense) to the one for $\theta_U = 0$ given by \eqref{temp38}. Notice that even for geodesic motion of the observer, in the presence of curvature it is impossible to have both $\theta_U = 0$ and $\theta_V = 0$ simultaneously throughout the motion in general.

Due to the shortcomings, We conclude that \eqref{temp39} is not a suitable candidate for zero velocity, and abandon the physical expectations of a relatively static object suggested at the start of our investigation. From what we have learned, the clear approach to overcome the asymmetry issue is to impose the requirement $\theta_U = \theta_V$. That is, we can define relatively static observers as observers that measure the same angle between the directions of outgoing and incoming bouncing photons moving between them. It should be clear that this equality refers to the simultaneity constructed with the connecting geodesics discussed throughout. (The constancy of $D$ ensures mutual simultaneity along the connecting geodesic.) Equating \eqref{thetau} and \eqref{thetav} leads directly to our newest condition on $V^\alpha$,
\begin{equation}
	\bar V^\alpha = U^\alpha,
\end{equation}
which is obviously symmetric; compare to \eqref{vbarzerofermi} and \eqref{temp38}. By means of \eqref{ddot} and \eqref{taudot}, for this motion we have
\begin{equation}
	\dot D^\alpha = D^\beta \dot U_\beta U^\alpha +\frac{1}{2}R^\beta_{DUD}H^\alpha_\beta + \mathcal{O}(D^3)
\end{equation}
as the propagation rule for the position vector, and by \eqref{thetau} and \eqref{thetav} the measured angles are
\begin{equation}
	\theta_U = \theta_V = \frac{1}{3}\left| \left( D^2\ddot U^\beta + R^\beta_{DUD} \right)H^\alpha_\beta \right| + \mathcal{O}(D^3).
\end{equation}
An important observation is that for this relative motion, differentiation and application of \eqref{partransdot} gives
\begin{equation}
	\bar{\dot V}^\alpha = \dot U^\alpha -\dot U^\beta D_\beta \dot U^\alpha + R^\alpha_{UDU} + \mathcal{O}(D^2),
\end{equation}
so although we have the symmetry in the 4-velocities, we see that if one of the observers is on geodesic motion, then in general the other is not. Further, by \eqref{kpm} and \eqref{temp41a}, the corresponding spacelike directions of the bouncing photons are
\begin{multline}
	E^{\pm\alpha} = \left(1 -\frac{D^2}{8}\dot U^\beta \dot U^\gamma H_{\beta\gamma}\right)\hat D^\alpha \\
    + \left(\frac{D}{2}\dot U^\beta - \frac{D}{4}D^\gamma \dot U_\gamma \dot U^\beta + \frac{D}{6}R^\beta_{UDU} \right)H^\alpha_\beta \\
	\pm \frac{1}{6}\left( D^2\ddot U^\beta + R^\beta_{DUD} \right)H^\alpha_\beta+ \mathcal{O}(D^3),  \label{temp55}
\end{multline}
where the $E^{+\alpha}$ and $E^{-\alpha}$ are the directions of the outgoing and incoming photons, respectively (see Figure \ref{fig:sd}). Differentiating and substituting for $\hat D^\alpha$ and $\dot{\hat D}^\alpha$, we get
\begin{equation}
	\dot E^{\pm\alpha} = E^{\pm\beta} \dot U_\beta U^\alpha + \frac{D}{2}\left(\ddot U^\beta+ R^\beta_{EUE} \right)H^\alpha_\beta + \mathcal{O}(D^2), \label{temp56}
\end{equation}
and again we see that these optical directions will be changing with respect to a local Fermi-Walker frame. Thus, in general, and even for geodesic motion of an observer (with $U^\alpha$), under this definition a static neighbor (with $\bar V^\alpha=U^\alpha$) will be visible from a direction that is rotating relative to a local non-rotating frame.\\

Defining the outgoing optical direction as $E^\alpha_o$ ($=E^{+ \alpha}$), and the incoming optical direction as $E^\alpha_i$ ($=E^{- \alpha}$), by means of \eqref{eibar} we can test that a bouncing photon leaving in direction $E^\alpha_o$ at some time $t$, will come back to the observer from the direction $E^\alpha_i$ at a later time $t+\delta t_{RT}$ (after a round trip). We can express the parallel transport ($\bar E^\alpha_i$) of $E^\alpha_i(t+\delta t_{RT})$ from $x^\alpha(t+\delta t_{RT})$ to $x^\alpha(t)$ in terms of $E^\alpha_i(t)$ and its derivatives at $x^\alpha(t)$ by use of \eqref{abartdeltat}. Then $E^\alpha_i(t)$ and its derivatives can be expressed in terms of $E^\alpha_o(t)$ with help of \eqref{temp55} and \eqref{temp56}, producing a relationship between $\bar E^\alpha_i$ and $E^\alpha_o$ that can be compared with \eqref{eibar}. With $\delta t_{RT}= 2D+\mathcal{O}(D^2)$, we find
\begin{align}
	\bar E^\alpha_i &= E^\alpha_i + \dot E^\alpha_i \delta t_{RT} + \frac{1}{2}\ddot E^\alpha_i \delta t_{RT}^2+... \nonumber\\
	&= E^\alpha_o -\frac{1}{3}\left( D^2\ddot U^\beta + R^\beta_{DUD} \right)H^\alpha_\beta + \dot E^\alpha_o \delta t_{RT} \nonumber\\
    &\hspace{5cm}+ \frac{1}{2}\ddot E^\alpha_o \delta t_{RT}^2+... \nonumber \\
	&= E^\alpha_o-\frac{1}{3}\left( D^2\ddot U^\beta + D^2 R^\beta_{EUE} \right)H^\alpha_\beta + E^\beta_o \dot U_\beta U^\alpha \delta t_{RT} \nonumber\\
    &\hspace{1.5cm}+\frac{D}{2}\left(\ddot U^\beta+ R^\beta_{EUE} \right)H^\alpha_\beta \delta t_{RT} \nonumber \\
	&\hspace{2cm} + \frac{1}{2}\left( E^\beta_o \ddot U_\beta U^\alpha + E^\beta_o \dot U_\beta \dot U^\alpha \right)\delta t_{RT}^2+... \nonumber \\
	&= E^\alpha_o+ E^\beta_o \dot U_\beta U^\alpha \delta t_{RT} + \frac{1}{2}\bigg( E^\beta_o \ddot U_\beta U^\alpha + E^\beta_o \dot U_\beta \dot U^\alpha  \nonumber \\
    &\hspace{0.7cm}+\frac{1}{3}\left( \ddot U^\beta + R^\beta_{EUE} \right)H^\alpha_\beta \bigg)\delta t_{RT}^2+\mathcal{O}(\delta t_{RT}^3).
\end{align}
The above is in agreement with \eqref{eibar} for the time interval of a round trip corresponding to the distance to the object. By \eqref{eibar}, photons coming out of $E^\alpha_o$ at any time, will eventually return to the observer from $E^\alpha_i$ after bouncing from a distance $D$ away, as required. Thus, we have found a consistent and symmetric alternative to the zero Fermi velocity, based on which we can define an alternative relative velocity. The primary shortcoming of this symmetric zero velocity definition, is that stationary objects will not have constant Fermi coordinates.\\

Much of our analysis so far relied on optical directions, yet for both cases of Fermi static and symmetrically static we see that the direction of visibility of a non-moving object is in general rotating relative to the Fermi frame. Naturally, we should also seek for a definition of being relatively static by requiring that the optical directions of the object will not rotate while also maintaining constant distance. For a neighboring object with 4-velocity of the general type $\bar V^\alpha = U^\alpha+\mathcal{O}(D^2)$ as in the cases investigated above, equations \eqref{kpm} and \eqref{temp41a} lead to
\begin{multline}
	E^{\pm\alpha} = \left(1 -\frac{D^2}{8}\dot U^\beta \dot U^\gamma H_{\beta\gamma}\right)\hat D^\alpha \\
    + \left(\frac{D}{2}\dot U^\beta - \frac{D}{4}D^\gamma \dot U_\gamma \dot U^\beta + \frac{D}{6}R^\beta_{UDU} \right)H^\alpha_\beta\\
	\pm \left( \bar V^\beta+ \frac{1}{6}\left( D^2\ddot U^\beta + R^\beta_{DUD} \right)\right)H^\alpha_\beta+ \mathcal{O}(D^3),
\end{multline}
and
\begin{equation}
	\dot E^{\pm\alpha} = \dot{\hat D}^\alpha + \frac{D}{2}\dot U^\beta \dot U^\gamma H_{\beta\gamma}U^\alpha + \frac{D}{2}\ddot U^\beta H^\alpha_\beta + \mathcal{O}(D^2). \label{temp59}
\end{equation}
To the given order of accuracy, the derivatives of both $E^{\pm\alpha}$ are the same, and also the difference between $E^{\pm\alpha}$ is in second order, so we may drop the $\pm$ sign where the meaning is clear. In particular, for zero Fermi relative velocity objects, the time derivative of the optical directions is
\begin{equation}
	\dot E^{\alpha} = E^\beta \dot U_\beta U^\alpha + \frac{D}{2}\ddot U^\beta H^\alpha_\beta + \mathcal{O}(D^2).
\end{equation}

Now, if we require of a neighboring object that its optical directions remain constant relative to a local Fermi-Walker frame, then setting $\dot E^\alpha= E^\beta \dot U_\beta U^\alpha$ in \eqref{temp59} yields the following condition on $\dot D^\alpha$,
\begin{equation}
	\dot D^{\alpha} = D^\beta \dot U_\beta U^\alpha - \frac{1}{2}D^2\ddot U^\beta H^\alpha_\beta + \mathcal{O}(D^3).
\end{equation}
(The projection operators $H^\alpha_\beta$ may refer to either $\hat D^\alpha$ or $E^\alpha$ at times, but at the order where they appear in the above relationships and the given accuracy, there is no distinction between the two.) Combining the above with \eqref{ddot}, we get
\begin{equation}
	\bar V^\alpha = U^\alpha - \frac{1}{2}\left(D^2\ddot U^\beta+R^\beta_{DUD}\right) H^\alpha_\beta + \mathcal{O}(D^3).
\end{equation}
These are the 4-velocity and the corresponding derivative of the position vector of a neighboring object that will be visible from a direction which will not rotate relative to the Fermi frame. The latter is a fair expectation of a relatively static object, however the resulting condition is clearly non-symmetric. These outcomes reveal that in general it is impossible to have two neighboring observers that both see each other from directions that are locally non-rotating within their respective frames throughout the motions. Even for geodesic motion, the presence of curvature may not allow such a possibility.\\

We summarize the investigation of this section with the following three candidates for relatively static objects.\\

\paragraph*{Fermi static:}
\begin{equation}
	\dot D^{\alpha} = D^\beta \dot U_\beta U^\alpha, \label{temp63}
\end{equation}
\begin{equation}
	\bar V^\alpha = U^\alpha - \frac{1}{2}R^\beta_{DUD}H^\alpha_\beta + \mathcal{O}(D^3),
\end{equation}
\begin{equation}
	\dot E^{\alpha} = E^\beta \dot U_\beta U^\alpha + \frac{D}{2}\ddot U^\beta H^\alpha_\beta + \mathcal{O}(D^2).
\end{equation}
A neighboring object with zero Fermi relative velocity will be static in a sense that its distance from the observer remains constant in time, and its physical position vector (interpreted as a geodesic wire) will not be rotating relative to a local Fermi frame. These objects will have constant Fermi coordinates that can be constructed along the timelike curve of the observer. However, in general, the static neighbors will not make the same conclusions on the motion of the observer itself, and so this concept of being static is not symmetric. Furthermore, for general motion of the observer, the direction from which the object is seen may rotate over time relative to the local Fermi frame.\\

\paragraph*{Symmetrically static:}
\begin{equation}
	\dot D^{\alpha} = D^\beta \dot U_\beta U^\alpha+\frac{1}{2}R^\beta_{DUD}H^\alpha_\beta+ \mathcal{O}(D^3), \label{temp66}
\end{equation}
\begin{equation}
	\bar V^\alpha = U^\alpha,
\end{equation}
\begin{equation}
	\dot E^{\alpha} = E^{\beta} \dot U_\beta U^\alpha + \frac{D}{2}\left(\ddot U^\beta+ R^\beta_{EUE} \right)H^\alpha_\beta + \mathcal{O}(D^2).
\end{equation}
Two observers with 4-velocities that equal when parallel transported along the (unique) connecting spacelike geodesic are relatively static in the sense that they will make equivalent measurements on each other. In particular, the angle between the optical directions that each can measure are equal, which guided us to this condition. Such two observers are referred to as kinematically comoving in \cite{bolos}. However, we note that while this definition of being relatively static is symmetric for two observers, this symmetry does not in general extend to a collection of more than two observers. If each of two arbitrary observers is relatively static to some third observer under this definition, then in general the two observers may not be relatively static to each other. Further, both the physical position vectors and optical directions of such relatively static observers will be rotating relative to a local Fermi frame. In case of no curvature, for any motion of the observer, the Fermi and the symmetric definition of being relatively static coincide.\\

\paragraph*{Optically static:}
\begin{equation}
	\dot D^{\alpha} = D^\beta \dot U_\beta U^\alpha - \frac{1}{2}D^2\ddot U^\beta H^\alpha_\beta + \mathcal{O}(D^3), \label{temp69}
\end{equation}
\begin{equation}
	\bar V^\alpha = U^\alpha - \frac{1}{2}\left(D^2\ddot U^\beta+R^\beta_{DUD}\right) H^\alpha_\beta + \mathcal{O}(D^3),
\end{equation}
\begin{equation}
	\dot E^{\alpha} = E^{\beta} \dot U_\beta U^\alpha.
\end{equation}
An object may also be considered relatively static if its distance from the observer remains constant and the direction from which it is visible is not rotating relative to a local Fermi frame. For distant objects, this definition of being static is particularly appealing, since it gives a practical criterion when the direction of the physical position vector is not available. However, for general motion and curvature, this definition of being relatively static is not symmetric. We see that in case of geodesic motion, or simply with no jerk, the Fermi and optical conditions for being relatively static coincide, and neither the position vector or the optical direction will rotate. Only when there is no curvature or jerk do all three definitions coincide.

\subsection{\label{sec:relvel}Defining Relative Velocity}

Ability to identify relatively static objects allows for a clear definition of relative velocity. Given a moving object close to an observer, the instantaneous relative velocity of the object can be defined as the deviation of its position vector over time from that of a static object at the same location. In this section we define relative velocity with respect to the three different cases of relatively static objects of section \ref{sec:relsta}.

Our reasoning is as follows. Consider the 3-dimensional relative velocity vector to be identified by an observer with 4-velocity $U^\alpha$ of a neighboring object with 4-velocity $V^\alpha$, based on observing changes in the physical position vector of the object. The position vector is constructed according to the process outlined in section \ref{sec:phys}, and its derivative is related to $V^\alpha$ through \eqref{ddot}. Let $D^\alpha(t)$ be the position vector of the object at time $t$, and $\bar D^\alpha(t+\delta t)$ be the position vector at time $t+\delta t$ parallel transported back to $t$. Let $D^\alpha_s(t+\delta t)$ be the position vector of a static object at $t+\delta t$, that coincides with $D^\alpha(t)$ (when $\delta t =0$), and let $\bar D^\alpha_s(t+\delta t)$ be its parallel transport to $t$. The propagation rule that determines $D^\alpha_s(t+\delta t)$ is given by one of the definitions of stationarity proposed in section \ref{sec:relsta}. We define the instantaneous relative velocity $\vec v^\alpha(t)$ at time $t$ as the rate of the developing difference between the position of the moving object and that of a coincident static one,
\begin{align}
	\nonumber \vec v^\alpha(t) &= \lim\limits_{\delta t \to 0}\frac{\bar D^\alpha(t+\delta t)-\bar D^\alpha_s(t+\delta t)}{\delta t}\\
	\nonumber &= \lim\limits_{\delta t \to 0}\frac{\dot D^\alpha(t)\delta t-\dot D^\alpha_s(t)\delta t+\mathcal{O}(\delta t^2)}{\delta t}\\
	&= \dot D^\alpha(t)-\dot D^\alpha_s(t), \label{temp72}
\end{align}
where we made use of \eqref{abartdeltat} to express the parallel transports. The first important aspect that we notice in this definition is that the velocity $\vec v^\alpha$ belongs to the space frame of $U^\alpha$, as required. This is due to the fact that $D^\alpha_s=D^\alpha$ at $t$, which means that $\dot D^\alpha_s U_\alpha = -D^\alpha_s \dot U_\alpha = -D^\alpha \dot U_\alpha = \dot D^\alpha U_\alpha$, and therefore $\vec v^\alpha U_\alpha = \dot D^\alpha U_\alpha - \dot D^\alpha_s U_\alpha = 0$. Next, we can manipulate the expression \eqref{ddot} for $\dot D^\alpha$ as follows,
\begin{multline}
	\dot D^\alpha = \dot \tau \left(\bar V^\alpha-\gamma U^\alpha +\gamma U^\alpha\right)-U^\alpha\\
    +\frac{1}{6}\left( \dot\tau \left( \bar V^\beta -\gamma U^\beta + \gamma U^\beta \right) +2U^\beta \right)R^\alpha_{D\beta D}+\mathcal{O}(D^3)\\
	= \dot \tau \bar V^\beta h^\alpha_\beta+\frac{1}{6}\dot \tau \bar V^\gamma h^\beta_\gamma R^\alpha_{D\beta D}+\left( \dot \tau \gamma -1 \right)U^\alpha+\frac{1}{2}R^\alpha_{DUD}\\
    +\mathcal{O}(D^3),
\end{multline}
and with \eqref{taudot},
\begin{align}
	\nonumber \dot \tau \gamma &= 1+D^\alpha \dot U_\alpha+\frac{1}{6}\bigg( \frac{1}{\gamma}\left( \bar V^\alpha -\gamma U^\alpha +\gamma U^\alpha \right)\\
    &\hspace{4cm}+2U^\alpha \bigg)R_{\alpha DUD}+\mathcal{O}(D^3)\nonumber\\
	&=\! 1\!+\!D^\alpha \dot U_\alpha\!+\!\frac{1}{6}\dot \tau \bar V^\beta h^\alpha_\beta R_{UD\alpha D}\!+\!\frac{1}{2}R_{UDUD}\!+\!\mathcal{O}(D^3),
\end{align}
\begin{align}
	\nonumber \dot D^\alpha &= D^\beta \dot U_\beta U^\alpha+\dot \tau \bar V^\beta h^\alpha_\beta+\frac{1}{6}\dot \tau \bar V^\gamma H^\beta_\gamma R^\alpha_{D\beta D}\\
    &\!\!\!\!\!\!\!+\!\frac{1}{6}\dot \tau \bar V^\gamma H^\beta_\gamma R_{UD\beta D}U^\alpha\!+\!\frac{1}{2}R^\alpha_{DUD}\!+\!\frac{1}{2}R_{UDUD}U^\alpha \!+\!\mathcal{O}(\!D^3)\nonumber\\
	&= D^\beta \dot U_\beta U^\alpha+\dot \tau \bar V^\beta h^\alpha_\beta+\frac{1}{6}\dot \tau \bar V^\gamma R^\epsilon_{D\beta D}H^\beta_\gamma H^\alpha_\epsilon\nonumber\\
    &\hspace{3.3cm}+\frac{1}{2}R^\beta_{DUD} H^\alpha_\beta+\mathcal{O}(D^3). \label{temp75}
\end{align}

Let $V^\alpha_s$ be the corresponding 4-velocity of the static object that coincides with the moving object. Regardless of the definition established in the previous section, as deduced in appendix \ref{app:neigh}, it is sufficient to assume that a static object has constant distance to the observer to establish that $\bar V^\alpha_s=U^\alpha+\mathcal{O}(D)$. Thus, for a static object $\bar V^\beta_s h^\alpha_\beta=\bar V^\beta_s H^\alpha_\beta= 0 +\mathcal{O}(D)$, and we have
\begin{equation}
	\dot D^{\alpha}_s = D^\beta \dot U_\beta U^\alpha+\dot \tau_s \bar V^\beta_s H^\alpha_\beta+\frac{1}{2}R^\beta_{DUD} H^\alpha_\beta+\mathcal{O}(D^3),
\end{equation}
which further simplifies to
\begin{equation}
	\dot D^{\alpha}_s = D^\beta \dot U_\beta U^\alpha+\bar V^\beta_s H^\alpha_\beta+\frac{1}{2}R^\beta_{DUD} H^\alpha_\beta+\mathcal{O}(D^3) \label{temp77}
\end{equation}
for the case where $\bar V^\alpha_s=U^\alpha+\mathcal{O}(D^2)$, in which we are interested. With equations \eqref{temp77} for $\dot D^\alpha_s$ and the general \eqref{temp75} for $\dot D^\alpha$, we proceed to express $\vec v^\alpha$ given by \eqref{temp72},
\begin{equation}
	\vec v^\alpha = \dot \tau \bar V^\beta h^\alpha_\beta+\frac{1}{6}\dot \tau \bar V^\gamma R^\epsilon_{D\beta D}H^\beta_\gamma H^\alpha_\epsilon-\bar V^\beta_s H^\alpha_\beta+\mathcal{O}(D^3). \label{tempp78}
\end{equation}
Thus, with reference to static objects having a 4-velocity $V^\alpha_s$, the relative velocity of a moving object is given by the above equation. For the three cases of interest from section \ref{sec:relsta}, we have the following definitions.\\

\paragraph*{Fermi relative velocity:}
\begin{equation}
	\bar V^\alpha_s = U^\alpha - \frac{1}{2}R^\beta_{DUD}H^\alpha_\beta + \mathcal{O}(D^3), \label{temp79}
\end{equation}
\begin{equation}
	\vec v^\alpha = \dot \tau \bar V^\beta h^\alpha_\beta+\frac{1}{6}\dot \tau \bar V^\gamma R^\epsilon_{D\beta D}H^\beta_\gamma H^\alpha_\epsilon+\frac{1}{2}R^\beta_{DUD} H^\alpha_\beta+\mathcal{O}(D^3). \label{temp80}
\end{equation}

This is the relative velocity of an object within the Fermi frame of an observer. It expresses how the Fermi space coordinates of the object change in time. For zero velocity, Fermi coordinates would remain constant. If Fermi coordinates are used to study kinematics within the space of an observer, then the velocities of moving objects would precisely correspond to the Fermi relative velocity defined above. Since an observer can construct Fermi coordinates and measure the change of these coordinates with respect to proper time, Fermi velocity is directly measurable. However, if one observer sees another with zero Fermi velocity and constant Fermi coordinates, the other observer will not in general see the first observer at constant Fermi coordinates in its own frame.\\

\paragraph*{Symmetric relative velocity:}
\begin{equation}
	\bar V^\alpha_s = U^\alpha, \label{temp81}
\end{equation}
\begin{equation}
	\vec v^\alpha = \dot \tau \bar V^\beta h^\alpha_\beta+\frac{1}{6}\dot \tau \bar V^\gamma R^\epsilon_{D\beta D}H^\beta_\gamma H^\alpha_\epsilon+\mathcal{O}(D^3). \label{temp82}
\end{equation}

This relative velocity is symmetric in the sense that it will be the same (but opposite) for two nearby observers making measurements on each other. If one observer measures a zero relative velocity of the other, then the other will measure the same. The above is clearly different to the Fermi relative velocity, and two observers that measure zero symmetric velocity of each other will in general drift in each others Fermi frames. (This drift will be symmetric.) The usefulness of this definition of velocity is not yet clear, but it is the natural alternative to Fermi velocity which retains symmetry in the general case of curvature and motion. We can compare this velocity to the \textit{kinematic} relative velocity defined in \cite{bolos}, given by
\begin{equation}
	\vec v^\alpha = \frac{1}{\gamma}\bar V^\beta h^\alpha_\beta,
\end{equation}
which is also symmetric. The main difference in the two definitions is that \eqref{temp82} is directly measurable through the Fermi frame while the \textit{kinematic} relative velocity is purely a mathematical construction. Since Fermi coordinates can be constructed and symmetrically static objects can be identifies, the deviation from these objects can be measured directly. Therefore, we can calculate the velocity through \eqref{temp82} as well as measure it directly. \textit{Kinematic} relative velocity can only be calculated but not measured. Furthermore, even without curvature, when the object and observer are separated by space, the acceleration of the observer can strongly affect the observed velocity of the object because of relativistic frame change. This effect is present in \eqref{temp82} through $\dot\tau$, but is not accounted for in \textit{Kinematic} relative velocity.\\

\paragraph*{Optical relative velocity:}
\begin{equation}
	\bar V^\alpha_s = U^\alpha - \frac{1}{2}\left(D^2\ddot U^\beta+R^\beta_{DUD}\right) H^\alpha_\beta + \mathcal{O}(D^3), \label{temp84}
\end{equation}
\begin{multline}
	\vec v^\alpha = \dot \tau \bar V^\beta h^\alpha_\beta+\frac{1}{6}\dot \tau \bar V^\gamma R^\epsilon_{D\beta D}H^\beta_\gamma H^\alpha_\epsilon\\
    +\frac{1}{2}\left(D^2\ddot U^\beta+R^\beta_{DUD}\right) H^\alpha_\beta+\mathcal{O}(D^3). \label{temp85}
\end{multline}

This relative velocity is a measure of the deviation of the position vector of a moving object from that of an optically static coincident object. Objects that have zero optical relative velocity will be visible from a direction that is non-rotating relative to a local Fermi frame, and their (physical) distance $D$ will remain constant in time. While this optical velocity was developed through considering optical directions, it is still based on the (physical) position vector $D^\alpha$, an issue that we address in detail below. In case of $\ddot U^\beta h^\alpha_\beta=0$ (constant acceleration), optical relative velocity coincides with the Fermi relative velocity. Only when there is no jerk or curvature are the three relative velocities equivalent.\\

When no physical contact with distant objects can be made, and only optical signals can be used to make measurements on distance and motion, then the concept of optical velocity is most significant. However, we quickly find a discomforting flaw with the above definition. It turns out that when an object is in pure radial motion away from the observer, according to the above definition of optical relative velocity, then the direction in which it appears to the observer is rotating in the local Fermi frame. This is contrary to what one would expect from objects with purely radial optical velocity. Evidently the reference to physical distance and direction in the above definition turns out to be the source of the problem. Not only that we have the conceptual issue mentioned, but also such quantities may not be available in situations where optical measurements are relevant. Consequently, we will refine the definition of optical velocity to be based on quantities that are accessible from optical measurements only. Of main importance will be the concept of indirectly measurable optical distances, and for its richness we leave the discussion of proper optical velocity to its own section.

\section{\label{sec:opt}Proper Optical Velocity}

In this section we will define the optical velocity based on an optical position vector and its rate of change. A significant issue with the definition given by \eqref{temp85} arises when one considers a purely radially moving object. We start this section by exposing this issue and the main flaw with the definition itself. Based on the analysis we then redefine optical velocity accordingly, and discuss the concept of optical distances in detail. The process is driven by well defined physically measurable quantities, with the aim to define a measurable optical velocity and express it in terms of general system parameters.

\subsection{Introduction and the need for modification}

Consider an object with purely radial motion away from the observer according to the notation and the definitions of section \ref{sec:relvel}. With either definition of velocity, for purely radial motion the restriction is $\dot {\hat D}^\alpha=\dot {\hat D}^\alpha_s$, where $D^\alpha = D\hat D^\alpha$ is the physical position vector of the object, $D^\alpha_s = D_s\hat D^\alpha_s$ is the position vector of a coincident static object, and $\dot {D}^\alpha_s \,(=D_s\dot {\hat D}^\alpha_s=D\dot {\hat D}^\alpha_s)$ is given by \eqref{temp63}, \eqref{temp66}, or \eqref{temp69}. While the change of direction is restricted to that of a static object (at the given distance, $D$), the distance itself is allowed to change. Thus,
\begin{equation}
	\dot D^\alpha_r = \dot D \hat D^\alpha + D\dot {\hat D}^\alpha = \dot D \hat D^\alpha + D\dot {\hat D}^\alpha_s = \dot D \hat D^\alpha + \dot {D}^\alpha_s, \label{tempp86}
\end{equation}
where the subscript indicates purely radial motion. Let $\vec v^\alpha_r$ be the corresponding velocity (under any definition) to an object with this radial motion, then by \eqref{temp72}
\begin{equation}
	\vec v^\alpha_r = \dot D^\alpha_r - \dot D^\alpha_s = \dot D \hat D^\alpha,
\end{equation}
exactly as expected for the radial velocity!

Optical relative velocity is entirely based on the fact that static objects will be visible from a direction that is not rotating in the local Fermi frame, but we will show that the direction of visibility will in general be rotating for radially moving objects in this definition. Let $K^\alpha$ be a past pointing null connecting vector (with normalized affine parametrization), that represents the geodesic on which photons travel to the observer $U^\alpha$ at the event of observation. $K^\alpha$ is decomposed as $K^\alpha=\omega_U(E^\alpha-U^\alpha)$, where $\omega_U=K^\alpha U_\alpha$ and $U^\alpha E_\alpha=0$. We will find the derivative of $E^\alpha$ (the visibility direction of the object) and show that under the given circumstances $E^\alpha$ is rotating in the local Fermi frame. For simplicity, we take the case of relatively low and constant recession speed, so that $\dot D \ll 1$ and $\ddot D = 0$.

From \eqref{kpm}
\begin{multline}
	K^{\alpha}=D^{\alpha}+\bar V^{\alpha}\delta \tau+\frac{1}{2}\bar{\dot V}^{\alpha}\delta \tau^{2}+\frac{1}{6}\bar{\ddot V}^{\alpha}\delta \tau^{3}+\frac{1}{6}R^{\alpha}_{D \bar V D}\delta \tau\\
    -\frac{1}{3}R^{\alpha}_{\bar V D \bar V}\delta \tau^{2}+ ...\,, \label{temp88}
\end{multline}
and so
\begin{multline}
	E^{\alpha}=U^\alpha + \frac{D}{\omega_U}\hat D^{\alpha}+\bar V^{\alpha}\frac{\delta\tau}{\omega_U}+\frac{1}{2}\bar{\dot V}^{\alpha}\frac{\delta\tau}{\omega_U}\delta\tau+\frac{1}{6}\bar{\ddot V}^{\alpha}\frac{\delta\tau}{\omega_U}\delta \tau^{2}\\
    +\frac{1}{6}R^{\alpha}_{D \bar V D}\frac{\delta\tau}{\omega_U} -\frac{1}{3}R^{\alpha}_{\bar V D \bar V}\frac{\delta\tau}{\omega_U}\delta \tau + ...\,, \label{temp4}
\end{multline}
where $\delta\tau$ is negative for the case of past pointing $K^\alpha$. $D$, $\omega_U$, and $\delta\tau$ are all related through the conditions $K^\alpha K_\alpha = 0$ and $D^\alpha U_\alpha = 0$, but we must replace the parallel transports of $V^\alpha$ and its derivatives before applying the conditions. To this end, recall the expression for $\dot D^\alpha$ in terms of $\bar V^\alpha$ given by \eqref{ddot} and its derivation in appendix \ref{app:deri}. If we take the point of view that $D^\alpha(t)$ is given along the trajectory of $U^\alpha$ a priori, and solve for the corresponding $\bar V^\alpha$ (or $V^\alpha$, either from scratch of by flipping \eqref{ddot}), then the result is
\begin{equation}
	\dot\tau\bar V^\alpha = U^\alpha +\dot D^\alpha -\frac{1}{6}\left(3R^\alpha_{DUD}+R^\alpha_{D\dot D D}\right) +\mathcal{O}(D^3)
\end{equation}
where now $\dot\tau$ is set from the requirement that $V^\alpha V_\alpha=-1$, so
\begin{multline}
	\dot \tau^2 = 1+2\dot U^\alpha D_\alpha - |\dot D^\alpha|^2 + R_{UDUD} + \frac{4}{3}R_{\dot DDUD} +\frac{1}{3}R_{\dot DD\dot DD}\\
    +\mathcal{O}(D^3).
\end{multline}
(We have used the fact $\dot D^\alpha U_\alpha= -\dot U^\alpha D_\alpha$, and we must to pay attention to $|\dot D^\alpha|\ne \dot D$.) For the case of interest $\dot D^\alpha = \dot D^\alpha_r$ given by \eqref{tempp86}, and with $D^\alpha_s$ given by \eqref{temp69}, to second order in $D$ and $\dot D$,
\begin{align}
	\dot\tau\bar V^\alpha &= U^\alpha +\dot D^\alpha_r -\frac{1}{2}R^\alpha_{DUD}+\mathcal{O}(D^3, D^2\dot D)\nonumber \\
	&= U^\alpha +\dot D\hat D^\alpha + D^\beta \dot U_\beta U^\alpha - \frac{1}{2}D^2\ddot U^\beta H^\alpha_\beta -\frac{1}{2}R^\alpha_{DUD}\nonumber\\
    &\hspace{4cm}+\mathcal{O}(D^3, D^2\dot D),
\end{align}
\begin{align}
	\dot \tau^2 &= 1+2\dot U^\alpha D_{\alpha} - |\dot D^\alpha_r|^2 + R_{UDUD} + \mathcal{O}(D^3, D^2\dot D)\nonumber \\
	&= 1+2\dot U^\alpha D_{\alpha} - \left( \dot D^2 -(D^\beta \dot U_\beta)^2 \right) + R_{UDUD}\nonumber\\
    &\hspace{4cm}+ \mathcal{O}(D^3, D^2\dot D),
\end{align}
and
\begin{equation}
	\dot \tau = 1+\dot U^\alpha D_\alpha - \frac{1}{2}\dot D^2 + \frac{1}{2}R_{UDUD}+\mathcal{O}(D^3, D^2\dot D).
\end{equation}
Differentiating and making use of \eqref{partransdot}, we get
\begin{equation}
	\ddot \tau = \ddot U^\alpha D_\alpha + \dot U^\alpha \hat D_\alpha\dot D +\mathcal{O}(D^2,D\dot D),
\end{equation}
\begin{equation}
	\bar {\dot V}^\alpha = \dot U^\alpha +\dot D\hat D^\beta\dot U_\beta U^\alpha - D^\beta \dot U_\beta \dot U^\alpha + R^\alpha_{UDU}+ \mathcal{O}(D^2,D\dot D),
\end{equation}
and
\begin{equation}
	\bar {\ddot V}^\alpha = \ddot U^\alpha + \mathcal{O}(D,\dot D).
\end{equation}
Applying the condition $K^\alpha K_\alpha = 0$ to \eqref{temp88} and making use of the above gives
\begin{multline}
	\delta \tau = -D\bigg(1-\dot D+\frac{1}{2}\dot U^\alpha D_\alpha -\frac{1}{8}(\dot U^\alpha D_\alpha)^2+ \frac{1}{2}\dot D^2- \frac{1}{24}\dot U^2D^2\\
    -\frac{1}{6}D\ddot U^\alpha D_\alpha+ \frac{1}{3}R_{UDUD} \bigg) + \mathcal{O}(D,\dot D,{}^4).
\end{multline}
The inner product of \eqref{temp88} with $U^\alpha$ gives a second equation for $\delta \tau$, $D$ and $\omega_U$, which we use together with the above to find
\begin{multline}
	\frac{\delta\tau}{\omega_U} = -1+\frac{1}{2}\dot D^2-\frac{1}{2}\dot DD^\alpha\dot U_\alpha+\frac{1}{6}D^2\dot U^2-\frac{1}{6}R_{UDUD}\\
    + \mathcal{O}(D,\dot D,{}^3),
\end{multline}
and
\begin{multline}
	\frac{D}{\omega_U} = 1-\frac{1}{2}\dot U^\alpha D_\alpha+\dot D + \frac{3}{8}(\dot U^\alpha D_\alpha)^2-\frac{1}{8}D^2\dot U^2 +\frac{1}{6}D\ddot U^\alpha D_\alpha\\
    +\frac{1}{2}\dot D\dot U^\alpha D_\alpha -\frac{1}{6}R_{UDUD} + \mathcal{O}(D,\dot D,{}^3).
\end{multline}
Therefore, by \eqref{temp4} and what is established above
\begin{align}
	E^{\alpha} &= \left( 1-\frac{D^2}{8}\dot U^\gamma\dot U^\beta H_{\gamma\beta}+ \dot D\dot U^\beta D_\beta \right)\hat D^\alpha\nonumber\\
	&\qquad +\bigg( \frac{D}{2}\left(1-\frac{1}{2}\dot U^\gamma D_\gamma-\dot D\right)\dot U^\beta+\frac{D^2}{3}\ddot U^\beta\nonumber\\
    &+\frac{D}{6}R^\beta_{UDU}+\frac{1}{3}R^\beta_{DUD}\bigg)H^\alpha_\beta + \mathcal{O}(D,\dot D,{}^3),
\end{align}
and its derivative is found to be
\begin{equation}
	\dot E^\alpha = E^\beta\dot U_\beta U^\alpha+\frac{1}{2}\dot D \dot U^\beta H^\alpha_\beta+ \mathcal{O}(D,\dot D,{}^2).
\end{equation}
(At this order $H^\alpha_\beta$ can be considered with reference to either $\hat D^\alpha$ or $E^\alpha$.) This shows that the visibility direction of the radially moving object will rotate relative to that of a (optically) static object. The rate of this rotation is given by the second term above. Even for the case of small $\dot D$, we found this undesired outcome at the lowest order. Since the strongest effect (at this order) is also directly dependent on the acceleration of the observer, we can easily envision a process in which this outcome takes place by using Rindler's accelerating frame; but we won't digress.

It is clear that the reason for this flaw is entirely due to the fact that the definition of optical velocity is based on the physical position vector $D^\alpha$, on which we also based how we treated radial motion and defined stationarity. A proper definition of optical velocity must involve an optical position vector instead, which would have the direction of $E^\alpha$ and a magnitude that corresponds to a suitable optical distance. Our analysis thus far reveals that the simplest quantity which can be used as a distance measure instead of $D$ is $\omega_U$. It is naturally optical since it is associated with $K^\alpha$. We see from the construction that the magnitude of the projection of $K^\alpha$ onto the space of $U^\alpha$ represents a length measure of $K^\alpha$ and, therefore, the distance to the object (if under the normalized affine parametrization). This projected magnitude is given by $\omega_U$, since $K^\alpha$ is null. For this we will call $\omega_U$ the projected optical distance to the object, and consider it as the simplest optical distance to combine with the optical direction $E^\alpha$. What we define as $\omega_U$ has been considered as a measure of distance in the literature before. For example, in \cite{bolos6,bolos} the author calls it the \textit{affine distance} (a term that we prefer to use for the Fermi distance which we have here as $D$), also see \cite{kermack,ellis}. However, at this point we cannot consider $\omega_U$ as measurable, as opposing to $D$; $\omega_U$ is a mathematical parameter that can be calculated from the events involved, while $D$ is a geodesic distance that can be measured directly within the Fermi frame. In the following section we discuss in full detail the well known optical distances based on the angular size and the luminosity of a distant object, and develop general expressions for them in terms of the fundamental parameters of any setup. We will then use our findings to properly define an optical velocity that could be determined through direct optical measurements by an observer, that is suitable for distant objects on which only optical measurements can be made.

\subsection{\label{sec:optdis}Optical Distances}

\subsubsection{\label{sec:fund}Fundamentals}

\begin{figure}
  \centering
  \includegraphics[width=\columnwidth]{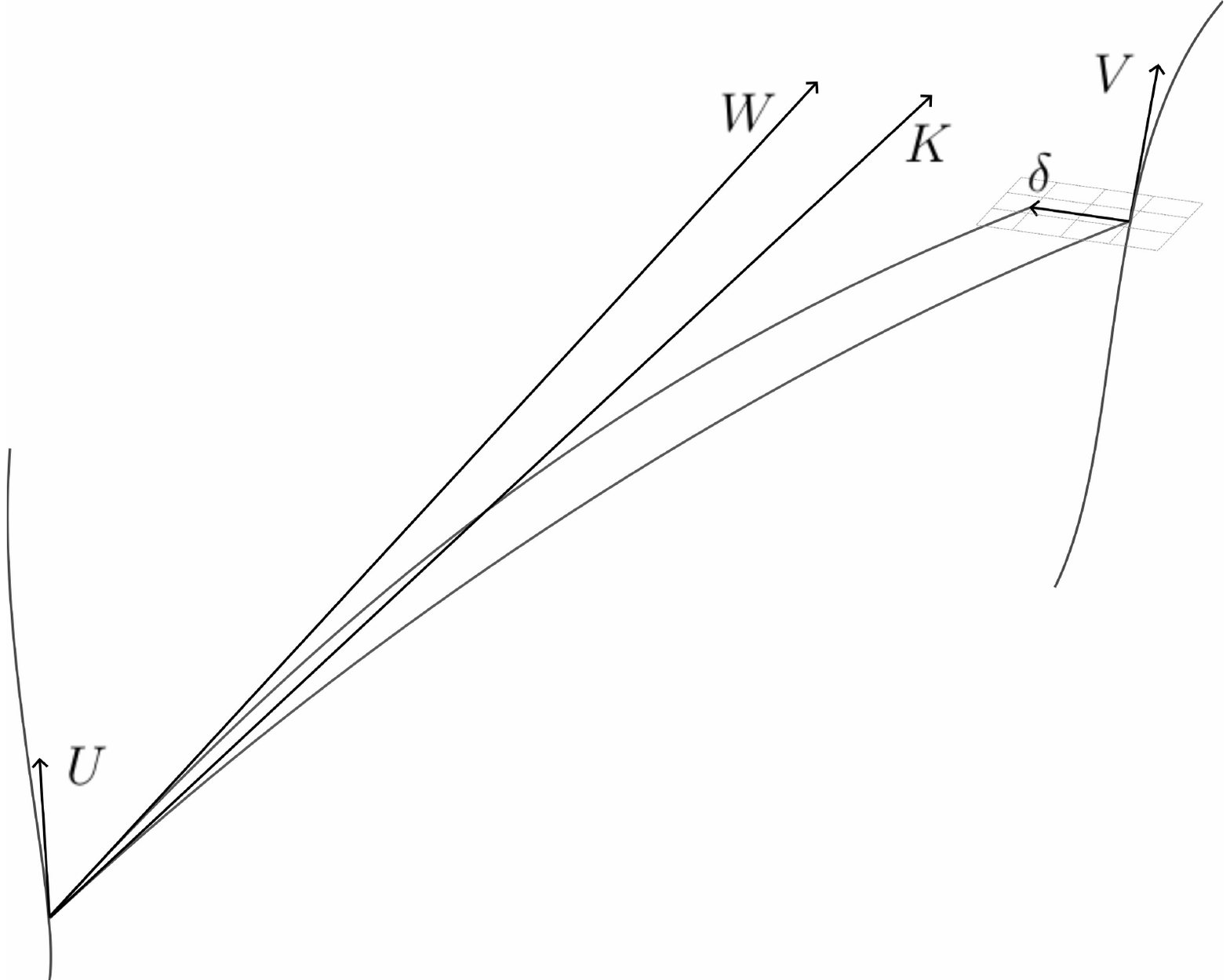}
  \caption{Two neighboring null geodesics starting a an event on the timelike curve with 4-velocity $U^\alpha$ and terminating at two nearby events within the same simultaneity slice of an observer with 4-velocity $V^\alpha$. $K$ and $W$ are tangents to the null geodesics at the event of $U^\alpha$, and $\delta$ is tangent to the spacelike geodesic connecting the events within the frame of $V^\alpha$. All three geodesics are parametrized with normalized affine parametrization, as depicted by the length of their tangents on the diagram.} \label{fig:delta}
\end{figure}

Consider the setup of Figure \ref{fig:delta}, where $K^\alpha$ and $W^\alpha = K^\alpha + \delta K^\alpha$ are tangents to null geodesics with the normalized affine parametrization, connecting to two neighboring events that are in the same simultaneity subspace of some observer $V^\alpha$, and are a short distance apart. With $\delta^\alpha$ being the tangent of the spacelike geodesic connecting the two events with the normalized affine parametrization, and $\bar \delta^\alpha$ its parallel transport to the event of $K^\alpha$, by equation \eqref{omegaoflambdau} in appendix \ref{app:deri},
\begin{align}
	\bar \delta^\alpha &=W^\alpha- K^\alpha+\frac{1}{3}R^\alpha_{WKW} + \frac{1}{6}R^\alpha_{KWK}+...\nonumber \\
	&= \delta K^\alpha-\frac{1}{6}R^\alpha_{K\delta K K} + \frac{1}{3}R^\alpha_{\delta K K \delta K}+... \label{temp86}
\end{align}
While $K^\alpha$ is null, its smallness is given by $\omega_U = -K^\alpha U_\alpha$ for some observer $U^\alpha$. $\delta K^\alpha$ is guaranteed to be spacelike in this construction, and the above expression is accurate to $...+\mathcal{O}(\omega_U^3,|\delta K|^3)$.

In this setup, for a given null vector $K^\alpha$ that connects to the event of $V^\alpha$, the vector $\delta K^\alpha$ has $n-2$ free parameters that fully describe it, where $n$ is the dimension of the manifold. $\delta K^\alpha$ is subject to the following two constraints:
\begin{gather}
	\left( K^\alpha +\delta K^\alpha \right) \left( K_\alpha +\delta K_\alpha \right) = 0,\\
	\implies |\delta K|^2 = -2 K^\alpha \delta K_\alpha, \label{temp89}
\end{gather}
since $K^\alpha$ and $W^\alpha$ are null; and
\begin{equation}
	\delta K^\alpha\bar V_\alpha = \frac{1}{6}R_{\bar V K\delta K K} - \frac{1}{3}R_{\bar V \delta K K \delta K}+\mathcal{O}(\omega_U^3,|\delta K|^3), \label{vcond}
\end{equation}
since $\bar V^\alpha \bar \delta_\alpha = V^\alpha \delta_\alpha = 0$, where $\bar V^\alpha$ is the parallel transport of $V^\alpha$ along the null geodesic. The last constraint means that the tangent $W^\alpha$ of the neighbouring photon must be set so that it leads to an event in the same simultaneity subspace with respect to $V^\alpha$.

Let $\theta_U$ be the measurable angle between $K^\alpha$ and $W^\alpha$ by the observer $U^\alpha$. Then by \eqref{angleformula}
\begin{align}
	\cos(\theta_U) &= 1+\frac{K^\alpha W_\alpha}{(U^\alpha K_\alpha)(U^\alpha W_\alpha)}\nonumber\\
	&= 1+\frac{K^\alpha(K_\alpha+\delta K_\alpha)}{(U^\alpha K_\alpha)(U^\alpha K_\alpha +U^\alpha \delta K_\alpha)}, \label{temp22}
\end{align}
and we get
\begin{equation}
	|\delta K|^2 = -2 K^\alpha \delta K_\alpha = \omega_U^2 \theta_U^2 \left(1-\frac{\delta K^\alpha U_\alpha}{\omega_U}\right) +\mathcal{O}(\theta_U^4). \label{temp23}
\end{equation}
Working in four dimensions, we can regard $\theta_U$ as one of the two free parameters that determine $\delta K^\alpha$; it has a clear physical interpretation. $\delta K^\alpha$ can be decomposed as follows:
\begin{equation}
	\delta K^\alpha = a U^\alpha + k K^\alpha + b \hat B^\alpha,
\end{equation}
where $\hat B^\alpha$ is a unit spacelike vector perpendicular to both $K^\alpha$ and $U^\alpha$. The scalar coefficients $a,k,b$ are determined from the inner products $\delta K^\alpha K_\alpha$, $\delta K^\alpha U_\alpha$, and $\delta K^\alpha \delta K_\alpha$. We get
\begin{gather}
	a=\frac{1}{2}\omega_U\theta_U^2\left(1-\frac{\delta K^\alpha U_\alpha}{\omega_U}\right) +\mathcal{O}(\theta_U^3),\\
	k = -\frac{\delta K^\alpha U_\alpha}{\omega_U}-\frac{1}{2}\theta_U^2\left(1-\frac{\delta K^\alpha U_\alpha}{\omega_U}\right)+\mathcal{O}(\theta_U^3),\\
	b = \omega_U\theta_U\left(1-\frac{\delta K^\alpha U_\alpha}{\omega_U}\right)+\mathcal{O}(\theta_U^3),
\end{gather}
and therefore
\begin{align}
	\nonumber	\delta K^\alpha &= \frac{1}{2}\omega_U\theta_U^2\left(1-\frac{\delta K^\alpha U_\alpha}{\omega_U}\right) U^\alpha \\
    &\hspace{1.5cm}-\left(\frac{\delta K^\alpha U_\alpha}{\omega_U}+\frac{1}{2}\theta_U^2\left(1-\frac{\delta K^\alpha U_\alpha}{\omega_U}\right)\right) K^\alpha\nonumber\\
	&\hspace{1cm} + \omega_U\theta_U\left(1-\frac{\delta K^\alpha U_\alpha}{\omega_U}\right) \hat B^\alpha+\mathcal{O}(\theta_U^3) \label{deltak} \\
	&= -\frac{\delta K^\alpha U_\alpha}{\omega_U} K^\alpha + \omega_U\theta_U\left(1-\frac{\delta K^\alpha U_\alpha}{\omega_U}\right) \hat B^\alpha\nonumber\\
    &\hspace{5cm}+\mathcal{O}(\theta_U^2). \label{deltak2}
\end{align}
When expressed in this form we see that for given $K^\alpha$ and $U^\alpha$, $\delta K^\alpha$ is fully determined by the measurable angle $\theta_U$ and the unit vector $\hat B^\alpha$ within the space of $U^\alpha$ and perpendicular to $K^\alpha$. For a choice of angle $\theta_U$ and orientation $\hat B^\alpha$, the quantity $\frac{\delta K^\alpha U_\alpha}{\omega_U}$ is determined from the second constraint on $\delta K^\alpha$, $\bar V^\alpha \bar \delta_\alpha = V^\alpha \delta_\alpha = 0$. This turns out to be the only place where the vector $V^\alpha$ enters the expression for $\delta K^\alpha$. Thus, for given $K^\alpha$ and $U^\alpha$, $\theta_U$ and $\hat B^\alpha$ determine the outgoing direction and therefore the full path of the neighboring photon to $K^\alpha$, the quantity $\frac{\delta K^\alpha U_\alpha}{\omega_U}$, which depends on $V^\alpha$, is only involved in setting the particular event on the neighboring null geodesic that happens to be within the same simultaneity slice with respect to observer $V^\alpha$.

Before applying the second constraint on $\delta K^\alpha$ to get an expression for $\frac{\delta K^\alpha U_\alpha}{\omega_U}$, we note the following. First, clearly $\delta K^\alpha \rightarrow 0$ as $\theta_U \rightarrow 0$, so we expect $\delta K^\alpha U_\alpha$ to be small and proportional to $\theta_U$. Second, the distance between the neighboring events in the space of $V^\alpha$ is given by
\begin{widetext}
\begin{align}
	|\delta| &= |\bar \delta| = \sqrt{|\delta K|^2-\frac{1}{3}R_{\delta K K \delta K K}+\mathcal{O}(\omega_U^5,\theta_U^3)}\nonumber \\
	&= \sqrt{\omega_U^2\theta_U^2\left(1-\frac{\delta K^\alpha U_\alpha}{\omega_U}\right)-\frac{1}{3}\omega_U^2\theta_U^2\left(1-\frac{\delta K^\alpha U_\alpha}{\omega_U}\right)^2R_{\hat B K \hat B K}+\mathcal{O}(\omega_U^5,\theta_U^3)}\nonumber \\
	&=\omega_U\theta_U\sqrt{\left(1-\frac{\delta K^\alpha U_\alpha}{\omega_U}\right)}\left(1-\frac{1}{6}\left(1-\frac{\delta K^\alpha U_\alpha}{\omega_U}\right)R_{\hat B K \hat B K}\right) + \mathcal{O}(\omega_U^4,\theta_U^2) \label{temp102} \\
	&= \omega_U\theta_U\sqrt{\left(1-\frac{\delta K^\alpha U_\alpha}{\omega_U}\right)} + \mathcal{O}(\omega_U^3,\theta_U^2), \label{temp103}
\end{align}
\end{widetext}
and its dependence on $V^\alpha$ comes only through the $\frac{\delta K^\alpha U_\alpha}{\omega_U}$ term. It is also well known (\cite{sef,ellis2,sachs}) that the distance between very close photon trajectories (such as members of a narrow beam) is independent of the motion of the observer that measures it, which further points to the smallness of $\frac{\delta K^\alpha U_\alpha}{\omega_U}$. Of course, if the small angle $\theta_U$ is considered infinitesimal, then no matter the value of $\frac{\delta K^\alpha U_\alpha}{\omega_U}$ (which depends on $V^\alpha$), $\theta_U$ can always be made smaller to make $\frac{\delta K^\alpha U_\alpha}{\omega_U} \ll 1$, since $\delta K^\alpha \rightarrow 0$ as $\theta_U \rightarrow 0$. However, if $\theta_U$ is considered very small but finite, then simple analysis reveals that even in flat space the range of $\frac{\delta K^\alpha U_\alpha}{\omega_U}$ is $(-\infty,1)$ and it cannot be neglected without some assumptions on $V^\alpha$. Merely assuming that $|\delta| \ll 1$ (narrow beam) is clearly insufficient to justify the removal of $\frac{\delta K^\alpha U_\alpha}{\omega_U}$ from the above expression. In appendix \ref{app:condistance} we analyze this situation and derive the required condition on $V^\alpha$ under which we can justifiably conclude that $\frac{\delta K^\alpha U_\alpha}{\omega_U} \ll 1$ and can be omitted from the distance expression.

It follows that as long as
\begin{equation}
	\frac{\frac{1}{2}\theta_U^2 E^\alpha \bar V_\alpha-\theta_U\hat B^\alpha \bar V_\alpha}{\gamma- E^\alpha \bar V_\alpha} \ll 1 \label{temp104}
\end{equation}
(which is not guaranteed), where $K^\alpha = \omega_U\left(U^\alpha+ E^\alpha\right)$ and $\gamma = -U^\alpha\bar V_\alpha$, to lowest order the condition $V^\alpha\delta_\alpha=0$ gives
\begin{align}
	\frac{\delta K^\alpha U_\alpha}{\omega_U}&=-\theta_U\frac{\hat B^\alpha \bar V_\alpha}{\gamma- E^\alpha \bar V_\alpha}+\mathcal{O}(\omega_U^2,\theta_U^2)\nonumber \\
	&=-\theta_U\frac{\omega_U}{\omega_V}\hat B^\alpha \bar V_\alpha+\mathcal{O}(\omega_U^2,\theta_U^2), \label{temp33}
\end{align}
where $\omega_V = -K^\alpha \bar V_\alpha = \omega_U\left( \gamma- E^\alpha \bar V_\alpha\right)$, see equation \eqref{deltaku}. With the above expression for $\frac{\delta K^\alpha U_\alpha}{\omega_U}$, we have 
\begin{align}
	\delta K^\alpha &= \theta_U\frac{\omega_U}{\omega_V}\hat B^\beta \bar V_\beta K^\alpha + \omega_U\theta_U \hat B^\alpha+\mathcal{O}(\omega_U^2,\theta_U^2)\nonumber \\
	&=\omega_U\theta_U\left(\frac{1}{\omega_V}\bar V_\beta K^\alpha+\delta^\alpha_\beta\right)\hat B^\beta+\mathcal{O}(\omega_U^2,\theta_U^2), \label{temp108}
\end{align}
\begin{align}
	\bar{\delta}^\alpha &= \delta K^\alpha-\frac{1}{6}R^\alpha_{K\delta K K} + \frac{1}{3}R^\alpha_{\delta K K \delta K}+\mathcal{O}(\omega_U^4,\theta_U^3)\\
	&= \delta K^\alpha-\frac{1}{6}R^\alpha_{K\delta K K}+\mathcal{O}(\omega_U^4,\theta_U^2)\nonumber \\
	&= \left(\delta^\alpha_\beta-\frac{1}{6}R^\alpha_{K\beta K}\right) \delta K^\beta+\mathcal{O}(\omega_U^4,\theta_U^2), \label{temp111}
\end{align}
and
\begin{equation}
	|\delta| = \omega_U\theta_U\left(1-\frac{1}{6}R_{\hat BK\hat BK}\right)+\mathcal{O}(\omega_U^4,\theta_U^2).
\end{equation}
Notice how $V^\alpha$ affects both $\delta K^\alpha$ and $\bar{\delta}^\alpha$, while $|\delta|$ is independent of $V^\alpha$ to first order in $\theta_U$ (as long as \eqref{temp104} is satisfied). From here on we will work only with the lowest order in $\theta_U$.

Thus, for a null trajectory given by $K^\alpha$, a neighboring null path is fully determined by $\theta_U$ and $\hat B^\alpha$, and the distance between the paths can be considered observer independent at any event. In four dimensional spacetime, the unit vector $\hat B^\alpha$ belongs to a two dimensional plane, so $\theta_U$ and $\hat B^\alpha$ is a two parameter pair, which can be considered as a vector in this plane, that determine $\delta K^\alpha$ and consequently $\bar\delta^\alpha$. Therefore we have a map form a two dimensional space orthogonal to $U^\alpha$ and $K^\alpha$ to another two dimensional space that is orthogonal to $\bar V^\alpha$ and $K^\alpha$. The goal now is to establish the relationship between a solid angle $d\Omega_U$, measured by observer $U^\alpha$ of a diverging narrow photon beam, and the corresponding (observer independent) cross-sectional area $dA$ of the beam at a distant event. This will allow us to produce expressions for the angular diameter distance and the luminosity distance, which we can use in defining optical velocity.

\subsubsection{\label{sec:solid}Cross-sectional Areas and Solid angles}

Consider a pair of vectors $\theta_1\hat B^\alpha_1$ and $\theta_2\hat B^\alpha_2$ that span the domain two dimensional plane, and the corresponding $\delta^\alpha_1$ and $\delta^\alpha_2$ that span the range two dimensional plane. The areas generated by each pair and the local metric can be related through the transformations derived above. While the area generated by $\delta^\alpha_1$ and $\delta^\alpha_2$ is physical and directly measurable, the area generated by $\theta_1\hat B^\alpha_1$ and $\theta_2\hat B^\alpha_2$ is mathematical, but related to the measurable solid angle opening. Once we establish the connection between the two areas, we will define the measurable solid angle from first principles without reference to any coordinates, and relate it to the mathematical area generated by $\theta_1\hat B^\alpha_1$ and $\theta_2\hat B^\alpha_2$.

The transformations given by equations \eqref{temp108} and \eqref{temp111} can be combined into one, thereby removing any reference to $\delta K^\alpha$. However we would like to avoid doing this due to the fact that the transformation $\theta_U\hat B^\alpha \rightarrow \delta K^\alpha$ given by \eqref{temp108} is singular. As we shall see, the existence of a Jacobian and an inverse make the derivation and the final relationship between the areas neat and clear. So instead we will first relate the areas corresponding to $(\theta_1\hat B^\alpha_1,\theta_2\hat B^\alpha_2)$ and $(\delta K^\alpha_1,\delta K^\alpha_2)$, which can be done in a simple way, and then make use of the non-singular properties of \eqref{temp111} to establish the more sophisticated connection between the areas corresponding to $(\delta K^\alpha_1,\delta K^\alpha_2)$ and $(\delta^\alpha_1,\delta^\alpha_2)$. For $i,j = 1,2$,
\begin{align}
	&g_{\alpha \beta}\delta K^\alpha_i \delta K^\beta_j\nonumber\\
    &= \!g_{\alpha \beta}\omega_U\theta_i\left(\!\frac{1}{\omega_V}\bar V_\gamma K^\alpha\!+\!\delta^\alpha_\gamma\!\right)\hat B^\gamma_i \omega_U\theta_j\left(\!\frac{1}{\omega_V}\bar V_\epsilon K^\beta\!+\!\delta^\beta_\epsilon\!\right)\hat B^\epsilon_j\nonumber\\
    &\hspace{6cm}+\mathcal{O}(\omega_U^3,\theta_U^3)\nonumber \\
	&= \omega_U^2g_{\alpha \beta}\left(\theta_i\hat B^\alpha_i\right)\left(\theta_j\hat B^\beta_j\right)+\mathcal{O}(\omega_U^3,\theta_U^3), \label{temp113}
\end{align}
(where $\theta_U$ in $\mathcal{O}(\omega_U^3,\theta_U^3)$ stands for either $\theta_i$ or $\theta_j$) and since these inner products are just related by a scalar, it is a very simple matter to relate the corresponding areas. For a collection of linearly independent vectors in a metric space, the subvolume (of dimension equal or smaller than the space) that corresponds to these vectors is easily obtained through taking the determinant of the metric confined to the subspace and expressed in terms of the given vectors. (Or in different words, through taking the determinant of the matrix generated from the inner products of all the participating vectors). In particular, for two vectors $X^\alpha$ and $Y^\alpha$, the corresponding area, $Area(X^\alpha,Y^\alpha)$, is given by
\begin{equation}
	Area(X^\alpha,Y^\alpha) = \sqrt{det(g,X,Y)} = \sqrt{g_{XX}g_{YY}-g_{XY}^2}.
\end{equation}
And since
\begin{align}
	&det(g,\delta K_1,\delta K_2) = g_{\delta K_1\delta K_1}g_{\delta K_2\delta K_2}-g_{\delta K_1\delta K_2}^2\nonumber\\
	&\hspace{1cm}= \omega_U^4\left(\theta_1^2g_{\hat B_1\hat B_1}\theta_2^2g_{\hat B_2\hat B_2}-\theta_1^2\theta_2^2g_{\hat B_1\hat B_2}^2\right)\!+\!\mathcal{O}(\omega_U^5,\theta_U^5)\nonumber \\
	&\hspace{1cm}= \omega_U^4 det(g,\theta_1\hat B_1,\theta_2\hat B_2)+\mathcal{O}(\omega_U^5,\theta_U^5),
\end{align}
we have
\begin{align}
	Area(\delta K^\alpha_1,\delta K^\alpha_2) &= \sqrt{det(g,\delta K_1,\delta K_2)}\nonumber\\
	&= \omega_U^2\sqrt{det(g,\theta_1\hat B_1,\theta_2\hat B_2)+\mathcal{O}(\omega_U,\theta_U^5)}\nonumber\\
	&=\omega_U^2Area(\theta_1\hat B^\alpha_1,\theta_2\hat B^\alpha_2)+\mathcal{O}(\omega_U^3,\theta_U^3). \label{temp121}
\end{align}

The relationship between $Area(\bar\delta^\alpha_1,\bar\delta^\alpha_2)$ and $Area(\delta K^\alpha_1,\delta K^\alpha_2)$ is not as easy to establish due to the fact that $g_{\alpha\beta}\bar\delta^\alpha_i\bar\delta^\beta_j$ is not simply related to $g_{\alpha\beta}\delta K^\alpha_i\delta K^\beta_j$ by a scalar as above, but will now depend on orientation as well. The transformation of the $\delta K^\alpha_i$'s to the $\bar\delta^\alpha_i$'s is given by \eqref{temp111}, but the $\delta K^\alpha_i$'s are confined to a two dimensional subspace, so we cannot simply use the Jacobian of the transformation to relate the subvolumes. Also, while the $\hat B^\alpha_i$'s are perpendicular to $U^\alpha$ and $K^\alpha$, the $\delta K^\alpha_i$'s are perpendicular to $\bar V^\alpha-\frac{1}{6}R^\alpha_{K\bar VK}$ and $K^\alpha$ (to first order in $\theta_i$'s), due to the restrictions given by \eqref{temp89} and \eqref{vcond}. With this in mind, the general problem can be stated as follows: given a transformation from an event in a metric space to another event in a metric space of the same dimension, together with one or more vectors that define an orthogonal subspace with respect to the metric at the original event, what is the relationship between a subvolume in the domain and that of its image under the transformation and the given restrictions? Clearly this relationship must depend not only on the transformation itself (which usually comes through the Jacobian) but also on the vectors that confine the transformation to a subspace. We dedicate appendix \ref{app:sub} to address this issue in detail. For the case at hand we have (see equation \eqref{temp322})
\begin{multline}
	\frac{det(g,\bar{\delta_1},\bar{\delta_2})}{det(g,\delta K_1,\delta K_2)} = \frac{1}{2} \frac{J^2}{det(g,K,L)} g^{\rho\mu} g^{\nu\epsilon} T^{-1}\,^\alpha_\rho T^{-1}\,^\gamma_\mu \\
    \times T^{-1}\,^\beta_\nu T^{-1}\,^\lambda_\epsilon (K_\alpha L_\beta - K_\beta L_\alpha)(K_\gamma L_\lambda - K_\lambda L_\gamma), \label{temp122}
\end{multline}
where $T^\alpha_\beta$ is the transformation, $K^\alpha$ and $L^\alpha$ are vectors that restrict the subspace, and $J$ is the Jacobian of the transformation, the determinant of $T^\alpha_\beta$. The above result is general, but our transformation $T^\alpha_\beta$ and the vector $L^\alpha$ are only accurate to second order in $\omega_U$. In particular, $T^\alpha_\beta =\delta^\alpha_\beta - \frac{1}{6}R^\alpha_{K\beta K}+\mathcal{O}(\omega_U^3,\theta_U)$, $L^\alpha=\bar V^\alpha-\frac{1}{6}R^\alpha_{K\bar VK}+\mathcal{O}(\omega_U^3,\theta_U)$, and $T^{-1}\,^\alpha_\beta = \delta^\alpha_\beta + \frac{1}{6}R^\alpha_{K\beta K}+\mathcal{O}(\omega_U^3,\theta_U)$. For this specific situation, we first find that
\begin{align}
	\nonumber g^{\rho\mu} & g^{\nu\epsilon} T^{-1}\,^\alpha_\rho T^{-1}\,^\gamma_\mu T^{-1}\,^\beta_\nu T^{-1}\,^\lambda_\epsilon \\
    &\hspace{2cm}\times(K_\alpha L_\beta - K_\beta L_\alpha)(K_\gamma L_\lambda - K_\lambda L_\gamma)\nonumber\\
	\nonumber	&\hspace{0.2cm}=\left(g^{\alpha\gamma} +\frac{1}{6}g^{\alpha\rho}R^\gamma_{K\rho K}+\frac{1}{6}g^{\gamma\rho}R^\alpha_{K\rho K}\right) \\
    &\hspace{1.5cm}\times \left(g^{\beta\lambda} +\frac{1}{6}g^{\beta\rho}R^\lambda_{K\rho K}+\frac{1}{6}g^{\lambda\rho}R^\beta_{K\rho K}\right)\nonumber\\
	&\hspace{0.5cm} \times\left(K_\alpha L_\beta - K_\beta L_\alpha\right)\left(K_\gamma L_\lambda - K_\lambda L_\gamma\right) + \mathcal{O}(\omega_U^5,\theta_U) \label{temp123} \\
	&= g_{KK}g_{LL}-g_{LK}g_{KL}-g_{KL}g_{LK}+g_{LL}g_{KK} \nonumber\\
    &\hspace{2cm}+ \frac{2}{3}g_{KK}R_{LKLK} + \mathcal{O}(\omega_U^5,\theta_U) \label{temp124} \\
	&=2det(g,K,L) + \mathcal{O}(\omega_U^5,\theta_U), \label{temp125}
\end{align}
so we get
\begin{align}
	&\frac{det(g,\bar{\delta_1},\bar{\delta_2})}{det(g,\delta K_1,\delta K_2)} \nonumber\\
    &\hspace{1.5cm}= \frac{1}{2} \frac{J^2}{det(g,K,L)} \left(2det(g,K,L) + \mathcal{O}(\omega_U^5,\theta_U)\right)\nonumber\\
	&\hspace{1.5cm}=J^2 + \mathcal{O}(\omega_U^3,\theta_U).
\end{align}
And therefore,
\begin{equation}
	\frac{Area(\bar \delta^\alpha_1,\bar \delta^\alpha_2)}{Area(\delta K^\alpha_1,\delta K^\alpha_2)} = J + \mathcal{O}(\omega_U^3,\theta_U). \label{temp128}
\end{equation}
Interestingly, for this particular case where $K^\alpha$ is null and also a part of the transformation itself, the relationship between the areas reduces back to the Jacobian of the transformation, which is by no means necessary. We notice how in going from \eqref{temp123} to \eqref{temp124} above, all the curvature terms either become zero due to the antisymmetry of the Riemann tensor, or get multiplied by $g_{KK}$, which rids the relationship from the extra terms whenever $K^\alpha$ is null. As we shall see, increasing the accuracy of $T^\alpha_\beta$ to higher orders of $\omega_U$ (with more curvature related terms together with more $K^\alpha$'s) will still lead to the same disappearance of extra terms, and only leave the Jacobian present.

In appendix \ref{app:det} we find an expression for the determinant of a matrix given by $M^\alpha_\beta=\delta^\alpha_\beta + \delta M^\alpha_\beta$, where $\delta M^\alpha_\beta \ll 1$. The relevant result is (see equation \eqref{temp276})
\begin{multline}
	det(M_{(\beta)}^{(\alpha)})=1+\delta M_{\alpha}^{\alpha}+\frac{1}{2}(\delta M_{\alpha}^{\alpha})(\delta M_{\beta}^{\beta})\\
    -\frac{1}{2}(\delta M_{\alpha}^{\beta})(\delta M_{\beta}^{\alpha})+...
\end{multline}
For our transformation $T^\alpha_\beta =\delta^\alpha_\beta - \frac{1}{6}R^\alpha_{K\beta K} + \mathcal{O}(\omega_U^3,\theta_U)$,
\begin{equation}
	det(T_{(\beta)}^{(\alpha)})=1-\frac{1}{6}R^\alpha_{K\alpha K}+ \mathcal{O}(\omega_U^3,\theta_U). \label{temp130}
\end{equation}
At this point we cannot make use of the higher order terms in the determinant that come from $\frac{1}{2}(\delta M_{\alpha}^{\alpha})(\delta M_{\beta}^{\beta})-\frac{1}{2}(\delta M_{\alpha}^{\beta})(\delta M_{\beta}^{\alpha})$ in the above expression, since it goes beyond the accuracy of $T^\alpha_\beta$ itself, and consequently will give an incorrect result. Thus, we find
\begin{equation}
	J=1-\frac{1}{6}R_{KK}+ \mathcal{O}(\omega_U^3,\theta_U), \label{temp131}
\end{equation}
where $R_{\alpha\beta}=R^\rho_{\alpha\rho\beta}$ is the Ricci curvature tensor. It is worth noting that in empty space with or without the cosmological constant $\Lambda$, the above expression for the Jacobian is just $1$ to the given order of accuracy (for this reason we later refine it further to expose curvature effects in vacuum).

With the above results we proceed to define the measurable solid angle by an observer with 4-velocity vector $U^\alpha$ from first principles. Clearly $dA=Area(\delta^\alpha_1,\delta^\alpha_2)=Area(\bar \delta^\alpha_1,\bar \delta^\alpha_2)$, since the inner products involved in determining the area are preserved under parallel transport. We may think of $dA$ as the cross-sectional area of the narrow photon beam, which can be considered observer independent. The event along the beam where $dA$ is measured will be referred to as the screen in the following definition. Accordingly, the measurable solid angle of a diverging narrow beam of photons is defined as the ratio of the cross-sectional area of the beam divided by the square of the distance from the event of origin to the screen, in the limit that the distance goes to zero. That is
\begin{equation}
	d\Omega_U = \lim\limits_{(distance) \to 0}\frac{dA}{(distance)^2}, \label{temp61}
\end{equation}
where we still need to make clear the type of distance measure to be used in the above definition. Among such possible distances there is the physical (Fermi) distance to the screen at the event of emission measured in the frame of $U^\alpha$, as well as the physical distance to the screen at the event of reception measured in the frame of $U^\alpha$. In addition we also have the non-measurable projected optical distance $\omega_U$. Conveniently, these distances are equivalent in the limit, and all go to zero together and at the same rate. For example, the relationship between the physical distance in the frame of $U^\alpha$ at the time of reception $D$ and $\omega_U$ (these two distances make the most sense to use here) is found to be
\begin{equation}
	D=\omega_U\left(1-\frac{1}{2}\dot U^\alpha K_\alpha\right)+\mathcal{O}(\omega_U^3),
\end{equation}
which is established in appendix \ref{app:deri} (see equation \eqref{temp67}). So we may as well use $\omega_U$ as the distance in the above definition for the current situation. Therefore, by \eqref{temp128}, \eqref{temp131}, and \eqref{temp121}
\begin{align}
	d\Omega_U &= \lim\limits_{\omega_U \to 0}\frac{dA}{\omega_U^2}\nonumber\\
	&= \lim\limits_{\omega_U \to 0}\frac{J+\mathcal{O}(\omega_U^3,\theta_U)}{\omega_U^2}Area(\delta K^\alpha_1,\delta K^\alpha_2)\nonumber\\
	&= \lim\limits_{\omega_U \to 0}\frac{1-\frac{1}{6}R_{KK}+\mathcal{O}(\omega_U^3,\theta_U)}{\omega_U^2}\nonumber\\
    &\hspace{1.5cm}\times\left(\omega_U^2 Area(\theta_1 \hat B^\alpha_1,\theta_2 \hat B^\alpha_2)+\mathcal{O}(\omega_U^3,\theta_U^3)\right)\nonumber\\
	&= Area(\theta_1 \hat B^\alpha_1,\theta_2 \hat B^\alpha_2) + \mathcal{O}(\theta_U^3). \label{temp136}
\end{align}
This gives us a clear interpretation of the mathematical area $Area(\theta_1 \hat B^\alpha_1,\theta_2 \hat B^\alpha_2)$, it is the measurable solid angle by the observer $U^\alpha$. It is clear that the definition of solid angles above can be extended to narrow beams that converge to the observer as well. With this association, \eqref{temp128} and \eqref{temp121} give
\begin{equation}
	\frac{dA}{\omega_U^2 d\Omega_U}=J+\mathcal{O}(\omega_U^3,\theta_U), \label{temp138}
\end{equation}
and finally
\begin{equation}
	\frac{dA}{d\Omega_U}=\omega_U^2 \left(1-\frac{1}{6}R_{KK}\right)+\mathcal{O}(\omega_U^5,\theta_U). \label{temp139}
\end{equation}
This is the relationship that we were seeking. It will be used to establish expressions for the angular diameter distance and the luminosity distance. Additionally, notice that we immediately get the solid angle aberration relationship from \eqref{temp138}. Indeed, as made clear through the derivation, the right hand side of \eqref{temp138} would not depend on the vector $U^\alpha$ at any level of accuracy. This means that for any other observer $U'^\alpha$ at the event of emission we get
\begin{equation}
	\frac{dA}{\omega_U^2 d\Omega_U}=\frac{dA}{\omega_{U'}^2 d\Omega_{U'}}, \label{temp140}
\end{equation}
which gives
\begin{equation}
	\frac{d\Omega_{U'}}{d\Omega_U}=\frac{\omega_U^2}{\omega_{U'}^2} = \frac{(K^\alpha U_\alpha)^2}{(K^\alpha U'_\alpha)^2}, \label{temp141}
\end{equation}
as is well known in the literature, \cite{sef}. Furthermore, while it is rarely stated explicitly, the above aberration relationship holds only for observers that are not in extreme relative motion when the solid angles are not infinitesimal. The precise condition under which the above is true is derived in appendix \ref{app:relab} and is the same for small (non-solid) angles. Of course, if the solid angles are considered infinitesimal, then they can always be adjusted to ensure the aberration relationship holds, but with small and finite solid angles the extra condition must be satisfied. In the present derivation of the above relationship, the restriction on the relative motion between $U^\alpha$ and $U'^\alpha$ comes through the condition on $\bar V^\alpha$, \eqref{temp104}, under which $dA$ can be considered observer independent, and consequently canceled out in going from \eqref{temp140} to \eqref{temp141}.

\subsubsection{\label{sec:exp}Expressing Optical Distances}

We are finally ready to turn attention to the angular diameter distance and the luminosity distance, beginning with a brief review of the concepts. These well known optical distances are determined from measurements of angular sizes and brightnesses of distant sources. They are closely related by the reciprocity relation (see \cite{ether,ellis2,sef}), which we will state and verify. The reciprocity theorem is true for any spacetime and theory of gravity, but its derivation conveniently avoids the need for general expressions of the two distances involved by exploiting the antisymmetries of the Riemann tensor. As far as we are aware, there is no explicit general expression for either of these distances in the literature. We will present such expressions and ensure that they satisfy the reciprocity relation.

\begin{figure}
  \centering
  \includegraphics[width=\columnwidth]{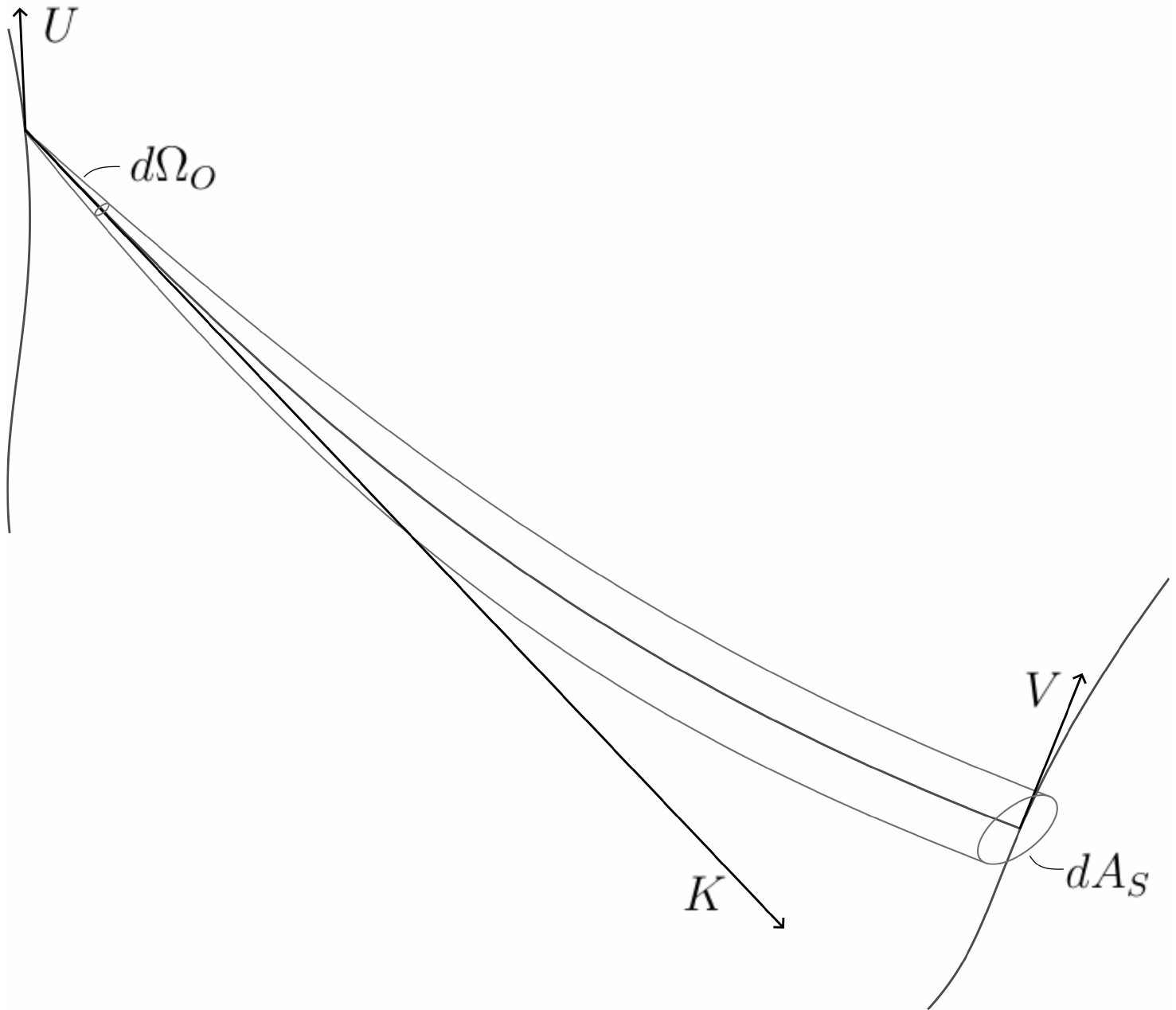}
  \caption{A narrow beam of light converging to an event on the timelike curve with 4-velocity $U^\alpha$. $d\Omega_O$ is the solid angle measured by the observer $U^\alpha$ and $dA_S$ is the corresponding cross-sectional area of the beam as measured by the source with 4-velocity $V^\alpha$. If $dA_S$ is the `size' of the object, then $d\Omega_O$ is its observed `angular size'. The main null geodesic that connects the source event to the observer is represented by its tangent $K^\alpha$ at the observer event; it fully determines the geodesic path under the normalized affine parametrization.} \label{fig:ad}
\end{figure}

Consider the setup in Figure \ref{fig:ad} and the given definitions. Now the vector $K^\alpha$ is past pointing, and the emitter of the light signal is the observer with 4-velocity $V^\alpha$, to which we will refer as the source. The observer with 4-velocity $U^\alpha$ receives the signal and makes measurements in order to establish the two optical distances. The reasoning behind the angular diameter distance $D_{A}$ is the following. For a source of cross-sectional area $dA_S$, the size of the corresponding solid angle image $d\Omega_O$ that arrives at an observer will depend on how far the source is. In analogy with flat space, the angular diameter distance is defined as
\begin{equation}
	D_A = \sqrt{\frac{dA_S}{d\Omega_O}},
\end{equation}
and can be determined by the observer if the size of the source is known a priori. In the same way we can also define the reciprocal distance $D_R$, based on a diverging beam from a point like source,
\begin{equation}
	D_R = \sqrt{\frac{dA_O}{d\Omega_S}}. \label{temp143}
\end{equation}
In this case the cone of rays in Figure \ref{fig:ad} is flipped, with $dA_O$ and $d\Omega_S$ being the cross-sectional area at the observer and the corresponding solid angle opening at the source. This geometrically defined distance is closely related to the luminosity distance $D_L$. Note that what we call the reciprocal distance here is sometimes referred to as the corrected luminosity distance (as in \cite{sef}). In \cite{ellis2} Ellis calls $D_R$ (his $r_G$) the galaxy area distance, and later defines the corrected luminosity distance that is different to $D_R$. To avoid confusion, we will not use the term corrected luminosity distance here.

Unlike the angular diameter distance $D_A$, the reciprocal distance $D_R$ cannot be determined by the observer by simply measuring out some $dA_O$, since the solid angle $d\Omega_S$ at the source that corresponds to it would generally be unknown. The reasoning behind the luminosity distance, $D_L$, is that for an isotropically radiating source of known luminosity, the radiation energy flux through a fixed area $dA_O$ at the observer will depend on how far the source is. Based on this reasoning we can define $D_L$ with analogy to flat space in terms of the measured flux and total known luminosity, and then relate these quantities to $dA_O$ and $d\Omega_S$. This will allow us to establish a connection between $D_L$ and $D_R$.

Let $L_S$ be the the total luminosity of a given source, and $F_O$ the radiation energy flux at a distant observer, then the luminosity distance is defined through analogy with flat space where $L_S = (4\pi D_L^2) F_O$, so that
\begin{equation}
	D_L = \sqrt{\frac{L_S}{4\pi F_O}}, \label{temp144}
\end{equation}
which, unlike $D_R$, can be determined from knowing $L_S$ a priori and measuring the local flux $F_O$.

Now suppose that a burst of isotropic radiation is emitted from the source with a total energy $E_S$ and over a proper time interval $d\tau_S$. Let $dE_S$ correspond to the fraction of energy that leaves the source in the form of photons distributed over a solid angle $d\Omega_S$, then
\begin{equation}
	\frac{d\Omega_S}{4 \pi} = \frac{dE_S}{E_S}. \label{temp145}
\end{equation}
These photons eventually arrive at a distant observer being distributed over an area $dA_O$ that corresponds to $d\Omega_S$. Let $dE_O$ be the radiation energy measured by the observer within this area over a proper time interval $d\tau_O$ that corresponds to the full burst interval $d\tau_S$. For a monochromatic signal, since the number of photons is conserved, $dE_S \propto \omega_S$ and $dE_O \propto \omega_O$ with the same proportionality constant, where $\omega_S$ and $ \omega_O$ are the respective frequencies at the source and observer. Consequently we have
\begin{equation}
	\frac{dE_O}{dE_S} = \frac{\omega_O}{\omega_S}, \label{temp146}
\end{equation}
and since the frequency shift is the same for all frequencies (the ratio $\frac{\omega_O}{\omega_S}$ is the same for all photons), the above holds true for non-monochromatic light also, only the frequency ratio should then be replaced by $(1+z)^{-1}$, where $z$ is the redshift parameter.

The total luminosity $L_S$ and the measurable flux $F_O$ are by definition related to the energies $E_S$ and $d E_O$ through
\begin{equation}
	E_S = L_S d\tau_S \label{temp147}
\end{equation}
and 
\begin{equation}
	dE_O = F_O dA_O d\tau_O. \label{temp148}
\end{equation}
Combining equations \eqref{temp143}, \eqref{temp144}, \eqref{temp145}, \eqref{temp146}, \eqref{temp147}, and \eqref{temp148}, we find
\begin{align}
	D_R^2 &= \frac{dA_O}{d\Omega_S}\nonumber\\
	&= \frac{dA_O E_S}{4\pi dE_S}\nonumber\\
	&= \frac{dA_O E_S\omega_O}{4\pi dE_O\omega_S}\nonumber\\
	&= \frac{L_S d\tau_S\omega_O}{4\pi F_O d\tau_O \omega_S}\nonumber\\
	&= D_L^2\frac{\omega_O^2}{\omega_S^2}. \label{temp153}
\end{align}
(Since clearly $\frac{d\tau_S}{d\tau_O}=\frac{\omega_O}{\omega_S}$; the frequency shift is itself an outcome of the time interval correspondence.) Again, the above is also true for non-monochromatic light, in which case the ratio $\frac{\omega_O^2}{\omega_S^2}$ would be replaced with $(1+z)^{-2}$.

Thus, when no other means of measuring the distance to an object are available, knowing the size ($dA_S$) and brightness ($L_S$) of the object allows the indirect measurement of the optical distances $D_A$ and $D_L$ through measuring $d\Omega_O$ and $F_O$. The reciprocity theorem is a purely geometrical result that relates $D_R$ and $D_A$, it states that (\cite{ether,ellis2,sef})
\begin{equation}
	D_R = D_A \frac{\omega_S}{\omega_O} = D_A (1+z), \label{reciprocity}
\end{equation}
which yields the following relationship for $D_A$ and $D_L$ by means of \eqref{temp153}
\begin{equation}
	D_L = D_A \frac{\omega_S^2}{\omega_O^2} = D_A (1+z)^2. \label{temp155}
\end{equation}
We emphasis again that the reciprocity theorem as well as \eqref{temp155} are well known, and what is missing in the literature are independent general expressions for the two measurable optical distances in terms of the fundamental parameters of the setup, which we are now in a position to present.\\

Adopting equation \eqref{temp139} to the setup in Figure \ref{fig:ad} and the definitions of $D_A$ and $D_R$, we have
\begin{equation}
	D_A^2 = \frac{dA_S}{d\Omega_O} = \omega_O^2 \left(1-\frac{1}{6}R_{KK} \right)+\mathcal{O}(\omega_O^5),
\end{equation}
and 
\begin{equation}
	D_R^2 = \frac{dA_O}{d\Omega_S} = \omega_S^2 \left(1-\frac{1}{6} R_{K_S K_S}^S\right)+\mathcal{O}(\omega_S^5),
\end{equation}
where $K^\alpha$ is past pointing and parametrized with the normalized affine parametrization so that $\omega_O=K^\alpha U_\alpha$ and $\omega_S=-K^\alpha_S V_\alpha$ are small; $K^\alpha_S = -\bar K^\alpha$ with $\bar K^\alpha$ being the parallel transport of $K^\alpha$ along the connecting geodesic to the source (which has 4-velocity $V^\alpha$). $R^S_{\alpha\beta}$ is the Ricci tensor evaluated at the source. Notice how the independence of the cross-sectional areas $dA_S$ and $dA_O$ on observer motion is clearly reflected in the above expressions, with only the 4-velocity that is associated with the solid angle being present. Let $\bar K_S^\alpha$ and $\bar R^S_{\alpha\beta}$ be the parallel transports of the tensors to the observer event along the connecting null geodesic. Then by the properties of the parallel transport $R^S_{K_SK_S}= \bar R^S_{\bar K_S \bar K_S}$. Clearly $\bar K^\alpha_S = - K^\alpha$, and by the generalized Taylor expansion for tensors (see \eqref{abartdeltat}, with the normalized affine parametrization)
\begin{equation}
	\bar R_{\alpha\beta}^S = R_{\alpha\beta} + \nabla_K R_{\alpha\beta} + \frac{1}{2}\nabla_K\nabla_K R_{\alpha\beta} + \mathcal{O}(\omega_O^3),
\end{equation}
where the terms on the right side are evaluated at the observer, and so
\begin{equation}
	\bar R_{\bar K_S\bar K_S}^S = R_{KK} + \nabla_K R_{KK} + \frac{1}{2}\nabla_K\nabla_K R_{KK} + \mathcal{O}(\omega_O^5), \label{temp156}
\end{equation}
which gives
\begin{equation}
	D_R^2 = \omega_S^2 \left(1-\frac{1}{6} R_{KK}\right)+\mathcal{O}(\omega^5),
\end{equation}
where $\omega$ in $\mathcal{O}(\omega^5)$ stands for the smallness in both $\omega_S$ and $\omega_O$. Finally, with reference to curvature and vectors at the observer event, and with the redshift $1+z = \frac{\omega_S}{\omega_O}$, we have the following results
\begin{equation}
	D_A = \omega_O \left(1-\frac{1}{12}R_{KK}\right)+\mathcal{O}(\omega_O^4), \label{temp175}
\end{equation}
\begin{align}
	D_R &= \omega_S \left(1-\frac{1}{12}R_{KK}\right)+\mathcal{O}(\omega^4)\\
	&= (1+z)\omega_O \left(1-\frac{1}{12}R_{KK}\right)+\mathcal{O}(\omega_O^4),
\end{align}
and by \eqref{temp153}
\begin{align}
	D_L &= \frac{\omega_S^2}{\omega_O} \left(1-\frac{1}{12}R_{KK}\right)+\mathcal{O}(\omega^4)\\
	&= (1+z)^2\omega_O \left(1-\frac{1}{12}R_{KK}\right)+\mathcal{O}(\omega_O^4). \label{temp179}
\end{align}
(Assuming $z$ is not extreme.) The null vector $K^\alpha$ in these expressions, which explicitly appears in $R_{KK}$ and $\omega_O \;(= \omega_U = K^\alpha U_\alpha)$, is assumed to either be given in some form by the setup or it can be obtained from the trajectories of the observer and source through the method described in appendix \ref{app:deri}. Thus, in any spacetime and coordinate system, given the world lines of observer and source, and an event of measurement for the observer, one can construct $K^\alpha$ and obtain an expression for the distances above. These expressions are general and independent of the field equations, but they are limited in accuracy due to the expansion in $\omega_O$, which we already called the projected optical distance for intuitive reasons. It can be easily verified that to this order in $\omega_O$ (as should be for all orders) $D_A$ and $D_R$ clearly satisfy the reciprocity relation \eqref{reciprocity}. Of central importance in the above is the effect of curvature on the optical distances. Curvature terms enter the above expressions explicitly through the Ricci tensor in the bracket, but can also contribute implicitly through the determination of $K^\alpha$ and $\omega_O$ for a given configuration, since curvature affects the null trajectory itself. Here, $\omega_O$ should not be thought of as measurable, but is mainly a geometrical quantity; only the ratio $\frac{\omega_S}{\omega_O}$ (or redshift) can be considered measurable. It satisfies intuition to see that the explicit and strongest contribution of curvature comes from of the Ricci tensor. As one would expect, it is the local density of matter and energy that causes extra convergence (or less divergence) of geodesics, and which ends up contributing to the most prominent effect of curvature on light ray bundles.

It must be stressed that the concept of photons traveling on null geodesics is merely used to simplify the explanations and build intuitive understanding of the outcomes. The above results are fully based on the fact that the propagation of electromagnetic radiation is governed by Maxwell's equations in curved spacetime. In most literature on the subject (for example \cite{sef}), the WKB approximation is the main method in establishing that the null vector $K^\alpha$ is normal to the wavefronts of outgoing radiation and other results that follow (also see chapter 6 in \cite{ellis2}).

\subsubsection{\label{sec:vac}The Case of Vacuum}

For purpose of defining proper optical velocity (as well as the acceleration) the accuracy in the above expressions of optical distances would suffice. However, in the case where the observer and source are in vacuum, the explicit curvature term disappears from the expressions, even in presence of the cosmological constant $\Lambda$. In practice, these situations may be of most interest, and it would be revealing to see how curvature affects ray bundles in empty space. To achieve this we must modify the accuracy of the above expressions to include curvature related terms of higher orders in $\omega_O$ that do not disappear in vacuum. These remaining terms would obviously be related to the Weyl components of the Riemann curvature tensor.

Armed with the fact that the distance between neighboring rays within an infinitesimally thin ray bundle is observer independent, we can refine our expression for $\frac{dA}{d\Omega_U}$ in equation \eqref{temp139} as follows. First, based on the derivation of the expression for $\delta K^\alpha$ given by \eqref{temp108} (also see \eqref{deltak2} and \eqref{temp33}), we see that as long as we keep the accuracy in $\theta_U$ to first order, any higher orders terms of $\omega_U$ in this expression will only appear within the coefficient of $K^\alpha$. For this reason these extra terms of $\omega_U$ will vanish from the inner product given by \eqref{temp113}, making it exact in $\omega_U$ at this order in $\theta_U$. Consequently, the area relationship given by \eqref{temp121} is exact in $\omega_U$ also. Therefore, \eqref{temp121} and \eqref{temp136} give
\begin{align}
	Area(\delta K^\alpha_1,\delta K^\alpha_2) &= \omega_U^2Area(\theta_1\hat B^\alpha_1,\theta_2\hat B^\alpha_2)+\mathcal{O}(\theta_U^3) \nonumber\\
    &= \omega_U^2 d\Omega_U +  \mathcal{O}(\theta_U^3), \label{temp163}
\end{align}
so we are able to replace $Area(\delta K^\alpha_1,\delta K^\alpha_2)$ in \eqref{temp128} without affecting the accuracy involving $\omega_U$.

Next we must modify the relationship between $\bar\delta^\alpha$ and $\delta K^\alpha$ given by \eqref{temp111}, and get a more accurate transformation tensor $T^\alpha_\beta$ to be used in relating $Area(\bar\delta^\alpha_1,\bar\delta^\alpha_2)$ to $Area(\delta K^\alpha_1,\delta K^\alpha_2)$, as prescribed by \eqref{temp122} and \eqref{temp128}. To this end, consider the connecting vector $\delta^\alpha(\lambda)$ as a function of the affine parameter $\lambda$ on the null geodesic with tangent $K^\alpha = \frac{dx^\alpha}{d\lambda}$. $\delta^\alpha(\lambda)$ is tangent to the geodesic connecting $x^\alpha(\lambda)$ to a nearby event $x^\alpha(\nu)$ on a neighboring null geodesic with an affine parameter $\nu$ and tangent $W^\alpha=\frac{dx^\alpha}{d\nu}$. The magnitude of $\delta^\alpha(\lambda)$ is the small geodesic distance between the nearby events, and the associations of the parameters $\lambda$ and $\nu$ is determined with respect to some observer (or congruence of observers) in mind, such that $\delta^\alpha(\lambda)$ will be in that observer's space frame at a given event $x^\alpha(\lambda)$. We make use of the crucial fact that this distance is observer independent to conclude that this construction allows any convenient choice of $\nu(\lambda)$. Such reparametrizations will not affect the area relations in which we are interested.

For the situation in Figure \ref{fig:delta}, let $\lambda_0=\nu_0=0$ at the event where the null geodesics intersect, with $K^\alpha = \frac{dx^\alpha}{d\lambda}\big|_0$ and $W^\alpha = \frac{dx^\alpha}{d\nu}\big|_0$. By equation \eqref{abartdeltat} in appendix \ref{app:deri}, the parallel transport of $\delta^\alpha(\lambda)$ from $x^\alpha(\lambda)$ to $x^\alpha(\lambda=0)$ along the geodesic is
\begin{equation}
	\bar\delta^\alpha(\lambda) = \delta^\alpha_0 + \nabla_K \delta^\alpha_0\lambda + \frac{1}{2}\nabla_K^2 \delta^\alpha_0\lambda^2 + \frac{1}{3!}\nabla_K^3 \delta^\alpha_0\lambda^3 + ... \label{temppp86}
\end{equation}
(Where $\nabla_K^2 = \nabla_K\nabla_K$ and so on.) As discussed in appendix \ref{app:deri}, reparametrization of $\lambda$ alters $K^\alpha$ in a way that keeps $K^\alpha\lambda$ the same. With the normalized affine parametrization, where for a particular event of interest along the geodesic $\lambda$ is set to $1$, the `smallness' of $\lambda$ is absorbed in the components of $K^\alpha$, which will eventually manifest itself in the parameter $\omega_U=-K^\alpha U_\alpha$ for some $U^\alpha$. The expression for $\bar\delta^\alpha$ in \eqref{temp86} and \eqref{temp111} assumes normalized affine parametrization in both $K^\alpha$ and $W^\alpha \;(=K^\alpha+\delta K^\alpha)$, which we adopt for the above expansion and set $\frac{d\nu}{d\lambda}=1$. The neighboring event to $x^\alpha(\lambda=1)$ where $\nu=1$ is set by any choice of observer $V^\alpha$ at $x^\alpha(\lambda=1)$ as described in the beginning of this section (see equations \eqref{vcond}, \eqref{temp108}), but this choice will not affect the length of $\bar\delta^\alpha$. With this in mind it is clear that $\bar\delta^\alpha(\lambda=1)$ given by the above is defined in the exact same way as $\bar\delta^\alpha$ in \eqref{temp86} and \eqref{temp111}, so we can use \eqref{temppp86} to establish a more accurate expression for $\bar\delta^\alpha$. To this end we find expressions for higher order derivatives of the connecting vector in appendix \ref{app:high}.

In the above expression, $\delta^\alpha_0=0$ since $\lambda_0$ is the event of intersection; and therefore by equation \eqref{temp321}, $\nabla_K\delta^\alpha_0 = \frac{d\nu}{d\lambda}W^\alpha-K^\alpha = W^\alpha-K^\alpha = \delta K^\alpha$ is exact. Thus, with $\delta^\alpha_0=0$ and $\nabla_K\delta^\alpha_0 = \delta K^\alpha$, the higher order derivatives of $\delta^\alpha$ at the event of intersection are given by (see appendix \ref{app:high})
\begin{equation}
	\nabla^2_K \delta^\alpha_0 = 0 \;\;\; (exactly),
\end{equation}
\begin{equation}
	\nabla^3_K \delta^\alpha_0 = -R^\alpha_{K(\delta K) K}+\mathcal{O}(\delta K^2),
\end{equation}
\begin{equation}
	\nabla^4_K \delta^\alpha_0 = -2\nabla_K R^\alpha_{K(\delta K) K}+\mathcal{O}(\delta K^2),
\end{equation}
and
\begin{equation}
	\nabla^5_K \delta_0^\alpha = -3 \nabla_K^2 R^\alpha_{K(\delta K) K}+R^\alpha_{K\beta K}R^\beta_{K(\delta K) K}+\mathcal{O}(\delta K^2).
\end{equation}
The smallness of $\delta K^\alpha$ can be encompassed by the small measurable angle $\theta_U$ between the null geodesics, with reference to some observer at the event (see equations \eqref{temp22}, \eqref{temp23}, \eqref{temp108}), and therefore $\mathcal{O}(\delta K^2)$ can be replaced with $\mathcal{O}(\theta_U^2)$ to make the approximation clear. As we shall see shortly, the reason for going up to the fifth order derivative is because this is where the lowest order non-vanishing curvature term in vacuum comes from.

Putting the above expressions together, we find
\begin{align}
	\bar\delta^\alpha &= \delta^\alpha_0 + \nabla_K \delta^\alpha_0 + \frac{1}{2}\nabla_K^2 \delta^\alpha_0 + \frac{1}{3!}\nabla_K^3 \delta^\alpha_0 + \frac{1}{4!}\nabla_K^4 \delta^\alpha_0\nonumber\\
    &\hspace{5cm}+ \frac{1}{5!}\nabla_K^5 \delta^\alpha_0 + \mathcal{O}(\omega^6)\nonumber\\
	&=\delta K^\alpha - \frac{1}{6}R^\alpha_{K(\delta K) K} - \frac{1}{12}\nabla_K R^\alpha_{K(\delta K) K}\nonumber\\
    &\hspace{0.3cm}- \frac{3}{5!}\nabla_K^2 R^\alpha_{K(\delta K) K} + \frac{1}{5!}R^\alpha_{K\beta K}R^\beta_{K(\delta K) K} + \mathcal{O}(\omega_U^6,\theta_U^2)\nonumber\\
	&=\bigg( \delta^\alpha_\beta - \frac{1}{6}R^\alpha_{K\beta K} - \frac{1}{12}\nabla_K R^\alpha_{K\beta K} - \frac{3}{5!}\nabla_K^2 R^\alpha_{K\beta K}\nonumber\\
    &\hspace{2.6cm}+ \frac{1}{5!}R^\alpha_{K\rho K}R^\rho_{K\beta K} \bigg)\delta K^\beta + \mathcal{O}(\omega_U^6,\theta_U^2)\nonumber\\
	&= T^\alpha_\beta\delta K^\beta + \mathcal{O}(\omega_U^6,\theta_U^2).
\end{align}
Compare this expression to \eqref{temp111}. Thus, we have modified the accuracy of the transformation $T^\alpha_\beta$ in \eqref{temp122}, which we will now use to refine the relationship between the areas $Area(\bar\delta^\alpha_1,\bar\delta^\alpha_2)$ and $Area(\delta K^\alpha_1,\delta K^\alpha_2)$ in \eqref{temp128}.

With $T^\alpha_\beta$ given above, and the corresponding $T^{-1}\,^\alpha_\beta$, as we've found in going from \eqref{temp122} to \eqref{temp125}, all terms involving the vector $L^\alpha$ vanish. This is entirely due to the antisymmetries of the Riemann tensor and the fact that $K^\alpha$ is null; and it is clear that this will be true for any higher accuracy of the transformation $T^\alpha_\beta$. Thus, \eqref{temp122} leads to
\begin{equation}
	\frac{det(g,\bar{\delta_1},\bar{\delta_2})}{det(g,\delta K_1,\delta K_2)} = J^2 + \mathcal{O}(\theta_U),
\end{equation}
and
\begin{equation}
	\frac{Area(\bar \delta^\alpha_1,\bar \delta^\alpha_2)}{Area(\delta K^\alpha_1,\delta K^\alpha_2)} = J + \mathcal{O}(\theta_U). \label{temp171}
\end{equation}
Again, for this particular case, the areas are simply related by the Jacobian of the transformation, which is not obvious. Defining $\delta T^\alpha_\beta$ such that $T^\alpha_\beta = \delta^\alpha_\beta + \delta T^\alpha_\beta$, then by equation \eqref{temp276}
\begin{align}
	det(T_{(\beta)}^{(\alpha)}) &= 1+\delta T_{\alpha}^{\alpha}+\frac{1}{2}(\delta T_{\alpha}^{\alpha})(\delta T_{\beta}^{\beta})-\frac{1}{2}(\delta T_{\beta}^{\alpha})(\delta T_{\alpha}^{\beta})+...\nonumber\\
	&= 1 - \frac{1}{6}R_{KK} - \frac{1}{12}\nabla_K R_{KK} - \frac{3}{5!}\nabla_K^2 R_{KK}\nonumber\\
    &\hspace{0.2cm}+ \frac{1}{5!}R^\alpha_{K\beta K}R^\beta_{K\alpha K} + \frac{1}{2} \left(\!- \frac{1}{6}R_{KK}\!\right)\left(\!- \frac{1}{6}R_{KK}\!\right)\nonumber\\
    &\hspace{0.2cm}- \frac{1}{2}\left(- \frac{1}{6}R^\alpha_{K\beta K}\right)\left(- \frac{1}{6}R^\beta_{K\alpha K}\right) + \mathcal{O}(\omega_U^5,\theta_U)\nonumber\\
	&= 1 - \frac{1}{6}R_{KK} - \frac{1}{12}\nabla_K R_{KK} - \frac{1}{40}\nabla_K^2 R_{KK}\nonumber\\
    &\hspace{0.2cm}+ \frac{1}{72} R_{KK}^2 - \frac{1}{180}R^\alpha_{K\beta K}R^\beta_{K\alpha K} + \mathcal{O}(\omega_U^5,\theta_U). \label{temp172}
\end{align}
Compare the above expression to \eqref{temp130}. Finally, with \eqref{temp171}, \eqref{temp172}, \eqref{temp163}, and since $dA=Area(\delta^\alpha_1,\delta^\alpha_2)=Area(\bar \delta^\alpha_1,\bar \delta^\alpha_2)$, we find
\begin{multline}
	\frac{dA}{d\Omega_U} = \omega_U^2 \bigg(1 - \frac{1}{6}R_{KK} - \frac{1}{12}\nabla_K R_{KK} - \frac{1}{40}\nabla_K^2 R_{KK} \\
    + \frac{1}{72} R_{KK}^2 - \frac{1}{180}R^\alpha_{K\beta K}R^\beta_{K\alpha K}\bigg) + \mathcal{O}(\omega_U^7,\theta_U). \label{temp173}
\end{multline}
This result is general for any null vector $K^\alpha$, whether it is future or past pointing, but it assumes normalized affine parametrization of the corresponding null geodesic, which means that $K^\alpha$ holds information of both events where $d\Omega$ and $dA$ are measured.

Adopting the refined ratio of $\frac{dA}{d\Omega_U}$ to the setup in Figure \ref{fig:ad} will yield more accurate expressions for optical distances in which curvature terms will remain even in vacuum. From the definitions of $D_A$ and $D_R$,
\begin{multline}
	D_A^2 = \frac{dA_S}{d\Omega_O} = \omega_O^2 \bigg(1 - \frac{1}{6}R_{KK} - \frac{1}{12}\nabla_K R_{KK}\\
    - \frac{1}{40}\nabla_K^2 R_{KK}  + \frac{1}{72} R_{KK}^2 - \frac{1}{180}R^\alpha_{K\beta K}R^\beta_{K\alpha K} \bigg)\\
    +\mathcal{O}(\omega_O^7),
\end{multline}
and 
\begin{multline}
	D_R^2 = \frac{dA_O}{d\Omega_S} = \omega_S^2 \bigg(1 - \frac{1}{6}R_{K_S K_S}^S - \frac{1}{12}\nabla_{K_S} R_{K_SK_S}^S\\
    - \frac{1}{40}\nabla_{K_S}^2 R_{K_SK_S}^S +\frac{1}{72} R_{K_SK_S}^{S\,2} - \frac{1}{180}R^{\alpha\,(S)}_{K_S\beta K_S}R^{\beta\,(S)}_{K_S\alpha K_S} \bigg)\\
    +\mathcal{O}(\omega_S^7),
\end{multline}
where $S$ refers to tensors at the source event. Let the barred tensors be their parallel transports along the null geodesic to the observer event. Then by the properties of the parallel transport and by means of the generalized Taylor expansion, we convert the above expression for $D_R$ to refer to the observer event as follows. In addition to $R_{K_S K_S}^S$ already given by equation \eqref{temp156} (and $\bar K_S^\alpha=-K^\alpha$), we have
\begin{equation}
	\overline{\nabla_{\alpha} R}_{\beta\gamma}^S = \nabla_{\alpha} R_{\beta\gamma} + \nabla_K \nabla_{\alpha} R_{\beta\gamma} +\mathcal{O}(\omega_O^2),
\end{equation}
\begin{equation}
	\overline{\nabla_{\bar K_S} R}_{\bar K_S\bar K_S}^S =  - \nabla_{K} R_{KK} - \nabla_K\nabla_{K} R_{KK} +\mathcal{O}(\omega_O^5),
\end{equation}
\begin{equation}
	\overline{\nabla^2_{\bar K_S} R}_{\bar K_S\bar K_S}^S = \nabla^2_{K} R_{KK} +\mathcal{O}(\omega_O^5),
\end{equation}
and
\begin{equation}
	\bar R^{\alpha\,(S)}_{\bar K_S\beta \bar K_S}\bar R^{\beta\,(S)}_{\bar K_S\alpha \bar K_S} = R^{\alpha}_{K\beta K} R^{\beta}_{K\alpha K}+\mathcal{O}(\omega_O^5).
\end{equation}
Therefore,
\begin{align}
	D_R^2 &= (1+z)^2\omega_O^2 \bigg(\!1 \!-\! \frac{1}{6}\left(\!R_{KK} \!+\! \nabla_K R_{KK} \!+\! \frac{1}{2}\nabla_K^2 R_{KK}\!\right) \nonumber\\
    &\hspace{1cm} + \frac{1}{12}\left( \nabla_{K} R_{KK} + \nabla_{K}^2 R_{KK} \right) \!-\! \frac{1}{40}\nabla_{K}^2 R_{KK}\nonumber\\
    &\hspace{1.2cm} +\frac{1}{72} R_{KK}^{2} - \frac{1}{180}R^{\alpha}_{K\beta K}R^{\beta}_{K\alpha K} \bigg)+\mathcal{O}(\omega_O^7) \nonumber\\
	&= (1+z)^2\omega_O^2 \!\left(\!1 \!-\! \frac{1}{6}R_{KK} \!-\! \frac{1}{12}\nabla_{K} R_{KK} \!-\! \frac{1}{40}\nabla_{K}^2 R_{KK} \right. \nonumber\\
	& \hspace{0.3cm} + \left. \frac{1}{72} R_{KK}^{2} - \frac{1}{180}R^{\alpha}_{K\beta K}R^{\beta}_{K\alpha K} \right)+\mathcal{O}(\omega_O^7).
\end{align}
Finally,
\begin{multline}
	D_A = \omega_O \bigg(1 - \frac{1}{12}R_{KK} - \frac{1}{24}\nabla_K R_{KK}\\
    - \frac{1}{80}\nabla_K^2 R_{KK} + \frac{1}{288} R_{KK}^2 - \frac{1}{360}R^\alpha_{K\beta K}R^\beta_{K\alpha K} \bigg)\\
    +\mathcal{O}(\omega_O^6),
\end{multline}
\begin{multline}
	D_R = (1+z)\omega_O \bigg(1 - \frac{1}{12}R_{KK} - \frac{1}{24}\nabla_K R_{KK} \\
    - \frac{1}{80}\nabla_K^2 R_{KK} + \frac{1}{288} R_{KK}^2 - \frac{1}{360}R^\alpha_{K\beta K}R^\beta_{K\alpha K} \bigg)\\
    +\mathcal{O}(\omega_O^6),
\end{multline}
and
\begin{multline}
	D_L = (1+z)^2\omega_O \bigg(1 - \frac{1}{12}R_{KK} - \frac{1}{24}\nabla_K R_{KK}\\
    - \frac{1}{80}\nabla_K^2 R_{KK} + \frac{1}{288} R_{KK}^2 - \frac{1}{360}R^\alpha_{K\beta K}R^\beta_{K\alpha K} \bigg)\\
    +\mathcal{O}(\omega_O^6).
\end{multline}
Once again, we immediately verify that the reciprocity relation is satisfied and arrive at the same conclusions as before. In vacuum, with or without $\Lambda$, the above distances become
\begin{equation}
	D_A = \omega_O \left(1 - \frac{1}{360}R^\alpha_{K\beta K}R^\beta_{K\alpha K} \right)+\mathcal{O}(\omega_O^6),
\end{equation}
\begin{equation}
	D_R = (1+z)\omega_O \left(1 - \frac{1}{360}R^\alpha_{K\beta K}R^\beta_{K\alpha K} \right)+\mathcal{O}(\omega_O^6),
\end{equation}
and
\begin{equation}
	D_L = (1+z)^2\omega_O \left(1 - \frac{1}{360}R^\alpha_{K\beta K}R^\beta_{K\alpha K} \right)+\mathcal{O}(\omega_O^6).
\end{equation}
Given any spacetime, an event of observation, 4-velocity of the observer, and the trajectory of a source, the above distances can be determined. These expressions would work well for relatively short distances, slowly varying metric components along the path of light, or high symmetry spacetimes. In cases where the metric changes significantly throughout the path, the above must be used with caution, as higher order terms may be relevant.

\subsection{\label{sec:defopt}Defining Optical Velocity}

With similar reasoning as in section \ref{sec:relvel} (see equation \eqref{temp72}), we define the optical velocity as the rate of change of the optical position vector of a given object with respect to that of a stationary object. Let $D_O$ represent the optical distance, so that $D_O E^\alpha$ is the optical position vector. A static object is one for which $\dot D_O = 0$ and $\dot E^\alpha = E^\beta \dot U_\beta U^\alpha$; its optical distance remains constant and its visibility direction is not rotating in the frame of the observer. Then using the optical position vector $D_O E^\alpha$ instead of the physical position vector $D^\alpha$ in the derivation of \eqref{temp72}, we have
\begin{equation}
	\vec v^\alpha_O = \left(D_O E^\alpha\right)^{\bullet} - D_O E^\beta \dot U_\beta U^\alpha. \label{temp204}
\end{equation}
Note that as opposed to the $\bar V^\alpha$ in section \ref{sec:relvel}, in the context of optical measurements the relevant 4-velocity of the object is that at the event to which $K^\alpha$ connects, not $D^\alpha$. So $\bar V^\alpha$ is the parallel transport of $V^\alpha$ from where the object's worldline intersects the past light cone of the observer, to the event of observation, along the connecting null geodesic, see Figure \ref{fig:sim}. We will establish a general expression for the above definition of optical velocity in terms of the fundamental vectors involved, their derivatives, and curvature components.

The derivative of the connecting vector $K^\alpha$ is found in the same way as that for $D^\alpha$ (see appendix \ref{app:deri}, equation \eqref{ddotap}),
\begin{equation}
	\dot K^\alpha = \dot\tau\bar V^\alpha - U^\alpha + \frac{1}{6}\left(\dot\tau \bar V^\beta + 2U^\beta\right)R^\alpha_{K\beta K}+\mathcal{O}(\omega_U^3), \label{temp205}
\end{equation}
where now $\bar V^\alpha$ is as described above, and $\dot \tau$ is established from the condition that $K^\alpha$ remains null throughout the motion. That is $(K^\alpha K_\alpha)^{\bullet}=2\dot K^\alpha K_\alpha = 0$, so we get
\begin{equation}
	\dot\tau = \frac{\omega_U}{\omega_V}, \label{temp206}
\end{equation}
as expected for this setup. Observe that $(\omega_U E^\alpha)^{\bullet} = (K^\beta h^\alpha_\beta)^{\bullet}$, so by means of \eqref{temp205}, $\omega_U$ would be the simplest optical distance to use in \eqref{temp204}; but since it is not measurable and we have the other options for $D_O$ expressed in term of $\omega_U$, we can proceed as follows,
\begin{align}
	\vec v^\alpha_O &= \left(\frac{D_O}{\omega_U}\omega_U E^\alpha\right)^{\bullet} - \frac{D_O}{\omega_U}\omega_U E^\beta \dot U_\beta U^\alpha\nonumber\\
	&= \left(\!\frac{D_O}{\omega_U}\!\right)^{\bullet}\omega_U E^\alpha \!+\! \frac{D_O}{\omega_U}\left(\!\dot K^\beta h^\alpha_\beta \!+\! K^\beta\dot h^\alpha_\beta \!-\! \omega_U E^\beta \dot U_\beta U^\alpha\!\right)\nonumber\\
	&= \left(\frac{D_O}{\omega_U}\right)^{\bullet}K^\beta h^\alpha_\beta + \frac{D_O}{\omega_U}\left(\dot K^\beta h^\alpha_\beta + \omega_U \dot U^\alpha\right)\nonumber\\
	&= \left(\frac{D_O}{\omega_U}\right)^{\bullet}K^\beta h^\alpha_\beta + \frac{D_O}{\omega_U}\Bigg( \frac{\omega_U}{\omega_V}\bar V^\beta\nonumber\\
    &\hspace{1.5cm}+\frac{1}{6}\left(\frac{\omega_U}{\omega_V} \bar V^\gamma + 2U^\gamma\right)R^\beta_{K\gamma K}+\omega_U \dot U^\beta \Bigg)h^\alpha_\beta\nonumber\\
    &\hspace{5cm}+\mathcal{O}(\omega_U^3). \label{temp207}
\end{align}
Now, $\bar V^\alpha = \gamma U^\alpha + \bar V^\beta h^\alpha_\beta$, and taking the inner product with $K^\alpha$ gives
\begin{equation}
	\frac{\omega_V}{\omega_U} = \gamma + \bar V^\alpha E_\alpha, \label{temp208}
\end{equation}
so
\begin{equation}
	\frac{\omega_U}{\omega_V}\bar V^\alpha = \left(1-\frac{\omega_U}{\omega_V}\bar V^\beta E_\beta\right) U^\alpha + \frac{\omega_U}{\omega_V}\bar V^\beta h^\alpha_\beta.
\end{equation}
($\gamma=-\bar V^\alpha U_\alpha$ is still the generalized relativistic $\gamma$ factor as before, but now with reference to a different event for $V^\alpha$.) Therefore,
\begin{align}
	\vec v^\alpha_O &= \left(\frac{D_O}{\omega_U}\right)^{\bullet}K^\beta h^\alpha_\beta \!+\! \left(\frac{D_O}{\omega_U}\right)\frac{\omega_U}{\omega_V} \bar V^\beta h^\alpha_\beta\!+\!\left(\frac{D_O}{\omega_U}\right)\omega_U \dot U^\alpha\nonumber\\
	&\hspace{1.3cm} +\frac{1}{6}\left(\frac{D_O}{\omega_U}\right)\frac{\omega_U}{\omega_V} \left( \bar V^\epsilon h^\gamma_\epsilon -\bar V^\epsilon E_\epsilon U^\gamma\right)R^\beta_{K\gamma K} h^\alpha_\beta\nonumber\\
    &\hspace{3.3cm}+ \frac{1}{2}\left(\frac{D_O}{\omega_U}\right)R^\beta_{KUK} h^\alpha_\beta+\mathcal{O}(\omega_U^3)\nonumber\\
	&= \left(\frac{D_O}{\omega_U}\right)^{\bullet}K^\beta h^\alpha_\beta \!+\! \left(\frac{D_O}{\omega_U}\right)\frac{\omega_U}{\omega_V} \bar V^\beta h^\alpha_\beta\!+\!\left(\frac{D_O}{\omega_U}\right)\omega_U \dot U^\alpha\nonumber\\
	&\hspace{0.1cm} +\!\frac{1}{6}\left(\frac{D_O}{\omega_U}\right)\frac{\omega_U}{\omega_V}\bar V^\epsilon H^\gamma_\epsilon R^\beta_{K\gamma K} h^\alpha_\beta \!+\! \frac{1}{2}\left(\frac{D_O}{\omega_U}\right)R^\beta_{KUK} h^\alpha_\beta\nonumber\\
    &\hspace{5.5cm}+\mathcal{O}(\omega_U^3). \label{temp210}
\end{align}
This is the general expression for optical velocity that we were seeking. It can now be easily adopted to the optical distances $D_A$ and $D_L$ by means of \eqref{temp175} and \eqref{temp179}. It is obvious that under this definition of velocity, the visibility direction of an object in purely radial motion will not be rotating, since this is the very definition of radial motion when using the optical position vector. That is, in this case $\dot E^\alpha_r = E^\beta \dot U_\beta U^\alpha$ by definition, so
\begin{equation}
	\left(D_O E^\alpha\right)^{\bullet}_{r} = \dot D_O E^\alpha + E^\beta \dot U_\beta U^\alpha,
\end{equation}
which gives
\begin{equation}
	\vec v^\alpha_{O\,r} = \left(D_O E^\alpha\right)^{\bullet}_{r} - D_O E^\beta \dot U_\beta U^\alpha = \dot D_O E^\alpha,
\end{equation}
as expected for purely radial velocity.

With the angular diameter distance $D_A$ as the optical distance $D_O$ in \eqref{temp210}, and with the help of equation \eqref{temp175}, we have
\begin{equation}
	\left(\frac{D_A}{\omega_U}\right) = 1-\frac{1}{12}R_{KK}+\mathcal{O}(\omega_U^3), \label{temp213}
\end{equation}
and
\begin{align}
	\left(\frac{D_A}{\omega_U}\right)^{\bullet} &= -\frac{1}{6}R_{\dot KK}+\mathcal{O}(\omega_U^2) \nonumber\\
    &= -\frac{1}{6} \frac{\omega_U}{\omega_V}R_{\bar VK}+\frac{1}{6}R_{UK}+\mathcal{O}(\omega_U^2). \label{temp214}
\end{align}
So
\begin{align}
	\vec v^\alpha_{O\,(D_A)} &= \left(-\frac{1}{6} \frac{\omega_U}{\omega_V}R_{\bar VK}+\frac{1}{6}R_{UK}\right)K^\beta h^\alpha_\beta \nonumber\\
    &\hspace{1cm}+ \left(1-\frac{1}{12}R_{KK}\right)\frac{\omega_U}{\omega_V} \bar V^\beta h^\alpha_\beta+\omega_U \dot U^\alpha\nonumber\\
	&\hspace{0.2cm} +\frac{1}{6}\frac{\omega_U}{\omega_V}\bar V^\epsilon H^\gamma_\epsilon R^\beta_{K\gamma K} h^\alpha_\beta + \frac{1}{2}R^\beta_{KUK} h^\alpha_\beta+\mathcal{O}(\omega_U^3) \nonumber\\
	&= \frac{\omega_U}{\omega_V}\bar V^\beta h^\alpha_\beta + \omega_U \dot U^\alpha -\frac{1}{12}\frac{\omega_U}{\omega_V}R_{KK} \bar V^\beta h^\alpha_\beta\nonumber\\
    &\hspace{0.7cm}-\frac{1}{6} \frac{\omega_U}{\omega_V}R_{\bar VK}K^\beta h^\alpha_\beta + \frac{1}{6}\frac{\omega_U}{\omega_V}\bar V^\epsilon H^\gamma_\epsilon R^\beta_{K\gamma K} h^\alpha_\beta \nonumber\\
	&\hspace{0.2cm} + \frac{1}{6}R_{UK}K^\beta h^\alpha_\beta + \frac{1}{2}R^\beta_{KUK} h^\alpha_\beta+\mathcal{O}(\omega_U^3). \label{temp215}
\end{align}
In a situation where the rotation of the visibility direction and the angular diameter distance can be experimentally determined, the optical velocity $\vec v^\alpha_{O\,(D_A)}$ is fully established by these two measurements. The above expression allows us to calculate this velocity with a given model for the setup, it is fully general and coordinate independent. As discussed in section \ref{sec:exp}, the ratio $\frac{\omega_V}{\omega_U}$ can be considered measurable, since this is the redshift ($\frac{\omega_V}{\omega_U}=1+z$).

For the luminosity distance $D_L$, equation \eqref{temp179} gives
\begin{equation}
	\left(\frac{D_L}{\omega_U}\right) = \left(\frac{\omega_V}{\omega_U}\right)^2 \left(1-\frac{1}{12}R_{KK}\right)+\mathcal{O}(\omega_U^3),
\end{equation}
and
\begin{multline}
	\left(\frac{D_L}{\omega_U}\right)^{\bullet} = 2\left(\frac{\omega_V}{\omega_U}\right) \left(\frac{\omega_V}{\omega_U}\right)^{\bullet} \left(1-\frac{1}{12}R_{KK}\right)\\
    + \left(\frac{\omega_V}{\omega_U}\right)^2 \left(-\frac{1}{6} \frac{\omega_U}{\omega_V}R_{\bar VK}+\frac{1}{6}R_{UK}\right)+\mathcal{O}(\omega_U^2). \label{temp217}
\end{multline}
Differentiating and making use of \eqref{temp205}, \eqref{temp206}, \eqref{temp208} and \eqref{partransdot},
\begin{align}
	\left(\frac{\omega_V}{\omega_U}\right)^{\bullet} &= \frac{\dot \omega_V \omega_U - \dot \omega_U \omega_V}{\omega_U^2} \nonumber\\
	&= \frac{1}{\omega_U}\left(\!\dot K^\alpha \bar V_\alpha\! +\! K^\alpha \dot{\bar V}_\alpha \!-\! \frac{\omega_V}{\omega_U}\left(\!\dot K^\alpha U_\alpha \!+\!K^\alpha\dot U_\alpha\!\right)\!\right)\nonumber\\
	&= \frac{1}{\omega_U}\Bigg(\!-\!\frac{\omega_U}{\omega_V}\!+\!\gamma \!+\! \frac{1}{6}\left(\!\frac{\omega_U}{\omega_V}\bar V^\alpha+2U^\alpha\!\right)\!R_{\bar V K\alpha K}\nonumber\\
    &\hspace{0.5cm}+ \frac{\omega_U}{\omega_V}K^\alpha \bar{\dot V}_\alpha +\frac{1}{2}R_{K\bar V UK}+\frac{1}{2}\frac{\omega_U}{\omega_V}R_{K\bar V \bar VK} \nonumber\\
	&\hspace{0.2cm}-\! \frac{\omega_V}{\omega_U}\!\bigg(\!\!-\!\frac{\omega_U}{\omega_V}\gamma \!+\!1\!+\!  \frac{1}{6}\!\left(\!\frac{\omega_U}{\omega_V}\bar V^\alpha\!+\!2U^\alpha\!\right)\!R_{U K\alpha K}\nonumber\\
    &\hspace{3.5cm}+  K^\alpha\dot U_\alpha\bigg)\Bigg) + \mathcal{O}(\omega_U^2)\nonumber\\
	&= \frac{1}{\omega_U}\bigg(2\gamma -\frac{\omega_V}{\omega_U}\left(1+K^\alpha \dot U_\alpha+\frac{1}{3}R_{UKUK}\right)\nonumber\\
    &\;-\frac{\omega_U}{\omega_V}\left(1-K^\alpha \bar {\dot V}_\alpha+\frac{1}{3}R_{\bar VK\bar VK}\right) \!-\! \frac{1}{3}R_{\bar V KUK}\bigg)\nonumber\\
    &\hspace{4.5cm}+\mathcal{O}(\omega_U^2). \label{temp218}
\end{align}
The zero order term in the bracket can be simplified by use of the angle between $\bar V^\beta h^\alpha_\beta$ and $E^\alpha$; let it be $\alpha$. Then by \eqref{temp208}
\begin{align}
	2\gamma\left(\frac{\omega_V}{\omega_U}\right)-&\left(\frac{\omega_V}{\omega_U}\right)^2-1=\gamma^2-1-\left(\bar V^\alpha E_\alpha\right)^2\nonumber\\
	&=\left(\gamma v-\bar V^\alpha E_\alpha\right) \left(\gamma v+\bar V^\alpha E_\alpha \right)\nonumber\\
	&=\left(\gamma v-\gamma v \cos(\alpha)\right) \left(\gamma v+\gamma v \cos(\alpha) \right)\nonumber\\
	&=\gamma^2v^2\sin^2(\alpha)\nonumber\\
	&=\left(\frac{\omega_V}{\omega_U}\right)^2\left(1-\frac{\omega_U}{\omega_V}\bar V^\alpha E_\alpha\right)^2v^2\sin^2(\alpha)\nonumber\\
	&=\left(\frac{\omega_V}{\omega_U}\right)^2\left(\frac{v\sin(\alpha)}{1+v\cos(\alpha)}\right)^2
\end{align}
($v$ is defined through $\gamma$ by $\gamma=(1-v^2)^{-\sfrac{1}{2}}$), and we have
\begin{multline}
	\left(\frac{\omega_V}{\omega_U}\right)^{\bullet} = \frac{1}{\omega_U}\frac{\omega_V}{\omega_U}\Bigg(\left(\frac{v\sin(\alpha)}{1+v\cos(\alpha)}\right)^2 \\
    - \left(K^\alpha \dot U_\alpha+\frac{1}{3}R_{UKUK}\right) +\left(\frac{\omega_U}{\omega_V}\right)^2\left(K^\alpha \bar {\dot V}_\alpha-\frac{1}{3}R_{\bar VK\bar VK}\right)\\
    -\frac{\omega_U}{\omega_V}\frac{1}{3}R_{\bar V KUK}\Bigg) +\mathcal{O}(\omega_U^3).
\end{multline}
In principle, the zero order term in the bracket could be very large. However, for non-extreme redshift, and also with the condition on $V^\alpha$ given by equations \eqref{temp180} or \eqref{temp104}, we can assume that the term is moderate. Because of this we can drop the second order $R_{KK}$ term in \eqref{temp217}, since we are only interested in up to first order terms in that expression.

And so, with $D_L$ taking place of the optical distance $D_O$ in \eqref{temp210}, we get
\begin{align}
	&\vec v^\alpha_{O\,(D_L)} = 2\left(\frac{\omega_V}{\omega_U}\right)\Bigg(2\gamma -\frac{\omega_V}{\omega_U}\left(1+K^\alpha \dot U_\alpha+\frac{1}{3}R_{UKUK}\right) \nonumber\\
	&\hspace{0.8cm} -\frac{\omega_U}{\omega_V}\left(1-K^\alpha \bar {\dot V}_\alpha+\frac{1}{3}R_{\bar VK\bar VK}\right) -\frac{1}{3}R_{\bar V KUK}\Bigg)E^\alpha\nonumber\\
	&\hspace{1cm} + \left(\frac{\omega_V}{\omega_U}\right)^2 \left(-\frac{1}{6} \frac{\omega_U}{\omega_V}R_{\bar VK}+\frac{1}{6}R_{UK}\right)K^\beta h^\alpha_\beta \nonumber\\
    &\hspace{0.2cm}+ \left(\frac{\omega_V}{\omega_U}\right)^2 \left(1-\frac{1}{12}R_{KK}\right)\frac{\omega_U}{\omega_V} \bar V^\beta h^\alpha_\beta+\left(\frac{\omega_V}{\omega_U}\right)^2\omega_U \dot U^\alpha \nonumber\\
    &\hspace{2.5cm}+\frac{1}{6}\left(\frac{\omega_V}{\omega_U}\right)^2\frac{\omega_U}{\omega_V}\bar V^\epsilon H^\gamma_\epsilon R^\beta_{K\gamma K} h^\alpha_\beta\nonumber\\
    &\hspace{3.5cm}+ \frac{1}{2}\left(\frac{\omega_V}{\omega_U}\right)^2R^\beta_{KUK} h^\alpha_\beta+\mathcal{O}(\omega_U^3) \nonumber\\
	&\hspace{1cm}= \frac{\omega_V}{\omega_U}\bar V^\beta h^\alpha_\beta + \left(\frac{\omega_V}{\omega_U}\right)^2 \omega_U \dot U^\alpha \nonumber\\
    &\hspace{0.2cm}+ 2\left(\!\frac{\omega_V}{\omega_U}\!\right)\!\left(\!2\gamma \!-\!\frac{\omega_V}{\omega_U}\left(1\!+\!K^\alpha \dot U_\alpha\right) \!-\!\frac{\omega_U}{\omega_V}\left(\!1\!-\!K^\alpha \bar {\dot V}_\alpha\!\right)\! \right)\!E^\alpha    \nonumber\\
	&\hspace{0.2cm} +\frac{2}{3}\left(\frac{\omega_V}{\omega_U}\right)^2\Bigg(\frac{1}{4}\omega_U R_{UK} -\frac{1}{4}\frac{\omega_U}{\omega_V}\omega_U R_{\bar V K}   - R_{UKUK} \nonumber\\
    &\hspace{0.2cm}- \left(\frac{\omega_U}{\omega_V}\right)^2R_{\bar VK\bar VK}-\left(\frac{\omega_U}{\omega_V}\right) R_{UK\bar VK}\Bigg) E^\alpha \nonumber\\
	&\hspace{0.2cm} -\frac{1}{12}\frac{\omega_V}{\omega_U}R_{KK} \bar V^\beta h^\alpha_\beta + \frac{1}{6}\frac{\omega_V}{\omega_U}\bar V^\epsilon H^\gamma_\epsilon R^\beta_{K\gamma K} h^\alpha_\beta\nonumber\\
    &\hspace{2cm}+ \frac{1}{2}\left(\frac{\omega_V}{\omega_U}\right)^2R^\beta_{KUK} h^\alpha_\beta+\mathcal{O}(\omega_U^3). \label{velld}
\end{align}
This is the general expression for optical velocity with respect to luminosity distance. It can be determined from measurements and also calculated through the above expression with a model for a given physical setup. Notice that
\begin{align}
	&\vec v^\alpha_{O\,(D_L)} = \left(\frac{\omega_V}{\omega_U}\right)^2\vec v^\alpha_{O\,(D_A)} \nonumber\\
    &\hspace{0.2cm}+\! 2\left(\!\frac{\omega_V}{\omega_U}\!\right)\!\left(\!2\gamma \!-\!\frac{\omega_V}{\omega_U}\!\left(\!1\!+\!K^\alpha \dot U_\alpha\!\right) \!-\!\frac{\omega_U}{\omega_V}\!\left(\!1\!-\!K^\alpha \bar {\dot V}_\alpha\!\right)\!\! \right)\!E^\alpha    \nonumber\\
	&\hspace{1.8cm}-\frac{2}{3}\left(\frac{\omega_V}{\omega_U}\right)^2\Bigg(  R_{UKUK} + \left(\frac{\omega_U}{\omega_V}\right)^2R_{\bar VK\bar VK}\nonumber\\
    &\hspace{2.2cm}+ \left(\frac{\omega_U}{\omega_V}\right) R_{UK\bar VK}\Bigg) E^\alpha+\mathcal{O}(\omega_U^3),
\end{align}
so in addition to the redshift factor, there is a difference between the two optical velocities that happens to be in the direction of visibility $E^\alpha$, and is caused by relative motion (given by $\gamma$ and the frequency ratios), the 4-accelerations, and curvature.

The explicit effects of curvature for both velocities come in as second order terms (as expected). However, the expressions are considerably simplified when reducing one order of accuracy. Then we get,
\begin{equation}
	\vec v^\alpha_{O\,(D_A)} = \frac{\omega_U}{\omega_V}\bar V^\beta h^\alpha_\beta + \omega_U \dot U^\alpha +\mathcal{O}(\omega_U^2), \label{temp221}
\end{equation}
\begin{multline}
	\vec v^\alpha_{O\,(D_L)} = \frac{\omega_V}{\omega_U}\bar V^\beta h^\alpha_\beta + \left(\frac{\omega_V}{\omega_U}\right)^2 \omega_U \dot U^\alpha\\
	+\!2\left(\!\frac{\omega_V}{\omega_U}\!\right)\left(\!2\gamma \!-\!\frac{\omega_V}{\omega_U}\left(1\!+\!K^\alpha \dot U_\alpha\right) \!-\!\frac{\omega_U}{\omega_V}\left(\!1\!-\!K^\alpha \bar {\dot V}_\alpha\!\right) \!\!\right)\!E^\alpha\\
    +\mathcal{O}(\omega_U^2), \label{temp222}
\end{multline}
and 
\begin{multline}
	\vec v^\alpha_{O\,(D_L)} = \left(\frac{\omega_V}{\omega_U}\right)^2\vec v^\alpha_{O\,(D_A)}\\
    +\! 2\left(\!\frac{\omega_V}{\omega_U}\!\right)\left(\!2\gamma \!-\!\frac{\omega_V}{\omega_U}\left(1\!+\!K^\alpha \dot U_\alpha\right) \!-\!\frac{\omega_U}{\omega_V}\left(\!1\!-\!K^\alpha \bar {\dot V}_\alpha\!\right) \!\right)\!E^\alpha\\
    +\mathcal{O}(\omega_U^2).
\end{multline}

A striking difference between the expressions for optical velocities and the physical relative velocities defined in section \ref{sec:relvel} is the explicit appearance of the acceleration vectors. While for physical velocities $\dot U^\alpha$ came only through $\dot\tau$ (see \eqref{temp80}, \eqref{temp82}, \eqref{temp85}, and \eqref{taudot}), which only scales the velocity, for the case of optical velocities the acceleration is explicit even in first order, and is able to affect both magnitude and direction of the velocity itself. We find a peculiar phenomenon due to the explicit appearance of $\dot U^\alpha$ in \eqref{temp221} and \eqref{temp222} for a very simple setup, which we will briefly discuss. In the special case where $\bar V^\alpha=U^\alpha$ at the moment of measurement, and the observer is accelerating directly away from the source, so that $\dot U^\alpha = -\dot U E^\alpha$, \eqref{temp221} gives
\begin{equation}
	\vec v^\alpha_{O\,(D_A)} = -\omega_U \dot U E^\alpha +\mathcal{O}(\omega_U^2).
\end{equation}
Thus, when the observer is momentarily at rest with the source but accelerates away, the optical velocity appears to be in the negative direction of the source. So the source appears to move closer. This is only a temporary effect, since $\bar V^\alpha$ would quickly gain a component in the positive $E^\alpha$ direction which will cancel out the $\dot U$ term and grow more positive. Therefore, in the process, the distant object will first appear to move towards the observer as the acceleration begins, and only after some time will appear to move away as one would have expected. The reason for this strange effect is entirely due to relativistic aberration of light and can be explained as follows. The optical velocity $\vec v^\alpha_{O\,(D_A)}$ is with respect to the angular diameter distance $D_A$, which is based on the apparent solid angle size of the distant object. A different observer that moves away from the source and coincident with $U^\alpha$ at the event of measurement will measure a smaller frequency of the converging light bundle, and therefore a larger solid angle, as given by \eqref{temp141}. So to the coincident moving observer the distant object appears closer. Thus, before the acceleration $\dot U^\alpha$ results in a gain of distance (or increase in $\omega_U$) that will make the object appear smaller (assuming $\dot V^\alpha=0$), it results in a change of frames, to one in which the object appears larger and closer. (It is possible to have an increasing acceleration so that the terms in \eqref{temp221} will always cancel out leading to a stationary looking object for an observer that accelerates away.) Of course, if instead the observer accelerates towards the object, then the effect would be opposite, the object will appear to move farther away before it appears to move closer, due to the same reason of aberration. This particular effect is exclusive to optical velocity with respect to the angular diameter distance.

For the same case with $\bar V^\alpha=U^\alpha$, $\dot U^\alpha = -\dot U E^\alpha$ (and $\dot V^\alpha=0$) in \eqref{temp222}, we get
\begin{align}
	\vec v^\alpha_{O\,(D_L)} &= -\omega_U \dot UE ^\alpha + 2\left( -K^\alpha \dot U_\alpha \right)E^\alpha+\mathcal{O}(\omega_U^2)\nonumber \\
	&= -\omega_U \dot UE ^\alpha + 2\left( \omega_U \dot U \right)E^\alpha+\mathcal{O}(\omega_U^2)\nonumber \\
	&= \omega_U \dot UE ^\alpha+\mathcal{O}(\omega_U^2),
\end{align}
so the initial acceleration causes an opposite effect to what it does for $\vec v^\alpha_{O\,(D_A)}$. Thus, due to the explicit $\dot U^\alpha$ term, an observer that is momentarily stationary but accelerating away from the source will observe a quick recession of the source even before any significant distance (or increase in $\omega_U$) is gained. This effect is not as strange as the one for $\vec v^\alpha_{O\,(D_A)}$, and can be understood as follows. The luminosity distance is based on the measurable energy flux by an observer, which depends on its motion. So before there is any significant distance gained, due to which the object will eventually become less bright, there is a gain in relative speed that results in a shift of frequency. Since the frequency decreases for a receding observer, the measured energy flux decreases as well, and the object appears less bright and farther away due to gain in speed only. The same spike in apparent optical velocity will be observed when the acceleration of the observer is towards the object; where the object would appear to get brighter and closer before any significant motion by the observer is taking place.\\

Before ending this section we make some observations that will help in comparing optical velocities and the physical relative velocities previously derived. First, consider the special case where $\bar V^\alpha=U^\alpha$ at the event of measurement, and both observer and object are in geodesic motion. Then only the symmetric relative velocity given by \eqref{temp82} turns out to be zero (as expected), while the Fermi relative velocity given by \eqref{temp80} (and the optical relative velocity given by \eqref{temp85}) ends up being
\begin{equation}
	\vec v^\alpha = \frac{1}{2}R^\beta_{DUD}H^\alpha_\beta+\mathcal{O}(D^3).
\end{equation}
In this case the two proper optical velocities are given by
\begin{align}
	\vec v^\alpha_{O\,(D_A)} &= \frac{1}{2}R^\beta_{KUK}h^\alpha_\beta+\mathcal{O}(\omega_U^3)\nonumber\\
	&= \frac{1}{2}R^\beta_{KUK}H^\alpha_\beta+\frac{1}{2}R_{UKUK}E^\alpha+\mathcal{O}(\omega_U^3),
\end{align}
and
\begin{align}
	\vec v^\alpha_{O\,(D_L)} &= \frac{1}{2}R^\beta_{KUK}h^\alpha_\beta -2R_{UKUK}E^\alpha +\mathcal{O}(\omega_U^3)\nonumber\\
	&=\frac{1}{2}R^\beta_{KUK}H^\alpha_\beta -\frac{3}{4}R_{UKUK}E^\alpha +\mathcal{O}(\omega_U^3).
\end{align}
So in addition to the perpendicular drift induced by the curvature, in the optical case we also have a radial component to the observed velocity caused by curvature. It must be kept in mind that the $\bar V^\alpha$ in the condition $\bar V^\alpha=U^\alpha$ is not the same in both physical and optical cases.

Next, consider the limit case where the observer and object are at the same event at the time of measurement. Then $\bar V^\alpha\,(=V^\alpha)$ is the same for both types of velocities, $D=\omega_U=0$, and $\bar V^\beta h^\alpha_\beta = |\bar V^\beta h^\alpha_\beta|E^\alpha = \gamma v E^\alpha$ (we could use $-E^\alpha$ instead, leading to the same results). Equations \eqref{temp80}, \eqref{temp82} and \eqref{temp85} (with \eqref{taudot}) all give
\begin{equation}
	\vec v^\alpha = \frac{1}{\gamma}\bar V^\beta h^\alpha_\beta = v E^\alpha,
\end{equation}
as expected (clearly $\hat D^\alpha = E^\alpha$ here). With help of \eqref{temp208}, we can replace the redshift factor
\begin{equation}
	\frac{\omega_V}{\omega_U} = \gamma + \bar V^\alpha E_\alpha = \gamma(1+v),
\end{equation}
so from \eqref{temp221} and \eqref{temp222} we find
\begin{equation}
	\vec v^\alpha_{O\,(D_A)} = \frac{1}{\gamma(1+v)}\bar V^\beta h^\alpha_\beta = \frac{v}{1+v} E^\alpha, \label{temp231}
\end{equation}
and
\begin{equation}
	\vec v^\alpha_{O\,(D_L)} = \gamma(1+v)\bar V^\beta h^\alpha_\beta = \gamma^2(1+v) v E^\alpha = \frac{v}{1-v} E^\alpha. \label{temp232}
\end{equation}
Apparently in this special case proper optical velocity is modified by a factor of $(1\pm v)^{-1}$, depending on the type of optical distance used. For small velocities the effect is negligible, but when $v \to 1$, $|\vec v^\alpha_{O\,(D_A)}|$ goes only up to $\frac{1}{2}$, while $|\vec v^\alpha_{O\,(D_L)}|$ becomes infinitely large. To understand the reason for the $(1+v)^{-1}$ factor in \eqref{temp231} we recall that a small angular diameter distance is the same as the projected optical distance, and examine the case of a fast moving object with $v \approx 1$. A simple Minkowski diagram reveals that for the same time interval, the projected optical distance to the null line would be half the length of the physical distance. This explains the $\frac{1}{2}$ speed limit and the reason why $|\vec v^\alpha_{O\,(D_A)}|$ should always be smaller than the actual speed $v$ in this case of coincidence. The reason for the $(1-v)^{-1}$ factor in \eqref{temp232} comes from the fact that $|\vec v^\alpha_{O\,(D_L)}|$ is based on the changing brightness of the moving object. For very fast moving objects this brightness is strongly affected by the headlight effect as well as redshift. When the coincident object takes a small step away from the observer, due to its velocity, the isotropic radiation coming from it will concentrate in the direction away from the observer, and whatever will travel back to the observer will be redshifted. So the brightness of a fast traveling object can reduce very fast, and in the limit $v \to 1$ it will reduce to zero
in an infinitesimally short time. This explains the reason for $|\vec v^\alpha_{O\,(D_L)}|$ to have an limit of infinity, and the reason why $|\vec v^\alpha_{O\,(D_L)}|$ must always be larger than $v$ for receding objects.

Finally, it is important to recognize that except for special cases the optical velocities defined in this section are by no means symmetric. We find that the relationships between $\bar V^\alpha$ and $U^\alpha$ for objects that appear stationary to the observer, for both definitions of optical velocity. Taking $\vec v^\alpha_{O\,(D_A)} = 0$ in \eqref{temp215} and solving for $\frac{\omega_U}{\omega_V}\bar V^\beta h^\alpha_\beta$, we get
\begin{equation}
	\frac{\omega_U}{\omega_V}\bar V^\beta h^\alpha_\beta = -\omega_U\dot U^\alpha -\frac{1}{2}R^\beta_{KUK}h^\alpha_\beta+\mathcal{O}(\omega_U^3).
\end{equation}
By \eqref{temp208} we have that $\frac{\omega_V}{\omega_U}=\gamma+\bar V^\beta h^\alpha_\beta E_\alpha$, which leads to
\begin{equation}
	\bar V^\beta h^\alpha_\beta = -\gamma\omega_U\dot U^\alpha +\gamma\omega_U\dot U^\beta K_\beta \dot U^\alpha -\frac{1}{2}\gamma R^\beta_{KUK}h^\alpha_\beta+\mathcal{O}(\omega_U^3).
\end{equation}
Since $\gamma^2 = 1+|\bar V^\beta h^\alpha_\beta|^2$, we find that $\gamma = 1+\frac{1}{2}\omega_U^2\dot U^2 +\mathcal{O}(\omega_U^3)$, and finally
\begin{multline}
	\bar V^\alpha = \left(1+\frac{1}{2}\omega_U^2\dot U^2\right) U^\alpha -\omega_U\dot U^\alpha +\omega_U\dot U^\beta K_\beta \dot U^\alpha\\
    -\frac{1}{2}R^\beta_{KUK}h^\alpha_\beta+\mathcal{O}(\omega_U^3). \label{temp235}
\end{multline}
This is the required 4-velocity of the source at the time of emission, in order for it to appear stationary (with respect to the angular diameter distance) to an observer with 4-velocity $U^\alpha$ and 4-acceleration $\dot U^\alpha$. As for luminosity distance, setting $\vec v^\alpha_{O\,(D_L)} = 0$ in \eqref{velld} and following the same method,
\begin{align}
	&\bar V^\beta h^\alpha_\beta = -\frac{\omega_V}{\omega_U}\omega_U\dot U^\alpha\nonumber\\
    &-2\left(2\gamma-\frac{\omega_V}{\omega_U}\left(1+K^\beta\dot U_\beta\right)-\frac{\omega_U}{\omega_V}\left(1-K^\beta\bar{\dot V}_\beta\right)\right)E^\alpha\nonumber\\
	&\hspace{1cm} +2\frac{\omega_V}{\omega_U} R_{UKUK}E^\alpha  -\frac{1}{2}\frac{\omega_V}{\omega_U} R^\beta_{KUK}h^\alpha_\beta+\mathcal{O}(\omega_U^3) \nonumber \\
	&\hspace{0.8cm}= -\frac{\omega_V}{\omega_U}\omega_U\dot U^\alpha -2\Bigg(\frac{\omega_U}{\omega_V}\left(\gamma^2-1-\left(\bar V^\beta E_\beta\right)^2\right)\nonumber\\
    &\hspace{3.5cm}-\frac{\omega_V}{\omega_U}K^\beta\dot U_\beta+\frac{\omega_U}{\omega_V}K^\beta\bar{\dot V}_\beta\Bigg)E^\alpha\nonumber \\
	&\hspace{1cm}+2\gamma R_{UKUK}E^\alpha  -\frac{1}{2}\gamma R^\beta_{KUK}h^\alpha_\beta+\mathcal{O}(\omega_U^3).
\end{align}
With this we find that $\gamma^2=1+\mathcal{O}(\omega_U^2)$, and therefore $\frac{\omega_V}{\omega_U}=1+\bar V^\alpha E_\alpha=1+\bar V^\beta h^\alpha_\beta E_\alpha=1+\mathcal{O}(\omega_U)$, so
\begin{multline}
	\bar V^\beta h^\alpha_\beta = -\omega_U\dot U^\alpha -\omega_U\bar V^\beta E_\beta \dot U^\alpha  +2\left( K^\beta\dot U_\beta-K^\beta\bar{\dot V}_\beta\right)E^\alpha \\
	  -2\left(\gamma^2-1-\left(\bar V^\beta E_\beta\right)^2-\bar V^\gamma E_\gamma \left(K^\beta \dot U_\beta+K^\beta\bar{\dot V}_\beta\right)\right)E^\alpha\\
	+2R_{UKUK}E^\alpha  -\frac{1}{2}R^\beta_{KUK}h^\alpha_\beta+\mathcal{O}(\omega_U^3), \label{temp237}
\end{multline}
which helps establish $\bar V^\alpha E_\alpha=K^\alpha\dot U_\alpha-2K^\alpha\bar{\dot V}_\alpha+\mathcal{O}(\omega_U^2)$. We also have that $\dot K^\alpha = \bar V^\beta h^\alpha_\beta-\bar V^\beta E_\beta U^\alpha +\mathcal{O}(\omega_U^2)$, and $\dot \gamma = 0 +\mathcal{O}(\omega^2)$, so the derivative $\bar{\dot V}^\alpha$ is found as follows,
\begin{align}
	\bar V^\alpha &= \gamma U^\alpha+\bar V^\beta h^\alpha_\beta\nonumber \\
	\dot{\bar V}^\alpha &= \dot \gamma U^\alpha +\gamma \dot U^\alpha +\left(\bar V^\beta h^\alpha_\beta\right)^{\bullet}   \nonumber\\
	\frac{\omega_U}{\omega_V} \bar{\dot V}^\alpha +R^\alpha_{UUK} &= \dot U^\alpha -\dot \omega_U\dot U^\alpha -\omega_U\ddot U^\alpha \nonumber\\
    &+\!2\left(\!\dot K^\beta\dot U_\beta\!+\!K^\beta \ddot U_\beta\! -\! \dot K^\beta \bar{\dot V}_\beta \!-\!K^\beta\dot{\bar{\dot V}}_\beta\!\right)\!E^\alpha\nonumber\\
	&\qquad+2\left( K^\beta\dot U_\beta-K^\beta\bar{\dot V}_\beta\right)\dot E^\alpha +\mathcal{O}(\omega_U^2)  \nonumber\\
	\frac{\omega_U}{\omega_V} \bar{\dot V}^\alpha  &= \dot U^\alpha -\left(\dot K^\beta U_\beta+K^\beta \dot U_\beta\right)\dot U^\alpha -\omega_U\ddot U^\alpha \nonumber\\
    &+2\left(K^\beta \ddot U_\beta  -K^\beta\bar{\ddot V}_\beta\right) +R^\alpha_{UKU}  +\mathcal{O}(\omega_U^2) \nonumber\\
	\bar{\dot V}^\alpha  &= \frac{\omega_V}{\omega_U}\left(\dot U^\alpha -\omega_U\ddot U^\alpha +R^\alpha_{UKU}\right)+\mathcal{O}(\omega_U^2)\nonumber\\
	&= \dot U^\alpha -\dot U^\beta K_\beta \dot U^\alpha -\omega_U\ddot U^\alpha +R^\alpha_{UKU}\nonumber\\
    &\hspace{3.5cm}+\mathcal{O}(\omega_U^2).
\end{align}
Using this expression for $\bar{\dot V}^\alpha$ and replacing $\bar V^\alpha E_\alpha$ in \eqref{temp237} gives
\begin{multline}
	\!\!\bar V^\beta h^\alpha_\beta \!=\! -\omega_U\dot U^\alpha \!+\!\omega_UK^\beta\dot U_\beta \dot U^\alpha\! -\!2\!\left(\!\gamma^2\!-\!1\!-\!\omega_U\ddot U^\beta K_\beta\!\right)\!E^\alpha \\
    - \frac{1}{2}R^\beta_{KUK}h^\alpha_\beta+\mathcal{O}(\omega_U^3).
\end{multline}
It is easy to find that $\gamma^2 = 1+\omega_U^2\dot U^2 +\mathcal{O}(\omega_U^3)$, so the final result is
\begin{align}
	\bar V^\alpha &= \left(1+\frac{1}{2}\omega_U^2\dot U^2\right)U^\alpha-\omega_U\dot U^\alpha +\omega_UK^\beta\dot U_\beta \dot U^\alpha \nonumber\\
    &\;+2\left(\omega_U\ddot U^\beta K_\beta-\omega_U^2\dot U^2\right)E^\alpha - \frac{1}{2}R^\beta_{KUK}h^\alpha_\beta+\mathcal{O}(\omega_U^3)\nonumber\\
	&= \left(1+\frac{1}{2}\omega_U^2\dot U^2\right)U^\alpha-\omega_U\dot U^\alpha +\omega_UK^\beta\dot U_\beta \dot U^\alpha \nonumber\\
    &\qquad+2\omega_U^2 \ddot U^\beta E_\beta E^\alpha - \frac{1}{2}R^\beta_{KUK}h^\alpha_\beta+\mathcal{O}(\omega_U^3). \label{tempp147}
\end{align}
This is the required 4-velocity for an object to appear static, with respect to the luminosity distance. Note the appearance of the jerk $\ddot U^\alpha$ of the observer in this condition, and the similarity of the above to equation \eqref{temp235}. Thus, in addition to the fact that optical velocity is not symmetric due to its very definition, the above analysis also shows that if one observer perceives another as stationary for an extended duration of time, the other will not make the same conclusion in general. Furthermore, as \eqref{temp235} and \eqref{tempp147} suggest, it is possible for the observer to perceive the object as stationary with respect to one optical distance, but moving radially with respect to another.

\section{\label{sec:acc}Acceleration and the Generalized Geodesic Deviation Equation}

With the hard work already done, we are in the position to define the observed acceleration for each type of velocity. Under any definition discussed above, the velocity vector of a moving object in the frame of an observer with 4-velocity $U^\alpha$ can be defined in each event along the world line of the observer, producing the velocity vector field $\vec v^\alpha(t)$ that is confined to the space of the observer, such that $U^\alpha(t)\vec v_\alpha(t)=0$ for all proper time $t$. Accordingly, we define the observed acceleration of an object as the Fermi derivative of the velocity vector, which is the rate of change of $\vec v^\alpha(t)$ within the space frame of $U^\alpha$.
\begin{equation}
	\vec a^\alpha = \dot{\vec v}^\alpha - \vec v^\beta \dot U_\beta U^\alpha,
\end{equation}
and since $U^\alpha(t)\vec v_\alpha(t)=0 \implies \dot U^\alpha(t)\vec v_\alpha(t)=-U^\alpha(t)\dot{\vec v}_\alpha(t)$, we have
\begin{equation}
	\vec a^\alpha = \dot{\vec v}^\alpha + \dot{\vec v}^\beta U_\beta U^\alpha = \dot{ \vec v}^\beta h^\alpha_\beta.
\end{equation}
The task now is to differentiate the velocity expressions along the observer's world line and project the results with $h^\alpha_\beta$.

For the three velocities defined through the physical position vector $D^\alpha$, we differentiate \eqref{tempp78} and adopt it to each case.
\begin{align}
	\dot{\vec v}^\beta h^\alpha_\beta &= \ddot\tau \bar V^\beta h^\alpha_\beta + \dot\tau \dot{\bar V}^\beta h^\alpha_\beta + \dot \tau \bar V^\gamma \left( \dot U_\gamma U^\beta +U_\gamma \dot U^\beta \right) h^\alpha_\beta\nonumber\\
	&\qquad + \frac{1}{6}\dot\tau \left(\bar V^\beta-\gamma U^\beta\right)\left(R^\epsilon_{\dot D\beta D} + R^\epsilon_{D\beta \dot D}\right)h^\alpha_\epsilon \nonumber\\
    &\hspace{3.7cm}- \left(\bar{V}^\gamma_S h^\beta_\gamma\right)^{\bullet}h^\alpha_\beta+\mathcal{O}(D^2)\nonumber\\
	&= \ddot\tau \bar V^\beta h^\alpha_\beta + \dot\tau \left(\dot \tau\bar{\dot V}^\beta+\frac{1}{2}R^\beta_{\bar V U D}+\frac{1}{2}\dot\tau R^\beta_{\bar V \bar V D}\right) h^\alpha_\beta \nonumber\\
    &\quad + \dot \tau \bar V^\beta U_\beta \dot U^\alpha- \left(\bar{V}^\gamma_S h^\beta_\gamma\right)^{\bullet}h^\alpha_\beta + \frac{1}{6}\dot\tau \bigg(\dot\tau R^\epsilon_{\bar V\bar V D}\nonumber\\
    &\quad -\dot\tau\gamma R^\epsilon_{\bar V U D}- R^\epsilon_{U\bar V D}+\gamma R^\epsilon_{U U D} - \dot\tau\gamma R^\epsilon_{DU \bar V} \nonumber\\
    &\hspace{3.8cm} -R^\epsilon_{D \bar V U}\bigg)h^\alpha_\epsilon +\mathcal{O}(D^2)\nonumber\\
	&= \dot\tau^2\bar{\dot V}^\beta h^\alpha_\beta - \left(1+\dot U^\beta D_\beta\right) \dot U^\alpha +\ddot\tau \bar V^\beta h^\alpha_\beta \nonumber\\
    &\quad- \left(\bar{V}^\gamma_S h^\beta_\gamma\right)^{\bullet}h^\alpha_\beta - \bigg(\frac{2}{3}\dot\tau^2 R^\beta_{\bar V D \bar V} +\frac{1}{3}\dot\tau R^\beta_{\bar V D U}\nonumber\\
    &\quad\quad-\frac{1}{6}\dot\tau R^\beta_{UD\bar V}+ \frac{1}{6}R^\beta_{UDU}\bigg)h^\alpha_\beta+\mathcal{O}(D^2).
\end{align}
(The $H^\alpha_\beta$'s appearing in \eqref{tempp78} have been converted from $h^\alpha_\beta$ without changing the result. It is advantageous to convert the $H^\alpha_\beta$'s back to $h^\alpha_\beta$ before differentiating to save algebra and maintain correctness for the given accuracy.) In the above $\dot \tau$ and $\ddot\tau$ are given by \eqref{taudot} and \eqref{taudotdot}, respectively; and the 4-velocities $\bar{V}^\alpha_S$ of static observers under each definition are given by \eqref{temp79}, \eqref{temp81}, and \eqref{temp84}. We get the following results for the three cases.\\

\paragraph*{Fermi relative acceleration:}
\begin{align}
	\left(\bar{V}^\gamma_S h^\beta_\gamma\right)^{\bullet}h^\alpha_\beta &= \left(-\frac{1}{2}R^\gamma_{DUD}h^\beta_\gamma\right)^{\bullet}h^\alpha_\beta +\mathcal{O}(D^2)\nonumber\\
	&= \left(\dot \tau R^\beta_{\bar V DU}-\frac{1}{2}\dot\tau R^\beta_{UD\bar V}-\frac{1}{2}R^\beta_{UDU}\right)h^\alpha_\beta\nonumber\\
    &\hspace{4cm}+\mathcal{O}(D^2).
\end{align}
Therefore,
\begin{align}
	\vec a^\alpha &=  \dot\tau^2\bar{\dot V}^\beta h^\alpha_\beta - \left(1+\dot U^\beta D_\beta\right) \dot U^\alpha +\ddot\tau \bar V^\beta h^\alpha_\beta \nonumber\\
	&- \left(\frac{2}{3}\dot\tau^2 R^\beta_{\bar V D \bar V} \!+\!\frac{4}{3}\dot\tau R^\beta_{\bar V D U}\!-\!\frac{2}{3}\dot\tau R^\beta_{UD\bar V}\!- \!\frac{1}{3}R^\beta_{UDU}\!\right)\!h^\alpha_\beta\nonumber\\
    &\hspace{5.5cm}+\mathcal{O}(D^2).
\end{align}\\

\paragraph*{Symmetric relative acceleration:}
\begin{equation}
	\left(\bar{V}^\gamma_S h^\beta_\gamma\right)^{\bullet}h^\alpha_\beta = \left(0\right)^{\bullet}h^\alpha_\beta=0.
\end{equation}
Therefore,
\begin{align}
	\vec a^\alpha &=  \dot\tau^2\bar{\dot V}^\beta h^\alpha_\beta - \left(1+\dot U^\beta D_\beta\right) \dot U^\alpha +\ddot\tau \bar V^\beta h^\alpha_\beta\nonumber\\
	&- \left(\frac{2}{3}\dot\tau^2 R^\beta_{\bar V D \bar V} +\frac{1}{3}\dot\tau R^\beta_{\bar V D U}-\frac{1}{6}\dot\tau R^\beta_{UD\bar V}+ \frac{1}{6}R^\beta_{UDU}\!\right)\!h^\alpha_\beta\nonumber\\
    &\hspace{5cm}+\mathcal{O}(D^2).
\end{align}\\

\paragraph*{Optical relative acceleration (w.r.t physical position vector):}
\begin{align}
	&\left(\bar{V}^\gamma_S h^\beta_\gamma\right)^{\bullet}h^\alpha_\beta = \left(\!-\frac{1}{2}D^2\ddot U^\gamma H^\beta_\gamma\!-\!\frac{1}{2}R^\gamma_{DUD}h^\beta_\gamma\!\right)^{\bullet}\!h^\alpha_\beta\!+\!\mathcal{O}(D^2)\nonumber\\
	&\quad= -D\dot D\ddot U^\beta H^\alpha_\beta + \left(\!\dot \tau R^\beta_{\bar V DU}\!-\!\frac{1}{2}\dot\tau R^\beta_{UD\bar V}\!-\!\frac{1}{2}R^\beta_{UDU}\!\right)\!h^\alpha_\beta\nonumber\\
    &\hspace{6.9cm}+\mathcal{O}(D^2)\nonumber\\
	&\quad= \!-\dot\tau\bar V^\gamma D_\gamma \ddot U^\beta H^\alpha_\beta \!+\!\! \left(\!\!\dot \tau R^\beta_{\bar V DU}\!-\!\frac{1}{2}\dot\tau R^\beta_{UD\bar V}\!-\!\frac{1}{2}R^\beta_{UDU}\!\!\right)\!h^\alpha_\beta\nonumber\\
    &\hspace{5.8cm}+\mathcal{O}(D^2).
\end{align}
Therefore,
\begin{align}
	\vec a^\alpha &=  \dot\tau^2\bar{\dot V}^\beta h^\alpha_\beta \!-\! \left(\!1\!+\!\dot U^\beta D_\beta\!\right) \!\dot U^\alpha \!+\!\ddot\tau \bar V^\beta h^\alpha_\beta \!+\!\dot\tau\bar V^\gamma D_\gamma \ddot U^\beta H^\alpha_\beta\nonumber\\
	&\;-\! \left(\!\frac{2}{3}\dot\tau^2 R^\beta_{\bar V D \bar V} \!+\!\frac{4}{3}\dot\tau R^\beta_{\bar V D U}\!-\!\frac{2}{3}\dot\tau R^\beta_{UD\bar V}\!- \!\frac{1}{3}R^\beta_{UDU}\!\right)\!h^\alpha_\beta\nonumber\\
    &\hspace{5cm}+\mathcal{O}(D^2).
\end{align}\\

For any of the above definitions, when the object is momentarily static so that the corresponding $\bar V^\alpha$ is given by \eqref{temp79}, \eqref{temp81}, or \eqref{temp84}, and in addition both observer and object travel on geodesics, so that $\dot U^\alpha = \bar{\dot V}^\alpha=0$, the accelerations reduce to
\begin{equation}
	\vec a^\alpha = -R^\alpha_{UDU}+\mathcal{O}(D^2),
\end{equation}
the well known geodesic deviation equation. This relationship is usually derived for a congruence of geodesics, so that $\bar V^\alpha=U^\alpha +\mathcal{O}(D)$, see for example \cite{rindler,synge}; and the result is expressed for $\ddot D^\alpha$, which is equivalent to our $\vec a^\alpha$ under the given circumstances. The method that we used to arrive at the above relationship allows us to generalize it by setting $\dot U^\alpha = \bar{\dot V}^\alpha=0$ in each of the acceleration expressions while allowing $\bar V^\alpha$ to remain general. The most interesting result is that for the Fermi relative acceleration, since then $\vec a^\alpha=\ddot D^\alpha$ even for arbitrary $\bar V^\alpha$, and the outcome can be considered as a generalization of the geodesic deviation equation. (Optical and Fermi relative accelerations are also equal in this case.) Thus, for two observers in geodesic motion that come sufficiently close, the acceleration of the connecting vector between them is the Fermi relative acceleration given by,
\begin{multline}
	\ddot D^\alpha = \frac{2}{3\gamma^2}\left(\frac{1}{\gamma}\bar V^\epsilon +2U^\epsilon\right)R_{\epsilon DU\bar V} \bar V^\beta h^\alpha_\beta \\
    - \left(\frac{2}{3\gamma^2} R^\beta_{\bar V D \bar V} +\frac{4}{3\gamma} R^\beta_{\bar V D U}-\frac{2}{3\gamma} R^\beta_{UD\bar V}- \frac{1}{3}R^\beta_{UDU}\right)h^\alpha_\beta\\
    +\mathcal{O}(D^2).
\end{multline}
The above does not rely on a congruence of geodesics and can be used for any relative orientation of $V^\alpha$ and $U^\alpha$. That is, while the geodesic deviation equation is applicable to `nearly parallel' observers (see \cite{synge2}), the generalized geodesic deviation equation is applicable to any observers in geodesic motion. Furthermore, it can be shown that the above is also applicable to any types of geodesics (the derivation is identical), but requires a modified interpretation of the connecting vector and its rate of change.\\

For the proper optical velocities of section \ref{sec:defopt}, differentiating and projecting \eqref{temp207} gives
\begin{multline}
	\dot{\vec v}^\beta_O h^\alpha_\beta = \left(\frac{D_O}{\omega_U}\right)^{\bullet\bullet}K^\beta h^\alpha_\beta +\left(\frac{D_O}{\omega_U}\right)^{\bullet}\dot K^\beta h^\alpha_\beta\\
    +2\left(\frac{D_O}{\omega_U}\right)^{\bullet}\omega_U\dot U^\alpha + \left(\frac{D_O}{\omega_U}\right)^{\bullet}\frac{\omega_U}{\omega_V}\bar V^\beta h^\alpha_\beta \\
    +\frac{D_O}{\omega_U}\Bigg(\left(\frac{\omega_U}{\omega_V}\right)^{\bullet}\bar V^\beta+\frac{\omega_U}{\omega_V}\bigg(\dot\tau\bar{\dot V}^\beta+\frac{1}{2}R^\beta_{\bar VUK}\\
    +\frac{1}{2}\dot\tau R^\beta_{\bar V\bar VK}\bigg)\Bigg) h^\alpha_\beta -\frac{D_O}{\omega_U}\frac{\omega_U}{\omega_V}\gamma\dot U^\alpha\\
    +\frac{1}{6}\frac{D_O}{\omega_U}\left(\frac{\omega_U}{\omega_V}\bar V^\gamma+2U^\gamma\right)\left(R^\beta_{\dot K\gamma K}+R^\beta_{K\gamma \dot K}\right)h^\alpha_\beta\\
    +\frac{D_O}{\omega_U}\left(\dot\omega_U\dot U^\alpha+\omega_U\ddot U^\beta h^\alpha_\beta\right)+\mathcal{O}(\omega_U^2),
\end{multline}
where in this case $\dot\tau$ is given by \eqref{temp206} ($=\frac{\omega_U}{\omega_V}$). With the help of \eqref{temp218} we have
\begin{multline}
	\left(\frac{\omega_U}{\omega_V}\right)^{\bullet} \!\!= \!\left(\frac{\omega_U}{\omega_V}\right)^2\!\frac{1}{\omega_U}\!\Bigg(\!2\gamma \!-\!\frac{\omega_V}{\omega_U}\!\left(\!1\!+\!K^\alpha \dot U_\alpha\!+\!\frac{1}{3}R_{UKUK}\!\right)\\
    -\frac{\omega_U}{\omega_V}\left(1-K^\alpha \bar {\dot V}_\alpha+\frac{1}{3}R_{\bar VK\bar VK}\right) -\frac{1}{3}R_{\bar V KUK}\Bigg)\\
    +\mathcal{O}(\omega_U^2).
\end{multline}
Therefore,
\begin{multline}
	\vec a^\alpha_{O} = \left(\frac{D_O}{\omega_U}\right)^{\bullet\bullet}\!\!K^\beta h^\alpha_\beta \!+\!\left(\frac{D_O}{\omega_U}\right)^{\bullet}\!\dot K^\beta h^\alpha_\beta\!+\!2\left(\frac{D_O}{\omega_U}\right)^{\bullet}\!\omega_U\dot U^\alpha\\
    + \left(\frac{D_O}{\omega_U}\right)^{\bullet}\frac{\omega_U}{\omega_V}\bar V^\beta h^\alpha_\beta +\frac{D_O}{\omega_U}  \left(\frac{\omega_U}{\omega_V}\right)^2\frac{1}{\omega_U}\Bigg(2\gamma \\
    -\frac{\omega_V}{\omega_U}\left(1+K^\alpha \dot U_\alpha+\frac{1}{3}R_{UKUK}\right)\\
    -\frac{\omega_U}{\omega_V}\left(1-K^\alpha \bar {\dot V}_\alpha+\frac{1}{3}R_{\bar VK\bar VK}\right) -\frac{1}{3}R_{\bar V KUK}\Bigg) \bar V^\beta h^\alpha_\beta\\
    +\frac{D_O}{\omega_U}\left(\frac{\omega_U}{\omega_V}\right)^2\left(\bar{\dot V}^\beta-\frac{1}{2}\frac{\omega_V}{\omega_U}R^\beta_{\bar VKU}-\frac{1}{2}R^\beta_{\bar VK\bar V}\right) h^\alpha_\beta\\ -2\frac{D_O}{\omega_U}\frac{\omega_U}{\omega_V}\gamma\dot U^\alpha -\frac{1}{6}\frac{D_O}{\omega_U}\Bigg(\left(\frac{\omega_U}{\omega_V}\right)^2 R^\beta_{\bar VK\bar V} \\
    + 5\frac{\omega_U}{\omega_V}R^\beta_{\bar VKU}-4\frac{\omega_U}{\omega_V}R^\beta_{UK\bar V}-2R^\beta_{UKU}\Bigg)h^\alpha_\beta \\
    +\frac{D_O}{\omega_U}\left(\left(1+\dot U^\beta K_\beta\right)\dot U^\alpha+\omega_U\ddot U^\beta h^\alpha_\beta\right)+\mathcal{O}(\omega_U^2).
\end{multline}
The above can be adopted to $D_O$ being either the angular diameter distance or the luminosity distance. To this end, we must establish the second derivatives of $\frac{D_A}{\omega_U}$ and $\frac{D_L}{\omega_U}$ by differentiating \eqref{temp214} and \eqref{temp217}, and then substitute for $\frac{D_O}{\omega_U}$ and its derivatives in the above for each type of distance. The task is straightforward but the explicit results are very long and not particularly illuminating, so we leave the above as the general expression for proper optical acceleration, with reference to the already established relationships for $\frac{D_A}{\omega_U}$ and $\frac{D_L}{\omega_U}$.\\

Of the velocities defined, the ones that are particularly useful for realistic observations and experiments are the Fermi relative velocity and the proper optical velocities. These are based on real measurements that can be made directly by the observer. Therefore, the corresponding accelerations to these velocities are of main interest.

\section{\label{sec:dis}Discussion and Conclusion}

We make a final comparison between the distances and velocities defined in this work and the ones mentioned in the introduction. For distances and velocities with respect to spacelike simultaneity we have the following. Our definition of the Fermi distance $D$ coincides with all such definitions in the literature, and since it is the magnitude of the physical position vector $D^\alpha$ given by \eqref{DofA}, it can be established for a given setup at any event through the method described in section \ref{sec:phys}. What is defined as the \textit{kinematic} relative velocity in \cite{bolos}, in our notation is given by
\begin{equation}
	\vec v^\alpha = \frac{1}{\gamma}h^\alpha_\beta \bar V^\beta,
\end{equation}
where $\bar V^\alpha$ is the parallel transport with respect to the spacelike connecting geodesic. For a given model the above can be calculated at an event by constructing the connecting geodesic and the parallel transport $\bar V^\alpha$ through the methods outlined in appendix \ref{app:deri}. While this velocity is a well defined mathematical measure of relative motion, and can be indirectly found through other measurements, in itself it is not a velocity that can be directly observed. On the other hand, what is defined as Fermi velocity can be observed within an extended frame. As is the case for the Fermi distance, the definition of Fermi velocity seems to coincide through the literature. We have solved the problem of finding a general expression for it (as posed by \cite{bolos3}), which is given by equation \eqref{temp80}, and which explicitly reveals effects of curvature and relative motion on velocity measurements. Our definitions of symmetric and optical relative velocities given by equations \eqref{temp82} and \eqref{temp85} are also based on spacelike simultaneity but are not found in other works on the subject. They may be of use when analyzing local relative motion as these can also be measured; they coincide with Fermi velocity in the limit of small distance, or in case of no curvature or jerk effects.

For distances and velocities with respect to the past light cone, sometimes referred to as lightlike simultaneity, \cite{bolos6}, we have the following comparisons. What we call the projected optical distance (given by $|U^\alpha K_\alpha|$, see section \ref{sec:opt}) is referred to as `affine distance' in \cite{bolos,bolos6}, and the concept is used in other sources, for example \cite{kermack}. While it provides a relative mathematical measure of how far a photon has traveled along a null geodesic relative to an observer, it cannot be observed directly and can only be found through calculation. For distances based on optical observations, the angular diameter distance and the luminosity distance play a more important role when considering possible measurements. We provided general expressions for these measurable distances, which can predict observations given any model. We then developed relationships between all three optical distances, and defined the proper optical velocity based on the rate of change of the optical position vector.

The \textit{spectroscopic} relative velocity of \cite{bolos} is defined similar to the \textit{kinematic} and also expressed through an equation of the above form, but with $\bar V^\alpha$ being the parallel transport with respect to the past null connecting geodesic (see Figure \ref{fig:sim} and its description). It can be calculated in the same way by utilizing some of the results derived in appendix \ref{app:deri}, but is also a velocity that cannot be directly measured, and also does not account for acceleration and curvature effects for the same reasons as the \textit{kinematic} relative velocity. The relative velocity defined in \cite{synge2} is exactly the \textit{spectroscopic} multiplied by a factor of $\gamma$, to which the same comments apply. Finally, the \textit{astrometric} relative velocity is a step in the direction of measurability analogous to the Fermi relative velocity. Its definition is equivalent to our general definition of proper optical velocity, but particularly with respect to the projected optical distance. Therefore, the general expression for the \textit{astrometric} relative velocity is provided by (see equation \eqref{temp210})
\begin{multline}
	\vec v^\alpha_O \!=\! \frac{\omega_U}{\omega_V} \bar V^\beta h^\alpha_\beta\!+\!\omega_U \dot U^\alpha\!+\!\frac{1}{6}\frac{\omega_U}{\omega_V}\bar V^\epsilon H^\gamma_\epsilon R^\beta_{K\gamma K} h^\alpha_\beta \!+\! \frac{1}{2}R^\beta_{KUK} h^\alpha_\beta\\
    +\mathcal{O}(\omega_U^3).
\end{multline}
However, the advantage of the proper optical velocity (with respect to angular diameter distance or luminosity distance) over the one given above is in the fact that it can be established directly through measuring the change in visibility direction and optical distance. While the direction of visibility, angular diameter distance, and luminosity distance, are directly measurable, the projected optical distance is not. So the optical position vector can always be determined and its rate of change can be measured, where as the velocity given by the above can only be calculated.

We can consider the most important results presented in this work to be the following: the general expression for measurable intersection angles and aberration relationships of section \ref{sec:ang}; the general expression for Fermi velocity and acceleration of sections \ref{sec:vel} and \ref{sec:acc}; the general expressions for the angular diameter distance and the luminosity distance of section \ref{sec:opt}; and the general expressions for proper optical velocity with respect to the optical distances as well as the corresponding accelerations of sections \ref{sec:opt} and \ref{sec:acc}. From the four relative velocities presented in \cite{bolos} and studied thereafter, we propose reducing the set to three, keeping only the Fermi relative velocity from the original set, as it is the only one which can be calculated and compared to direct observations. In addition to Fermi velocity, the set should include the proper optical velocities of section \ref{sec:opt}, which are the most relevant ones for astronomical measurements. The general expressions for these velocities and their corresponding accelerations are fundamental for understanding kinematics in curved spacetime and allow for direct calculations.

We should clarify the meaning of smallness in the various approximations involving distance and time throughout this work. While the distances and time lapses may be very large in some coordinates and units, the assumption is that the combination of curvature terms with these quantities is overall very small. Specifically, it is the combination of the Riemann tensor components with the square of the distances (or time) that is assumed to be small. This means that the distances (or time lapses) in the approximations must be small in comparison to distances over which curvature effects are significant. Therefore, if it is assumed that curvature only weakly affects the phenomena under investigation, and it is sufficient to only expose the strongest contributions of curvature, then the truncations in the approximations are justified. In particular, for the simple case of de Sitter spacetime, the combined curvature terms with the distances will be of the form $\Lambda D^2$, where $\Lambda$ is the cosmological constant and $D$ is the distance in the approximation (or time lapse). So as long as the distance is much smaller than the cosmological horizon, the truncations are justified. For the simple case of Schwarzschild, the combined terms would be of the form $\frac{m}{r^3}D^2$, where $m$ is the mass parameter and $r$ is the Schwarzschild radial coordinate. So even when $D$ (the Fermi distance between observers for example) is comparable to $r$ (the coordinate location of observation event), as long as $r \gg m$ the truncations are justified. In cases when curvature effects are strong or when the lowest order terms disappear, most approximations we have used can be easily adjusted to higher accuracy. If the curvature changes drastically throughout the path of the connecting geodesic (null or not), then higher order terms must be considered, since the derivatives of the Riemann tensor would strongly contribute to the outcome.

The material presented in this work along with the method of analysis can be extended to further investigations. For example, the concept of `radial velocity', discussed at length in the literature, \cite{klein2,nord,terk,terk2,but}, can be made more rigorous by means of our general expressions. The results for velocity, acceleration, and optical distances can be applied to known cosmological models or other spacetimes, and be used in studying some interesting setups. Furthermore, the general expressions for acceleration will be useful in clarifying the decompositions of observed acceleration in attempts to study the so called `dark force', as has been recently done in some sources, see \cite{zhang,ho} for example. As for the general case, an interesting question to ask is, to what extent can an observer determine the Fermi distance and relative velocity of a distant object through purely optical measurements? In other words, for example, knowing the measurable optical distances and velocities, can an observer calculate the (current) Fermi distance to the object? To get the general relationship for that would require connecting the expressions for the Fermi distance to the optical quantities through the 4-velocity of the object and its derivatives, which should be possible. Additionally, equations \eqref{ddot}, \eqref{taudot} and the expression for Fermi relative velocity reveal that for non-zero acceleration of the observer, some objects can have relative velocities with corresponding speeds exceeding the speed of light; while there are also events where the relative velocity is zero for any 4-velocity of an object located there. These outcomes bring about the concept of emergent horizons due to acceleration, extensively discussed by Rindler, see \cite{rindler}. In fact, our results suggest that we can generalize the concept of Rindler horizons to arbitrary motion and curvature, and discuss the properties of the subspace constituted by all such (frozen) events for a given general timelike curve of an observer. Finally, our expansion of the geometric exponential map, given by \eqref{AofL}, allows for a natural construction of Fermi coordinates around any timelike curve; and since we can always improve the accuracy of \eqref{AofL}, we can construct higher order terms for the metric expressed in Fermi coordinates. This method of finding higher order terms is more intuitive and mathematically simpler than what is commonly found in the literature, see for example \cite{man,mtw,lini}, and might shed light on a simple algorithmic way for generating higher order terms in the metric expression.

\begin{widetext}

\appendix
\section{\label{app:deri}Derivations I}

\paragraph*{\textbf{Derivation of $\Lambda^\alpha = A^\alpha + \frac{1}{2}\Gamma^{\alpha}_{\beta\gamma}A^\beta A^\gamma + \frac{1}{6}(\partial_\beta\Gamma^{\alpha}_{\gamma\epsilon} + \Gamma^{\alpha}_{\beta\rho}\Gamma^{\rho}_{\gamma\epsilon})A^\beta A^\gamma A^\epsilon + ...$}}\textbf{:}~\\

Consider two nearby events $x^\alpha_t$ and $x^\alpha_\tau$ with a unique geodesic connecting them. Let this geodesic be parametrized by an affine parameter $\lambda$ and have tangent $\Lambda^\alpha=\frac{dx^{\alpha}}{d\lambda}$, with $x^{\alpha}(\lambda=0)=x^{\alpha}_t$ and $x^{\alpha}(\lambda)=x^{\alpha}_\tau$ for some small $\lambda$.
\begin{equation}
	x^{\alpha}_\tau = x^{\alpha}(\lambda) = x^{\alpha}(\lambda=0) +\frac{dx^{\alpha}}{d\lambda}\lambda + \frac{1}{2}\frac{d^2x^{\alpha}}{d\lambda^2}\lambda^2 + \frac{1}{6}\frac{d^3x^{\alpha}}{d\lambda^3}\lambda^3 + ...
\end{equation}
Due to the geodesic nature of $x^{\alpha}(\lambda)$,
\begin{equation}
	\frac{d^2x^{\alpha}}{d\lambda^2}=\frac{d}{d\lambda}\Lambda^\alpha=\Lambda^\beta\partial_\beta\Lambda^\alpha=-\Gamma^{\alpha}_{\beta\gamma}\Lambda^\beta \Lambda^\gamma=-\Gamma^{\alpha}_{\Lambda\Lambda}
\end{equation}
and
\begin{equation}
	\frac{d^3x^{\alpha}}{d\lambda^3}=-\frac{d}{d\lambda}(\Gamma^{\alpha}_{\beta\gamma}\Lambda^\beta \Lambda^\gamma)= -\Lambda^\epsilon\partial_\epsilon\Gamma^{\alpha}_{\beta\gamma}\Lambda^\beta \Lambda^\gamma + 2\Gamma^{\alpha}_{\beta\rho}\Gamma^{\rho}_{\gamma\epsilon}\Lambda^\beta \Lambda^\gamma \Lambda^\epsilon= -\partial_\Lambda\Gamma^{\alpha}_{\Lambda\Lambda} + 2\Gamma^{\alpha}_{\Lambda\rho}\Gamma^{\rho}_{\Lambda\Lambda}.
\end{equation}
Therefore,
\begin{equation}
	x^{\alpha}_\tau =  x^{\alpha}_t + \Lambda^\alpha \lambda - \frac{1}{2}\Gamma^{\alpha}_{\Lambda\Lambda}\lambda^2 - \frac{1}{6}(\partial_\Lambda\Gamma^{\alpha}_{\Lambda\Lambda} - 2\Gamma^{\alpha}_{\Lambda\rho}\Gamma^{\rho}_{\Lambda\Lambda})\lambda^3 + ...
\end{equation}
Defining $A^\alpha = x^{\alpha}_\tau -  x^{\alpha}_t$ for brevity, noting it is not a vector,
\begin{equation}
	A^\alpha = \Lambda^\alpha \lambda - \frac{1}{2}\Gamma^{\alpha}_{\Lambda\Lambda}\lambda^2 - \frac{1}{6}(\partial_\Lambda\Gamma^{\alpha}_{\Lambda\Lambda} - 2\Gamma^{\alpha}_{\Lambda\rho}\Gamma^{\rho}_{\Lambda\Lambda})\lambda^3 + ...
\end{equation}
Inverting for $\Lambda^\alpha\lambda$ (with the assumption of smallness in all $A^\alpha$ and $\Lambda^\alpha\lambda$),
\begin{align}
	\Lambda^\alpha\lambda &= A^\alpha + \frac{1}{2}\Gamma^{\alpha}_{\beta\gamma}A^\beta A^\gamma + \frac{1}{6}(\partial_\beta\Gamma^{\alpha}_{\gamma\epsilon} + \Gamma^{\alpha}_{\beta\rho}\Gamma^{\rho}_{\gamma\epsilon})A^\beta A^\gamma A^\epsilon + ...\nonumber\\
	&=A^\alpha + \frac{1}{2}\Gamma^{\alpha}_{AA} + \frac{1}{6}(\partial_A\Gamma^{\alpha}_{AA} + \Gamma^{\alpha}_{A\rho}\Gamma^{\rho}_{AA}) + ...
\end{align}
Reparametrization of $\lambda$ to another affine parameter that is also zero at $x^\alpha_t$ does not affect the combination $\Lambda^\alpha\lambda$ ($=\frac{dx^\alpha}{d\lambda}\lambda$). In fact, $|\Lambda|\lambda$ is the small metric distance between the two events. If it is not null, then the arc length parametrization makes the latter claim clear (in which case $|\Lambda|=\pm 1$). For an equivalent approach to either case, null or not, we can always parametrize so that $x^{\alpha}(\lambda=1)=x^{\alpha}_\tau$. Then $|\Lambda|$ is the small (or null) metric distance between the events. This reasoning allows finding Synge's world function for any metric, see \cite{synge2}. With this parametrization
\begin{equation}
	A^\alpha = \Lambda^\alpha - \frac{1}{2}\Gamma^{\alpha}_{\Lambda\Lambda} - \frac{1}{6}(\partial_\Lambda\Gamma^{\alpha}_{\Lambda\Lambda} - 2\Gamma^{\alpha}_{\Lambda\rho}\Gamma^{\rho}_{\Lambda\Lambda}) + ...\,, \label{AofL}
\end{equation}
and
\begin{equation}
	\Lambda^\alpha =A^\alpha + \frac{1}{2}\Gamma^{\alpha}_{AA} + \frac{1}{6}(\partial_A\Gamma^{\alpha}_{AA} + \Gamma^{\alpha}_{A\rho}\Gamma^{\rho}_{AA}) + ... \label{LofA}
\end{equation}
Thus, for any close by events $x^{\alpha}_t$ and $x^{\alpha}_\tau$ the above is an expression for the tangent $\Lambda^\alpha$ to the connecting geodesic at $x^{\alpha}_t$, with magnitude that equals the metric distance between the events. This unique parametrization of the connecting geodesic, often called the normalized affine parametrization, was used by Synge in \cite{synge2}, and by Hawking and Ellis in \cite{hawk}. We will make frequent use of it throughout. Finally, \eqref{AofL} and \eqref{LofA} are in fact third order expansions of the geometric exponential map and its inverse, respectively. They can be used to quickly build local Fermi coordinates without use of the geodesic deviation equation and the accuracy can easily be extended to higher orders as the situation requires.\\

\paragraph*{\textbf{Derivation of $[\tau \to t]^\alpha_\beta=\delta^\alpha_\beta + \Gamma^\alpha_{\Lambda\beta} + \frac{1}{2}\left(\partial_\Lambda\Gamma^\alpha_{\Lambda\beta} - \Gamma^\alpha_{\beta\rho}\Gamma^\rho_{\Lambda\Lambda} + \Gamma^\alpha_{\Lambda\rho}\Gamma^\rho_{\Lambda\beta}\right)+...$}}\textbf{:}\\

Consider a vector $A^\alpha$ at $x^\alpha_\tau$ and its parallel transport $\bar A^\alpha$ to $x^\alpha_t$, along the connecting geodesic discussed above. Let $\tilde A^\alpha(\lambda)$ be a constant vector field on the geodesic ($\lambda$ an affine parameter), such that $\tilde A^\alpha(\lambda=0)=\bar A^\alpha$ and $\tilde A^\alpha(\lambda)=A^\alpha$ for the small value of $\lambda$ at which $x^\alpha(\lambda)=x^\alpha_\tau$.
\begin{equation}
	A^{\alpha} = \tilde A^{\alpha}(\lambda) = \tilde A^{\alpha}(\lambda=0) + \frac{d\tilde A^{\alpha}}{d\lambda}\lambda + \frac{1}{2}\frac{d^2\tilde A^{\alpha}}{d\lambda^2}\lambda^2 + ...
\end{equation}
Since $\tilde A^\alpha$ is a constant vector,
\begin{equation}
	\frac{d\tilde A^{\alpha}}{d\lambda} = \Lambda^\beta\partial_\beta\tilde A^\alpha = -\Gamma^\alpha_{\beta\gamma}\Lambda^\beta\tilde A^{\gamma}=-\Gamma^\alpha_{\Lambda\tilde A},
\end{equation}
\begin{align}
	\frac{d^2\tilde A^{\alpha}}{d\lambda^2} &= - \Lambda^\epsilon\partial_\epsilon\Gamma^\alpha_{\beta\gamma}\Lambda^\beta\tilde A^{\gamma} + \Gamma^\alpha_{\beta\gamma}\Gamma^\beta_{\Lambda\Lambda}\tilde A^{\gamma} + \Gamma^\alpha_{\beta\gamma}\Lambda^\beta\Gamma^\gamma_{\Lambda \tilde A}\nonumber\\
	&= - \partial_\Lambda\Gamma^\alpha_{\Lambda\tilde A} + \Gamma^\alpha_{\tilde A\rho}\Gamma^\rho_{\Lambda\Lambda} + \Gamma^\alpha_{\Lambda \rho}\Gamma^\rho_{\Lambda \tilde A}.
\end{align}
Therefore,
\begin{align}
	\nonumber A^{\alpha} &= \bar A^{\alpha} - \Gamma^\alpha_{\Lambda \bar A}\lambda - \frac{1}{2}\left(\partial_\Lambda\Gamma^\alpha_{\Lambda\bar A} - \Gamma^\alpha_{\bar A\rho}\Gamma^\rho_{\Lambda\Lambda} - \Gamma^\alpha_{\Lambda \rho}\Gamma^\rho_{\Lambda \bar A}\right) \lambda^2 + ...\\
	&= \bar A^{\beta}\left(\delta^\alpha_\beta - \Gamma^\alpha_{\Lambda\beta}\lambda - \frac{1}{2}\left(\partial_\Lambda\Gamma^\alpha_{\Lambda\beta} - \Gamma^\alpha_{\beta\rho}\Gamma^\rho_{\Lambda\Lambda} - \Gamma^\alpha_{\Lambda\rho}\Gamma^\rho_{\Lambda\beta}\right)\lambda^2\right) + ... \label{x0toxtau}
\end{align}
Inverting, assuming smallness in all $\Lambda^\alpha\lambda$,
\begin{equation}
	\bar A^{\alpha} = A^{\beta}\left(\delta^\alpha_\beta + \Gamma^\alpha_{\Lambda\beta}\lambda + \frac{1}{2}\left(\partial_\Lambda\Gamma^\alpha_{\Lambda\beta} - \Gamma^\alpha_{\beta\rho}\Gamma^\rho_{\Lambda\Lambda} + \Gamma^\alpha_{\Lambda\rho}\Gamma^\rho_{\Lambda\beta}\right)\lambda^2\right) + ...
\end{equation}
Again, reparametrizing so that $x^{\alpha}(\lambda=1)=x^{\alpha}_\tau$, which leaves $\Lambda^\alpha\lambda$ unchanged and makes $|\Lambda|$ the metric distance between the events (assumed to be small and allowed to be null).
\begin{equation}
	\bar A^{\alpha} = A^{\beta}\left(\delta^\alpha_\beta + \Gamma^\alpha_{\Lambda\beta} + \frac{1}{2}\left(\partial_\Lambda\Gamma^\alpha_{\Lambda\beta} - \Gamma^\alpha_{\beta\rho}\Gamma^\rho_{\Lambda\Lambda} + \Gamma^\alpha_{\Lambda\rho}\Gamma^\rho_{\Lambda\beta}\right)\right) + ...
\end{equation}
Defining $[\tau \to t]^\alpha_\beta$ as the parallel transport operator, we have
\begin{equation}
	[\tau \to t]^\alpha_\beta=\delta^\alpha_\beta + \Gamma^\alpha_{\Lambda\beta} + \frac{1}{2}\left(\partial_\Lambda\Gamma^\alpha_{\Lambda\beta} - \Gamma^\alpha_{\beta\rho}\Gamma^\rho_{\Lambda\Lambda} + \Gamma^\alpha_{\Lambda\rho}\Gamma^\rho_{\Lambda\beta}\right)+... \label{partrans}
\end{equation}
The components of this operator are expressed in terms of the tangent and the metric components at $x^\alpha_t$ by choice, but it operates on vectors at $x^\alpha_\tau$ and parallel transports them to $x^\alpha_t$. The tangent to the connecting geodesic $\Lambda^\alpha$ is given by \eqref{LofA} also in terms of metric components at $x^\alpha_t$ and with the normalized affine parametrization.\\

\paragraph*{\textbf{Derivation of $\dot D^\alpha = \dot \tau \bar V^\alpha - U^\alpha + \frac{1}{6}\left(\dot \tau \bar V^\beta + 2U^\beta\right)R^\alpha_{D\beta D}+\mathcal{O}(D^3)$}}\textbf{:}\\

With reference to the construction presented above, consider the case where the connecting geodesic is spacelike, and with the requirement that the tangent $\Lambda^\alpha$ at $x^\alpha_t$ is orthogonal to $U^\alpha$. In this case we relabel $\Lambda \to D$. The requirement $D^\alpha U_\alpha=0$ applied to \eqref{LofA} gives a relationship between $\tau$ and $t$. For a region in which the geodesics are sufficiently close, this $D^\alpha(t)$ is a uniquely constructed as a spacelike vector field within the space frame of $U^\alpha$ that connects $x^\alpha(t)$ to the corresponding $x^\alpha(\tau(t))$.

Differentiating \eqref{LofA},
\begin{align}
	\dot D^\alpha &= U^\beta \nabla_\beta D^\alpha\nonumber\\
	&= U^\beta \partial_\beta D^\alpha + \Gamma^\alpha_{\beta \gamma} U^\beta D^\gamma\nonumber\\
	&= U^\beta \partial_\beta \left( A^\alpha + \frac{1}{2}\Gamma^{\alpha}_{\epsilon\gamma}A^\epsilon A^\gamma + ... \right) + \Gamma^\alpha_{\beta \gamma} U^\beta \left( A^\gamma + \frac{1}{2}\Gamma^{\gamma}_{\epsilon\mu}A^\epsilon A^\mu + ... \right)\nonumber\\
	&=\! \partial_U A^\alpha \!+\! \frac{1}{2} \partial_U \Gamma^{\alpha}_{AA} \!+\! \Gamma^{\alpha}_{A \beta}\partial_U A^\beta \!+\! \frac{1}{6}(\partial_\beta\Gamma^{\alpha}_{AA} \!+\! \Gamma^{\alpha}_{\beta\rho}\Gamma^{\rho}_{AA})\partial_U A^\beta\!+\! \frac{1}{3}(\partial_\beta\Gamma^{\alpha}_{A \gamma} \!+\! \Gamma^{\alpha}_{\beta\rho}\Gamma^{\rho}_{A \gamma}) A^\beta \partial_U A^\gamma\! +\! \Gamma^\alpha_{UA} \!+\! \frac{1}{2}\Gamma^\alpha_{U \rho}\Gamma^{\rho}_{AA} \!+ ...\,,
\end{align}
where
\begin{align}
	\partial_U A^\alpha = U^\beta \partial_\beta A^\alpha &=\frac{dx^\beta}{dt}\frac{\partial}{\partial x^\beta}A^\alpha\nonumber\\
	&=\frac{d}{dt}(x^\alpha(\tau)-x^\alpha(t))\nonumber\\
	&=\frac{d\tau}{dt}\frac{dx^\alpha(\tau)}{d\tau}-\frac{dx^\alpha(t)}{dt}\nonumber\\
	&=\dot\tau V^\alpha - U^\alpha,
\end{align}
so
\begin{align}
	\dot D^\alpha &= \dot\tau V^\alpha - U^\alpha + \frac{1}{2} \partial_U \Gamma^{\alpha}_{AA} + \dot\tau\Gamma^{\alpha}_{AV} - \Gamma^{\alpha}_{AU} + \frac{1}{6}\dot\tau(\partial_\beta\Gamma^{\alpha}_{AA} + \Gamma^{\alpha}_{\beta\rho}\Gamma^{\rho}_{AA})V^\beta - \frac{1}{6}(\partial_U\Gamma^{\alpha}_{AA} + \Gamma^{\alpha}_{U\rho}\Gamma^{\rho}_{AA})\nonumber\\
	& \hspace{3cm}+ \frac{1}{3}\dot\tau(\partial_\beta\Gamma^{\alpha}_{A V} + \Gamma^{\alpha}_{\beta\rho}\Gamma^{\rho}_{A V}) A^\beta - \frac{1}{3}(\partial_\beta\Gamma^{\alpha}_{A U} + \Gamma^{\alpha}_{\beta\rho}\Gamma^{\rho}_{A U}) A^\beta + \Gamma^\alpha_{UA} + \frac{1}{2}\Gamma^\alpha_{U \rho}\Gamma^{\rho}_{AA} + ...
\end{align}
It is safe to define $D^\alpha=D\hat D^\alpha$, where $D=|D|$ is the (small) spacelike distance between the events, so it is evident that \eqref{AofL} is expressed up to terms of order $D^3$, and we can make reference to $D$ in the accuracy of our expressions. Replacing the remaining $A^\alpha$'s with $D^\alpha$'s by using \eqref{AofL},
\begin{align}
	\dot D^\alpha &= \dot\tau V^\alpha - U^\alpha + \frac{1}{2} \partial_U \Gamma^{\alpha}_{DD} + \dot\tau\Gamma^{\alpha}_{V\beta}\left(D^\beta-\frac{1}{2}\Gamma^\beta_{DD} \right) + \frac{1}{6}\dot\tau\left(\partial_\beta\Gamma^{\alpha}_{DD} + \Gamma^{\alpha}_{\beta\rho}\Gamma^{\rho}_{DD}\right)V^\beta - \frac{1}{6}\left(\partial_U\Gamma^{\alpha}_{DD} + \Gamma^{\alpha}_{U\rho}\Gamma^{\rho}_{DD}\right)\nonumber\\
	&\hspace{5cm}+ \frac{1}{3}\dot\tau\left(\partial_D\Gamma^{\alpha}_{DV} + \Gamma^{\alpha}_{D\rho}\Gamma^{\rho}_{DV}\right) - \frac{1}{3}\left(\partial_D\Gamma^{\alpha}_{DU}+\Gamma^{\alpha}_{D\rho}\Gamma^{\rho}_{DU}\right)+\frac{1}{2}\Gamma^\alpha_{U \rho}\Gamma^{\rho}_{DD}+\mathcal{O}(D^3)\nonumber\\
	&= \dot\tau V^\beta \left( \delta^\alpha_\beta + \Gamma^\alpha_{D \beta} + \frac{1}{6}\partial_\beta\Gamma^\alpha_{DD} + \frac{1}{3}\partial_D\Gamma^\alpha_{D\beta} - \frac{1}{3}\Gamma^\alpha_{\beta\rho}\Gamma^\rho_{DD} +  \frac{1}{3}\Gamma^\alpha_{D\rho}\Gamma^\rho_{D\beta}\right)\nonumber\\
	&\hspace{6.5cm}- U^\alpha + \frac{1}{3} \partial_U \Gamma^{\alpha}_{DD} - \frac{1}{3} \partial_D \Gamma^{\alpha}_{UD} + \frac{1}{3}\Gamma^\alpha_{U \rho}\Gamma^{\rho}_{DD} - \frac{1}{3}\Gamma^\alpha_{D\rho}\Gamma^{\rho}_{UD} +\mathcal{O}(D^3)\nonumber\\
	&= \dot\tau V^\beta \left( \delta^\alpha_\beta + \Gamma^\alpha_{D \beta} + \frac{1}{6}R^\alpha_{D\beta D} + \frac{1}{2}\partial_D\Gamma^\alpha_{D\beta} - \frac{1}{2}\Gamma^\alpha_{\beta\rho}\Gamma^\rho_{DD} +  \frac{1}{2}\Gamma^\alpha_{D\rho}\Gamma^\rho_{D\beta}\right) - U^\alpha + \frac{1}{3}R^\alpha_{DUD} +\mathcal{O}(D^3),
\end{align}
where $R^\alpha_{\beta\gamma\epsilon}$ is the Riemann curvature tensor, defined by
\begin{equation}
	R^\alpha_{\beta\gamma\epsilon} = \partial_\gamma \Gamma^{\alpha}_{\epsilon\beta} - \partial_\epsilon \Gamma^{\alpha}_{\gamma\beta} + \Gamma^\alpha_{\gamma \rho}\Gamma^{\rho}_{\epsilon\beta} - \Gamma^\alpha_{\epsilon\rho}\Gamma^{\rho}_{\gamma\beta}. \label{riem}
\end{equation}
Using \eqref{partrans} and defining $\bar V^\alpha = [\tau \to t]^\alpha_\beta V^\beta$,
\begin{equation}
	\dot D^\alpha = \dot \tau \bar V^\alpha - U^\alpha + \frac{1}{6}\left(\dot\tau \bar V^\beta + 2U^\beta\right)R^\alpha_{D\beta D}+\mathcal{O}(D^3). \label{ddotap}
\end{equation}

An expression for $\dot \tau$ can be established by differentiating the condition $D^\alpha U_\alpha=0$ and applying to the above. $\dot D^\alpha U_\alpha = -D^\alpha \dot U_\alpha$ yields
\begin{equation}
	\dot\tau \gamma = 1+ D^\alpha \dot U_\alpha+ \frac{1}{6}\left(\frac{1}{\gamma} \bar V^\alpha + 2U^\alpha\right)R_{\alpha DUD}+\mathcal{O}(D^3), \label{taudotap}
\end{equation}
where $\gamma=-\bar V^\alpha U_\alpha$, the generalized relativistic gamma factor.\\

\paragraph*{\textbf{Derivation of $\dot{\bar A}^\alpha = \dot{\tau} \bar{\dot A}^\alpha+\frac{1}{2}R^\alpha_{\bar A UD}+\frac{1}{2}\dot \tau R^\alpha_{\bar A \bar V D}+\mathcal{O}(D^2)$}}\textbf{:}\\

Consider vector field $A^\alpha(\tau)$ on $x^\alpha(\tau)$ and its covariant derivative $\dot A^\alpha(\tau)$ with respect to $\tau$. Let $\bar A^\alpha(t)$ and $\bar{\dot A}^\alpha(t)$ be their parallel transports to the corresponding $x^\alpha(t)$ along the connecting spacelike geodesic discussed above.
\begin{equation}
	\bar A^\alpha(t) = [\tau \to t]^\alpha_\beta A^\beta(\tau(t)),
\end{equation}
where by \eqref{partrans}
\begin{equation}
	[\tau \rightarrow t]^\alpha_\beta=\delta^\alpha_\beta + \Gamma^\alpha_{D\beta} + \frac{1}{2}(\partial_D\Gamma^\alpha_{D\beta} - \Gamma^\alpha_{\beta\rho}\Gamma^\rho_{DD} + \Gamma^\alpha_{D\rho}\Gamma^\rho_{D\beta})+\mathcal{O}(D^3).
\end{equation}
Differentiating,
\begin{align}
	\dot{\bar A}^\alpha &= \nabla_U \bar A^\alpha = \partial_U \bar A^\alpha + \Gamma^\alpha_{\bar A U}\nonumber\\
	&= \partial_U [\tau \to t]^\alpha_\beta A^\beta + [\tau \to t]^\alpha_\beta \dot \tau \partial_V A^\beta + \Gamma^\alpha_{\beta U}[\tau \to t]^\beta_\gamma A^\gamma\nonumber\\
	&= \partial_U [\tau \to t]^\alpha_\beta A^\beta + [\tau \to t]^\alpha_\beta \dot \tau (\dot A^\beta-\tilde \Gamma^\beta_{AV}) + \Gamma^\alpha_{\beta U}[\tau \to t]^\beta_\gamma A^\gamma,
\end{align}
where
\begin{equation}
	\tilde \Gamma^\alpha_{\beta\gamma}(\tau(t)) = \Gamma^\alpha_{\beta\gamma}(t) + \partial_D \Gamma^\alpha_{\beta\gamma}(t) + \mathcal{O}(D^2).
\end{equation}
The derivative of the operator is found to be
\begin{equation}
	\partial_U [\tau \to t]^\alpha_\beta = \dot \tau\Gamma^\alpha_{\beta \bar V}-\Gamma^\alpha_{\beta U} + \dot{\tau}\partial_D\Gamma^\alpha_{\beta \bar V}+ \dot\tau\Gamma^\alpha_{D\rho}\Gamma^\rho_{\beta \bar V}- \Gamma^\alpha_{U\rho}\Gamma^\rho_{\beta D}  +\frac{1}{2}\dot\tau R^\alpha_{\beta \bar V D} +\frac{1}{2}R^\alpha_{\beta U D}+\mathcal{O}(D^2).
\end{equation}
Therefore, combining the results we have
\begin{equation}
	\dot{\bar A}^\alpha = \dot \tau \bar{\dot A}^\alpha+\frac{1}{2}R^\alpha_{\bar A U D} +\frac{1}{2}\dot\tau R^\alpha_{\bar A \bar V D} +\mathcal{O}(D^2). \label{partransdot}
\end{equation}\\

\paragraph*{\textbf{Derivation of $\ddot D^\alpha = \dot{\tau}^2\bar{\dot V}^\alpha -\dot U^\alpha + \ddot\tau \bar V^\alpha+R^\alpha_{UUD} + \frac{2}{3}(\dot \tau \bar V^\beta-U^\beta)(\dot \tau \bar V^\gamma+2U^\gamma))R^\alpha_{\beta\gamma D}+\mathcal{O}(D^2)$}}\textbf{:}\\

Differentiating \eqref{ddotap} and \eqref{taudotap}, and making use of \eqref{partransdot},
\begin{align}
	\ddot D^\alpha &= \ddot\tau \bar V^\alpha + \dot\tau\dot{\bar V}^\alpha - \dot U^\alpha + \frac{1}{6}(\dot\tau \bar V^\beta + 2U^\beta)R^\alpha_{\epsilon\beta\gamma}(\dot D^\epsilon D^\gamma + D^\epsilon\dot D^\gamma)+\mathcal{O}(D^2)\nonumber\\
	&=\ddot\tau \bar V^\alpha + \dot\tau^2\bar{\dot V}^\alpha+\frac{1}{2}\dot \tau R^\alpha_{\bar VUD} +\frac{1}{2}\dot \tau^2R^\alpha_{\bar V \bar V D}-\dot U^\alpha +\frac{1}{6}(\dot \tau \bar V^\epsilon-U^\epsilon)(\dot \tau \bar V^\beta+2U^\beta)R^\alpha_{\epsilon\beta D}+\frac{1}{2}\dot \tau R^\alpha_{DU\bar V}+\mathcal{O}(D^2)\nonumber\\
	&=\ddot\tau \bar V^\alpha + \dot\tau^2\bar{\dot V}^\alpha - \dot U^\alpha + \dot \tau R^\alpha_{\bar VUD}+\frac{1}{2}\dot \tau R^\alpha_{\epsilon\bar VD}(\dot \tau \bar V^\epsilon-U^\epsilon)+\frac{1}{6}(\dot \tau \bar V^\epsilon-U^\epsilon)(\dot \tau \bar V^\beta+2U^\beta)R^\alpha_{\epsilon\beta D}+\mathcal{O}(D^2)\nonumber\\
	&=\ddot\tau \bar V^\alpha + \dot\tau^2\bar{\dot V}^\alpha - \dot U^\alpha +R^\alpha_{UUD}+\frac{2}{3}(\dot \tau \bar V^\beta-U^\beta)(\dot \tau \bar V^\gamma+2U^\gamma)R^\alpha_{\beta\gamma D}+\mathcal{O}(D^2), \label{ddotdotap}
\end{align}
and
\begin{align}
	\ddot \tau \gamma &= -\dot \tau \dot \gamma +\dot D^\alpha\dot U_\alpha+D^\alpha \ddot U_\alpha+\frac{1}{6}(\dot \tau \bar V^\alpha + 2 U^\alpha)R_{\alpha\beta U \gamma}(\dot D^\beta D^\gamma+D^\beta\dot D^\gamma)+\mathcal{O}(D^2)\nonumber\\
	&=\dot \tau(\dot U^\alpha \bar V_\alpha+\dot{\bar V}^\alpha U_\alpha)+\dot \tau \bar V^\alpha\dot U_\alpha+D^\alpha \ddot U_\alpha+\frac{1}{6}(\dot \tau \bar V^\alpha+2U^\alpha)(\dot \tau \bar V^\beta-U^\beta)R_{\alpha\beta UD}+\frac{1}{6}\dot \tau (\dot \tau \bar V^\alpha+2U^\alpha)R_{\alpha DU\bar V}+\mathcal{O}(D^2)\nonumber\\
	&= 2\dot \tau\bar V^\alpha \dot U_\alpha+\dot \tau^2\bar{\dot V}^\alpha U_\alpha+D^\alpha \ddot U_\alpha+\frac{2}{3}\dot \tau (\dot \tau \bar V^\alpha+2U^\alpha)R_{\alpha DU\bar V}+\mathcal{O}(D^2). \label{taudotdotap}
\end{align}\\

\paragraph*{\textbf{Derivation of $[t \rightarrow t_0]^\alpha_\beta = \delta^\alpha_\beta + \Gamma^\alpha_{U\beta}\delta t + \frac{1}{2}\left(\partial_U\Gamma^\alpha_{U\beta} - \Gamma^\alpha_{\beta\rho}\Gamma^\rho_{UU} + \Gamma^\alpha_{U\rho}\Gamma^\rho_{U\beta}+\Gamma^\alpha_{\dot U \beta}\right)\delta t^2+...$}}\textbf{:}\\

Consider a vector $A^\alpha$ at $x^\alpha(t)$ and its parallel transport $\bar A^\alpha$ to $x^\alpha(t_0)$ along the curve. Let $\tilde A^\alpha(t)$ be a constant vector field on the curve, such that $\tilde A^\alpha(t_0)=\bar A^\alpha$ and $\tilde A^\alpha(t)=A^\alpha$, and let $\delta t = t-t_0$.
\begin{align}
	A^{\alpha} = \tilde A^{\alpha}(t) &= \tilde A^{\alpha}(t_0) + \frac{d\tilde A^{\alpha}}{dt}\delta t + \frac{1}{2}\frac{d^2\tilde A^{\alpha}}{dt^2}\delta t^2 + ...\nonumber\\
	&= \bar A^{\alpha} + \partial_U \tilde A^{\alpha}\delta t + \frac{1}{2}\partial^2_U\tilde A^{\alpha}\delta t^2 + ...\,,
\end{align}
where
\begin{equation}
	\partial_U \tilde A^\alpha=-\Gamma^\alpha_{U\tilde A},
\end{equation}
and
\begin{equation}
	\partial^2_U \tilde A^\alpha=\partial_U(\partial_U \tilde A^\alpha)= -\partial_U\Gamma^\alpha_{U \tilde A} + \Gamma^\alpha_{\tilde A \rho}\Gamma^\rho_{UU} + \Gamma^\alpha_{U \rho}\Gamma^\rho_{U\tilde A}-\Gamma^\alpha_{\dot U \tilde A},
\end{equation}
under the condition $\nabla_U \tilde A^\alpha=0$. Therefore,
\begin{align}
	\nonumber A^{\alpha} &= \bar A^{\alpha} -\Gamma^\alpha_{U\bar A}\delta t + \frac{1}{2}\left(-\partial_U\Gamma^\alpha_{U \bar A} + \Gamma^\alpha_{\bar A \rho}\Gamma^\rho_{UU} + \Gamma^\alpha_{U \rho}\Gamma^\rho_{U\bar A}-\Gamma^\alpha_{\dot U \bar A}\right)\delta t^2 + ...\\
	&= \bar A^\beta \left( \delta^\alpha_\beta - \Gamma^\alpha_{U\beta}\delta t - \frac{1}{2}\left(\partial_U\Gamma^\alpha_{U\beta} - \Gamma^\alpha_{\beta\rho}\Gamma^\rho_{UU} - \Gamma^\alpha_{U\rho}\Gamma^\rho_{U\beta}+\Gamma^\alpha_{\dot U \beta}\right)\delta t^2 \right) + ... \label{x0toxt}
\end{align}
Inverting,
\begin{equation}
	\bar A^{\alpha} = A^{\beta}\left(\delta^\alpha_\beta + \Gamma^\alpha_{U\beta}\delta t + \frac{1}{2}\left(\partial_U\Gamma^\alpha_{U\beta} - \Gamma^\alpha_{\beta\rho}\Gamma^\rho_{UU} + \Gamma^\alpha_{U\rho}\Gamma^\rho_{U\beta}+\Gamma^\alpha_{\dot U \beta}\right)\delta t^2\right) + ...
\end{equation}
Defining $[t \to t_0]^\alpha_\beta$ as the parallel transport operator, we have
\begin{equation}
	[t \to t_0]^\alpha_\beta=\delta^\alpha_\beta + \Gamma^\alpha_{U\beta}\delta t + \frac{1}{2}\left(\partial_U\Gamma^\alpha_{U\beta} - \Gamma^\alpha_{\beta\rho}\Gamma^\rho_{UU} + \Gamma^\alpha_{U\rho}\Gamma^\rho_{U\beta}+\Gamma^\alpha_{\dot U \beta}\right)\delta t^2+... \label{partransu}
\end{equation}
The components of this operator are expressed in terms of the Cauchy data of the curve and the metric components at $x^\alpha(t_0)$ by choice, but it operates on vectors at $x^\alpha(t)$ and parallel transports them to $x^\alpha(t_0)$ along the curve.\\

\paragraph*{\textbf{Derivation of $\bar{\Omega}^{\alpha} = \Lambda^\alpha - U^{\alpha}\delta t-\frac{1}{2}\dot U^{\alpha}\delta t^2 - \frac{1}{6}\ddot U^{\alpha}\delta t^3 + \frac{1}{3}R^{\alpha}_{\Lambda U \Lambda}\delta t + \frac{1}{6}R^{\alpha}_{U \Lambda U}\delta t^2 + ...$}}\textbf{:}\\
\begin{figure}
  \centering
  \includegraphics[width=8cm]{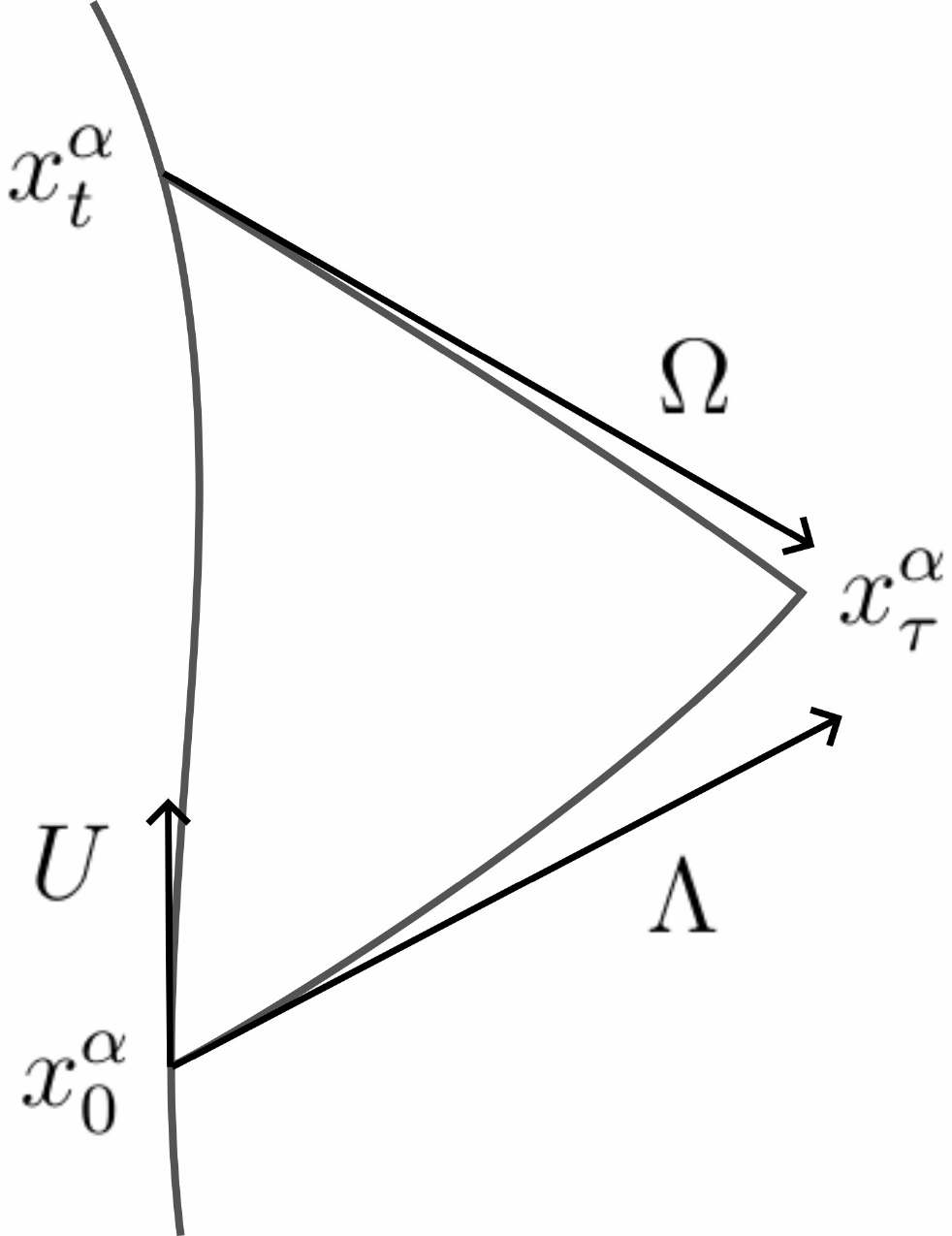}
  \caption{A timelike curve, not necessarily a geodesic, with 4-velocity $U^\alpha$, and two geodesics connecting the events $x^\alpha_0$ and $x^\alpha_t$ on the timelike curve to an event $x^\alpha_\tau$ not on the curve. $\Lambda$ and $\Omega$ are the tangents to the geodesics at events $x^\alpha_0$ and $x^\alpha_t$, respectively, parametrized with the normalized affine parametrization, depicted by their lengths on the diagram.} \label{fig:omega}
\end{figure}

Consider the setup in Figure \ref{fig:omega} and the given definitions. Let $A^\alpha= x^\alpha_\tau-x^\alpha_0$ and $B^\alpha= x^\alpha_\tau-x^\alpha_t$. With the normalized affine parametrization as described above, by equation \eqref{LofA}
\begin{equation}
	\Lambda^\alpha =A^\alpha + \frac{1}{2}\Gamma^{\alpha}_{AA} + \frac{1}{6}(\partial_A\Gamma^{\alpha}_{AA} + \Gamma^{\alpha}_{A\rho}\Gamma^{\rho}_{AA}) + ...\,, \label{ap-lambdaofa}
\end{equation}
and
\begin{equation}
	\Omega^\alpha =B^\alpha + \frac{1}{2}\tilde{\Gamma}^{\alpha}_{BB} + \frac{1}{6}(\partial_B\tilde{\Gamma}^{\alpha}_{BB} + \tilde{\Gamma}^{\alpha}_{B\rho}\tilde{\Gamma}^{\rho}_{BB}) + ...\,, \label{ap-omegaofb}
\end{equation}
where
\begin{equation}
	\tilde{\Gamma}^{\alpha}_{\beta \gamma}=\Gamma^{\alpha}_{\beta \gamma}(t)= \Gamma^{\alpha}_{\beta \gamma}(t_0) + \partial_U \Gamma^{\alpha}_{\beta \gamma}(t_0)\delta t +... \label{ap-gammatilde}
\end{equation}
Let
\begin{equation}
	\bar \Omega^\alpha =\Omega^\beta[t \to t_0]^\alpha_\beta, \label{ap-omegabar}
\end{equation}
where $[t \to t_0]^\alpha_\beta$ is the parallel transport operator from $x^\alpha_t$ to $x^\alpha_0$ along the curve, given by \eqref{partransu}. Furthermore,
\begin{align}
	x^\alpha_t &=x^\alpha_0 + \frac{dx^\alpha}{dt}\delta t + \frac{1}{2}\frac{d^2 x^\alpha}{dt^2}\delta t^2 + \frac{1}{6}\frac{d^3 x^\alpha}{dt^3}\delta t^3 + ...\nonumber\\
	&=x^\alpha_0 + U^\alpha \delta t + \frac{1}{2}\partial_U U^\alpha \delta t^2 + \frac{1}{6}\partial^2_U U^\alpha \delta t^3 + ...\,,
\end{align}
where
\begin{equation}
	\partial_U U^\alpha=\dot U^\alpha -\Gamma^\alpha_{UU}, \label{ap-pu}
\end{equation}
and
\begin{equation}
	\partial^2_U U^\alpha=\partial_U(\partial_U U^\alpha)=\ddot U^\alpha-3\Gamma^\alpha_{\dot U U}-\partial_U\Gamma^\alpha_{UU}+2\Gamma^\alpha_{U\rho}\Gamma^\rho_{UU}. \label{ap-p2u}
\end{equation}
Therefore,
\begin{align}
	\nonumber B^\alpha &= x^\alpha_\tau - x^\alpha_0 - U^\alpha \delta t - \frac{1}{2}\partial_U U^\alpha \delta t^2 - \frac{1}{6}\partial^2_U U^\alpha \delta t^3 + ...\\
	&= A^\alpha - U^\alpha \delta t - \frac{1}{2}\partial_U U^\alpha \delta t^2 - \frac{1}{6}\partial^2_U U^\alpha \delta t^3 + ... \label{ap-bofa}
\end{align}
Combining equations \eqref{ap-omegabar}, \eqref{ap-omegaofb}, \eqref{ap-gammatilde}, \eqref{ap-bofa}, \eqref{ap-pu}, \eqref{ap-p2u}, \eqref{ap-lambdaofa}, and \eqref{partransu}, the expression for $\bar \Omega^\alpha$ simplifies to
\begin{equation}
	\bar \Omega^{\alpha}=\Lambda^{\alpha}-U^{\alpha}\delta t-\frac{1}{2}\dot U^{\alpha}\delta t^2-\frac{1}{6}\ddot U^{\alpha}\delta t^3+\frac{1}{3}R^{\alpha}_{\Lambda U \Lambda}\delta t+\frac{1}{6}R^{\alpha}_{U \Lambda U}\delta t^2 + ... \label{omegaoflambdau}
\end{equation}\\

\paragraph*{\textbf{Derivation of $\bar A^\alpha(t,t+\delta t) = A^\alpha(t) + \dot A^\alpha(t) \delta t + \frac{1}{2}\ddot A^\alpha(t) \delta t^2+...$}}\textbf{:}\\

Consider a vector field $A^\alpha(t)$ on the curve $x^\alpha(t)$, and let
\begin{equation}
	\bar A^\alpha(t_0,t) = [t \rightarrow t_0]^\alpha_\beta A^\beta(t),
\end{equation}
were $[t \rightarrow t_0]^\alpha_\beta$ is the parallel transport from $x^\alpha(t)$ to $x^\alpha(t_0)$, along the curve itself. Note that $\bar A^\alpha(t_0,t)$ is a vector at $t_0$ for all $t$. So
\begin{equation}
	\dot{\bar A}^\alpha(t_0,t) = \lim_{\delta t\to 0}\frac{\bar A^\alpha(t_0,t+\delta t)-\bar A^\alpha(t_0,t)}{\delta t}=\lim_{\delta t\to 0}[t \rightarrow t_0]^\alpha_\beta\frac{[t+\delta t \rightarrow t]^\beta_\gamma A^\gamma(t+\delta t)-A^\beta(t)}{\delta t},
\end{equation}
since $[t+\delta t \rightarrow t_0]^\alpha_\beta = [t \rightarrow t_0]^\alpha_\gamma[t+\delta t \rightarrow t]^\gamma_\beta $. Now with \begin{equation}
	\dot A^\beta(t)=\lim_{\delta t\to 0}\frac{[t+\delta t \rightarrow t]^\beta_\gamma A^\gamma(t+\delta t)-A^\beta(t)}{\delta t},
\end{equation}
we find
\begin{equation}
	\dot{\bar A}^\alpha(t_0,t) = [t \rightarrow t_0]^\alpha_\beta \dot A^\beta(t),
\end{equation}
where $\dot{\bar A}^\alpha(t_0,t)$ is the ordinary derivative, while $\dot A^\alpha(t)$ is the covariant (or intrinsic in this case) derivative along the curve. In the same way
\begin{equation}
	\ddot{\bar A}^\alpha(t_0,t) = [t \rightarrow t_0]^\alpha_\beta \ddot A^\beta(t),
\end{equation}
where again we have ordinary differentiation on the left and covariant on the right. This association is clearly true for all orders. Therefore,
\begin{align}
	  \bar A^\alpha(t_0,t) &= \bar A^\alpha (t_0,t_0)+\dot{\bar A}^\alpha(t_0,t_0)(t-t_0)+\frac{1}{2}\ddot{\bar A}^\alpha(t_0,t_0)(t-t_0)^2+...\nonumber\\
	&= A^\alpha (t_0)+\dot{A}^\alpha(t_0)(t-t_0)+\frac{1}{2}\ddot{A}^\alpha(t_0)(t-t_0)^2+...\,,
\end{align}
or
\begin{equation}
	\bar A^\alpha(t,t+\delta t) = A^\alpha(t) + \dot A^\alpha(t) \delta t + \frac{1}{2}\ddot A^\alpha(t) \delta t^2+... \label{abartdeltat}
\end{equation}
It is clear that the above derivation also applies to tensors of any rank. This is a generalized version of the Taylor expansion about an event with respect to any curve in an arbitrary metric space. When the curve is null or spacelike the dot in the above expression should be replaced with $\nabla$ and the tangent to the curve. And for a curve connecting the two events of interest under the normalized affine parametrization, the expansion parameter ($\delta t$ in the above) is just equal $1$ with the smallness being contained in the tangent vector itself.\\

\paragraph*{\textbf{Derivation of $A^\alpha_{U\Lambda} = A^\beta \left( \delta^\alpha_\beta + \frac{1}{2}R^\alpha_{\beta \Lambda U}\delta t \right) + ...$}}\textbf{:}\\

Let $A^\alpha$ be a vector at an event $x_0$ on the timelike curve of an observer with 4-velocity $U^\alpha$, and $\Lambda^\alpha$ be the 
tangent to the connecting geodesic to some neighboring event $x_\tau$, with normalized affine parametrization as described above, see Figure \ref{fig:omega}. With $\Omega^\alpha$ as defined on the diagram, and its parallel transport $\bar \Omega^\alpha$ given by \eqref{omegaoflambdau}, we connect the events $x_0$, $x_t$ and $x_\tau$, and define $A^\alpha_{U\Lambda}$ as the parallel transport of $A^\alpha$ around the triangle. The parallel transport operator from $x_0$ to $x_t$ along the timelike curve is given by (see equation \eqref{x0toxt})
\begin{equation}
	[x_0 \to x_t]^\alpha_\beta= \delta^\alpha_\beta - \Gamma^\alpha_{U\beta}\delta t - \frac{1}{2}\left(\partial_U\Gamma^\alpha_{U\beta} - \Gamma^\alpha_{\beta\rho}\Gamma^\rho_{UU} - \Gamma^\alpha_{U\rho}\Gamma^\rho_{U\beta}+\Gamma^\alpha_{\dot U \beta}\right)\delta t^2 + ... \label{x0toxt2}
\end{equation}
The parallel transport along the connecting geodesic from $x_\tau$ to $x_0$ is given by \eqref{partrans},
\begin{equation}
	[x_\tau \to x_0]^\alpha_\beta=\delta^\alpha_\beta + \Gamma^\alpha_{\Lambda\beta} + \frac{1}{2}\left(\partial_\Lambda\Gamma^\alpha_{\Lambda\beta} - \Gamma^\alpha_{\beta\rho}\Gamma^\rho_{\Lambda\Lambda} + \Gamma^\alpha_{\Lambda\rho}\Gamma^\rho_{\Lambda\beta}\right)+...
\end{equation}
In both cases above, the terms are evaluated at $x_0$. As for the transport from $x_t$ to $x_\tau$, we can use $\Omega^\alpha$ at $x_t$ in an analogous way to $\Lambda^\alpha$ at $x_0$, and the same process that led to \eqref{x0toxtau} gives
\begin{equation}
	[x_t \to x_\tau]^\alpha_\beta=\delta^\alpha_\beta - \tilde \Gamma^\alpha_{\Omega\beta} - \frac{1}{2}\left(\partial_\Omega\tilde \Gamma^\alpha_{\Omega\beta} - \tilde \Gamma^\alpha_{\beta\rho}\tilde \Gamma^\rho_{\Omega\Omega} - \tilde \Gamma^\alpha_{\Omega\rho}\tilde \Gamma^\rho_{\Omega\beta}\right)+...\,,
\end{equation}
but this time the terms are evaluated at $x_t$. To relate the terms in the above to the event $x_0$ we use $\Omega^\alpha = \bar \Omega^\beta [x_0 \to x_t]^\alpha_\beta$, where $[x_0 \to x_t]^\alpha_\beta$ is given by \eqref{x0toxt2} and $\bar \Omega^\alpha$ by \eqref{omegaoflambdau}, and
\begin{equation}
	\tilde \Gamma^\alpha_{\beta\gamma} = \Gamma^\alpha_{\beta\gamma} + \partial_U \Gamma^\alpha_{\beta\gamma}\delta t + ...
\end{equation}
Finally, the combination of the three parallel transports is found to be
\begin{equation}
	[x_0 \to x_t \to x_\tau \to x_0]^\alpha_\beta=\delta^\alpha_\beta + \frac{1}{2}R^\alpha_{\beta \Lambda U}\delta t +...\,, \label{tritrans}
\end{equation}
and therefore
\begin{equation}
	A^\alpha_{U\Lambda} = A^\beta \left( \delta^\alpha_\beta + \frac{1}{2}R^\alpha_{\beta \Lambda U}\delta t \right) + ... \label{tritransa}
\end{equation}
The direction of transport around the triangle matters; $A^\alpha_{U\Lambda}$ is the result of transporting in the $U^\alpha$ direction first and transporting back from the $\Lambda^\alpha$ direction last. The antisymmetry of the Riemann tensor makes clear that the change gained by moving around will be opposite in sign if traversed in an opposite direction, as expected; and looping around twice but in opposite directions will result in no change. Importantly, only the curvature term remains and the non-geodesicity of the timelike curve plays no role within this order of accuracy. (The components of $\Lambda^\alpha$ and $\delta t$ are small, so the expression is correct to second order.) In fact, none of the connecting curves of the three events need to be a geodesic for the result to hold for the small triangle. The Riemann term resulting in the established relationship is the smallest order curvature effect, which happens to appear at the highest order we are interested in, and clearly, in the absence of curvature, any accelerations of the connecting curves will have no effect on the parallel transport round trip. (So any acceleration terms in the round trip transport will necessarily get combined with curvature terms, and appear at even higher orders.) Thus, the above result is fully general for any two small vectors at an event, which give rise to a triangle through the (geodesic) construction described above.\\

\paragraph*{\textbf{Derivation of $\Phi^{\alpha} = \Lambda^\alpha + \bar V^{\alpha}\delta \tau + \frac{1}{2}\bar{\dot V}^{\alpha}\delta \tau^2 + \frac{1}{6}\bar{\ddot V}^{\alpha}\delta \tau^3 + \frac{1}{6}R^{\alpha}_{\Lambda \bar V \Lambda}\delta \tau - \frac{1}{3}R^{\alpha}_{\bar V \Lambda \bar V}\delta \tau^2 + ...$}}\textbf{:}\\

Consider an analogous situation to that of Figure \ref{fig:omega}, but for the neighboring timelike curve with 4-velocity $V^\alpha$ and proper time $\tau$. The events $x_0$, $x_t$, and $x_\tau$ get relabeled to $x_0$, $x_\tau$, and $x_t$, respectively (with $x_0$ and $x_\tau$ now being on the same timelike curve). With analogous definition for the vector $\Omega^\alpha$ as on the figure, let $\Phi^\alpha = -\Omega^\beta[x_\tau \to x_t]^\alpha_\beta$ be the tangent of the connecting geodesic to $x_\tau$ at $x_t$. The aim is to establish an expression for $\Phi^\alpha$ at $x_t$ for the given setup. Identical to the result \eqref{omegaoflambdau}, we have at the event $x_0$
\begin{equation}
	\bar \Omega^{\alpha}=\Lambda^{\alpha}-V^{\alpha}\delta \tau-\frac{1}{2}\dot V^{\alpha}\delta \tau^2-\frac{1}{6}\ddot V^{\alpha}\delta \tau^3+\frac{1}{3}R^{\alpha}_{\Lambda V \Lambda}\delta \tau+\frac{1}{6}R^{\alpha}_{V \Lambda V}\delta \tau^2 + ...\,,
\end{equation}
where the $\Lambda^\alpha$ in the above is analogous to the one in the figure. Before proceeding, for the sake of consistency, we drop this definition of $\Lambda^\alpha$ and simply reuse the symbol for the tangent vector to the same connecting geodesic but at $x_t$ (so that $\Lambda^\alpha_{new}=-\Lambda^\beta_{old}[x_0 \to x_t]^\alpha_\beta$ and $\Lambda^\alpha_{old}=-\Lambda^\beta_{new}[x_t \to x_0]^\alpha_\beta = -\bar \Lambda^\alpha_{new}$). Then the above becomes
\begin{equation}
	\bar \Omega^{\alpha}=-\Lambda^\beta_{new}[x_t \to x_0]^\alpha_\beta-V^{\alpha}\delta \tau-\frac{1}{2}\dot V^{\alpha}\delta \tau^2-\frac{1}{6}\ddot V^{\alpha}\delta \tau^3+\frac{1}{3}R^{\alpha}_{\bar\Lambda V \bar\Lambda}\delta \tau-\frac{1}{6}R^{\alpha}_{V \bar\Lambda V}\delta \tau^2 + ...
\end{equation}
The parallel transport of this relationship to $x_t$ is given by
\begin{equation}
	\bar \Omega^{\beta}[x_0 \to x_t]^\alpha_\beta=-\Lambda^{\alpha}-\bar V^{\alpha}\delta \tau-\frac{1}{2}\bar{\dot V}^{\alpha}\delta \tau^2-\frac{1}{6}\bar{\ddot V}^{\alpha}\delta \tau^3+\frac{1}{3}R^{\alpha}_{\Lambda \bar V \Lambda}\delta \tau-\frac{1}{6}R^{\alpha}_{\bar V \Lambda \bar V}\delta \tau^2 + ...\,,
\end{equation}
where the curvature terms remain unaffected by the transformation in this order, and we dropped the subscript on $\Lambda^\alpha$ for brevity. Now, $\bar \Omega^{\beta}[x_0 \to x_t]^\alpha_\beta = \Omega^\gamma[x_\tau \to x_0]^\beta_\gamma[x_0 \to x_t]^\alpha_\beta = \Omega^\mu[x_\tau \to x_t]^\epsilon_\mu[x_t \to x_\tau]^\gamma_\epsilon[x_\tau \to x_0]^\beta_\gamma[x_0 \to x_t]^\alpha_\beta = -\Phi^\epsilon \left( \delta^\alpha_\epsilon +\frac{1}{2}R^\alpha_{\epsilon \Lambda \Phi} \right)$, by using equation \eqref{tritrans} adopted to the situation at hand and the definition for $\Phi^\alpha$ above. Therefore,
\begin{equation}
	-\Phi^\beta\left( \delta^\alpha_\beta +\frac{1}{2}R^\alpha_{\beta \Lambda \Phi}\delta \tau \right)=-\Lambda^{\alpha}-\bar V^{\alpha}\delta \tau-\frac{1}{2}\bar{\dot V}^{\alpha}\delta \tau^2-\frac{1}{6}\bar{\ddot V}^{\alpha}\delta \tau^3+\frac{1}{3}R^{\alpha}_{\Lambda \bar V \Lambda}\delta \tau-\frac{1}{6}R^{\alpha}_{\bar V \Lambda \bar V}\delta \tau^2 + ...\,,
\end{equation}
and
\begin{equation}
	\Phi^\alpha =\Lambda^{\alpha}+\bar V^{\alpha}\delta \tau+\frac{1}{2}\bar{\dot V}^{\alpha}\delta \tau^2+\frac{1}{6}\bar{\ddot V}^{\alpha}\delta \tau^3+\frac{1}{6}R^{\alpha}_{\Lambda \bar V \Lambda}\delta \tau-\frac{1}{3}R^{\alpha}_{\bar V \Lambda \bar V}\delta \tau^2 + ...\,, \label{temp62}
\end{equation}
which is the result we were seeking, and we interpret it as follows. Given two events $x_t$ and $x_0$ on neighboring timelike curves $x^\alpha(t)$ and $x^\alpha(\tau)$, respectively, which are connected by a geodesic, and for some interval $\delta \tau$ of proper time from $x_0$ along the curve, the tangent to the connecting geodesic from $x_t$ to the event $\delta \tau$ up the neighboring curve is given by the above relationship, where all the terms are evaluated at the event $x_t$.\\

\paragraph*{\textbf{Derivation of $D=\omega_U\left(1-\frac{1}{2}\dot U^\alpha K_\alpha\right)+\mathcal{O}(\omega_U^3)$}}\textbf{:}\\

Let $U^\alpha (=U^\alpha(t_0))$ be the 4-velocity of an observer at some event $x^\alpha_0=x^\alpha(t_0)$. At the same event, let $K^\alpha$ be a future pointing null vector that connects to a neighboring event $x^\alpha_1$ under the normalized affine parametrization. Let $D^\alpha(t)$ (with $t$ the proper time for $U^\alpha(t)$ and $t > t_0$) be the position vector in the space frame of $U^\alpha(t)$ pointing to the event $x^\alpha_1$ (as constructed in section \ref{sec:relsta}, see equation \eqref{outphoton}). The parallel transport of $D^\alpha(t_0+\delta t)$ from $x^\alpha(t_0+\delta t)$ back to $x^\alpha(t_0)$ along the timelike trajectory is given by (with help of \eqref{omegaoflambdau})
\begin{equation}
	\bar D^\alpha = K^\alpha - U^{\alpha}\delta t-\frac{1}{2}\dot U^{\alpha}\delta t^2-\frac{1}{6}\ddot U^{\alpha}\delta t^3+\frac{1}{3}R^{\alpha}_{K U K}\delta t+\frac{1}{6}R^{\alpha}_{U K U}\delta t^2 + ...
\end{equation}
Here $\delta t$ and the magnitude $D=|D^\alpha|$ are assumed to be small, while the smallness of the null vector $K^\alpha$ can be encompassed by the parameter $\omega_U = -U^\alpha K_\alpha$. This makes the above expression correct to third order combinations of $\delta t$ and $\omega_U$. Clearly the values of $\omega_U$, $\delta t$ and $D$ are connected in a way where the value of one sets the other two (at least when they are small and positive). The facts that $D^\alpha(t)$ is perpendicular to $U^\alpha(t)$ and that $K^\alpha$ is null will give rise to two equations in the three unknowns as expected. The parallel transport of $U^\alpha(t_0+\delta t)$ from $x^\alpha(t_0+\delta t)$ back to $x^\alpha(t_0)$ is given by (equation \eqref{abartdeltat})
\begin{equation}
	\bar U^\alpha = U^\alpha(t_0) + \dot U^\alpha(t_0) \delta t + \frac{1}{2}\ddot U^\alpha(t_0) \delta t^2+\mathcal{O}(\delta t^3).
\end{equation}
Since parallel transport preserves inner products and $D^\alpha(t)$ is in the space frame of $U^\alpha(t)$, we have $\bar D^\alpha\bar U_\alpha=D^\alpha(t)U_\alpha(t)=0$. Applying it to the above expressions and solving for $\delta t$ gives
\begin{equation}
	\delta t = \omega_U \left(1-K^\alpha \dot U_\alpha+\left(K^\alpha \dot U_\alpha\right)^2-\frac{1}{2}\omega_U K^\alpha \ddot U_\alpha - \frac{1}{6}\omega_U^2 \dot U^2 - \frac{1}{3}R_{UKUK}\right) + \mathcal{O}(\omega_U^4),
\end{equation}
which can be used to eliminate $\delta t$ in favor of $\omega_U$ in expressing the magnitude $D$. Since $D=|D^\alpha(t)|=|\bar D^\alpha|$, we find
\begin{align}
	D &= \omega_U \left(1-\frac{1}{2}K^\alpha \dot U_\alpha+\frac{3}{8}\left(K^\alpha \dot U_\alpha\right)^2-\frac{1}{6}\omega_U K^\alpha \ddot U_\alpha - \frac{1}{24}\omega_U^2 \dot U^2 - \frac{1}{6}R_{UKUK}\right) + \mathcal{O}(\omega_U^4)\\
	&= \omega_U \left(1-\frac{1}{2}K^\alpha \dot U_\alpha\right) + \mathcal{O}(\omega_U^3). \label{temp67}
\end{align}

\section{Analysis}

\subsection{\label{app:relab}Relativistic Aberration for Small Angles}

For small angles $\theta_U$ and $\theta_V$ ($\theta_U, \theta_V \ll 1$) equation \eqref{relab} reduces to
\begin{equation}
	\frac{\theta_V^2}{\theta_U^2}=\frac{(U^\alpha K_\alpha)(U^\alpha W_\alpha)}{(V^\alpha K_\alpha)(V^\alpha W_\alpha)}. \label{relabs}
\end{equation}
Defining
\begin{equation}
	1+x = \frac{\sfrac{V^\alpha W_\alpha}{V^\alpha K_\alpha}}{\sfrac{U^\alpha W_\alpha}{U^\alpha K_\alpha}},
\end{equation}
allows us to rewrite the above as
\begin{equation}
	\frac{\theta_V^2}{\theta_U^2} = \frac{(U^\alpha K_\alpha)^2}{(V^\alpha K_\alpha)^2(1+x)}, \;\; or \;\; \frac{(U^\alpha W_\alpha)^2(1+x)}{(V^\alpha W_\alpha)^2}.
\end{equation}
It is immediately evident that for either zero relative speed, $v$, or zero angle $\theta_U$, we get $x=0$, which implies that $x=0+\mathcal{O}(\theta_U)$. Indeed, the aberration relationship for small angles can be cast in a simpler form
\begin{equation}
	\frac{\theta_V}{\theta_U}=\frac{(U^\alpha K_\alpha)}{(V^\alpha K_\alpha)}=\frac{(U^\alpha W_\alpha)}{(V^\alpha W_\alpha)}, \label{relabsa}
\end{equation}
where only one of the participating null vectors is used. This form immediately leads to the well known aberration relationship for solid angles (given by \eqref{temp3}). However, there are several issues that are worth exploring in the situation where the measurable angles are considered small but not infinitesimal. As we will find, it is possible for the angles to be small in a situation of extreme relative velocity, with the approximate relationship given by \eqref{relabsa} being incorrect. We will see under which circumstances the above version of the aberration relationship for small angles holds true, and when one must resort to the more general version \eqref{relabs} for the correct expression.

With reference to the decompositions given by \eqref{temp297} and the expressions given by \eqref{temp299}, take $\alpha$ as the smaller of the two angles, so that $\beta=\alpha + \epsilon \theta_U$ for some $\epsilon \in [0,1]$. Dividing the ratios in \eqref{temp299} gives, to lowest orders in $\theta_U$,
\begin{equation}
	\frac{\sfrac{V^\alpha W_\alpha}{V^\alpha K_\alpha}}{\sfrac{U^\alpha W_\alpha}{U^\alpha K_\alpha}} = 1 + v \epsilon \theta_U \frac{\sin(\alpha)+\frac{1}{2}\cos(\alpha)\epsilon \theta_U}{1-v\cos(\alpha)},
\end{equation}
and this gives us an expression for $x$. As expected, if $\theta_U$ is treated as an infinitesimal quantity, it can always be considered small enough to ensure that $x \ll 1$. However, if $\theta_U$ is very small but of a prefixed value, then there can always be a small enough angle $\alpha$ and a large enough speed $v$ for which $x \gtrsim 1$. The only possible way to put a restriction on $x$ with the hope that $x \ll 1$ in the current setup, is to use the initial assumptions $\theta_U, \theta_V \ll 1$, which may or may not disallow speeds and orientations that make $x$ depart from being small. We proceed to investigate the connection between the magnitudes of $x$ and $\theta_V$.

Firstly, regardless of $\theta_V$, when $\alpha \gg \theta_U$, $x \ll 1$ for all $v$. Things only get interesting when $\epsilon \sim 1$ and $\alpha \lesssim \theta_U$, in which case we can express
\begin{equation}
	x=v \epsilon \theta_U \frac{\alpha+\frac{1}{2}\epsilon\theta_U}{1-v+\frac{1}{2}v\alpha^2},
\end{equation}
and
\begin{equation}
	\theta_V^2=\frac{\theta_U^2(1-v)(1+v)}{(1-v+\frac{1}{2}v\alpha^2)^2(1+x)}.
\end{equation}
As $v$ increases from zero the behavior of the above expressions is as follows:\\

- When $\theta_U^2 \ll 1-v \leq 1$, $x \ll 1$ and $\theta_V^2 \ll 1$.\\

- As $1-v \to \theta_U^2$, $x \sim 1$ and $\theta_V^2 \sim 1$.\\

- As $1-v \to \alpha^2$, $x \sim \frac{\theta_U^2}{\alpha^2}$ and $\theta_V^2 \sim 1$. (If $\alpha=0$, $x \to \infty, \; \theta_V^2 \sim 1$)\\

- When $0 < 1-v \ll \alpha^2$, $x \sim \frac{\theta_U^2}{\alpha^2}$ and $\theta_V^2 \sim \frac{1-v}{\alpha^2}$. (If $\alpha=0$, $x \to \infty, \; \theta_V^2 \sim 1$)\\
\\
Thus, we see that it is possible for $x$ to be of moderate value or even very large without violating the requirement that $\theta_V \ll 1$. This can happen when $0 < \alpha \lesssim \theta_U$ and the speed is extreme. This means that only equation \eqref{relabs} holds in general for small angles, while the simpler form \eqref{relabsa} requires an additional condition (on top of the assumption $\theta_U, \theta_V \ll 1$).

We also see that if $\alpha=0$ and we have perfect alignment, then the requirement $\theta_V \ll 1$ does not allow $x$ to be moderate or large. In that case, $\theta_V \ll 1$ forces $\frac{\theta_U^2}{1-v} \ll 1$, which means that $x \ll 1$. This is the only situation in which the smallness of the angles implies $x \ll 1$.

In conclusion, the assumption that $\theta_U, \theta_V \ll 1$ does not guarantee that $x \ll 1$, and therefore an extra condition must be imposed for the form \eqref{relabsa} to hold true. Otherwise, only the original form \eqref{relabs} is true for small angles in general. The obvious condition that must be satisfied for \eqref{relabsa} to be correct is $x \ll 1$, and without restrictions on $\epsilon$ it is equivalent to
\begin{equation}
	\frac{v\theta_U^2}{1-v\cos(\alpha)} \ll 1. \label{cond}
\end{equation}
The above is surely satisfied for $\alpha \gg \theta_U$, the situation of non-alignment of observer and light rays. Then, the angles and $x$ remain small for any speed. For near alignment, $0 < \alpha \lesssim \theta_U$, the above must be taken as an extra condition on the speed for \eqref{relabsa} to be true. Interestingly, for perfect alignment, $\alpha=0$, the above condition is implied by the smallness of the angles, which immediately excludes extreme velocities and therefore ensures that $x \ll 1$ and that \eqref{relabsa} is correct.

Overall, the condition given by \eqref{cond} would rarely be unsatisfied, and therefore may not seem necessary to impose. However, it is important to note that the aberration relationship in the simple from of \eqref{relabsa} (as well as that for solid angles \eqref{temp3}) is not always correct, and the circumstances under which it holds must be fully disclosed. As far as we are aware, \eqref{cond} is the first explicit statement of the condition that must be satisfied for the small and solid angle aberration relationships (\eqref{relabsa} and \eqref{temp3}) to work.

Finally, although $\alpha$ was taken as the smaller angle, it is clear that either of the two angles would do in the condition \eqref{cond}. In fact, without any assumptions on $\alpha$ and $\beta$, we could allow $-1 \leq \epsilon \leq 1$, and through a slightly more complicated process establish that the extremes of $x$ are $-1$ and $\infty$. These happen in case of perfect alignment with either of the two light rays. The same conclusions will follow of the necessity for an additional condition of the form \eqref{cond} for the case of near alignment (without any further restrictions on $\epsilon$).

\subsection{\label{app:neigh}Neighboring Objects at Constant Distance}

In this section we derive an expression for the proper time lapse $\delta \tau^\pm$, and for other related quantities needed in section \ref{sec:relsta}, under the condition of constant distance.\\

For the setup in Figure \ref{fig:sd} and the expression for $K^{\pm\alpha}$ given by \eqref{kpm}, $\delta \tau^\pm$ is set by the requirement $K^{\pm\alpha}K^{\pm}_\alpha=0$,
\begin{equation}
	0=D^2-\delta\tau^{\pm 2}+2D^\alpha\bar V_\alpha\delta\tau^\pm+D^\alpha\bar{\dot V}_\alpha\delta\tau^{\pm 2}+\frac{1}{3}D^\alpha\bar{\ddot V}_\alpha\delta\tau^{\pm 3}-\frac{1}{12}\dot V^2\delta\tau^{\pm 4}-\frac{1}{3}R_{D\bar V D\bar V}\delta\tau^{\pm 2}+...
\end{equation}
For the particular case where the distance to the object remains the same, we have $\dot D^\alpha D_\alpha=0$ and by equation \eqref{ddot} also
\begin{equation}
	D^\alpha \bar V_\alpha = 0 +\mathcal{O}(D^4).
\end{equation}
Differentiating the above with respect to the time $t$ twice and applying \eqref{ddot}, \eqref{partransdot} and \eqref{taudot}, we get
\begin{equation}
	D^\alpha \bar {\dot V}_\alpha = 1-\gamma^2+D^\alpha \dot U_\alpha \gamma^2-(D^\alpha \dot U_\alpha)^2+\frac{1}{3}\left( R_{DUDU}+R_{D \bar V D \bar V}+R_{D\bar V DU} \right) +\mathcal{O}(D^3), \label{dvdot}
\end{equation}
and
\begin{equation}
	D^\alpha \bar {\ddot V}_\alpha = D^\alpha\ddot U_\alpha-3\dot\gamma+\frac{1}{2}\left( R_{DU\bar V U}+R_{D \bar V \bar V U} \right) +\mathcal{O}(D^2).
\end{equation}
Notice that merely as a consequence of the constancy of the distance to the object (in case of non-extreme acceleration or jerk) we have that
\begin{equation}
	\gamma = 1 + \mathcal{O}(D), \;\;\;\;\;\; \dot{\tau} = 1+\mathcal{O}(D), \;\;\;\;\;\; \dot{\gamma}=0+\mathcal{O}(D),
\end{equation}
and since (by \eqref{partransdot})
\begin{equation}
	\dot \gamma = -\dot U^\alpha \bar V_\alpha - \dot \tau \bar{\dot V}^\alpha U_\alpha - \frac{1}{2}\left( \dot{\tau}R_{D \bar V \bar V U}+R_{D U \bar V U} \right) + \mathcal{O}(D^2),
\end{equation}
we also have
\begin{equation}
	\bar{\dot V}^\alpha U_\alpha = -\dot U^\alpha \bar V_\alpha + \mathcal{O}(D).
\end{equation}
These restrictions can be pushed one step further by recognizing that $\bar V^\alpha = \gamma U^\alpha + \sqrt{\gamma^2-1}\hat{V^\beta h^\alpha_\beta}=U^\alpha + \mathcal{O}(\sqrt{D})$, and since $D$ is constant, $\dot{\bar V}^\alpha = \dot U^\alpha+\mathcal{O}(\sqrt{D})$ and $\bar{\dot V}^\alpha = \dot U^\alpha+\mathcal{O}(\sqrt{D})$ (by \eqref{partransdot}). This means that $D^\alpha\bar{\dot V}_\alpha = D^\alpha\dot U_\alpha+\mathcal{O}(D\sqrt{D})$, and from which we must conclude that $\gamma = 1+\mathcal{O}(D\sqrt{D})$ by \eqref{dvdot}. Repeating the process with the adjusted restriction on $\gamma$ will yield a stronger restriction, particularly that $\gamma = 1+\mathcal{O}(D^\frac{7}{4})$, and the cycle continues with a clear pattern approaching $\gamma = 1+\mathcal{O}(D^2)$. In fact, it can be easily shown that if the difference $\gamma^2-1$ is assumed to have any terms larger than $\mathcal{O}(D^2)$ it leads to a contradiction in \eqref{dvdot}. Therefore, the restriction of constant distance to the object (assuming no extreme acceleration or jerk), gives rise to the following conditions,
\begin{equation}
	D^\alpha \bar {\dot V}_\alpha = 1-\gamma^2+D^\alpha \dot U_\alpha-(D^\alpha \dot U_\alpha)^2+ R_{DUDU} +\mathcal{O}(D^3), \label{temp78}
\end{equation}
\begin{equation}
	D^\alpha \bar {\ddot V}_\alpha = D^\alpha \ddot U_\alpha+\mathcal{O}(D^2),
\end{equation}
\begin{equation}
	\gamma = 1 + \mathcal{O}(D^2), \;\;\;\;\;\; \dot{\tau} = 1+\mathcal{O}(D), \;\;\;\;\;\; \dot{\gamma}=0+\mathcal{O}(D^2),
\end{equation}
\begin{equation}
	\bar V^\alpha = U^\alpha + \mathcal{O}(D), \;\;\;\;\;\; \bar{\dot V}^\alpha = \dot U^\alpha+\mathcal{O}(D), \;\;\;\;\;\; \bar{\ddot V}^\alpha=\ddot U^\alpha+\mathcal{O}(D), \label{dconstcondv}
\end{equation}
and
\begin{equation}
	\bar{\dot V}^\alpha U_\alpha = -\dot U^\alpha \bar V_\alpha + \mathcal{O}(D^2). \label{dconstcondvdot}
\end{equation}
(It can be shown from first principles that $\gamma=1+\mathcal{O}(D^n) \implies \dot{\gamma}=0+\mathcal{O}(D^n)$ for constant $D$, which is expected since $\gamma$ is a scalar on the curve.) With these we find
\begin{equation}
	\delta \tau^\pm=\pm\frac{D}{\gamma}\left( 1+\frac{1}{2}D^\alpha\dot U_\alpha-\frac{1}{8}(D^\alpha\dot U_\alpha)^2-\frac{1}{24}\dot U^2 D^2+\frac{1}{3} R_{DUDU} \right)+\frac{D^2}{6}D^\alpha \ddot U_\alpha+\mathcal{O}(D^4). \label{temp21}
\end{equation}

\subsection{\label{app:condistance}Distance Between Neighboring Null Geodesics}

In what follows we establish an expression for $\frac{\delta K^\alpha U_\alpha}{\omega_U}$ from the condition $V^\alpha \delta_\alpha=0$ for section \ref{sec:fund}, and set a condition on $V^\alpha$ which ensures that $\frac{\delta K^\alpha U_\alpha}{\omega_U} \ll 1$. Working to the highest accuracy the current analysis allows, the condition $V^\alpha \delta_\alpha=0$ requires $\delta K^\alpha$ to satisfy equation \eqref{vcond}. With $\delta K^\alpha$ given by \eqref{deltak} we have
\begin{equation}
	R_{\bar V K\delta K K}=\frac{1}{2}\omega_U\theta_U^2\left(1-\frac{\delta K^\alpha U_\alpha}{\omega_U}\right)R_{\bar V K UK} + \omega_U\theta_U\left(1-\frac{\delta K^\alpha U_\alpha}{\omega_U}\right)R_{\bar V K UK}+\mathcal{O}(\omega_U^4,\theta_U^3),
\end{equation}
and
\begin{multline}
	R_{\bar V \delta K K \delta K}=\omega_U^2\theta_U^2\left(1-\frac{\delta K^\alpha U_\alpha}{\omega_U}\right)^2R_{\bar V B KB} + \omega_U\theta_U\left(1-\frac{\delta K^\alpha U_\alpha}{\omega_U}\right)\frac{\delta K^\alpha U_\alpha}{\omega_U}R_{\bar V K BK}\\
	+\frac{1}{2}\omega_U\theta_U^2\left(1-\frac{\delta K^\alpha U_\alpha}{\omega_U}\right)\frac{\delta K^\alpha U_\alpha}{\omega_U}R_{\bar V K UK}+\mathcal{O}(\omega_U^4,\theta_U^3).
\end{multline}
For the sake of algebraic elegance we derive an expression for $\left(1-\frac{\delta K^\alpha U_\alpha}{\omega_U}\right)$ before getting one for $\frac{\delta K^\alpha U_\alpha}{\omega_U}$, and for brevity let $y=\left(1-\frac{\delta K^\alpha U_\alpha}{\omega_U}\right)$. Using the above in equation \eqref{vcond}, we get
\begin{equation}
	\delta K^\alpha\bar V_\alpha = \frac{1}{6}\omega_U\theta_U y \left(\left(2y-1\right)R_{\bar V K\hat BK}+\left(2y-1\right)\frac{1}{2}\theta_UR_{\bar V KUK}-2y\omega_U\theta_U R_{\bar V \hat BK\hat B}\right)+\mathcal{O}(\omega_U^4,\theta_U^3). \label{temp164}
\end{equation}
Evaluating the left hand side of the above,
\begin{equation}
	\delta K^\alpha\bar V_\alpha = -\frac{1}{2}\omega_U\theta_U^2 y\gamma+\left(1-y+\frac{1}{2}\theta_U^2y\right)\omega_V+\omega_U\theta_U y \hat B^\alpha \bar V_\alpha + \mathcal{O}(\theta_U^3),
\end{equation}
where $\gamma = -U^\alpha \bar V_\alpha$ and $\omega_V = -K^\alpha \bar V_\alpha$. Decomposing $K^\alpha = \omega_U\left(U^\alpha+ E^\alpha\right)$ (where $U^\alpha E_\alpha=0$), we have $\omega_V = -K^\alpha \bar V_\alpha = \omega_U\left( \gamma- E^\alpha \bar V_\alpha\right)$. Clearly, $E^\alpha$ represents the space direction in which the photon $K^\alpha$ travels in the frame of $U^\alpha$. Replacing $\omega_V$ in the above,
\begin{equation}
	\delta K^\alpha\bar V_\alpha = \omega_U\left(\gamma- E^\alpha \bar V_\alpha\right)\left(1-y\left(1+\frac{\frac{1}{2}\theta_U^2 E^\alpha \bar V_\alpha-\theta_U\hat B^\alpha \bar V_\alpha}{\gamma- E^\alpha \bar V_\alpha}\right)\right) + \mathcal{O}(\theta_U^3).
\end{equation}
Let us define
\begin{equation}
	x=\frac{\frac{1}{2}\theta_U^2 E^\alpha \bar V_\alpha-\theta_U\hat B^\alpha \bar V_\alpha}{\gamma- E^\alpha \bar V_\alpha},
\end{equation}
then combining the above with \eqref{temp164} we find
\begin{equation}
	y=\frac{1}{1+x}\left(1-\frac{1}{6}\frac{\theta_U y}{\gamma- E^\alpha \bar V_\alpha}\left(\left(2y-1\right)R_{\bar V K\hat BK}+\left(2y-1\right)\frac{1}{2}\theta_UR_{\bar V KUK}-2y\omega_U\theta_U R_{\bar V \hat BK\hat B}\right)\right)+ \mathcal{O}(\omega_U^3,\theta_U^3). \label{temp168}
\end{equation}
The above expression becomes much simpler in the absence of curvature or if we simply reduce accuracy by one order of $\omega_U$. Then we would have an explicit expression for $y$ and therefore $\frac{\delta K^\alpha U_\alpha}{\omega_U}$. However, at this stage we cannot neglect the curvature term without justification since it could in principle be very large for certain choices of $\bar V^\alpha$. In fact, it is already evident and will be clearer below that the range of $x$ is $(-1,\infty)$, so we see that $y$ cannot be taken as $1+\mathcal{O}(\theta_U)$ without some restriction on $V^\alpha$. As we shall shortly see, the curvature term departs from near zero for $V^\alpha$'s that also make $x$ depart from near zero, and the condition that keeps $x$ small also ensures that the curvature term remains small; and since this term contains extra $\omega_U$ factors it can be justifiably removed due to its smallness relative to that of $x$.

Decomposing $\bar V^\alpha$,
\begin{equation}
	\bar V^\alpha = \gamma U^\alpha +\left(\bar V^\beta  E_\beta\right) E^\alpha +\sqrt{\gamma^2 v^2-\left(\bar V^\beta  E_\beta\right)^2}\hat V^\alpha_\perp,
\end{equation}
where $\hat V^\alpha_\perp$ is a unit vector in the space frame of $U^\alpha$ that is perpendicular to the direction of the photon $ E^\alpha$, and $v=\frac{\sqrt{\gamma^2-1}}{\gamma}$. With $h^\alpha_\beta$ being the projection operator onto the frame of $U^\alpha$, it is easy to see that $|h^\alpha_\beta \bar V^\beta|=\gamma v$. Let $\alpha$ be the angle between $h^\alpha_\beta \bar V^\beta$ and $ E^\alpha$ in the frame of $U^\alpha$. Then
\begin{align}
	\bar V^\alpha &= \gamma U^\alpha +\gamma v \cos(\alpha) E^\alpha +\sqrt{\gamma^2 v^2-\gamma^2 v^2 \cos^2(\alpha)}\hat V^\alpha_\perp\nonumber\\
	&= \gamma U^\alpha +\gamma v \cos(\alpha) E^\alpha +\gamma v\sin(\alpha)\hat V^\alpha_\perp.
\end{align}
Since $\bar V^\alpha$ is the parallel transport of a 4-vector, $v$ and $\alpha$ are not measurable quantities and should be treated as purely mathematical; however, they do serve as relative measures of speed and alignment of $\bar V^\alpha$ with the photon $K^\alpha$. With these definitions and the above decomposition for $\bar V^\alpha$, $\hat B^\alpha \bar V_\alpha = \gamma v \sin(\alpha)\hat B_\alpha \hat V^\alpha_\perp= \epsilon \gamma v \sin(\alpha)$, for some $\epsilon \in [-1,1]$, and we get
\begin{equation}
	x=v\frac{\frac{1}{2}\theta_U^2\cos(\alpha)-\epsilon\theta_U\sin(\alpha)}{1-v\cos(\alpha)}.
\end{equation}
Notice the similarity to the $x$ defined in appendix \ref{app:relab}. Further, for the curvature terms in \eqref{temp168}
\begin{equation}
	R_{\bar V K \alpha K} = \gamma \left(1-v\cos(\alpha)\right)R_{UK\alpha K}+\gamma v \sin(\alpha)R_{\hat V_\perp K\alpha K},
\end{equation}
and 
\begin{equation}
	\omega_UR_{\bar V \hat B K\hat B} = \gamma \omega_U \left(1-v\cos(\alpha)\right)R_{U\hat BK\hat B}+\gamma v \cos(\alpha)R_{K\hat BK\hat B}+\omega_U\gamma v\sin(\alpha)R_{\hat V_\perp \hat BK\hat B}.
\end{equation}
Therefore,
\begin{align}
	&\frac{\theta_U}{\gamma- E^\alpha \bar V_\alpha} \left(\left(2y-1\right)R_{\bar V K\hat BK}+ \left(2y-1\right)\frac{1}{2}\theta_UR_{\bar V KUK}-2y\omega_U\theta_U R_{\bar V \hat BK\hat B}\right)\nonumber\\
	&\hspace{4cm}=\theta_U\left(\left(2y-1\right)R_{UK\hat BK}+\left(2y-1\right)\frac{1}{2}\theta_UR_{UKUK}-2y\omega_U\theta_UR_{U\hat BK\hat B}\right)\nonumber\\
	&\hspace{5cm}+v\frac{\theta_U\sin(\alpha)}{1-v\cos(\alpha)}\left(\left(2y-1\right)R_{\hat V_\perp K\hat BK}+\left(2y-1\right)\frac{1}{2}\theta_UR_{\hat V_\perp KUK}-2y\omega_U\theta_UR_{\hat V_\perp\hat BK\hat B}\right)\nonumber\\
	&\hspace{8cm}-2v\frac{\theta_U^2}{1-v\cos(\alpha)}y\cos(\alpha)R_{K\hat B K\hat B}+\mathcal{O}(\omega_U^3,\theta_U^3).
\end{align}
On the right hand side of the above, the first term remains small for any $v$ and $\alpha$, while the other two terms may become large when $\alpha \lesssim \theta_U$ and $1-v \lesssim \alpha^2$, which is exactly when $x$ becomes large. Expressed this way makes it clear that whatever magnitude $x$ may be, the curvature terms are smaller because of the extra $\omega_U$ factors. Particularly important is that when $x$ remains very small, the combined curvature term is much smaller.

We can now justifiably drop the curvature terms in the expression for $y$ by slightly reducing accuracy and making it explicit,
\begin{equation}
	y=\frac{1}{1+x}+\mathcal{O}(\omega_U^2,\theta_U^3).
\end{equation}
Thus, it is $x$ that dominates the expression for $y$, and a condition on $x$ will ensure the required smallness of $y$. Evidently, if we work with the higher order terms in $\omega_U$ then the equation that establishes $y$ is a quadratic. This means that there could be two potential solutions for $y$, and therefore two possible events on the neighboring null geodesic that are within the same simultaneity slice with respect to $V^\alpha$. While this may be interesting for cases of extreme curvature, it is not relevant for the present analysis. It is worth noting that if we worked with higher orders of $\omega_U$ in the expression for $\bar \delta^\alpha$, then we would have even more possible solutions for $y$. However, it is the smallest distance between the photons, $|\delta|$, that is of our concern, and the goal is to determine the conditions under which it can be considered observer independent when expressed to smallest order of $\theta_U$.

From the expression of $|\delta|$ given by equation \eqref{temp102} and \eqref{temp103}, we see that only when $\frac{\delta K^\alpha U_\alpha}{\omega_U} \ll 1$ $(\sim \theta_U)$ can it be justifiably removed from the right hand side, taking with it all the possible dependence of $|\delta|$ on $V^\alpha$. Therefore, for independence of $|\delta|$ on $V^\alpha$ to lowest order of $\theta_U$ we require that $y=1-\frac{\delta K^\alpha U_\alpha}{\omega_U}=1+\mathcal{O}(\theta_U)$, for which we must have $x \ll 1$. Thus, the required condition for independence is $x \ll 1$, which is equivalent to
\begin{equation}
	\frac{v\theta_U^2}{1-v\cos(\alpha)} \ll 1. \label{temp180}
\end{equation}
It is strikingly similar to the case of aberration for small angles (appendix \ref{app:relab}, equation \eqref{cond}), except here we have to interpret $v$ and $\alpha$ differently. The accompanying discussion of the behavior of $x$ in appendix \ref{app:relab} is also relevant to the present case. Clearly, most choices of $V^\alpha$ will satisfy the above, but if extreme relative motion is of relevance and the condition may be violated, then a modified expression for $|\delta|$ must be used with $1-\frac{\delta K^\alpha U_\alpha}{\omega_U} \ne 1 +\mathcal{O}(\theta_U)$. The analysis we presented makes it clear how such modified expressions (for $|\delta|$ and $\frac{\delta K^\alpha U_\alpha}{\omega_U}$) can be readily obtained, enabling us to deal with any possible relative motion. Finally, with the above condition satisfied we have
\begin{align}
	y=1-\frac{\delta K^\alpha U_\alpha}{\omega_U}&=\frac{1}{1+x}+\mathcal{O}(\omega_U^2,\theta_U^3)\nonumber\\
	&=1-x+\mathcal{O}(\omega_U^2,\theta_U^3),
\end{align}
and 
\begin{align}
	\frac{\delta K^\alpha U_\alpha}{\omega_U}&=x+\mathcal{O}(\omega_U^2,\theta_U^3)\nonumber\\
	&=\frac{\frac{1}{2}\theta_U^2 E^\alpha \bar V_\alpha-\theta_U\hat B^\alpha \bar V_\alpha}{\gamma- E^\alpha \bar V_\alpha}+\mathcal{O}(\omega_U^2,\theta_U^3)\\
	&=-\theta_U\frac{\hat B^\alpha \bar V_\alpha}{\gamma- E^\alpha \bar V_\alpha}+\mathcal{O}(\omega_U^2,\theta_U^2),
\end{align}
and since $\omega_V = -K^\alpha \bar V_\alpha = \omega_U\left( \gamma- E^\alpha \bar V_\alpha\right)$,
\begin{equation}
	\frac{\delta K^\alpha U_\alpha}{\omega_U}=-\theta_U\frac{\omega_U}{\omega_V}\hat B^\alpha \bar V_\alpha+\mathcal{O}(\omega_U^2,\theta_U^2). \label{deltaku}
\end{equation}
In ending this section we note that for the present case of distance between neighboring photons, as well as for the case of aberration for small angles (appendix \ref{app:relab}), requiring that the redshift is non-extreme ($\frac{\omega_U}{\omega_V} \sim 1$) elegantly leads to the desired results but is too restrictive of a condition. As made clear above, it is possible to have $\omega_U \gg \omega_V$, for example, while still having $\frac{\delta K^\alpha U_\alpha}{\omega_U} \ll 1$ as required.

\section{Derivations II}

\subsection{Inverse of a Matrix}

In this section we derive an explicit expression for the inverse of a square matrix in terms of the matrix itself and Levi-Civita symbols.\\

For a non-singular matrix $M^\alpha_\beta$, the determinant is neatly captured by use of the Levi-Civita symbol $\epsilon^{\alpha_1..\alpha_n}_{1..n}$ (or $\epsilon_{\alpha_1..\alpha_n}^{1..n}$), where $\alpha_1..\alpha_n$ is $\alpha_1,\alpha_2,..,\alpha_{n-1},\alpha_n$, $1..n$ is $1,2,..,n-1,n$ and $n$ is the dimension of the manifold. To be clear and to avoid confusion due to different conventions, let us be explicit in that the general $\epsilon^{\alpha_1..\alpha_n}_{\beta_1..\beta_n}$ symbol here is non-zero when all $\alpha_i$'s are different and all $\beta_i$'s are different. Clearly for non-zero value of $\epsilon^{\alpha_1..\alpha_n}_{\beta_1..\beta_n}$ there can be no repetition in the $\alpha$'s or in the $\beta$'s. The symbol is positive if the order of permutation between the $\alpha$'s and $\beta$'s is even, and negative if it is odd. It is easy to see that the Levi-Civita symbol $\epsilon^{\alpha_1..\alpha_n}_{1..n}$ (or $\epsilon_{\alpha_1..\alpha_n}^{1..n}$) is a tensor density of weight one (or negative one).
\begin{equation}
	det(M_{(\beta)}^{(\alpha)})=\epsilon^{\alpha_1..\alpha_n}_{1..n}M_{\alpha_1}^1..M_{\alpha_n}^n =\epsilon_{\alpha_1..\alpha_n}^{1..n}M^{\alpha_1}_1..M^{\alpha_n}_n= \frac{1}{n!}\epsilon^{\alpha_1..\alpha_n}_{1..n}M_{\alpha_1}^{\beta_1}..M_{\alpha_n}^{\beta_n}\epsilon_{\beta_1..\beta_n}^{1..n},
\end{equation}
\begin{equation}
	n = M_{\alpha_1}^{\beta_1} \frac{1}{(n-1)!det(M_{(\beta)}^{(\alpha)})} \epsilon^{\alpha_1..\alpha_n}_{1..n} M_{\alpha_2}^{\beta_2}..M_{\alpha_n}^{\beta_n}\epsilon_{\beta_1..\beta_n}^{1..n}.
\end{equation}
The above suggests that the inverse, $M^{-1}\,^{\alpha}_{\beta}$, of the matrix $M^{\alpha}_{\beta}$ is given by
\begin{equation}
	M^{-1}\,^{\alpha}_{\beta}=\frac{\epsilon^{\alpha\alpha_2..\alpha_n}_{1..n}M_{\alpha_2}^{\beta_2}..M_{\alpha_n}^{\beta_n}\epsilon_{\beta\beta_2..\beta_n}^{1..n}}{(n-1)!det(M_{(\beta)}^{(\alpha)})}, \label{temp272}
\end{equation}
which we verify as follows. Consider the product
\begin{equation}
	M^{\gamma}_{\alpha} \epsilon^{\alpha\alpha_2..\alpha_n}_{1..n}M_{\alpha_2}^{\beta_2}..M_{\alpha_n}^{\beta_n}\epsilon_{\beta\beta_2..\beta_n}^{1..n}.
\end{equation}
The free indices are $\gamma$ and $\beta$. If $\gamma$ and $\beta$ are the same, then for every possible permutation of $\beta_2..\beta_n$ the above product is just the determinant of the matrix. Since there are $(n-1)!$ such permutations, we have
\begin{equation}
	M^{\gamma}_{\alpha} \epsilon^{\alpha\alpha_2..\alpha_n}_{1..n}M_{\alpha_2}^{\beta_2}..M_{\alpha_n}^{\beta_n}\epsilon_{\beta\beta_2..\beta_n}^{1..n} = (n-1)!det(M_{(\beta)}^{(\alpha)}), \;\;\;\;\;\;\;\; (\gamma=\beta).
\end{equation}
When $\gamma$ and $\beta$ are not the same, then $\gamma$ will have to equal one of $\beta_2..\beta_n$ in every term of the sum (since these are the only other options). This means that there will be a repeating upper index of $M^\alpha_\beta$ in every term, and by the antisymmetry of the Levi-Civita symbol with the alphas, all the terms will cancel out in pairs giving a zero sum. (This is the same as having a repeating row or column in the matrix of which the determinant is taken.) Thus,
\begin{equation}
	M^{\gamma}_{\alpha} \epsilon^{\alpha\alpha_2..\alpha_n}_{1..n}M_{\alpha_2}^{\beta_2}..M_{\alpha_n}^{\beta_n}\epsilon_{\beta\beta_2..\beta_n}^{1..n} = (n-1)!det(M_{(\beta)}^{(\alpha)})\delta^\gamma_\beta,
\end{equation}
which proves the expression given by \eqref{temp272}.

The conclusion can easily be extended to any rank two tensor. In particular, for a metric tensor $g_{\alpha\beta}$ in an $n$ dimensional manifold, the determinant of the metric and the expression for its inverse are given by
\begin{equation}
	det(g_{(\alpha\beta)})=\epsilon^{\alpha_1..\alpha_n}_{1..n}g_{\alpha_1 1}..g_{\alpha_n n} = \frac{1}{n!}\epsilon^{\alpha_1..\alpha_n}_{1..n}g_{\alpha_1 \beta_1}..g_{\alpha_n \beta_n}\epsilon^{\beta_1..\beta_n}_{1..n},
\end{equation}
and
\begin{equation}
	g^{\alpha\beta}=\frac{\epsilon^{\alpha\alpha_2..\alpha_n}_{1..n}g_{\alpha_2 \beta_2}..g_{\alpha_n \beta_n}\epsilon^{\beta\beta_2..\beta_n}_{1..n}}{(n-1)!det(g_{(\alpha\beta)})}.
\end{equation}

\subsection{\label{app:sub}Subvolume Transformations}

In this section we derive relationships between volumes within subspaces under linear transformations.\\

Consider a transformation $T^\alpha_\beta$ from a point in an $n$ dimensional manifold with metric $g_{\alpha\beta}$, to a point in a different manifold of the same dimension with metric $g'_{\alpha\beta}$. Assume the transformation is non-singular and has a determinant given by $J=det(T^{(\alpha)}_{(\beta)})$. For any basis in the domain and its image in the range, the relationship between the corresponding volumes with respect to the two metrics is given by
\begin{equation}
	\frac{V'}{V}=\frac{\sqrt{det(g'_{(\alpha\beta)})}}{\sqrt{det(g_{(\alpha\beta)})}}J.
\end{equation}
The determinants on the right hand side are with reference to the original coordinates and the relationship is independent of basis. Consider now a vector $U^\alpha$ at the domain and the subspace perpendicular to it with respect to the metric $g_{\alpha\beta}$. With the transformation restricted to this subspace, we will now construct a relationship between the corresponding subvolumes for any basis on the subspace. We will then extend it to a subspace restricted by two vectors and adopt the results to pseudo Riemannian spacetime. 

Let $B_a^\alpha$, $a = 1,..,n-1$, be some basis on the subspace perpendicular to $U^\alpha$, so that $B_a^\alpha U_\alpha = 0$. Then for the full basis $U^\alpha,B_1^\alpha,..,B_{n-1}^\alpha$, we have a corresponding basis of 1-forms $dU_\alpha,dB^1_\alpha,..,dB^{n-1}_\alpha$, obtained under the prescription $dU(U)=1,\;dU(B_a)=0$, and $dB^a(U)=0,\;dB^a(B_b)=\delta^a_b$. The volume element $dV$ is
\begin{equation}
	dV = \sqrt{det(g_{(\alpha\beta)})}dx^1 \wedge .. \wedge dx^n = \sqrt{det(g,U,B_1,..,B_{n-1})}dU\wedge dB^1 \wedge .. \wedge dB^{n-1},
\end{equation}
and by the definition of the wedge product and the fact that $B_a^\alpha U_\alpha = 0$, we have
\begin{equation}
	\sqrt{det(g_{(\alpha\beta)})}\epsilon^{1..n}_{\alpha_1..\alpha_n}dx^{\alpha_1} \otimes .. \otimes dx^{\alpha_n} = |U|\sqrt{det(g,B_1,..,B_{n-1})}dU\wedge dB^1 \wedge .. \wedge dB^{n-1}.
\end{equation}
$det(g,X,..,Y)$ is the determinant of the square matrix obtained from the inner products of the vectors involved with respect to the metric. (Then $det(g_{(\alpha\beta)})$ is the same as $det(g,\partial_1,..,\partial_n)$.) Applying the ordered set $((empty),B_1,..,B_{n-1})$ to the each side of the above,
\begin{equation}
	\sqrt{det(g_{(\alpha\beta)})}\epsilon^{1..n}_{\alpha_1..\alpha_n}dx^{\alpha_1}B^{\alpha_2}_1..B^{\alpha_n}_{n-1} = |U|\sqrt{det(g,B_a)}dU. \label{temp281}
\end{equation}
(Where $det(g,B_a)=det(g,B_1,..,B_{n-1})$.) The 1-form $dU$ can be expressed in the basis $dx^\alpha$ as follows. Let $U^\downarrow$ be the \textit{flattened} version of the vector $U$ with respect to the metric, so that $U^\downarrow_\alpha = g_{\alpha\beta}U^\beta \; (= U_\alpha)$. $U^\downarrow$ can be expressed in either 1-form basis,
\begin{equation}
	U^\downarrow = U^\downarrow_\alpha dx^\alpha = U^\downarrow_U dU + U^\downarrow_a dB^a,
\end{equation}
where
\begin{equation}
	U^\downarrow_U = U^\downarrow (U) = g_{UU} = |U|^2,
\end{equation}
and
\begin{equation}
	U^\downarrow_a = U^\downarrow (B_a) = g_{\alpha\beta}U^\alpha B_a^\beta = 0.
\end{equation}
Therefore,
\begin{equation}
	U^\downarrow_\alpha dx^\alpha = |U|^2 dU,
\end{equation}
and
\begin{equation}
	dU = \frac{U_\alpha}{|U|^2} dx^\alpha.
\end{equation}
Inserting in \eqref{temp281} gives
\begin{equation}
	\sqrt{det(g_{(\alpha\beta)})}\epsilon^{1..n}_{\alpha\alpha_1..\alpha_{n-1}}B^{\alpha_1}_1..B^{\alpha_{n-1}}_{n-1}dx^{\alpha} = \frac{\sqrt{det(g,B_a)}}{|U|}U_\alpha dx^\alpha,
\end{equation}
and therefore,
\begin{equation}
	\frac{U^\alpha}{|U|} = \frac{\sqrt{det(g_{(\alpha\beta)})}}{\sqrt{det(g,B_a)}}g^{\alpha\beta}\epsilon^{1..n}_{\beta\beta_1..\beta_{n-1}}B^{\beta_1}_1..B^{\beta_{n-1}}_{n-1}. \label{temp288}
\end{equation}
The above can be viewed as a generalized version of the cross product; for a set of linearly independent $n-1$ vectors, the above produces a unit normal vector with respect to the metric. (Without a metric, we can get a 1-form that defines the subspace spanned by the set of vectors.) Squaring to eliminate $U^\alpha$, we get the following expression for $det(g,B_a)$
\begin{equation}
	det(g,B_a) = det(g_{(\alpha\beta)})g^{\alpha\beta}\epsilon^{1..n}_{\alpha\alpha_1..\alpha_{n-1}}\epsilon^{1..n}_{\beta\beta_1..\beta_{n-1}}B^{\alpha_1}_1..B^{\alpha_{n-1}}_{n-1} B^{\beta_1}_1..B^{\beta_{n-1}}_{n-1}.
\end{equation}

Let $\bar B^\alpha_a$ be the transformed basis vectors in the range of $T^\alpha_\beta$, where the metric is $g'_{\alpha\beta}$, then we also have
\begin{equation}
	det(g',\bar B_a) = det(g'_{(\alpha\beta)})g'^{\alpha\beta}\epsilon^{1..n}_{\alpha\alpha_1..\alpha_{n-1}}\epsilon^{1..n}_{\beta\beta_1..\beta_{n-1}}\bar B^{\alpha_1}_1..\bar B^{\alpha_{n-1}}_{n-1} \bar B^{\beta_1}_1..\bar B^{\beta_{n-1}}_{n-1}.
\end{equation}
Let $\bar g'_{\alpha\beta}$ be the pullback of the metric $g'_{\alpha\beta}$ to the domain, then by the definition of a pullback $\bar g'_{\alpha\beta}B^\alpha_a B^\beta_b = g'_{\alpha\beta}\bar B^\alpha_a \bar B^\beta_b$, which means that $det(\bar g',B_a)=det(g',\bar B_a)$, but also
\begin{equation}
	det(\bar g',B_a) = det(\bar g'_{(\alpha\beta)})\bar g'^{\alpha\beta}\epsilon^{1..n}_{\alpha\alpha_1..\alpha_{n-1}}\epsilon^{1..n}_{\beta\beta_1..\beta_{n-1}}B^{\alpha_1}_1..B^{\alpha_{n-1}}_{n-1}B^{\beta_1}_1..B^{\beta_{n-1}}_{n-1},
\end{equation}
where $\bar g'^{\alpha\beta}$ is the inverse of the pullback metric $\bar g'_{\alpha\beta}$. Since the transformation is non-singular, $\bar g'_{\alpha\beta}$ is invertible, and one would expect the pushforward of $\bar g'^{\alpha\beta}\;(=\bar g'^{\gamma\rho}T^\alpha_\gamma T^\beta_\rho)$ to coincide with the inverse $g'^{\alpha\beta}$ of the metric $g'_{\alpha\beta}$. Or equivalently, one would expect that the inverted pushforward of $g'^{\alpha\beta}$ given by $g'^{\gamma\rho}T^{-1}\,^\alpha_\gamma T^{-1}\,^\beta_\rho$ is the same as $\bar g'^{\alpha\beta}$. We confirm this suspicion by demonstrating that $\bar g'^{\gamma\rho}T^\alpha_\gamma T^\beta_\rho g'_{\epsilon\beta} = \delta^\alpha_\epsilon$.
\begin{equation}
	\bar g'^{\gamma\rho}T^\alpha_\gamma T_\rho^\beta g'_{\epsilon\beta} = T^\alpha_\gamma \bar g'^{\gamma\rho} T_\rho^\beta g'_{\beta\lambda}T^\lambda_\mu T^{-1}\,^\mu_\epsilon =  T^\alpha_\gamma \bar g'^{\gamma\rho} \bar g'_{\rho\mu} T^{-1}\,^\mu_\epsilon = \delta^\alpha_\epsilon.
\end{equation}
Thus, $\bar g'^{\gamma\rho}T^\alpha_\gamma T^\beta_\rho = g'^{\alpha\beta}$ and $\bar g'^{\alpha\beta} = g'^{\gamma\rho} T^{-1}\,^\alpha_\gamma T^{-1}\,^\beta_\rho$, and we find
\begin{align}
	det(g',\bar B_a) &= det(\bar g',B_a)\nonumber\\
	&= det(\bar g'_{(\alpha\beta)})\bar g'^{\alpha\beta}\epsilon^{1..n}_{\alpha\alpha_1..\alpha_{n-1}}\epsilon^{1..n}_{\beta\beta_1..\beta_{n-1}}B^{\alpha_1}_1..B^{\alpha_{n-1}}_{n-1}B^{\beta_1}_1..B^{\beta_{n-1}}_{n-1}\nonumber\\
	&= det(g'_{(\alpha\beta)}) J^2 g'^{\gamma\lambda}T^{-1}\,^\alpha_\gamma T^{-1}\,^\beta_\lambda\frac{U_\alpha}{|U|}\frac{U_\beta}{|U|}\frac{det(g,B_a)}{det(g_{(\alpha\beta)})}.
\end{align}
(It is easy to see from the definition of the determinant that $det(\bar g'_{(\alpha\beta)})=J^2 det(g'_{(\alpha\beta)})$.) Finally,
\begin{equation}
	\frac{det(g',\bar B_a)}{det(g,B_a)} = \frac{det(g'_{(\alpha\beta)})}{det(g_{(\alpha\beta)})} \frac{J^2}{|U|^2} g'^{\gamma\lambda}T^{-1}\,^\alpha_\gamma T^{-1}\,^\beta_\lambda U_\alpha U_\beta. \label{temp296}
\end{equation}
For any choice of basis $B_a$, the corresponding subvolume is given by $\sqrt{det(g,B_a)}$, so the ratio of the subvolumes under the transformation is given by the root of the right hand side. This ratio is basis independent, as it should be, but depends on the two metrics, the transformation itself, and the confining vector $U^\alpha$.\\

An interesting question arises in these circumstances, which we would briefly digress to before the next part: What would be the vector $V^\alpha$ in the range of $T^\alpha_\beta$ that is normal to the image of the subspace determined by $U^\alpha$ at the domain? Clearly, $V^\alpha$ will not simply be the pushforward of $U^\alpha$. We would like an expression for $V^\alpha$ in terms of $U^\alpha$, the transformation, and the two metrics. By \eqref{temp288},
\begin{align}
	\frac{V^\alpha}{|V|} &= \frac{\sqrt{det(g'_{(\alpha\beta)})}}{\sqrt{det(g',\bar B_a)}} g'^{\alpha\beta}\epsilon^{1..n}_{\beta\beta_1..\beta_{n-1}}\bar{B}^{\beta_1}_1..\bar{B}^{\beta_{n-1}}_{n-1}\nonumber\\
	&= \frac{\sqrt{det(g'_{(\alpha\beta)})}}{\sqrt{det(g',\bar B_a)}} g'^{\alpha\beta}\epsilon^{1..n}_{\beta\beta_1..\beta_{n-1}}T^{\beta_1}_{\alpha_1}..T^{\beta_{n-1}}_{\alpha_{n-1}}B^{\alpha_1}_1..B^{\alpha_{n-1}}_{n-1}.
\end{align}
The following is true for any covariant tensor $M$ of rank $k$.
\begin{equation}
	M_{[\alpha_1..\alpha_k]}=M_{[\beta_1..\beta_k]} \epsilon^{\rho_1..\rho_{n-k}..\beta_1..\beta_k}_{\rho_1..\rho_{n-k}\alpha_1..\alpha_k} \frac{1}{k!(n-k)!},
\end{equation}
where the square brackets mean antisymmetrization. For $k=n-1$, we have
\begin{equation}
	M_{[\alpha_1..\alpha_{n-1}]}=M_{[\beta_1..\beta_{n-1}
    ]} \epsilon^{\rho\beta_1..\beta_{n-1}}_{\rho\alpha_1..\alpha_{n-1}}\frac{1}{(n-1)!}. \label{temp300}
\end{equation}
By means of equations \eqref{temp300}, \eqref{temp272}, \eqref{temp288}, and \eqref{temp296},
\begin{align}
	\frac{V^\alpha}{|V|} &= \frac{\sqrt{det(g'_{(\alpha\beta)})}}{\sqrt{det(g',\bar B_a)}} g'^{\alpha\beta}\epsilon^{1..n}_{\beta\beta_1..\beta_{n-1}}T^{\beta_1}_{\alpha_1}..T^{\beta_{n-1}}_{\alpha_{n-1}} \epsilon^{\rho\alpha_1..\alpha_{n-1}}_{\rho\mu_1..\mu_{n-1}} B^{\mu_1}_1..B^{\mu_{n-1}}_{n-1}\frac{1}{(n-1)!}\nonumber\\
	&= \frac{\sqrt{det(g'_{(\alpha\beta)})}}{\sqrt{det(g',\bar B_a)}} g'^{\alpha\beta}\epsilon^{1..n}_{\beta\beta_1..\beta_{n-1}}T^{\beta_1}_{\alpha_1}..T^{\beta_{n-1}}_{\alpha_{n-1}} \epsilon^{\rho\alpha_1..\alpha_{n-1}}_{1..n} \epsilon^{1..n}_{\rho\mu_1..\mu_{n-1}} B^{\mu_1}_1..B^{\mu_{n-1}}_{n-1}\frac{1}{(n-1)!}\nonumber\\
	&= \frac{\sqrt{det(g'_{(\alpha\beta)})}}{\sqrt{det(g',\bar B_a)}} g'^{\alpha\beta} J T^{-1}\,^{\rho}_{\beta} \frac{\sqrt{det(g,B_a)}}{\sqrt{det(g_{(\alpha\beta)})}} \frac{U_\rho}{|U|}\nonumber\\
	&= \frac{1}{\sqrt{g'^{\gamma\epsilon}T^{-1}\,^\alpha_\gamma T^{-1}\,^\beta_\epsilon U_\alpha U_\beta}}g'^{\alpha\beta}T^{-1}\,^{\gamma}_{\beta} g_{\gamma\epsilon}U^\epsilon.
\end{align}
Thus, for any vector $U^\alpha$ in the domain the corresponding vector $V^\alpha$ in the range, as described, is given by the above (up to magnitude). Notice that
\begin{equation}
	V^\alpha \propto g'^{\alpha\beta}T^{-1}\,^{\gamma}_{\beta} g_{\gamma\epsilon}U^\epsilon,
\end{equation}
which suggests that we could have obtained the above result through an easier way. Indeed, if $V^\alpha$ is perpendicular to the image subspace, then clearly $g'_{\alpha\beta}V^\alpha$ acting on any transformed basis vector $\bar B^\alpha_a$ give zero, and by definition of the pullback $T^\alpha_\beta g'_{\alpha\gamma}V^\gamma$ acting on any basis vector $B^\alpha_a$ will give zero also. This means that the vector $g^{\epsilon\beta}T^\alpha_\beta g'_{\alpha\gamma}V^\gamma$ must be parallel to $U^\alpha$, which promptly yields the above.\\

Next we will derive a version of equation \eqref{temp296} for the case of two vectors $K^\alpha$ and $L^\alpha$ that confine the transformation $T^\alpha_\beta$ to an $n-2$ dimensional subspace. Let $B_a^\alpha$, $a=1,..,n-2$, be any basis for the subspace perpendicular to $K^\alpha$ and $L^\alpha$. Then, with the same reasoning as before, we identify a dual basis $dK,dL,dB^1,..,dB^{n-2}$ and proceed as follows,
\begin{align}
	dV = \sqrt{det(g_{(\alpha\beta)})}dx^1 \wedge .. \wedge dx^n &= \sqrt{det(g,K,L,B_a)}dK\wedge dL \wedge .. \wedge dB^{n-2},\nonumber\\
	\sqrt{det(g_{(\alpha\beta)})}\epsilon^{1..n}_{\alpha_1..\alpha_n}dx^{\alpha_1} \otimes .. \otimes dx^{\alpha_n} &= \sqrt{det(g,K,L)det(g,B_a)}dK\wedge dL \wedge .. \wedge dB^{n-2}\nonumber\\
	\sqrt{det(g_{(\alpha\beta)})}\epsilon^{1..n}_{\alpha_1..\alpha_n}dx^{\alpha_1} \otimes dx^{\alpha_2} B^{\alpha_3}_1..B^{\alpha_n}_{n-2} &= \sqrt{det(g,K,L)det(g,B_a)} dK \wedge dL.
\end{align}
The \textit{flattened} versions $K^\downarrow$, $L^\downarrow$ can be expressed in either basis,
\begin{align}
	K^\downarrow &= K_\alpha dx^\alpha = K^\downarrow_K dK + K^\downarrow_L dL + K^\downarrow_a dB^a,\\
	L^\downarrow &= L_\alpha dx^\alpha = L^\downarrow_K dK + L^\downarrow_L dL + L^\downarrow_a dB^a,
\end{align}
where
\begin{align}
	K^\downarrow_K &= K^\downarrow (K) = g_{KK},\\
	L^\downarrow_L &= L^\downarrow (L) = g_{LL},\\
	K^\downarrow_L = K^\downarrow (L) &= g_{KL} = L^\downarrow (K) = L^\downarrow_K,\\
	K^\downarrow_a &= L^\downarrow_a = 0.
\end{align}
Wedging,
\begin{align}
	K^\downarrow \wedge L^\downarrow = (K_\alpha dx^\alpha)\wedge(L_\beta dx^\beta) &= (g_{KK}dK+g_{KL}dL)\wedge(g_{KL}dK+g_{LL}dL),\nonumber\\
	(K_\alpha L_\beta - K_\beta L_\alpha)dx^\alpha \otimes dx^\beta &= det(g,K,L)dK\wedge dL.
\end{align}
Therefore,
\begin{equation}
	\sqrt{det(g_{(\alpha\beta)})}\epsilon^{1..n}_{\alpha\beta\mu_1..\mu_{n-2}} B^{\mu_1}_1..B^{\mu_{n-2}}_{n-2}dx^{\alpha} \otimes dx^{\beta} = \frac{\sqrt{det(g,B_a)}}{\sqrt{det(g,K,L)}} (K_\alpha L_\beta - K_\beta L_\alpha)dx^\alpha \otimes dx^\beta,
\end{equation}
and
\begin{equation}
	\frac{K_\alpha L_\beta - K_\beta L_\alpha}{\sqrt{det(g,K,L)}} = \frac{\sqrt{det(g_{(\alpha\beta)})}}{\sqrt{det(g,B_a)}}\epsilon^{1..n}_{\alpha\beta\mu_1..\mu_{n-2}} B^{\mu_1}_1..B^{\mu_{n-2}}_{n-2},
\end{equation}
compare to \eqref{temp288}. Squaring the 2-forms with respect to the metric gives
\begin{equation}
	\frac{g^{\alpha\gamma}g^{\beta\lambda}(K_\alpha L_\beta - K_\beta L_\alpha)(K_\gamma L_\lambda - K_\lambda L_\gamma)}{det(g,K,L)} =	\frac{det(g_{(\alpha\beta)})}{det(g,B_a)}g^{\alpha\gamma}g^{\beta\lambda}\epsilon^{1..n}_{\alpha\beta\mu_1..\mu_{n-2}}\epsilon^{1..n}_{\gamma\lambda\rho_1..\rho_{n-2}} B^{\mu_1}_1..B^{\mu_{n-2}}_{n-2}B^{\rho_1}_1..B^{\rho_{n-2}}_{n-2},
\end{equation}
\begin{equation}
	det(g,B_a) = \frac{1}{2}det(g_{(\alpha\beta)})g^{\alpha\gamma}g^{\beta\lambda}\epsilon^{1..n}_{\alpha\beta\mu_1..\mu_{n-2}}\epsilon^{1..n}_{\gamma\lambda\rho_1..\rho_{n-2}} B^{\mu_1}_1..B^{\mu_{n-2}}_{n-2}B^{\rho_1}_1..B^{\rho_{n-2}}_{n-2}.
\end{equation}
Similarly, for the transported basis we have
\begin{equation}
	det(g',\bar B_a) = \frac{1}{2}det(g'_{(\alpha\beta)})g'^{\alpha\gamma}g'^{\beta\lambda}\epsilon^{1..n}_{\alpha\beta\mu_1..\mu_{n-2}}\epsilon^{1..n}_{\gamma\lambda\rho_1..\rho_{n-2}} \bar{B}^{\mu_1}_1..\bar{B}^{\mu_{n-2}}_{n-2}\bar{B}^{\rho_1}_1..\bar{B}^{\rho_{n-2}}_{n-2},
\end{equation}
and for the pullback metric we have
\begin{equation}
	det(\bar g',B_a) = \frac{1}{2}det(\bar g'_{(\alpha\beta)}) \bar g'^{\alpha\gamma} \bar g'^{\beta\lambda} \epsilon^{1..n}_{\alpha\beta\mu_1..\mu_{n-2}}\epsilon^{1..n}_{\gamma\lambda\rho_1..\rho_{n-2}} B^{\mu_1}_1..B^{\mu_{n-2}}_{n-2}B^{\rho_1}_1..B^{\rho_{n-2}}_{n-2}.
\end{equation}
Therefore,
\begin{align}
	det(g',\bar B_a) &= det(\bar g',B_a)\nonumber \\
	&= \frac{1}{2}det(\bar g'_{(\alpha\beta)}) \bar g'^{\alpha\gamma} \bar g'^{\beta\lambda} \epsilon^{1..n}_{\alpha\beta\mu_1..\mu_{n-2}}\epsilon^{1..n}_{\gamma\lambda\rho_1..\rho_{n-2}} B^{\mu_1}_1..B^{\mu_{n-2}}_{n-2}B^{\rho_1}_1..B^{\rho_{n-2}}_{n-2}\nonumber\\
	&= \frac{1}{2}det(g'_{(\alpha\beta)}) J^2 g'^{\rho\mu} g'^{\nu\epsilon} T^{-1}\,^\alpha_\rho T^{-1}\,^\gamma_\mu T^{-1}\,^\beta_\nu T^{-1}\,^\lambda_\epsilon\frac{(K_\alpha L_\beta - K_\beta L_\alpha)(K_\gamma L_\lambda - K_\lambda L_\gamma)}{det(g,K,L)}\frac{det(g,B_a)}{det(g_{(\alpha\beta)})},
\end{align}
and finally,
\begin{equation}
	\frac{det(g',\bar B_a)}{det(g,B_a)} = \frac{1}{2}\frac{det(g'_{(\alpha\beta)})}{det(g_{(\alpha\beta)})} \frac{J^2}{det(g,K,L)} g'^{\rho\mu} g'^{\nu\epsilon} T^{-1}\,^\alpha_\rho T^{-1}\,^\gamma_\mu T^{-1}\,^\beta_\nu T^{-1}\,^\lambda_\epsilon (K_\alpha L_\beta - K_\beta L_\alpha)(K_\gamma L_\lambda - K_\lambda L_\gamma), \label{temp320}
\end{equation}
compare with \eqref{temp296}. The root of the right hand side is the ratio between the corresponding subvolumes under the restricted transformation that we were seeking. If the transformation is within the same event, then \eqref{temp296} and \eqref{temp320} respectively become
\begin{equation}
	\frac{det(g,\bar B_a)}{det(g,B_a)} = \frac{J^2}{|U|^2} g^{\gamma\lambda}T^{-1}\,^\alpha_\gamma T^{-1}\,^\beta_\lambda U_\alpha U_\beta, \label{tempp321}
\end{equation}
and
\begin{equation}
	\frac{det(g,\bar B_a)}{det(g,B_a)} = \frac{1}{2} \frac{J^2}{det(g,K,L)} g^{\rho\mu} g^{\nu\epsilon} T^{-1}\,^\alpha_\rho T^{-1}\,^\gamma_\mu T^{-1}\,^\beta_\nu T^{-1}\,^\lambda_\epsilon (K_\alpha L_\beta - K_\beta L_\alpha)(K_\gamma L_\lambda - K_\lambda L_\gamma). \label{temp322}
\end{equation}
In case of pseudo Riemannian manifold the derivations in this section must be done with a bit more care. However for a Lorentzian manifold the results \eqref{tempp321} and \eqref{temp322} remain exactly the same (unless $U^\alpha$ is null).

\subsection{\label{app:det}The Determinant of a Nearly Identity Matrix}

We derive an expansion of the determinant of a square matrix of the form
\begin{equation}
	M^\alpha_\beta = \delta^\alpha_\beta + \delta M^\alpha_\beta
\end{equation}
to second order in the small coefficients of the matrix $\delta M^\alpha_\beta$ ($ \ll 1$).\\

From the definition of the determinant and by basic counting, we find
\begin{align}
	det(M_{(\beta)}^{(\alpha)})&=\epsilon^{\alpha_1..\alpha_n}_{1..n}M_{\alpha_1}^1..M_{\alpha_n}^n =\epsilon_{\alpha_1..\alpha_n}^{1..n}M^{\alpha_1}_1..M^{\alpha_n}_n= \frac{1}{n!}\epsilon^{\alpha_1..\alpha_n}_{1..n}M_{\alpha_1}^{\beta_1}..M_{\alpha_n}^{\beta_n}\epsilon_{\beta_1..\beta_n}^{1..n}\nonumber\\
	&= \frac{1}{n!}\epsilon^{\alpha_1..\alpha_n}_{1..n}(\delta_{\alpha_1}^{\beta_1} + \delta M_{\alpha_1}^{\beta_1})..(\delta_{\alpha_n}^{\beta_n} + \delta M_{\alpha_n}^{\beta_n})\epsilon_{\beta_1..\beta_n}^{1..n}\nonumber\\
	\nonumber &= 1+\frac{1}{(n-1)!}\epsilon^{\alpha_1..\alpha_n}_{1..n}(\delta M_{\alpha_1}^{\beta_1})(\delta_{\alpha_2}^{\beta_2})..(\delta_{\alpha_n}^{\beta_n})\epsilon_{\beta_1..\beta_n}^{1..n}+\frac{1}{2(n-2)!}\epsilon^{\alpha_1..\alpha_n}_{1..n}(\delta M_{\alpha_1}^{\beta_1})(\delta M_{\alpha_2}^{\beta_2})(\delta_{\alpha_3}^{\beta_3})..(\delta_{\alpha_n}^{\beta_n})\epsilon_{\beta_1..\beta_n}^{1..n}+...\nonumber\\
	&= 1+\frac{1}{(n-1)!}\epsilon^{\alpha_1 \alpha_2..\alpha_n}_{\beta_1 \alpha_2..\alpha_n}\delta M_{\alpha_1}^{\beta_1} + \frac{1}{2(n-2)!}\epsilon^{\alpha_1 \alpha_2 \alpha_3..\alpha_n}_{\beta_1 \beta_2 \alpha_3..\alpha_n}(\delta M_{\alpha_1}^{\beta_1})(\delta M_{\alpha_2}^{\beta_2})+...\nonumber\\
	&= 1+\delta M_{\alpha}^{\alpha}+\frac{1}{2}(\delta M_{\alpha}^{\alpha})(\delta M_{\beta}^{\beta})-\frac{1}{2}(\delta M_{\alpha}^{\beta})(\delta M_{\beta}^{\alpha})+... \label{temp276}
\end{align}

\subsection{\label{app:high}Higher Order Derivatives of the Connecting Vector}

In this section we find higher order derivatives of the connecting vector $D^\alpha$ for neighboring timelike and null geodesics.

For two neighboring timelike trajectories with 4-velocities $U^\alpha$ and $V^\alpha$ as described in section \ref{sec:phys}, consider the case where both are geodesics and at a given initial event the position vector $D_0^\alpha$ in the frame of $U^\alpha$ and its derivative $\dot{D}_0^\alpha$ are both very small in magnitude. Since $\dot U^\alpha=0$, $\dot{D}^\alpha$ is spacelike and represents the Fermi (and optical) relative velocity as described in section \ref{sec:relvel}. Then to lowest orders in $D$ and $\dot D$, by \eqref{ddot} and \eqref{ddotdot}
\begin{equation}
	\dot D^\alpha = \dot \tau \bar V^\alpha-U^\alpha+\mathcal{O}(D^2),
\end{equation}
and
\begin{equation}
	\ddot D^\alpha = -R^\alpha_{UDU}+\mathcal{O}(D^2, D\dot D). \label{temp317}
\end{equation}
Since $\dot U^\alpha=0$ and $\dot V^\alpha=0$, we find from \eqref{taudotdot} that $\ddot \tau =0 +\mathcal{O}(D^2,D\dot D)$, so the only remaining term in the expression for $\ddot D^\alpha$ at this order is the curvature term. It comes in from the relationship between $\dot{\bar V}^\alpha$ and $\bar{\dot V}^\alpha$ as derived in appendix \ref{app:deri} (equation \eqref{partransdot}). This expression for $\ddot D^\alpha$ in the current setup is the well known geodesic deviation equation.

Given the fact that $\ddot D^\alpha$ is of order $D$ and is replaceable by means of \eqref{temp317}, we can easily establish any of the higher order derivatives of $D^\alpha$ in the direction of $U^\alpha$ to lowest orders in $D$ and $\dot D$. Proceeding, we find
\begin{equation}
	\dddot D^\alpha = -\dot R^\alpha_{UDU}-R^\alpha_{U\dot D U}+\mathcal{O}(D^2, \dot D^2 ,D\dot D),
\end{equation}
and
\begin{equation}
	\ddddot {D}^\alpha = -\ddot R^\alpha_{UDU}-2\dot R^\alpha_{U\dot D U}+R^\alpha_{U\beta U}R^\beta_{UDU}+\mathcal{O}(D^2, \dot D^2 ,D\dot D);
\end{equation}
and for $D_0=0$,
\begin{equation}
	\dddot{\ddot {D}}_0^\alpha = -3\ddot R^\alpha_{U\dot D_0 U}+R^\alpha_{U\beta U}R^\beta_{U\dot D_0 U}+\mathcal{O}(\dot D_0^2).
\end{equation}

The evolution of the connecting vector $D^\alpha$ can be similarly analyzed for the case of neighboring null geodesics. The interpretation of $D^\alpha$ becomes more sophisticated, however, since there is no natural observer for the frame to which this vector is restricted. This means that for a given event on one null geodesic, there is no unique event on the neighboring null geodesic to associate it with. Fortunately, as we've seen in section \ref{sec:fund} and appendix \ref{app:condistance}, for nearly parallel null geodesics, while $D^\alpha$ requires an observer to be uniquely set, its magnitude $D$ is observer independent (to first order in $D$ and $\dot D\; (\propto \theta)$ and for non-extreme motion, see appendix \ref{app:condistance}). This important property allows us to justifiably interpret the magnitude of the connecting vector $D^\alpha$ as the distance between the null geodesics, regardless of how the events on each are associated, and analyze its derivatives.

Let $x^\alpha(\lambda)$ and $x^\alpha(\nu)$ be neighboring null geodesics, where $\lambda$ and $\nu$ are affine parameters on each, with $K^\alpha = \frac{dx^\alpha}{d\lambda}$ and $W^\alpha = \frac{dx^\alpha}{d\nu}$ being their tangents. Then following the derivation in of equation \eqref{ddotap} in appendix \ref{app:deri}, the first derivative to lowest order is given by
\begin{equation}
	\nabla_K D^\alpha = \frac{d\nu}{d\lambda} \bar W^\alpha-K^\alpha+\mathcal{O}(D^2). \label{temp321}
\end{equation}
Unlike the case for timelike trajectories, here we are not restricted to any parametrizations $x^\alpha(\lambda)$, $x^\alpha(\nu)$, or association $\nu(\lambda)$. These will either be set with reference to a particular observer and setup in mind, or for mathematical simplicity. For the case where the geodesics are very close and nearly parallel at some initial event on $x^\alpha(\lambda)$, that is, for small magnitudes of $D_0$ and $D'_0$ ($D'=|\nabla_K D^\alpha|$), we have
\begin{equation}
	\nabla^2_K D^\alpha = \nabla_K\nabla_K D^\alpha = -R^\alpha_{KDK}+\mathcal{O}(D^2,DD'),
\end{equation}
which is obtained through an identical process as \eqref{ddotdot} and \eqref{temp317} for timelike geodesics. Again we arrive at the geodesic deviation equation, but for null geodesics. With the same reasoning as for the timelike case, to lowest orders in $D$ and $D'$, we find
\begin{equation}
	\nabla^3_K D^\alpha = -\nabla_K R^\alpha_{KDK}-R^\alpha_{K(\nabla_K D) K}+\mathcal{O}(D^2, D'^2 ,D D'),
\end{equation}
\begin{equation}
	\nabla^4_K {D}^\alpha = -\nabla_K^2 R^\alpha_{KDK}-2\nabla_K R^\alpha_{K(\nabla_K D) K}+R^\alpha_{K\beta K}R^\beta_{KDK}+\mathcal{O}(D^2, D'^2 ,D D'),
\end{equation}
and for the case $D_0=0$,
\begin{equation}
	\nabla^5_K D_0^\alpha = -3 \nabla_K^2 R^\alpha_{K(\nabla_K D_0) K}+R^\alpha_{K\beta K}R^\beta_{K(\nabla_K D_0) K}+\mathcal{O}(D'^2_0).
\end{equation}

\end{widetext}

\bibliography{general}

\end{document}